

\input epsf
\def\dofig#1#2{\vbox{\centerline{\epsfbox{#2}}}}

\def\varfig#1#2{\centerline{\epsfxsize=#1\epsfbox{#2}}}

\def\SkipFig#1#2{\vbox to 4in{\vss%
   \centerline{Scanned detector figure is too large for %
   electronic distribution.}\vss}}

\def\unredoffs{} \def\redoffs{\voffset=-.31truein\hoffset=-.59truein}
\def\speclscape{\special{ps: landscape}}
%
%
%
%
%
\newbox\leftpage \newdimen\fullhsize \newdimen\hstitle \newdimen\hsbody
\tolerance=1000\hfuzz=2pt
\catcode`@=11 
%
\def\PPTbig{
   \twelvepoint\unredoffs\baselineskip=16pt plus 2pt minus 1pt
   \hsbody=\hsize \hstitle=\hsize 
}
%
\def\PPTlittle{%
   \tenpoint\baselineskip=16pt plus 2pt minus 1pt \vsize=7truein
   \redoffs \hstitle=8truein\hsbody=4.75truein\fullhsize=10truein\hsize=\hsbody
   \output={\ifnum\pageno=0 
     \shipout\vbox{\speclscape{\hsize\fullhsize\makeheadline}
     \hbox to \fullhsize{\hfill\pagebody\hfill}}\advancepageno
   \else
     \almostshipout{\leftline{\vbox{\pagebody\makefootline}}}\advancepageno
   \fi}
}
\def\almostshipout#1{\if L\l@r \count1=1 \message{[\the\count0.\the\count1]}
      \global\setbox\leftpage=#1 \global\let\l@r=R
 \else \count1=2
  \shipout\vbox{\speclscape{\hsize\fullhsize\makeheadline}
      \hbox to\fullhsize{\box\leftpage\hfil#1}}  \global\let\l@r=L\fi}
%
\newcount\yearltd\yearltd=\year\advance\yearltd by -1900

%
%

\def\draftmode{\message{ DRAFTMODE }\def\draftdate{{\rm preliminary draft:
\number\month/\number\day/\number\yearltd\ \ \hourmin}}%
\headline={\hfil\draftdate}\writelabels\baselineskip=20pt plus 2pt minus 2pt
 {\count255=\time\divide\count255 by 60 \xdef\hourmin{\number\count255}
  \multiply\count255 by-60\advance\count255 by\time
  \xdef\hourmin{\hourmin:\ifnum\count255<10 0\fi\the\count255}}}
\def\nolabels{\def\wrlabeL##1{}\def\eqlabeL##1{}\def\reflabeL##1{}}
\def\writelabels{\def\wrlabeL##1{\leavevmode\vadjust{\rlap{\smash%
{\line{{\escapechar=` \hfill\rlap{\sevenrm\hskip.03in\string##1}}}}}}}%
\def\eqlabeL##1{{\escapechar-1\rlap{\sevenrm\hskip.05in\string##1}}}%
\def\reflabeL##1{\noexpand\llap{\noexpand\sevenrm\string\string\string##1}}}
\nolabels
%
\global\newcount\secno \global\secno=0
\global\newcount\meqno \global\meqno=1
\def\newsec#1{\global\advance\secno by1\message{(\the\secno. #1)}
\global\subsecno=0\eqnres@t\noindent{\bf\the\secno. #1}
\writetoca{{\secsym} {#1}}\par\nobreak\medskip\nobreak}
\def\eqnres@t{\xdef\secsym{\the\secno.}\global\meqno=1\bigbreak\bigskip}
\def\sequentialequations{\def\eqnres@t{\bigbreak}}\xdef\secsym{}
\global\newcount\subsecno \global\subsecno=0
\def\subsec#1{\global\advance\subsecno by1\message{(\secsym\the\subsecno. #1)}
\ifnum\lastpenalty>9000\else\bigbreak\fi
\noindent{\it\secsym\the\subsecno. #1}\writetoca{\string\quad
{\secsym\the\subsecno.} {#1}}\par\nobreak\medskip\nobreak}
\def\appendix#1#2{\global\meqno=1\global\subsecno=0\xdef\secsym{\hbox{#1.}}
\bigbreak\bigskip\noindent{\bf Appendix #1. #2}\message{(#1. #2)}
\writetoca{Appendix {#1.} {#2}}\par\nobreak\medskip\nobreak}
%
%
\def\eqnn#1{\xdef #1{(\secsym\the\meqno)}\writedef{#1\leftbracket#1}%
\global\advance\meqno by1\wrlabeL#1}
\def\eqna#1{\xdef #1##1{\hbox{$(\secsym\the\meqno##1)$}}
\writedef{#1\numbersign1\leftbracket#1{\numbersign1}}%
\global\advance\meqno by1\wrlabeL{#1$\{\}$}}
\def\eqn#1#2{\xdef #1{(\secsym\the\meqno)}\writedef{#1\leftbracket#1}%
\global\advance\meqno by1$$#2\eqno#1\eqlabeL#1$$}
%
\newskip\footskip\footskip14pt plus 1pt minus 1pt 
\def\footnotefont{\ninepoint}\def\f@t#1{\footnotefont #1\@foot}
\def\f@@t{\baselineskip\footskip\bgroup\footnotefont\aftergroup\@foot\let\next}
\setbox\strutbox=\hbox{\vrule height9.5pt depth4.5pt width0pt}
\global\newcount\ftno \global\ftno=0
\def\foot{\global\advance\ftno by1\footnote{$^{\the\ftno}$}}
%
\newwrite\ftfile
\def\footend{\def\foot{\global\advance\ftno by1\chardef\wfile=\ftfile
$^{\the\ftno}$\ifnum\ftno=1\immediate\openout\ftfile=foots.tmp\fi%
\immediate\write\ftfile{\noexpand\smallskip%
\noexpand\item{f\the\ftno:\ }\pctsign}\findarg}%
\def\footatend{\vfill\eject\immediate\closeout\ftfile{\parindent=20pt
\centerline{\bf Footnotes}\nobreak\bigskip\input foots.tmp }}}
\def\footatend{}
%
%
\global\newcount\refno \global\refno=1
\newwrite\rfile
\def\ref{[\the\refno]\nref}
\def\nref#1{\xdef#1{[\the\refno]}\writedef{#1\leftbracket#1}%
\ifnum\refno=1\immediate\openout\rfile=refs.tmp\fi
\global\advance\refno by1\chardef\wfile=\rfile\immediate
\write\rfile{\noexpand\item{#1\ }\reflabeL{#1\hskip.31in}\pctsign}\findarg}
\def\findarg#1#{\begingroup\obeylines\newlinechar=`\^^M\pass@rg}
{\obeylines\gdef\pass@rg#1{\writ@line\relax #1^^M\hbox{}^^M}%
\gdef\writ@line#1^^M{\expandafter\toks0\expandafter{\striprel@x #1}%
\edef\next{\the\toks0}\ifx\next\em@rk\let\next=\endgroup\else\ifx\next\empty%
\else\immediate\write\wfile{\the\toks0}\fi\let\next=\writ@line\fi\next\relax}}
\def\striprel@x#1{} \def\em@rk{\hbox{}}
\def\lref{\begingroup\obeylines\lr@f}
\def\lr@f#1#2{\gdef#1{\ref#1{#2}}\endgroup\unskip}
\def\semi{;\hfil\break}
\def\addref#1{\immediate\write\rfile{\noexpand\item{}#1}} 
\def\listrefs{\footatend\vfill\supereject\immediate\closeout\rfile\writestoppt
\baselineskip=14pt\centerline{{\bf References}}\bigskip{\frenchspacing%
\parindent=20pt\escapechar=` \input refs.tmp\vfill\eject}\nonfrenchspacing}
\def\startrefs#1{\immediate\openout\rfile=refs.tmp\refno=#1}
\def\xref{\expandafter\xr@f}\def\xr@f[#1]{#1}
\def\refs#1{\count255=1[\r@fs #1{\hbox{}}]}
\def\r@fs#1{\ifx\und@fined#1\message{reflabel \string#1 is undefined.}%
\nref#1{need to supply reference \string#1.}\fi%
\vphantom{\hphantom{#1}}\edef\next{#1}\ifx\next\em@rk\def\next{}%
\else\ifx\next#1\ifodd\count255\relax\xref#1\count255=0\fi%
\else#1\count255=1\fi\let\next=\r@fs\fi\next}
%

%
\newwrite\ffile\global\newcount\figno \global\figno=1
\def\fig{fig.~\the\figno\nfig}
\def\nfig#1{\xdef#1{fig.~\the\figno}%
\writedef{#1\leftbracket fig.\noexpand~\the\figno}%
\ifnum\figno=1\immediate\openout\ffile=figs.tmp\fi\chardef\wfile=\ffile%
\immediate\write\ffile{\noexpand\medskip\noexpand\item{Fig.\ \the\figno. }
\reflabeL{#1\hskip.55in}\pctsign}\global\advance\figno by1\findarg}
\def\xfig{\expandafter\xf@g}\def\xf@g fig.\penalty\@M\ {}
\def\figs#1{figs.~\f@gs #1{\hbox{}}}
\def\f@gs#1{\edef\next{#1}\ifx\next\em@rk\def\next{}\else
\ifx\next#1\xfig #1\else#1\fi\let\next=\f@gs\fi\next}
\newwrite\lfile
{\escapechar-1\xdef\pctsign{\string\%}\xdef\leftbracket{\string\{}
\xdef\rightbracket{\string\}}\xdef\numbersign{\string\#}}
\def\writedefs{\immediate\openout\lfile=labeldefs.tmp \def\writedef##1{%
\immediate\write\lfile{\string\def\string##1\rightbracket}}}
\newread\lfilein
%
\def\writestop{\def\writestoppt{\immediate\write\lfile{\string\pageno%
\the\pageno\string\startrefs\leftbracket\the\refno\rightbracket%
\string\def\string\secsym\leftbracket\secsym\rightbracket%
\string\secno\the\secno\string\meqno\the\meqno}\immediate\closeout\lfile}}
\def\writestoppt{}\def\writedef#1{}
\def\seclab#1{\xdef #1{\the\secno}\writedef{#1\leftbracket#1}\wrlabeL{#1=#1}}
\def\subseclab#1{\xdef #1{\secsym\the\subsecno}%
\writedef{#1\leftbracket#1}\wrlabeL{#1=#1}}
\newwrite\tfile \def\writetoca#1{}
\def\leaderfill{\leaders\hbox to 1em{\hss.\hss}\hfill}
\def\writetoc{\immediate\openout\tfile=toc.tmp
   \def\writetoca##1{{\edef\next{\write\tfile{\noindent ##1
   \string\leaderfill {\noexpand\number\pageno} \par}}\next}}}

%

\def\titlefont{
   \ifx\answ\bigans \sixteenpoint
      \else\twentypoint\fi}

\global\font\tencsc=cmcsc10
\global\font\twelvecsc=cmcsc10 scaled \magstep1

\def\authorfont{
   \ifx\answ\bigans\tenpoint\tencsc
   \else\twelvepoint\twelvecsc\fi}		

\def\abstractfont{
   \ifx\answ\bigans\tenpoint
   \else\twelvepoint\fi}			

%

\hyphenation{anom-aly anom-alies coun-ter-term coun-ter-terms}
\def\inv{^{\raise.15ex\hbox{${\scriptscriptstyle -}$}\kern-.05em 1}}

\def\Dsl{\,\raise.15ex\hbox{/}\mkern-13.5mu D} 
\def\dsl{\raise.15ex\hbox{/}\kern-.57em\partial}

\def\lspace{\ifx\answ\bigans{}\else\qquad\fi}
\def\lbspace{\ifx\answ\bigans{}\else\hskip-.2in\fi} 
\def\boxeqn#1{\vcenter{\vbox{\hrule\hbox{\vrule\kern3pt\vbox{\kern3pt
	\hbox{${\displaystyle #1}$}\kern3pt}\kern3pt\vrule}\hrule}}}
\def\mbox#1#2{\vcenter{\hrule \hbox{\vrule height#2in
		\kern#1in \vrule} \hrule}}  
%
 \def\CO{{\cal O}} 
   
\def\CL{{\cal L}}   
 \def\CR{{\cal R}}

\def\darr#1{\raise1.5ex\hbox{$\leftrightarrow$}\mkern-16.5mu #1}

\def\half{{\textstyle{1\over2}}} 
\def\roughly#1{\raise.3ex\hbox{$#1$\kern-.75em\lower1ex\hbox{$\sim$}}}

\catcode`@=11                           
\newskip\ttglue


\def\ninefonts{%
   \global\font\ninerm=cmr9
   \global\font\ninei=cmmi9
   \global\font\ninesy=cmsy9
   \global\font\nineex=cmex10
   \global\font\ninebf=cmbx9
   \global\font\ninesl=cmsl9
   \global\font\ninett=cmtt9
   \global\font\nineit=cmti9
   \skewchar\ninei='177
   \skewchar\ninesy='60
   \hyphenchar\ninett=-1
   \moreninefonts                               
   \gdef\ninefonts{\relax}}

\def\moreninefonts{\relax}                      



\def\elevenfonts{%
   \global\font\elevenrm=cmr10 scaled \magstephalf
   \global\font\eleveni=cmmi10 scaled \magstephalf
   \global\font\elevensy=cmsy10 scaled \magstephalf
   \global\font\elevenex=cmex10
   \global\font\elevenbf=cmbx10 scaled \magstephalf
   \global\font\elevensl=cmsl10 scaled \magstephalf
   \global\font\eleventt=cmtt10 scaled \magstephalf
   \global\font\elevenit=cmti10 scaled \magstephalf
   \global\font\elevenss=cmss10 scaled \magstephalf
   \skewchar\eleveni='177
   \skewchar\elevensy='60
   \hyphenchar\eleventt=-1
   \moreelevenfonts                            
   \gdef\elevenfonts{\relax}}%

\def\moreelevenfonts{\relax}                    

\def\twelvefonts{
   \global\font\twelverm=cmr10 scaled \magstep1
   \global\font\twelvei=cmmi10 scaled \magstep1
   \global\font\twelvesy=cmsy10 scaled \magstep1
   \global\font\twelveex=cmex10 scaled \magstep1
   \global\font\twelvebf=cmbx10 scaled \magstep1
   \global\font\twelvesl=cmsl10 scaled \magstep1
   \global\font\twelvett=cmtt10 scaled \magstep1
   \global\font\twelveit=cmti10 scaled \magstep1
   \global\font\twelvess=cmss10 scaled \magstep1
   \skewchar\twelvei='177
   \skewchar\twelvesy='60
   \hyphenchar\twelvett=-1
   \moretwelvefonts                             
   \gdef\twelvefonts{\relax}}

\def\moretwelvefonts{\relax}                    

\def\fourteenfonts{%
   \global\font\fourteenrm=cmr10 scaled \magstep2
   \global\font\fourteeni=cmmi10 scaled \magstep2
   \global\font\fourteensy=cmsy10 scaled \magstep2
   \global\font\fourteenex=cmex10 scaled \magstep2
   \global\font\fourteenbf=cmbx10 scaled \magstep2
   \global\font\fourteensl=cmsl10 scaled \magstep2
   \global\font\fourteenit=cmti10 scaled \magstep2
   \global\font\fourteenss=cmss10 scaled \magstep2
   \skewchar\fourteeni='177
   \skewchar\fourteensy='60
   \morefourteenfonts                           
   \gdef\fourteenfonts{\relax}}

\def\morefourteenfonts{\relax}                  

\def\sixteenfonts{%
   \global\font\sixteenrm=cmr10 scaled \magstep3
   \global\font\sixteeni=cmmi10 scaled \magstep3
   \global\font\sixteensy=cmsy10 scaled \magstep3
   \global\font\sixteenex=cmex10 scaled \magstep3
   \global\font\sixteenbf=cmbx10 scaled \magstep3
   \global\font\sixteensl=cmsl10 scaled \magstep3
   \global\font\sixteenit=cmti10 scaled \magstep3
   \skewchar\sixteeni='177
   \skewchar\sixteensy='60
   \moresixteenfonts                            
   \gdef\sixteenfonts{\relax}}

\def\moresixteenfonts{\relax}                   

\def\twentyfonts{%
   \global\font\twentyrm=cmr10 scaled \magstep4
   \global\font\twentyi=cmmi10 scaled \magstep4
   \global\font\twentysy=cmsy10 scaled \magstep4
   \global\font\twentyex=cmex10 scaled \magstep4
   \global\font\twentybf=cmbx10 scaled \magstep4
   \global\font\twentysl=cmsl10 scaled \magstep4
   \global\font\twentyit=cmti10 scaled \magstep4
   \skewchar\twentyi='177
   \skewchar\twentysy='60
   \moretwentyfonts                             
   \gdef\twentyfonts{\relax}}

\def\moretwentyfonts{\relax}                    

\def\twentyfourfonts{%
   \global\font\twentyfourrm=cmr10 scaled \magstep5
   \global\font\twentyfouri=cmmi10 scaled \magstep5
   \global\font\twentyfoursy=cmsy10 scaled \magstep5
   \global\font\twentyfourex=cmex10 scaled \magstep5
   \global\font\twentyfourbf=cmbx10 scaled \magstep5
   \global\font\twentyfoursl=cmsl10 scaled \magstep5
   \global\font\twentyfourit=cmti10 scaled \magstep5
   \skewchar\twentyfouri='177
   \skewchar\twentyfoursy='60
   \moretwentyfourfonts                         
   \gdef\twentyfourfonts{\relax}}

\def\moretwentyfourfonts{\relax}                



\def\tenmibfonts{
   \global\font\tenmib=cmmib10
   \global\font\tenbsy=cmbsy10
   \skewchar\tenmib='177
   \skewchar\tenbsy='60
   \gdef\tenmibfonts{\relax}}

\def\elevenmibfonts{
   \global\font\elevenmib=cmmib10 scaled \magstephalf
   \global\font\elevenbsy=cmbsy10 scaled \magstephalf
   \skewchar\elevenmib='177
   \skewchar\elevenbsy='60
   \gdef\elevenmibfonts{\relax}}

\def\twelvemibfonts{
   \global\font\twelvemib=cmmib10 scaled \magstep1
   \global\font\twelvebsy=cmbsy10 scaled \magstep1
   \skewchar\twelvemib='177
   \skewchar\twelvebsy='60
   \gdef\twelvemibfonts{\relax}}

\def\fourteenmibfonts{
   \global\font\fourteenmib=cmmib10 scaled \magstep2
   \global\font\fourteenbsy=cmbsy10 scaled \magstep2
   \skewchar\fourteenmib='177
   \skewchar\fourteenbsy='60
   \gdef\fourteenmibfonts{\relax}}

\def\sixteenmibfonts{
   \global\font\sixteenmib=cmmib10 scaled \magstep3
   \global\font\sixteenbsy=cmbsy10 scaled \magstep3
   \skewchar\sixteenmib='177
   \skewchar\sixteenbsy='60
   \gdef\sixteenmibfonts{\relax}}

\def\twentymibfonts{
   \global\font\twentymib=cmmib10 scaled \magstep4
   \global\font\twentybsy=cmbsy10 scaled \magstep4
   \skewchar\twentymib='177
   \skewchar\twentybsy='60
   \gdef\twentymibfonts{\relax}}

\def\twentyfourmibfonts{
   \global\font\twentyfourmib=cmmib10 scaled \magstep5
   \global\font\twentyfourbsy=cmbsy10 scaled \magstep5
   \skewchar\twentyfourmib='177
   \skewchar\twentyfourbsy='60
   \gdef\twentyfourmibfonts{\relax}}


\def\mib{
   \tenmibfonts
   \textfont0=\tenbf\scriptfont0=\sevenbf
   \scriptscriptfont0=\fivebf
   \textfont1=\tenmib\scriptfont1=\seveni
   \scriptscriptfont1=\fivei
   \textfont2=\tenbsy\scriptfont2=\sevensy
   \scriptscriptfont2=\fivesy}

\def\scrfonts{\relax}


\def\ninepoint{\ninefonts               
   \def\rm{\fam0\ninerm}%
   \textfont0=\ninerm\scriptfont0=\sevenrm\scriptscriptfont0=\fiverm
   \textfont1=\ninei\scriptfont1=\seveni\scriptscriptfont1=\fivei
   \textfont2=\ninesy\scriptfont2=\sevensy\scriptscriptfont2=\fivesy
   \textfont3=\nineex\scriptfont3=\nineex\scriptscriptfont3=\nineex
   \textfont\itfam=\nineit\def\it{\fam\itfam\nineit}%
   \textfont\slfam=\ninesl\def\sl{\fam\slfam\ninesl}%
   \textfont\ttfam=\ninett\def\tt{\fam\ttfam\ninett}%
   \textfont\bffam=\ninebf
   \scriptfont\bffam=\sevenbf
   \scriptscriptfont\bffam=\fivebf\def\bf{\fam\bffam\ninebf}%
   \def\mib{\relax}%
   \def\scr{\relax}%
   \tt\ttglue=.5emplus.25emminus.15em
   \normalbaselineskip=11pt
   \setbox\strutbox=\hbox{\vrule height 8pt depth 3pt width 0pt}%
   \normalbaselines\rm\singlespaced}%

\def\tenpoint{
   \def\rm{\fam0\tenrm}%
   \textfont0=\tenrm\scriptfont0=\sevenrm\scriptscriptfont0=\fiverm
   \textfont1=\teni\scriptfont1=\seveni\scriptscriptfont1=\fivei
   \textfont2=\tensy\scriptfont2=\sevensy\scriptscriptfont2=\fivesy
   \textfont3=\tenex\scriptfont3=\tenex\scriptscriptfont3=\tenex
   \textfont\itfam=\tenit\def\it{\fam\itfam\tenit}%
   \textfont\slfam=\tensl\def\sl{\fam\slfam\tensl}%
   \textfont\ttfam=\tentt\def\tt{\fam\ttfam\tentt}%
   \textfont\bffam=\tenbf
   \scriptfont\bffam=\sevenbf
   \scriptscriptfont\bffam=\fivebf\def\bf{\fam\bffam\tenbf}%
   \def\mib{%
      \tenmibfonts
      \textfont0=\tenbf\scriptfont0=\sevenbf
      \scriptscriptfont0=\fivebf
      \textfont1=\tenmib\scriptfont1=\seveni
      \scriptscriptfont1=\fivei
      \textfont2=\tenbsy\scriptfont2=\sevensy
      \scriptscriptfont2=\fivesy}%
   \def\scr{\scrfonts
      \global\textfont\scrfam=\tenscr\fam\scrfam\tenscr}%
   \tt\ttglue=.5emplus.25emminus.15em
   \normalbaselineskip=12pt
   \setbox\strutbox=\hbox{\vrule height 8.5pt depth 3.5pt width 0pt}%
   \normalbaselines\rm\singlespaced}%

\def\elevenpoint{\elevenfonts           
   \def\rm{\fam0\elevenrm}%
   \textfont0=\elevenrm\scriptfont0=\sevenrm\scriptscriptfont0=\fiverm
   \textfont1=\eleveni\scriptfont1=\seveni\scriptscriptfont1=\fivei
   \textfont2=\elevensy\scriptfont2=\sevensy\scriptscriptfont2=\fivesy
   \textfont3=\elevenex\scriptfont3=\elevenex\scriptscriptfont3=\elevenex
   \textfont\itfam=\elevenit\def\it{\fam\itfam\elevenit}%
   \textfont\slfam=\elevensl\def\sl{\fam\slfam\elevensl}%
   \textfont\ttfam=\eleventt\def\tt{\fam\ttfam\eleventt}%
   \textfont\bffam=\elevenbf
   \scriptfont\bffam=\sevenbf
   \scriptscriptfont\bffam=\fivebf\def\bf{\fam\bffam\elevenbf}%
   \def\mib{%
      \elevenmibfonts
      \textfont0=\elevenbf\scriptfont0=\sevenbf
      \scriptscriptfont0=\fivebf
      \textfont1=\elevenmib\scriptfont1=\seveni
      \scriptscriptfont1=\fivei
      \textfont2=\elevenbsy\scriptfont2=\sevensy
      \scriptscriptfont2=\fivesy}%
   \def\scr{\scrfonts
      \global\textfont\scrfam=\elevenscr\fam\scrfam\elevenscr}%
   \tt\ttglue=.5emplus.25emminus.15em
   \normalbaselineskip=13pt
   \setbox\strutbox=\hbox{\vrule height 9pt depth 4pt width 0pt}%
   \normalbaselines\rm\singlespaced}%

\def\twelvepoint{\twelvefonts\ninefonts 
   \def\rm{\fam0\twelverm}%
   \textfont0=\twelverm\scriptfont0=\ninerm\scriptscriptfont0=\sevenrm
   \textfont1=\twelvei\scriptfont1=\ninei\scriptscriptfont1=\seveni
   \textfont2=\twelvesy\scriptfont2=\ninesy\scriptscriptfont2=\sevensy
   \textfont3=\twelveex\scriptfont3=\twelveex\scriptscriptfont3=\twelveex
   \textfont\itfam=\twelveit\def\it{\fam\itfam\twelveit}%
   \textfont\slfam=\twelvesl\def\sl{\fam\slfam\twelvesl}%
   \textfont\ttfam=\twelvett\def\tt{\fam\ttfam\twelvett}%
   \textfont\bffam=\twelvebf
   \scriptfont\bffam=\ninebf
   \scriptscriptfont\bffam=\sevenbf\def\bf{\fam\bffam\twelvebf}%
   \def\mib{%
      \twelvemibfonts\tenmibfonts
      \textfont0=\twelvebf\scriptfont0=\ninebf
      \scriptscriptfont0=\sevenbf
      \textfont1=\twelvemib\scriptfont1=\ninei
      \scriptscriptfont1=\seveni
      \textfont2=\twelvebsy\scriptfont2=\ninesy
      \scriptscriptfont2=\sevensy}%
   \def\scr{\scrfonts
      \global\textfont\scrfam=\twelvescr\fam\scrfam\twelvescr}%
   \tt\ttglue=.5emplus.25emminus.15em
   \normalbaselineskip=14pt
   \setbox\strutbox=\hbox{\vrule height 10pt depth 4pt width 0pt}%
   \normalbaselines\rm\singlespaced}%

\def\fourteenpoint{\fourteenfonts\twelvefonts 
   \def\rm{\fam0\fourteenrm}%
   \textfont0=\fourteenrm\scriptfont0=\twelverm\scriptscriptfont0=\tenrm
   \textfont1=\fourteeni\scriptfont1=\twelvei\scriptscriptfont1=\teni
   \textfont2=\fourteensy\scriptfont2=\twelvesy\scriptscriptfont2=\tensy
   \textfont3=\fourteenex\scriptfont3=\fourteenex
      \scriptscriptfont3=\fourteenex
   \textfont\itfam=\fourteenit\def\it{\fam\itfam\fourteenit}%
   \textfont\slfam=\fourteensl\def\sl{\fam\slfam\fourteensl}%
   \textfont\bffam=\fourteenbf
   \scriptfont\bffam=\twelvebf
   \scriptscriptfont\bffam=\tenbf\def\bf{\fam\bffam\fourteenbf}%
   \def\mib{%
      \fourteenmibfonts\twelvemibfonts\tenmibfonts
      \textfont0=\fourteenbf\scriptfont0=\twelvebf
      \scriptscriptfont0=\tenbf
      \textfont1=\fourteenmib\scriptfont1=\twelvemib
      \scriptscriptfont1=\tenmib
      \textfont2=\fourteenbsy\scriptfont2=\tenbsy
      \scriptscriptfont2=\tenbsy}%
   \def\scr{\scrfonts
      \global\textfont\scrfam=\fourteenscr\fam\scrfam\fourteenscr}%
   \normalbaselineskip=17pt
   \setbox\strutbox=\hbox{\vrule height 12pt depth 5pt width 0pt}%
   \normalbaselines\rm\singlespaced}%

\def\sixteenpoint{\sixteenfonts\fourteenfonts\twelvefonts 
   \def\rm{\fam0\sixteenrm}%
   \textfont0=\sixteenrm\scriptfont0=\fourteenrm\scriptscriptfont0=\twelverm
   \textfont1=\sixteeni\scriptfont1=\fourteeni\scriptscriptfont1=\twelvei
   \textfont2=\sixteensy\scriptfont2=\fourteensy\scriptscriptfont2=\twelvesy
   \textfont3=\sixteenex\scriptfont3=\sixteenex\scriptscriptfont3=\sixteenex
   \textfont\itfam=\sixteenit\def\it{\fam\itfam\sixteenit}%
   \textfont\slfam=\sixteensl\def\sl{\fam\slfam\sixteensl}%
   \textfont\bffam=\sixteenbf%
   \scriptfont\bffam=\fourteenbf%
   \scriptscriptfont\bffam=\twelvebf\def\bf{\fam\bffam\sixteenbf}%
   \def\mib{%
      \sixteenmibfonts\fourteenmibfonts\twelvemibfonts
      \textfont0=\sixteenbf\scriptfont0=\fourteenbf
      \scriptscriptfont0=\twelvebf
      \textfont1=\sixteenmib\scriptfont1=\fourteenmib
      \scriptscriptfont1=\twelvemib
      \textfont2=\sixteenbsy\scriptfont2=\fourteenbsy
      \scriptscriptfont2=\twelvebsy}%
   \def\scr{\scrfonts
      \global\textfont\scrfam=\sixteenscr\fam\scrfam\sixteenscr}%
   \normalbaselineskip=20pt%
   \setbox\strutbox=\hbox{\vrule height 14pt depth 6pt width 0pt}%
   \normalbaselines\rm\singlespaced}%

\def\twentypoint{\twentyfonts\sixteenfonts\fourteenfonts 
   \def\rm{\fam0\twentyrm}%
   \textfont0=\twentyrm\scriptfont0=\sixteenrm\scriptscriptfont0=\fourteenrm
   \textfont1=\twentyi\scriptfont1=\sixteeni\scriptscriptfont1=\fourteeni
   \textfont2=\twentysy\scriptfont2=\sixteensy\scriptscriptfont2=\fourteensy
   \textfont3=\twentyex\scriptfont3=\twentyex\scriptscriptfont3=\twentyex
   \textfont\itfam=\twentyit\def\it{\fam\itfam\twentyit}%
   \textfont\slfam=\twentysl\def\sl{\fam\slfam\twentysl}%
   \textfont\bffam=\twentybf
   \scriptfont\bffam=\sixteenbf
   \scriptscriptfont\bffam=\fourteenbf\def\bf{\fam\bffam\twentybf}%
   \def\mib{%
      \twentymibfonts\sixteenmibfonts\fourteenmibfonts%
      \textfont0=\twentybf\scriptfont0=\sixteenbf%
      \scriptscriptfont0=\fourteenbf%
      \textfont1=\twentymib\scriptfont1=\sixteenmib%
      \scriptscriptfont1=\fourteenmib%
      \textfont2=\twentybsy\scriptfont2=\sixteenbsy%
      \scriptscriptfont2=\fourteenbsy}%
   \def\scr{\scrfonts
      \global\textfont\scrfam=\twentyscr\fam\scrfam\twentyscr}%
   \normalbaselineskip=24pt
   \setbox\strutbox=\hbox{\vrule height 17pt depth 7pt width 0pt}%
   \normalbaselines\rm\singlespaced}%

\def\twentyfourpoint{\twentyfourfonts\twentyfonts\sixteenfonts 
   \def\rm{\fam0\twentyfourrm}%
   \textfont0=\twentyfourrm\scriptfont0=\twentyrm\scriptscriptfont0=\sixteenrm
   \textfont1=\twentyfouri\scriptfont1=\twentyi\scriptscriptfont1=\sixteeni
   \textfont2=\twentyfoursy\scriptfont2=\twentysy\scriptscriptfont2=\sixteensy
   \textfont3=\twentyfourex\scriptfont3=\twentyfourex
      \scriptscriptfont3=\twentyfourex
   \textfont\itfam=\twentyfourit\def\it{\fam\itfam\twentyfourit}%
   \textfont\slfam=\twentyfoursl\def\sl{\fam\slfam\twentyfoursl}%
   \textfont\bffam=\twentyfourbf
   \scriptfont\bffam=\twentybf
   \scriptscriptfont\bffam=\sixteenbf\def\bf{\fam\bffam\twentyfourbf}%
   \def\mib{%
      \twentyfourmibfonts\twentymibfonts\sixteenmibfonts%
      \textfont0=\twentyfourbf\scriptfont0=\twentybf
      \scriptscriptfont0=\sixteenbf
      \textfont1=\twentyfourmib\scriptfont1=\twentymib
      \scriptscriptfont1=\sixteenmib
      \textfont2=\twentyfourbsy\scriptfont2=\twentybsy
      \scriptscriptfont2=\sixteenbsy}%
   \def\scr{\scrfonts
      \global\textfont\scrfam=\twentyfourscr\fam\scrfam\twentyfourscr}%
   \normalbaselineskip=28pt
   \setbox\strutbox=\hbox{\vrule height 19pt depth 9pt width 0pt}%
   \normalbaselines\rm\singlespaced}%




\def\singlespaced{
   \baselineskip=\normalbaselineskip            
}


%

\newskip\EnvTopskip     \EnvTopskip=\medskipamount      
\newskip\EnvBottomskip  \EnvBottomskip=\medskipamount   
\newskip\EnvLeftskip    \EnvLeftskip=2\parindent        
\newskip\EnvRightskip   \EnvRightskip=\parindent        
\newskip\EnvDelt@skip   \EnvDelt@skip=0pt               
\newcount\@envDepth     \@envDepth=\z@                  


\def\beginEnv#1{
   \begingroup                          
   \def\@envname{#1}
   \ifvmode\def\@isVmode{T}
   \else\def\@isVmode{F}\vskip 0pt\fi   
   \ifnum\@envDepth=\@ne\parindent=\z@\fi 
   \global\advance\@envDepth by \@ne    
   \EnvDelt@skip=\baselineskip          
   \advance\EnvDelt@skip by-\normalbaselineskip
   \@setenvmargins\EnvLeftskip\EnvRightskip 
   \setenvskip{\EnvTopskip}
   \vskip\skip@\penalty-500             
   }

\def\endEnv#1{
   \ifnum\@envDepth<1                   
      \emsg{> Tried to close ``#1'' environment, but no environment open!}%
      \begingroup                       
   \else                                
      \def\test{#1}
      \ifx\test\@envname\else           
         \emsg{> Miss-matched environments!}%
         \emsg{> Should be closing ``\@envname'' instead of ``\test''}%
      \fi                               %
   \fi                                  %
   \vskip 0pt                           
   \setenvskip\EnvBottomskip            
   \vskip\skip@\penalty-500             
   \xdef\@envtemp{\@isVmode}
   \endgroup                            
   \global\advance\@envDepth by -\@ne   
   \if F\@envtemp\vskip-\parskip\noindent\fi 
   }


\def\setenvskip#1{\skip@=#1 \divide\skip@ by \@envDepth}


\def\@setenvmargins#1#2{
   \advance \leftskip  by #1    \advance \displaywidth by -#1   %
   \advance \rightskip by #2    \advance \displaywidth by -#2   %
   \advance \displayindent by #1}                               %


\def\@eatpar{\futurelet\next\@testpar}
\def\@testpar{\ifx\next\par\let\@next=\@@eatpar\else\let\@next=\relax\fi\@next}
\long\def\@@eatpar#1{\relax}

%
%

\def\TeXexample{\beginEnv{TeXexample}
   \vskip\EnvDelt@skip                  
   \parskip=\z@ \parindent=\z@          
   \baselineskip=\normalbaselineskip    
   \def\par{\leavevmode\endgraf}
   \obeylines                           
   \catcode`|=\z@                       
   \ttverbatim                          
   \@eatpar}



\chardef\other=12

\def\ttverbatim{\begingroup                     
   \catcode`\(=\other \catcode`\)=\other        
   \catcode`\"=\other \catcode`\[=\other        %
   \catcode`\]=\other                           %
   \let\do=\uncatcode \dospecials               %
   \obeyspaces \obeylines                       
   \def\n{\vskip\baselineskip}
   \tt}                                         

\def\uncatcode#1{\catcode`#1=\other}            

{\obeyspaces\gdef {\ }}                        


\def\TeXquoteon{\catcode`\|=\active}            

{\TeXquoteon\obeylines                          
   \gdef|{\ifmmode\vert\else                    
     \ttverbatim \spaceskip=\ttglue             
     \let^^M=\ 
     \let|=\endgroup                            
     \fi}                                       
}


\def\raggedcenter{
    \flushenv                           
    \advance\leftskip\z@ plus4em        
    \advance\rightskip\z@ plus 4em      
    \spaceskip=.3333em \xspaceskip=.5em %
    \pretolerance=9999 \tolerance=9999  %
    \hyphenpenalty=9999 \exhyphenpenalty=9999   
    \@eatpar}                           %

\def\endraggedcenter{\endflushenv}              

\def\flushenv{
    \vskip \z@                          
    \bgroup                             
     \def\flushhmode{F}
     \parindent=\z@  \parfillskip=\z@}  %

\def\endflushenv{
   \ifhmode\endgraf\fi                          
   \if T\flushhmode \egroup\hss\fi              
   \egroup}                                     




\def\slashchar#1{\setbox0=\hbox{$#1$}           
   \dimen0=\wd0                                 
   \setbox1=\hbox{/} \dimen1=\wd1               
   \ifdim\dimen0>\dimen1                        
      \rlap{\hbox to \dimen0{\hfil/\hfil}}      
      #1                                        
   \else                                        
      \rlap{\hbox to \dimen1{\hfil$#1$\hfil}}   
      /                                         
   \fi}                                         %


\def\simge{
    \mathrel{\rlap{\raise 0.511ex
        \hbox{$>$}}{\lower 0.511ex \hbox{$\sim$}}}}

\def\simle{
    \mathrel{\rlap{\raise 0.511ex
        \hbox{$<$}}{\lower 0.511ex \hbox{$\sim$}}}}



\def\TeV{{\rm TeV}}                     
\def\GeV{{\rm GeV}}                     
\def\MeV{{\rm MeV}}                     

\def\mb{{\rm mb}}                       
\def\nb{{\rm nb}}                       
\def\pb{{\rm pb}}                       
\def\fb{{\rm fb}}                       

\def\cmsec{{\rm cm^{-2}s^{-1}}}         


\newcount\subsubsecno \subsubsecno=0		
\def\chapsym{}
\def\ch@psymdash{}
\def\ch@psymdot{}

\def\GEMdoc{%
   \openin\lfilein=labeldefs.tmp		
   \ifeof\lfilein\closein\lfilein		
   \else\closein\lfilein			
   \input labeldefs.tmp \relax\fi\writedefs				
   \twelvepoint\baselineskip=16pt
   \parskip=\medskipamount
   \GEMsuperrefstrue				
   \let\newchap=\GEMchapter
   \let\newsec=\GEMsection
   \let\subsec=\GEMsubsection
   \let\subsubsec=\GEMsubsubsection
   \let\subsubsubsec=\GEMsubsubsubsection
   \let\eqnres@t=\GEMeqnres@t
   \let\listrefs=\GEMlistrefs			
   \let\ref=\GEMref				
   \let\nref=\GEMnref				
   \let\xref=\GEMxref				
   \let\cite=\GEMcite				
   \let\nfig=\GEMnfig				
   \let\nfigwide=\GEMnfigwide			
   \footline={\hss\twelverm\ch@psymdash\folio\hss}
}

\newdimen\footlineoffset \footlineoffset=2pc	

\def\MIB{\mib}

\def\GEMchapter#1#2{
   \gdef\chapsym{#1}
   \gdef\ch@psymdash{#1--}
   \gdef\ch@psymdot{#1.}
   \vfill\supereject				
   \def\Hrulefill{ \leaders\hrule height1pt\hfill\ }%
   \hrule height0pt depth0pt			
   \vskip-10pt
   \centerline{\Hrulefill\ \lower6pt\hbox{\twentyfourpoint\chapsym}%
   \Hrulefill}%
   \vskip0pt
   \raggedcenter%
      {\sixteenpoint\bf\MIB
      #2
      \vskip0pt}				
   \endraggedcenter%
   \vskip3pc					
   \global\secno=0\global\subsecno=0		
   \global\subsubsecno=0\global\meqno=1		
}

\def\GEMsection#1{
   \bigbreak					
   \global\advance\secno by1
   \gdef\secsym{\ch@psymdot\the\secno.}
   \message{(\secsym #1)}
   \global\subsecno=0\global\subsubsecno=0	
   \global\meqno=1				
   \GEMhe@ding{\bf\MIB\secsym\ \ }{#1}
   \writetoca{{\secsym}{#1}}
}

\def\GEMsubsection#1{
   \bigbreak					
   \global\advance\subsecno by1			
   \gdef\subsecsym{\secsym\the\subsecno.}
   \message{(\subsecsym #1)}
   \global\subsubsecno=0
   \GEMhe@ding{\bf\MIB\subsecsym\ \ }{#1}
   \writetoca{\string\quad {\subsecsym} {#1}}
}

\def\GEMsubsubsection#1{
   \bigbreak					
   \global\advance\subsubsecno by1		
   \gdef\subsubsecsym{\subsecsym\the\subsubsecno.}
   \message{(\subsubsecsym #1)}
   \GEMhe@ding{\bf\MIB\subsubsecsym\ \ }{#1}
   \writetoca{\string\quad {\subsubsecsym} {#1}}
}%
\def\GEMsubsubsubsection#1{
   \bigbreak					
   \GEMhe@ding{}{\bf\MIB #1}
}%

\def\GEMhe@ding#1#2{
   \vbox{\raggedright\tolerance=10000		
      \setbox0=\hbox{#1}
      \dimen0=\wd0
      \hangindent=\dimen0 \hangafter=1
      {\noindent #1#2}
      \vskip-\parskip}\nobreak
}

\def\GEMlistrefs{%
   \footatend\relax\immediate\closeout\rfile\writestoppt
   \bigbreak
   \vbox{{\bf\noindent References}\medskip}
   \nobreak
   {\frenchspacing\parindent=20pt\escapechar=` \input refs.tmp\vskip0pt}%
   \nonfrenchspacing}


\def\GEMtable{%
   \hbox to\hsize\bgroup\hss\vbox\bgroup	
   \GEMpoint					
   \def~{\phantom{0}}
   \tabskip=0pt\halign\bgroup}			

\def\GEMpoint{\twelvepoint}

\def\endGEMtable{%
   \egroup\egroup\hss\egroup}			

\def\GEMrule{%
   \noalign{\vskip4pt\hrule\vskip4pt}}

\def\GEMrulerule{%
   \noalign{\vskip4pt\hrule\vskip1.5pt\hrule\vskip4pt}}



\newif\ifGEMsuperrefs	\GEMsuperrefstrue

\def\GEMcite#1{%
   \begingroup
      \let\@sf=\empty				
      \ifhmode\edef\@sf{
         \spacefactor\the\spacefactor}\/\fi     %
      \ifGEMsuperrefs                           
         $\relax{}^{#1}$\@sf
      \else {}~[{#1}]\@sf\fi
   \endgroup}

\def\GEMref{\GEMcite{\the\refno}\GEMnref}

\def\GEMnref#1{\xdef#1{\the\refno}\writedef{#1\leftbracket#1}%
\ifnum\refno=1\immediate\openout\rfile=refs.tmp\fi
\global\advance\refno by1\chardef\wfile=\rfile\immediate
\write\rfile{\noexpand\item{#1.\ }\reflabeL{#1\hskip.31in}\pctsign}\findarg}

\def\GEMxref#1{#1}


\def\GEMnfig#1#2#3{%
   \xdef#1{\ch@psymdash\the\figno}
   \writedef{#1\leftbracket#1}
   \topinsert
      #2
      \bigskip
      \GEMcaption{FIG.~#1}{#3}
   \endinsert
   \global\advance\figno by1}			

\newbox\c@pbox
\newcount\c@plines

\def\GEMnfigwide#1#2#3{%
   \xdef#1{\ch@psymdash\the\figno}
   \writedef{#1\leftbracket#1}
   \widetopinsert
      #2
      \bigskip
      \GEMcaption{Fig.~#1}{#3}
   \endinsert
   \global\advance\figno by1}			

\def\GEMcaption#1#2{
   \begingroup\GEMpoint				
   \global\setbox\c@pbox=\vbox{
      \noindent\hbox{\noindent #1.\ \ }{#2}
      \vskip0pt\relax				
      \global\c@plines=\prevgraf}
   \ifnum\@ne=\c@plines
      \global\setbox\c@pbox=\vbox{
      \noindent\hfil				
      \hbox{\noindent#1:\ \ }{#2}\hfil}\fi
   \centerline{\box\c@pbox}
   \endgroup					
}


\def\GEMepsf#1#2{%
   \ifx\undefined\epsfbox			
      \vskip#2\relax				
   \else					
      \epsfysize=#2				
      \epsfbox{#1}
   \fi
}

\catcode`@=12

\def\bigans{b }
\def\littleans{l }

\everyjob={%
   \message{ big or little (b/l)? }\read-1 to\answ
   \ifx\answ\bigans
      \message{This will come out unreduced.}
      \PPTbig
   \else
      \if\answ\littleans
         \message{This will come out reduced.}
         \PPTlittle
      \else
         \message{No format defined}
      \fi
   \fi}


\PPTbig


\def\loadssfonts{%
   \global\font\niness=cmss9
   \global\font\sevenss=cmss9 at 7pt
   \gdef\ninessfonts{\relax}%
}

\def\ninepointss{\ninefonts               
   \loadssfonts				  
   \def\rm{\fam0\niness}%
   \textfont0=\niness\scriptfont0=\sevenss\scriptscriptfont0=\fiverm
   \textfont1=\ninei\scriptfont1=\seveni\scriptscriptfont1=\fivei
   \textfont2=\ninesy\scriptfont2=\sevensy\scriptscriptfont2=\fivesy
   \textfont3=\nineex\scriptfont3=\nineex\scriptscriptfont3=\nineex
   \textfont\itfam=\nineit\def\it{\fam\itfam\nineit}%
   \textfont\slfam=\ninesl\def\sl{\fam\slfam\ninesl}%
   \textfont\ttfam=\ninett\def\tt{\fam\ttfam\ninett}%
   \textfont\bffam=\ninebf
   \scriptfont\bffam=\sevenbf
   \scriptscriptfont\bffam=\fivebf\def\bf{\fam\bffam\ninebf}%
   \def\mib{\relax}%
   \def\scr{\relax}%
   \tt\ttglue=.5emplus.25emminus.15em
   \normalbaselineskip=11pt
   \setbox\strutbox=\hbox{\vrule height 8pt depth 3pt width 0pt}%
   \normalbaselines\rm\singlespaced}%

%
%

\def\CL{{\cal L}}

\def\CO{{\cal O}}

\def\CQ{{\cal Q}}
\def\CR{{\cal R}}

%
%
\def\ts{\thinspace}
\def\ra{\rightarrow}
\def\ol{\overline}

\def\W+-{W^\pm}
\def\Z0{Z^0}

\def\Ms+-{M^2_\pm}

\def\M+-{M_\pm}

\def\kslash{\raise.15ex\hbox{/}\kern-.57em k}
\def\pslash{\raise.15ex\hbox{/}\kern-.57em p}
%
%
%
\def\simge{\mathrel{%
   \rlap{\raise 0.511ex \hbox{$>$}}{\lower 0.511ex \hbox{$\sim$}}}}
\def\simle{\mathrel{
   \rlap{\raise 0.511ex \hbox{$<$}}{\lower 0.511ex \hbox{$\sim$}}}}
\def\slashchar#1{\setbox0=\hbox{$#1$}           
   \dimen0=\wd0                                 
   \setbox1=\hbox{/} \dimen1=\wd1               
   \ifdim\dimen0>\dimen1                        
      \rlap{\hbox to \dimen0{\hfil/\hfil}}      
      #1                                        
   \else                                        
      \rlap{\hbox to \dimen1{\hfil$#1$\hfil}}   
      /                                         
   \fi}                                         %


%
\def\etmiss{\slashchar{E}_T}
\def\MeV{{\rm MeV}}
\def\gluino{{\widetilde g}}
\def\squark{{\widetilde q}}
\def\lsp{{\widetilde\chi_1^0}}
\def\ellell{\ell^+\ell^-}

\def\CL{{\cal L}}

\def\GeV{{\rm GeV}}
\def\Lhnc{\CL_{\rm HNC}}
\def\Liso{\CL_{\rm ISO}}
\def\TeV{{\rm TeV}}
\def\abseta{\vert \eta \vert}
\def\afb{A_{FB}}

\def\cmsec{{\rm cm^{-2}sec^{-1}}}
\def\cm{{\rm cm}}

\def\cstar{\cos \theta^*}
\def\ellm{\ell^-}
\def\ellpm{\ell^\pm}
\def\ellp{\ell^+}

\def\ee{e^+ e^-}

\def\et{E_T}
\def\fb{{\rm fb}}
\def\gev{{\rm GeV}}
\def\h{H^0}
\def\hgg{\h \ra \gamma \gamma}

\def\hl{10^{33} \ts {\rm cm}^{-2} \ts {\rm s}^{-1}}
\def\jet{{\rm jet}}
\def\jets{{\rm jets}}
\def\mb{{\rm mb}}
\def\mee{M_{e^+ e^-}}

\def\mmm{M_{\mu^+ \mu^-}}
\def\Mlm{M_{\ell \mu}}
\def\Mem{M_{e \mu}}
\def\Mmm{M_{\mu \mu}}

\def\mum{\mu^-}

\def\mup{\mu^+}
\def\nb{{\rm nb}}
\def\ol{\overline}

\def\pb{{\rm pb}}

\def\pt{p_T}

\def\ra{\rightarrow}

\def\rshat{\sqrt{\shat}}
\def\shat{\hat s}
\def\tev{{\rm TeV}}

\def\uhl{10^{34} \ts {\rm cm}^{-2} \ts {\rm s}^{-1}}

\def\mzp{M_{Z'}}
\def\gzp{\Gamma_{Z'}}

\def\Lhnc{\CL_{\rm HNC}}
\def\Liso{\CL_{\rm ISO}}
\def\abseta{\vert \eta \vert}
\def\afb{A_{FB}}
\def\cstar{\cos \theta^*}

\def\sigh{\hat \sigma}
\def\thw{\theta_W}

\def\half{{1 \over {2}}}
\def\thalf{\textstyle{{1 \over {2}}}}

\def\tfourth{\textstyle{{1 \over {4}}}}
\def\sscy{10\,\fb^{-1}}
\def\sscd{100\,\fb^{-1}}


\def\GEMlistrefs{%
   \footatend\relax\immediate\closeout\rfile\writestoppt
   \bigbreak
   \vbox{{\bf\noindent REFERENCES}\medskip}
   \nobreak
   {\frenchspacing\parindent=20pt\escapechar=` \input refs.tmp\vskip0pt}%
   \nonfrenchspacing}

\catcode`@=11
\def\GEMsection#1{
   \vfill\supereject				
   \global\advance\secno by1
   \gdef\secsym{\ch@psymdot\the\secno.}
   \message{(\secsym #1)}
   \global\subsecno=0\global\subsubsecno=0	
   \global\meqno=1				
   \GEMhe@ding{\bf\MIB\secsym\ \ }{#1}
   \writetoca{{\secsym}{#1}}
}
\catcode`@=12

\gdef\Emiss{\slashchar{E}}
\gdef\Dzero{D$\slashchar{0}$}
\gdef\cmsec{{\rm cm}^{-2}{\rm s}^{-1}}

\GEMdoc
\TeXquoteon					
\hyphenation{cal-o-rim-e-ter}

\def\doublecolumns{\relax}
\def\enddoublecolumns{\relax}

\def\widetopinsert{\topinsert}

\def\forceleft{\relax}
\def\forceright{\relax}


\rightline{\singlespaced\vbox{%
\hbox{BNL-61138}%
\hbox{BUHEP-94-31}%
\hbox{Fermilab Pub-94/392-E}%
}}

\bigskip\bigskip

\centerline{\fourteenpoint\bf SIMULATIONS OF SUPERCOLLIDER PHYSICS}

\bigskip\bigskip

\centerline{\bf K.D. Lane$^1$, F.E. Paige$^2$, T. Skwarnicki$^3$, and
W.J.~Womersley$^4$}

\medskip

\centerline{$^1$ Physics Department, Boston University, Boston, MA
02215}
\centerline{$^2$ Physics Department, Brookhaven National Laboratory,
Upton, NY 11973}
\centerline{$^3$ Physics Department, Southern Methodist University,
Dallas, TX 75275}
\centerline{$^4$ Fermi National Accelerator Laboratory,
Batavia, IL 60510}

\vskip1cm

\begingroup\narrower\narrower

\centerline{\bf ABSTRACT}

	The Standard Model of particle physics makes it possible to
simulate complete events for physics signatures and their backgrounds in
high energy collisions. Knowledge of how the produced particles interact
with the materials in a detector makes it possible to simulate the response
of any particular detector design to these events and so determine whether
the detector could observe the signal. The combination of these techniques
has played an important role in the design of new detectors, particularly
those for hadron supercolliders where the high rates and small signal cross
sections make the experiments very difficult. The technique is reviewed
here and illustrated using the simulations of the GEM detector proposed for
the Superconducting Super Collider. Although the simulations and results
described here are somewhat detector-specific, we believe that they can
serve as a useful model for this component of detector design for future
hadron supercolliders.

\vskip0pt\endgroup

\vskip1cm

\doublecolumns

\newsec{INTRODUCTION}

	The ``Standard Model'' of particle physics describes strong,
electromagnetic, and weak interactions in terms of an $SU(3)
\otimes SU(2) \otimes U(1)$ gauge theory. The theoretical elements of
the model have been in place for more than 20 years.%
\nref\sm{S.~L.~Glashow, Nucl.\ Phys.\ {\bf 22}, 579 (1961)\semi
S.~Weinberg, Phys.\ Rev.\ Lett.\ {\bf 19}, 1264 (1967)\semi
A.~Salam, in Proceedings of the 8th Nobel Symposium on Elementary
Particle
Theory, Relativistic Groups and Analyticity, edited by N.\ Svartholm
(Almquist and Wiksells, Stockholm, 1968), p.\ 367\semi
H.~Fritzsch, M.~Gell-Mann, H.~Leutwyler, Phys.\ Lett.\ {\bf 47B}, 365,
(1973)\semi
D.~Gross and F.~Wilczek, Phys.\ Rev.\ Lett.\ {\bf 30}, 1343 (1973)\semi
H.~D.~Politzer, Phys.\ Rev.\ Lett.\ {\bf 30}, 1346 (1973).}%
\nref\particles{R.~Cahn and G.~Goldhaber, {\it The Experimental
Foundations of Particle Physics}, (Cambridge University Press, 1989)}%
\cite{\sm,\particles}
Over that period a series of experimental results --- including the
discoveries of charm,%
\ref\EXPcharm{J.~J.~Aubert, et al., Phys.\ Rev.\ Lett.\ {\bf 33}, 1404
(1974)\semi
J.~E.~Augustin, et al., Phys.\ Rev.\ Lett.\ {\bf 33}, 1406 (1974)\semi
G.~Goldhaber, et al., Phys.\ Rev.\ Lett.\ {\bf 37}, 255 (1976).}
of the $\tau$-lepton and the $b$ quark,%
\ref\EXPthird{M.~L.~Perl et al., Phys.\ Rev.\ Lett.\ {\bf 35}, 1489
(1975)\semi
S.~W.~Herb et al., Phys.\ Rev.\ Lett.\ {\bf 39}, 252 (1977)}
and of the predicted $W^\pm$ and $Z^0$ bosons%
\ref\EXPwz{G.~Arnison, et al., Phys.\ Lett.\ {\bf 122B}, 103
(1983)\semi
M.~Banner, et al., Phys.\ Lett.\ {\bf 122B}, 476 (1983)\semi
P.~Bagnaia, et al., Phys.\ Lett.\ {\bf 129B}, 130 (1983).}
\ --- have provided increasing evidence of the correctness of the
Standard Model at present energy scales. More recently, the Standard
Model has withstood extremely stringent quantitative experimental
tests.%
\ref\smtests{For a recent review, see P.~Langacker, M.~-X.~Luo and
A.~K.~Mann, Rev.\ Mod.\ Phys.\ {\bf 64}, 87 (1992).}
Thus it is now accepted that it provides a good
description of particle physics down to the shortest distances yet
probed, about $10^{-16}\,\cm$.

	In the Standard Model the basic constituents of matter are
spin-$1\over2$ quarks and leptons. There are six types or ``flavors'' of
each. Under the $SU(2) \otimes U(1)$ group of
electroweak interactions, the left-handed fields are in three quark and
lepton doublets; the
right-handed ones are in singlets:
\eqna\Fermion
$$
\eqalignno
{
{\rm Leptons:}& \quad
\Biggl( {\nu_e \atop e} \Biggr) \quad
\Biggl( {\nu_\mu \atop \mu} \Biggr) \quad
\Biggl( {\nu_\tau \atop \tau} \Biggr) &\Fermion{a}\cr
{\rm Quarks:}& \quad
\Biggl( {u \atop d'} \Biggr) \quad
\Biggl( {c \atop s'} \Biggr) \quad
\Biggl( {t \atop b'} \Biggr) & \Fermion{b}\cr
}
$$
Under the $SU(3)$ group of strong interactions, which for historical reasons
is known as color, the quarks are triplets and the leptons are singlets.
The quarks and leptons can also be organized into three generations
identical except for mass. There is mixing between the generations of
quarks; the primed lower elements in Eq.~\Fermion{b} denote linear
combinations of the mass eigenstates. All the fermions definitely
have been found except for the top quark and the tau-neutrino.
Direct evidence for the top quark with mass
$m_t = 174\pm 10 \ts ^{+13}_{-12} \,\gev$ has recently been reported.%
\ref\topmass{F.~Abe, et al., The CDF Collaboration, Phys.\ Rev.\
Lett.\ {\bf 73}, 225 (1994); Phys.\ Rev.\ {\bf D50}, 2966,
(1994).\hfil\break At the time much of this work was done, the
experimental bound was $m_t > 95\,\GeV$.}
There is strong indirect evidence for the tau-neutrino.%
\nref\nutaumass{There is little doubt that $\nu_\tau$
exists and that it and the
$\tau^-$ form a standard lepton doublet. The 95\% limit on its mass is
$m_{\nu_\tau} < 35\,{\rm MeV}$.  See Ref.\ \PDG.}%
\nref\PDG{Particle Data Group, K.~Hikasa et al., Phys.\ Rev.\ {\bf
D50}, No.~3--I (1994).}%
\cite{\nutaumass, \PDG}
If the number of quark-lepton generations is equal to the number
$N_\nu$ of light neutrinos, then there are no more than these three.
The evidence for this comes from precision measurements of the $Z^0$
at LEP, which imply $N_\nu = 2.99 \pm 0.04$ in the Standard
Model.\cite{\PDG}

	The interactions of the quarks and leptons are invariant under
$SU(3) \otimes SU(2) \otimes U(1)$ locally at each point in
space-time. The requirement of local invariance implies that there is
one spin-one gauge boson for each generator of the symmetry
group, and it restricts their couplings so that the theory
is renormalizable. This means that cross sections do not grow more
rapidly than allowed by unitarity and that the form of the
interactions is insensitive to possible physics at much higher mass
scales. These gauge bosons are responsible for all known interactions
except gravity, for which there is no fully satisfactory quantum theory. The
eight generators of the $SU(3)$ color group correspond to the eight
massless gluons of Quantum Chromodynamics (QCD) that are responsible
for strong interactions.  Gauge invariance requires that these
interact. These self-interactions produce a potential energy
which grows linearly
with distance between isolated quarks or gluons, which are therefore
permanently confined into the observed hadrons such as the proton. At
short distances --- large energies and momentum transfers --- the same
interactions make
the QCD coupling weak, so that perturbation theory can be used to
describe the production of heavy particles or large-$p_T$ interactions
in hadronic collisions.  Because a gauge theory restricts the form of
the couplings and because calculations can be done in perturbation
theory, the Standard Model can be used to predict cross sections
even for new physics (provided, of course, that a model exists for the
couplings of the new physics to standard model degrees of freedom).

	The remaining four generators of $SU(2) \otimes U(1)$
correspond to the massless photon and the massive $W^\pm$ and $Z^0$
bosons responsible for electroweak interactions. If a gauge symmetry
is unbroken, then the gauge bosons and fermions associated with it
must be massless.  It is possible, however, to introduce new
interactions that break the symmetry ``spontaneously'' while preserving
the good high-energy behavior of the gauge theory.
This is known as the ``Higgs mechanism''.%
\ref\higgs{P.~W.~Anderson, Phys.\ Rev.\ {\bf 110}, 827 (1958); {\it
ibid.}, {\bf 130}, 439 (1963)\semi
Y.~Nambu, Phys.\ Rev.\ {\bf 117}, 648 (1959)\semi
J.~Schwinger, Phys.\ Rev.\ {\bf 125}, 397 (1962)\semi
P.~Higgs, Phys.\ Rev.\ Lett.\ {\bf 12}, 132 (1964)\semi
F.~Englert and R.~Brout, Phys.\ Rev.\ Lett.\ {\bf 13}, 321 (1964)\semi
G.~S.~Guralnik, C.~R.~Hagen and T.~W.~B.~Kibble, Phys.\ Rev.\ Lett.\
{\bf13}, 585 (1964).}
In the Standard
Model, both the $W^\pm$ and $Z^0$ bosons and the fermions get their
masses from an elementary scalar field, which is a doublet under
$SU(2)$ and which acquires a vacuum expectation value, thus breaking
the symmetry.

	While there is strong experimental evidence supporting the
gauge-theoretic part of the Standard Model, there is as yet no
evidence for or against the Higgs mechanism for
electroweak symmetry breaking. Nor is there data to indicate the mechanism
by which finite and unequal fermion masses are generated (flavor symmetry
breaking).
Understanding how these masses are produced is the central
problem in particle physics. Several scenarios have been proposed for
electroweak and flavor symmetry breaking:%
\ref\ehlq{E.~Eichten, I.~Hinchliffe, K.~Lane and C.~Quigg, Rev.\ Mod.\
Phys.\ {\bf 56}, 579 (1984).}

\medskip

\item{$\bullet$} Standard Higgs models, containing one or more
elementary Higgs boson multiplets.  These are generally complex weak
doublets. The minimal model has one doublet, with a single neutral
boson $H^0$.

\item{$\bullet$} Supersymmetry. In the minimal supersymmetric standard
model there are two Higgs doublets, and every known particle has a
superpartner.

\item{$\bullet$} Models of dynamical electroweak and flavor symmetry
breaking. The most studied proposal is
technicolor-plus-extended-technicolor, with one doublet or one family
of technifermions.

\item{$\bullet$} Composite models, in which quarks and leptons are
built of more fundamental constituents.

\medskip

        None of these proposals is fully satisfactory. In elementary
Higgs boson models, whether supersymmetric or not, there is no plausible
natural explanation of why electroweak symmetry breaking occurs and why it has
the observed scale. In non-supersymmetric models, the Higgs boson's
mass, $M_H$, and its vacuum expectation value, $v$, are unstable
against radiative corrections. There is no natural reason why these
two parameters should be very much less than the energy scale at which
the essential physics changes, e.g., a unification scale or the Planck
scale.%
\ref\natural{K.~G.~Wilson, unpublished; quoted in L.~Susskind, Phys.\
Rev.\
{\bf D20}, 2619 (1979)\semi
G.~'tHooft, in {\it Recent Developments in Gauge Theories}, edited by
G.~'tHooft, et al. (Plenum, New York, 1980).}
This instability does not arise in supersymmetric models, which is the
motivation for supersymmetry at the electroweak scale.%
\nref\SUSYref{S.~Dimopoulos and H.~Georgi, Nucl.\ Phys.\ {\bf B193},
153
(1981)\semi
A.H.~Chamseddine, R.~Arnowitt and P.~Nath, Phys.\ Rev.\ Lett.\ {\bf
49}, 970
(1982)\semi
L.~J.~Hall, J.~Lykken and S.~Weinberg, Phys.\ Rev.\ {\bf D27}, 2359
(1983).}%
\nref\HaberKane{For reviews of supersymmetry and its phenomenology,
see H.\ E.\ Haber and G.\ L.\ Kane, Phys.\ Rept.\ {\bf 117}, 75
(1985)\semi
S.~Dawson, E.~Eichten and C.~Quigg, Phys.\ Rev.\ {\bf D31}, 1581
(1985).}%
\cite{\SUSYref, \HaberKane}
Furthermore, elementary Higgs boson models are known to be
``trivial'', i.e., they cannot make sense as interacting field theories
with the cutoff taken to infinity.%
\ref\trivial{See, for
example, R.~Dashen and H.~Neuberger,
Phys.\ Rev.\ Lett.\ {\bf 50}, 1897 (1983)\semi J.~Kuti, L.~Lin and
Y.~Shen,
Phys.\ Rev.\ Lett.\ {\bf 61}, 678 (1988)\semi A.~Hasenfratz, et
al.\ Phys.\ Lett.\ {\bf B199}, 531 (1987)\semi G.~Bhanot and K.~Bitar,
Phys.\ Rev.\ Lett.\ {\bf 61}, 798 (1988).}
This means that they must be regarded as effective theories,
meaningful only below some cutoff $\Lambda$. Obviously, $\Lambda$ must
be somewhat greater than $M_H$ for the effective theory to make sense.
For a modest separation of these energies, $M_H < {\rm few} \times
\Lambda$, both perturbative and lattice gauge calculations give $M_H
\simle 650\,\gev$ in the minimal one-doublet model.%
\ref\footb{This is not a pressing issue in the popular ``minimal
supersymmetric standard model'' because the Higgs masses are relatively
low and, so, the cutoff $\Lambda$ may be very high indeed.}
Finally, elementary Higgs models provide no clue to the meaning of
flavor symmetry and the origin of its breaking.  The flavor-symmetry
breaking Yukawa couplings of the Higgs boson to fermions are arbitrary
free parameters.

Despite these apparent problems, the standard Higgs boson, $H^0$, charged
Higgses, $H^\pm$, and the supersymmetric partners of all the known
particles may exist and must be sought. However, if something like the
standard $\h$ is found and is heavier than about $700\,\gev$, experiments
must have the capacity to discover the additional, unspecified new physics
that surely exists in the same energy region, of order $1\,\tev$. One of
the basic considerations driving the high collision energy of the SSC was
the possibility that such physics exists.

        Dynamical theories of electroweak and flavor symmetries --- technicolor
and extended technicolor --- address these shortcomings of
the elementary Higgs boson models.%
\ref\TCref{S.~Weinberg, Phys.\ Rev.\ {\bf D13}, 974 (1976); {\it ibid}
{\bf
D19}, 1277 (1979)\semi
L.~Susskind, Phys.\ Rev.\ {\bf D20}, 2619 (1979)\semi
S.~Dimopoulous and L.~Susskind, Nucl.\ Phys.\ {\bf B155}, 237
(1979)\semi
E.~Eichten and K.~Lane, Phys.\ Lett.\ {\bf 90B}, 125 (1980).}
However, they do so at the price of introducing flavor-changing
neutral currents that are too large, and pseudo-Goldstone bosons
(technipions) that are too light.%
\ref\FCNCref{E.~Eichten and K.~Lane in Ref.\ \TCref\semi J.~Ellis,
M.~Gaillard,
D.~Nanopoulos and P.~Sikivie, Nucl.\ Phys.\ {\bf B182}, 529 (1981).}
Further, it is difficult to build realistic models with QCD-like dynamics
that are consistent with the precision tests of the electroweak interactions.
These difficulties have been mitigated, but only by invoking a
new form of strong technicolor dynamics.%
\nref\wtc{B.~Holdom, Phys.\ Rev.\ {\bf D24}, 1441 (1981); Phys.\ Lett.\
{\bf B150}, 301 (1985)\semi
T.~Appelquist, D.~Karabali and L.~C.~R. Wijewardhana, Phys.\ Rev.\
Lett.\
{\bf 57}, 957 (1986)\semi
T.~Appelquist and L.~C.~R.~Wijewardhana, Phys.\ Rev.\ {\bf D36}, 568
(1987) \semi
K.~Yamawaki, M.~Bando and K.~Matumoto, Phys.\ Rev.\ Lett.\
{\bf 56}, 1335 (1986) \semi
T.~Akiba and T.~Yanagida, Phys.~Lett.\ {\bf B169}, 432 (1986).}%
\nref\setc{T.~Appelquist, T.~Takeuchi, M.~B.~Einhorn,
L.~C.~R.~Wijewardhana,
Phys.\ Lett.\ {\bf B220}, 223 (1989); T.~Takeuchi, Phys. Rev. {\bf
D40},
2697 (1989)
\semi V.~A.~Miransky and K.~Yamawaki, Mod.\ Phys.\ Lett.\ {\bf A4}, 129
(1989)
\semi R.~S.~Chivukula, A.~G.~Cohen and K.~Lane, Nucl.\ Phys.\ {\bf
B343},
554 ( 1990).}%
\cite{\wtc,\setc}
Realistic models of
composite quarks and leptons are similarly difficult to construct.%
\ref\thooft{See G.~'tHooft in Ref.\ \natural.}
While no compelling models of dynamical electroweak and flavor
symmetry breaking exist, there are nevertheless model-independent
phenomenological programs for searching for them.\cite{\ehlq}

\bigskip

	While the origin of electroweak symmetry breaking is unknown,
the mass scale for the new physics responsible for it is set by
the vacuum expectation value of the Standard Model Higgs boson:
$$
v = 2^{-\tfourth} G_F^{-\thalf} = 246\,\gev \ts .
$$
The cross section for {\it any} new physics at this scale is set by
dimensional analysis and coupling strengths to be $\sigma
\sim 1\,\nb$ -- $1\,\fb$. So long as the nature of this new physics
remains unknown, the most promising way by far to explore this
mass and cross section scale is with a $pp$ (or $p\bar p$) supercollider.
That is because such a machine functions as a collider of broad-band beams of
quarks and gluons (thus allowing coupling to both color and electroweak quantum
numbers) and because these beams have the highest energy and luminosity
technically feasible today.
Clearly, it would have been easier to carry out the
high-mass searches at higher center of mass energy and
a luminosity of $\CL \sim \hl$ rather than $\uhl$.
Nevertheless, for the foreseeable future,
the capabilities of the CERN Large Hadron Collider (LHC),
$\sqrt{s} = 14\,\TeV$ and ${\cal L} = 10^{34}\,\cmsec$,
make it the best instrument for discovering the origin of electroweak
symmetry breaking.

	Hadron supercolliders such as the SSC or LHC present a
difficult experimental problem: the interesting cross sections are
very small compared to the total cross section of order $100\,\mb$ and small
even compared to the dijet cross section at a comparable mass. Thus,
careful study is needed to determine whether a given experimental design can
detect the signatures of models for new physics. Because these signatures
always manifest themselves as quarks and gluons (jets), charged leptons and
neutrinos (missing transverse energy, $\etmiss$), the
Standard Model allows simulation of events both for the new
physics and for its backgrounds in a reliable and predictive way.
Given a detector design and known facts about particle-matter
interactions, the response of the detector to these events can also be
simulated. Finally, the simulated data can be reconstructed and
analyzed just as for a real experiment to determine whether one could
observe a signal.

These in-depth simulations are necessitated by the long lead times and
large amounts of money involved in constructing a supercollider detector.
They have been made possible by the convergence of a reliable theoretical
framework (the Standard Model) with rapid increases in available computing
power. The simulations are an essential component of the design process,
providing feedback on the design's strengths and weaknesses. In this way,
the simulations have played and continue to play an important role in the
design of detectors for the SSC and LHC.

	This article reviews the simulations carried out by the GEM (Gammas,
Electrons, and Muons) Collaboration for its detector%
\ref\gemtdr{GEM Collaboration, {\it Technical Design Report},
SSCL--SR--1219(1993).}
at the SSC. Section~2 describes the tools available for simulation of
interactions and of the behavior of the resulting particles in a detector.
Section~3 briefly reviews the design proposed for the GEM detector at the
time of Technical Review. It is
essential to have a detailed engineering design to obtain realistic results,
and the details of results presented here are, strictly speaking,
valid principally for this design. However, any
well-developed design is likely to have
comparable strengths and problems, so the methods used for GEM and the
results obtained should be more generally useful.
Section~4 describes the tools developed for the simulation of the GEM
detector. The principal development was a fast all-detector
parameterization, called |gemfast|,
that allowed simulation of very large numbers of
events in a manageable amount of time. This was supplemented by the most
detailed possible simulations of the muon system and calorimetry
in those cases that the fast simulation
was insufficient to model important details and where the
importance of the physics to GEM's design was believed to warrant
the considerable extra effort.
In section~5, the bulk of this article, we describe the results obtained for
a variety of representative processes. These were chosen to impose stringent
tests on {\it all} areas of the GEM detector's design as well as for their
intrinsic physics interest. Finally, Section~6 summarizes the
results and conclusions. In publishing this document, it is our hope that
others will see it as describing a useful model of detector simulation. We
believe that that this model can be improved. We are also convinced
of the validity of the simulation philosophy that culminated in
|gemfast|. This gave GEM the capacity to simulate rapidly a very wide
variety of physics and background processes and to use the results of these
simulations to provide input for crucial design improvements.
Without |gemfast|, we would have
had to compromise either the breadth or the plausibility of our studies.

\newsec{SIMULATION TOOLS IN HIGH ENERGY PHYSICS}

	Since the Standard Model is a gauge theory, the coupling of
any new physics to known particles is determined by the
$SU(3) \otimes SU(2) \otimes U(1)$ representations of the new
particles. Furthermore, since the $SU(3)$ coupling becomes small at
large momentum transfer and the $SU(2)$ and $U(1)$ couplings are
intrinsically small, the cross sections for the production of new
particles can be calculated in perturbation theory provided that the
new physics does not involve nonperturbative dynamics. Even if the new
dynamics is nonperturbative, its coupling to ordinary matter remains
perturbative, and useful predictions can still be made, e.g., for
strong $WW$ scattering.%
\ref\Bagger{J.~Bagger, et al., FERMILAB--PUB--93--040-T, June 1993.}
All the Standard Model background cross sections can likewise be
calculated using perturbation theory.

	For testing a detector design, it is not sufficient to calculate
cross sections; it is necessary to generate complete events; event
generators are briefly reviewed in Section~2.1 below. It is also necessary
to simulate the interactions of each of the particles from these events with
the detector. One of the most generally useful programs for doing this is
GEANT,%
\ref\Geant{R.~Brun et al., GEANT Users' Guide, CERN/DD/78/2;
           GEANT2 Users' Guide, CERN/DD/EE/83/1;
           GEANT3 Users' Guide, CERN/DD/EE/84--1, Rev. 1987\semi
           K.~Lassila, CERN/CN/91/13.}
which contains tools for describing the geometry of the detector and
for propagating particles through it taking into account all the
possible interactions. CALOR89,%
\ref\Calor{T.~A.~Gabriel, et al., ORNL/TM--5619, 1977, ORNL/TM--11185 (1983);
also IEEE Trans. Nuc. Sci. {\bf 36}, 14 (1989).}
is less general but contains a better description of low-energy
neutrons and so is better suited to precision calculations of the
response of hadron calorimeters. These simulation tools are reviewed
in Section~2.2 below.

	While GEANT and CALOR89 are very powerful, they are too slow to
simulate the large number of events needed to determine the backgrounds for
new physics signatures in hadronic collisions. Hence it is necessary to
construct fast, parameterized simulations based on GEANT or CALOR89 results
for single particles and on test beam data. A sophisticated parameterized
simulation of the GEM detector, |gemfast|, is the basis of most of the
physics results herein. It is described in Section~4 below.

\subsec{Review of Event Generation}

	HERWIG,%
\ref\herwig{G.~Marchesini, et al., Comput.\ Phys.\ Commun.\ {\bf 67},
465 (1992).}
ISAJET,%
\ref\isajet{F.~E.~Paige and S.~D.~Protopopescu, BNL--38774 (1986).}
and PYTHIA%
\ref\pythia{H.~-U.~Bengtsson and T.~Sjostrand,
Comp.\ Phys.\ Commun.\ {\bf 46}, 43 (1987).}
are all programs widely used in particle physics to generate complete events
for a variety of processes. While these programs differ in many important
details, they share a common general approach, which is outlined here.

	The first step in producing a complete event is to generate an
initial hard scattering process using its perturbative QCD or electroweak
cross section.
Because the QCD effective coupling $\alpha_s(Q^2)$ becomes small at large
momentum transfer $Q^2$, the cross section $\hat\sigma$ for any hard process
involving quarks and gluons can be calculated in perturbation theory. Then
the $pp$ cross section is given by the convolution of $\hat\sigma$ with the
parton distributions $f_i(x,Q^2)$:
\eqn\EQparton{
\sigma = \int dx_1dx_2 \, \hat\sigma\left(\hat s,Q^2\right)
\delta\left(x_1x_2 -{\hat s \over s}\right) f_i\left(x_1,Q^2\right)
f_j\left(x_2,Q^2\right)\,.}
Beyond the leading order in $\alpha_s$, $\hat\sigma$ in general contains
infrared and collinear singularities. For inclusive cross sections, the
infrared singularities cancel between real and virtual graphs, and the
collinear singularities can be absorbed consistently into the parton
distributions $f_i(x,Q^2)$, producing their $Q^2$ dependence. To lowest
order, however, $\hat\sigma$ is free of any singularities, so Eq.~\EQparton\
can be used to generate an exclusive process using standard Monte Carlo
techniques. That is, the program selects the four-vectors for the initial
hard scattering.

	While Eq.~\EQparton\ is a good approximation for sufficiently
inclusive cross sections, it is a poor approximation for the structure of
events. The hard scattering process represented by $\hat\sigma$ involves
external quarks and gluons at least in the initial state and generally in
the final state as well. These are massless and carry color charge, so they
necessarily radiate gluons.  Since the coupling in QCD is dimensionless,
this radiation occurs on all $Q^2$ scales from that of the initial hard
scattering down to the confinement scale, ${\cal O}(1\,\GeV^2)$. In
particular, a massless quark or gluon can radiate a collinear massless gluon
while conserving energy and momentum, leading to a so-called mass or
collinear singularity. Because of these singularities, the higher order
corrections are of order $\alpha_s(Q^2)\log Q^2$, and they cannot be
neglected.

	Fortunately, the collinear singularities take a simple form. Each
additional radiation in which an external parton $i$ produces $j$ and $k$
introduces into the cross section a factor
\eqn\EQbranch{
{\alpha_s \over 2\pi p_i^2} P_{i \to jk} (z)
}
where $z = p_j/p_i$ is the momentum fraction carried by $j$ and $P_{i \to
jk}$ is one of the well-known Altarelli-Parisi functions.%
\ref\APfcns{G.~Altarelli and G.~Parisi, Nucl.\ Phys.\ {\bf B126}, 298
(1977).}
The dominant contribution comes from $p_{j,k}^2 \ll p_i^2 \ll Q^2$, so each
radiation is approximately independent. Because of this and because the
cross sections and not the amplitudes factorize, the quantum mechanical
radiation in the collinear limit can be treated as a classical Markov
branching process. The radiation of soft gluons also factorizes, just as it
does for QED. This makes it possible to write a Monte Carlo algorithm, known
as the branching approximation,\cite{\herwig,\isajet,\pythia} that correctly
generates all the $\alpha_s(Q^2) \log Q^2$ corrections to all orders in QCD
perturbation theory. Each branching leads to partons with smaller $p_i^2$.
Since the Monte Carlo algorithm uses exact, non-collinear kinematics, it also
generates a fair approximation to non-collinear radiation, i.e. to multijet
final states.

	For sufficiently small $p_i^2$ perturbation theory becomes invalid,
and one must resort to a model to convert partons into hadrons. Both ISAJET
and PYTHIA use fragmentation models based on low-energy data for this.
HERWIG carries perturbation theory down to $p_i^2 = {\cal O}(1\,\GeV^2)$ and
uses mainly two-body phase space for the final hadronization. It is also
necessary to model the hadronization of the spectator partons, which carry
most of the total energy; there is no reliable theory for this. Fortunately,
while the hadronization model is important for the details, perturbation
theory controls the main features.

	An alternative to the branching approximation is to use explicit
higher-order matrix elements. This approach produces a better approximation
to cross sections for events with several distinct jets, but it is difficult
to combine with parton showers to produce realistic events without double
counting. VECBOS%
\ref\vecbos{By F.~A.~Berends, H.~Kuijf, B.~Tausk, and W.~T.~Giele, Nucl.\
Phys.\ {\bf B357}, 32 (1991).}
contains the exact tree-level cross sections for $W + n\,{\rm jets}$ with
$n \le 4$, and has played an important role in searches for the top quark.
PAPAGENO%
\ref\papageno{I.~Hinchliffe, LBL--34372, (1993).}
contains a wide range of processes and has been used in some of the analyses
presented here.

\subsec{Review of GEANT and CALOR89}

      GEANT is a system of detector-description and simulation tools written
largely at CERN.  The first version, written in 1974, was a bare
framework which emphasized tracking of a few particles per event through a
relatively simple detector.  GEANT has been continuously developed
since then to handle much more complex detector systems, many hundreds of
particles per interaction, and greater detail in the physics simulated.
Because of its flexibility, good support on many hardware platforms, and years
of verification against data, GEANT has become a {\it de facto}
standard within high
energy physics and is used very extensively.

GEANT performs the following functions:

\item{$\bullet$}{\bf Define particle and material parameters.}
GEANT has internal databases of particle masses and lifetimes, properties
of common materials,
and interaction cross sections.  The user is able to define additional
materials, and create mixtures or compounds.

\item{$\bullet$}{\bf Geometry definition.}
The user must create a geometrical model of the detector.  GEANT will then
handle the tracking of particles through this model and their interactions
within it.  GEANT~3.15 provides a palette of 14 standard shapes for the user
which can be sized, positioned and rotated as desired.  The detector model
is a hierarchical tree in which each volume is contained within a `mother',
and each `mother' contains one or more `daughters.'

\item{$\bullet$}{\bf Compute physics processes.}
GEANT accepts lists of particles and vertices produced by event generators.
It contains routines to handle the decays of unstable particles, and the
interactions of particles with the detector.  The physics package computes
the probability of occurrence of a process by sampling the total cross
section and generates the final state after interaction.  It also handles
continuous or quasi-continuous processes such as energy loss. The processes
currently implemented for electrons and photons are:  multiple Coulomb
scattering, ionization and delta ray production, bremsstrahlung, positron
annihilation, $e^+e^-$ pair production, Compton scattering, photoelectric
effect, photofission of heavy elements, and Rayleigh scattering.  For muons,
decays in flight, multiple scattering, ionization and delta ray production,
bremsstrahlung, direct pair production and nuclear interactions are
implemented.  For hadrons, decays in flight, multiple scattering, ionization
and delta ray production, and hadronic interactions are included.  The
hadronic interactions are handled by parameterizations of experimental data
incorporated in the packages GHEISHA,%
\ref\Gheisha{H.~Fesefeldt, PITHA--86/05 (1986)}
HADRIN/NUCRIN%
\ref\Hadrin{K.~Hanssgen, et al., KMU--HEP--80--07 (Leipzig U., 1980);
K.~Hanssgen and J.~Ranft, Comp. Phys. Commun. {\bf 39}, 37 (1986).}
and FLUKA.
\ref\Fluka{P.~Aarnio, et al., FLUKA Users' Guide, CERN--TIS--RP--168 (1986),
     CERN--TIS--RP--190 (1987).}

\item{$\bullet$}{\bf Track particles through the detector.}
Particles from the primary interaction, and those generated within the
detector, are tracked through the detector model.
Tracking proceeds in steps; during each step, the particle may
decay or interact; otherwise, it loses energy through ionization and
continues to the next step.    Particles whose kinetic energy falls below a
cutoff value are stopped and their remaining energy dumped.

\item{$\bullet$}{\bf Simulation of detector response.}
The detector response is modeled in two stages.
Firstly the particles are tracked through the sensitive
detector elements; then,
after the whole event has been tracked, digitization routines
(provided by the user) are called to model the
response of the detector to the particles.  This two-step process enables
nonlinear effects (such as detectors able to respond only to the passage
of one particle) to be properly modeled.

\item{$\bullet$}{\bf Drawing.}
GEANT includes a drawing package to produce graphical output showing the
detector model and the trajectories of particles through it.  Tools are
provided for the user to
add markers for the locations of energy loss,
hit detector elements, etc.
The drawing package can also produce graphical representations of the
hierarchical tree of volumes or of the internal data structures of the
program.

Although GEANT is indispensable for general detector simulation,
we have also made use of the CALOR89%
\cite{\Calor}
package for calorimeter simulations.  We
have used CALOR89 both directly, and through an interface with GEANT.%
\ref\GeantCalor{C.~Zeitnitz and T.~A.~Gabriel, `The GEANT-CALOR Interface,'
in Proc. of 3rd Int. Conf. on
Calorimetry for High Energy Physics, Corpus Christi, 1992.}
CALOR89 provides a more detailed low-energy treatment of hadronic processes,
especially neutron transport, which is of interest in the prediction of
hadronic calorimeter resolutions.  It uses HETC with FLUKA%
\cite{\Fluka}
for general hadronic transport, MORSE%
\ref\Morse{S.~N.~Cramer, Applications Guide to the MORSE Monte Carlo code,
      ORNL/TM--9355, 1985.}
for neutrons with energy below 20 MeV, and EGS%
\ref\Egs{W.~R. Nelson, et al., The EGS4 Code System, SLAC--265 (1985).}
for electromagnetic particles.  CALOR89 contains its own geometry, tracking
and physics packages similar in concept to those of GEANT.

\newsec{THE GEM DETECTOR}

The GEM collaboration was formed in 1991 to develop a major detector for the
SSC. The physics objectives of GEM reflected the motivation for the SSC
itself: to study high transverse momentum physics --- exemplified by the
search for Higgs bosons --- and to search for new physics beyond the
Standard Model.  The GEM detector design was presented in a Technical Design
Report (TDR),\cite{\gemtdr}
and the simulations described in this paper are based on that design (See
Figure~\Oview).  The detector was designed as a general purpose device
emphasizing clean identification and high resolution measurement of gammas,
electrons, and muons, hence the name, GEM. High priority also was given in
the design to accurate measurement of jets and
missing transverse energy. The architecture and the detector
technologies chosen were intended to provide good performance even at the
highest luminosities that were contemplated for the SSC.  The design of the
GEM detector was based on the following principles:

\item{$\bullet$} Very precise electromagnetic calorimetry without a magnet
coil in front of it. Together with the redundant pointing capabilities of
the EM calorimeter and central tracker, this provides the best measurements
of photon and electron energies and allows the reconstruction of the masses
of narrow states with very high resolution.

\item{$\bullet$} A precise muon spectrometer in a large superconducting
solenoidal magnet, allowing measurement of the momenta of high energy muons
over a large solid angle with a minimum of multiple scattering. The muon
system is situated in a quiet environment, shielded by the thick calorimeter.
Muon measurement does not rely on central tracking, thereby providing a
robust stand-alone system for operation at the highest luminosities.

\item{$\bullet$} Hermetic hadronic calorimetry for the measurement of
jets and the reconstruction of missing energy.

\item{$\bullet$} Central tracking in a magnetic field with sufficiently low
occupancy to operate reliably at the highest luminosities anticipated at the
SSC ($\uhl$). The central tracker radius is 0.9~m, allowing for a compact
calorimeter and a large muon tracking volume. It was assumed in simulations
that the inner silicon tracking layers were removed at $\CL = \uhl$.

\nfig\Oview{\SkipFig{6.5in}{gemiso.eps}}{Perspective view of the GEM
detector.}

The performance specifications of the detector are given in Table~1.  The
reach of the GEM design for physics signals that were of
of major interest to the SSC is summarized in Figure~\Reach.

\topinsert
\GEMcaption{Table\ 1}{Top-level specifications for the GEM detector.}
\bigskip
\GEMtable
#\hfil\quad & #\hfil\cr
\GEMrulerule
Magnet\cr
\quad Central field 			& 0.8 T\cr
\quad Inner diameter			& 18 m\cr
\quad Length				& 31 m\cr
\GEMrule
Muon system \cr
\quad Coverage				& $0.1 < |\eta| < 2.5$\cr
\quad $\Delta p_T/p_T$ at $|\eta| = 0$, $p_T = 500\,\GeV$ & 5\%\cr
\quad $\Delta p_T/p_T$ at $|\eta| = 2.5$, $p_T = 500\,\GeV$ & 12\%\cr
\quad Charge separation ($\eta = 0$)	& $p \le 6.5\,\TeV$ at 95\% C.L.\cr
\GEMrule
Electromagnetic calorimeter\cr
\quad Coverage				& $|\eta| < 3$\cr
\quad Energy resolution 		& $6\hbox{--}8\% /\sqrt{E}
					\oplus 0.4\%$\cr
\quad Position resolution		& $4.4\,{\rm mm} / \sqrt{E}$\cr
\quad Pointing resolution		& $40\hbox{--}50\,{\rm mrad} /
					\sqrt{E} + 0.5\,{\rm mrad}$\cr
\GEMrule
Hadronic calorimeter\cr
\quad Coverage				& $|\eta| < 5.5$\cr
\quad Jet resolution 			& $60\% / \sqrt{E} \oplus 4\%$\cr
\GEMrule
Tracker\cr
\quad Coverage				& $|\eta| < 2.5$\cr
\quad Charge separation at 95\% C.L. ($\eta = 0$) & $p \le 600\,\GeV$\cr
\quad Momentum resolution\cr
\quad\quad at high momenta (measurement limited) &
					$\Delta p/p^2 = 1.2
					\times 10^{-3}\,\GeV^{-1}$\cr
\quad\quad at low momenta (multiple scattering limited) &
					$\Delta p/p = 3.5\%$\cr
\GEMrulerule
\endGEMtable
\endinsert

\nfig\Reach{\varfig{6.5in}{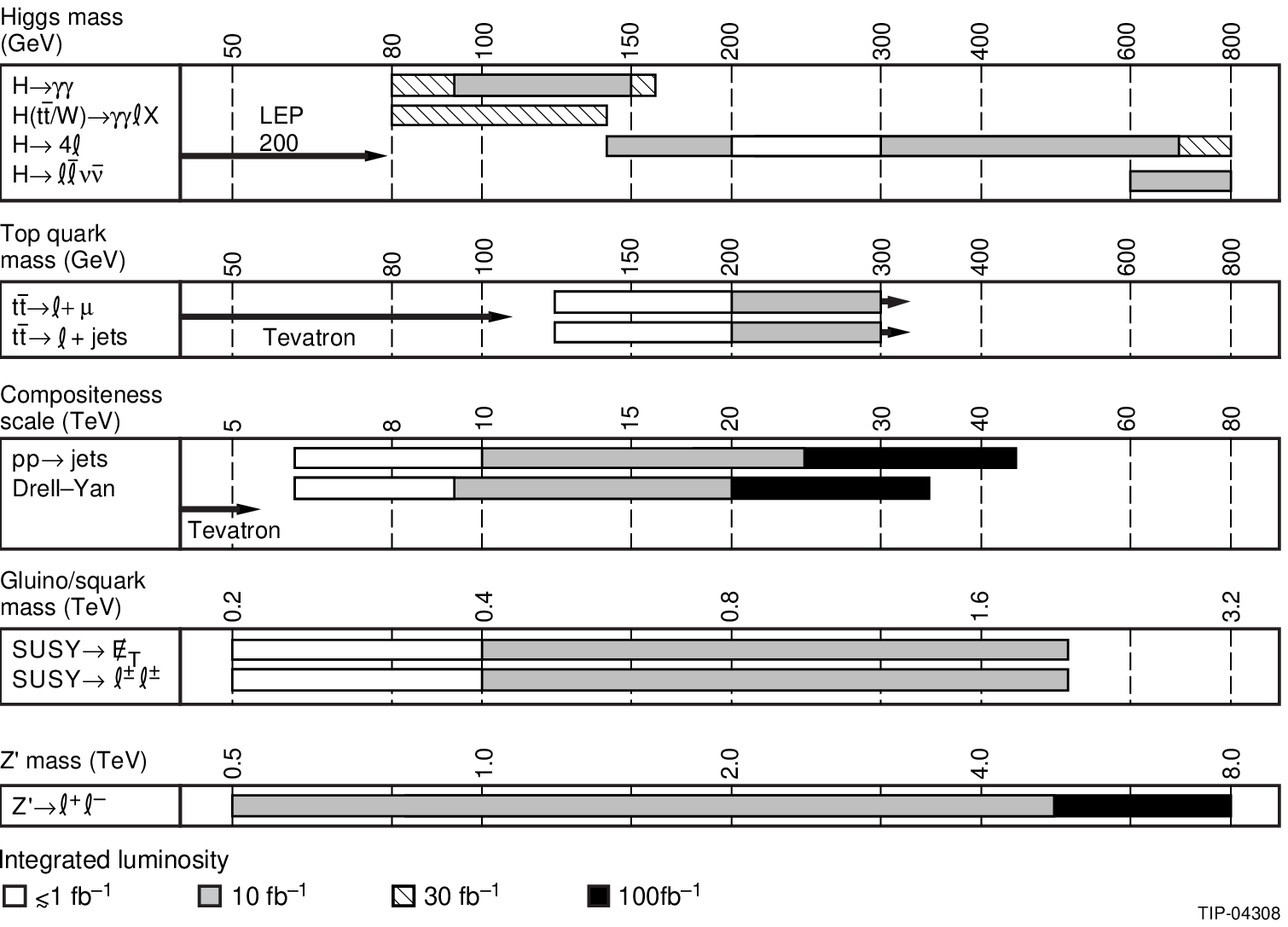}}{The physics reach of GEM.}

\subsec{Detector Design}

The detailed optimization of the design was determined by the physics
requirements, the practical ability to meet the necessary performance
specifications, and cost/schedule constraints.  Attention was also paid to
detector integration issues such as radiation shielding and the interaction
between the beamline and the inner radius of the detector.  It was also
intended that GEM should be complementary to the SDC detector, which used a
large tracker; GEM emphasized calorimetric measurement of gammas and
electrons, a precise and robust muon system, and unique capabilities at high
luminosity.

Detailed descriptions of the detector subsystems, including technical
features, implementation and integration issues, and studies of expected
performance are given in the Technical Design Report.\cite{\gemtdr} We
describe briefly some of the key features of each major subsystem below.

\subsubsec{Magnet}

A very large superconducting solenoid surrounded the detector elements, with
a nominal field of 0.8~T, an inner diameter of 18~m, and a length of 31~m.
The magnet provided a nearly uniform axial field in the region of the
central tracker, allowing measurements of the momenta and signs of charged
particles and helping with electron identification.  In the volume of
detector outside the calorimeters, the magnet provided a 0.8~T field for muon
momentum measurements.  The coil was in two halves, one on each side of the
central detector support (which carried the calorimeter and central
tracker).  In the forward region, conical iron poles were employed to shape
the field. These introduced a radial component to the forward field,
improving the resolution of the muon system in the forward direction.

\subsubsec{Muon System}

Muons were identified by their penetration through the calorimeter.  Muon
momentum was measured using the sagitta determined by three superlayers of
chambers between the calorimeter and the magnet; see Fig.~\Muon. Cathode
strip chambers (CSC's) were chosen, based on the need to obtain the required
spatial resolution, determination of the $z$ coordinate, triggering, and
occupancy, and the desire to use a single technology in both the endcaps and
barrel.  The resolution in the sagitta measurement varies as $BL^2$, where
$B$ is the magnetic field strength and $L$ is the lever arm of the
measurement.  Through the use of a large lever arm and a moderate field, a
precise momentum measurement was obtained:  $\Delta p_T/p_T = 5\%\ (12\%)$
at $\eta = 0\ (2.5)$ for $p_T = 500\,\GeV$.

\nfig\Muon{\SkipFig{6.6in}{gemmuon.eps}}{Quarter view of the GEM detector
showing the muon system, including shielding.}

The momentum resolution was limited at high momenta by the spatial
measurement errors (both inherent and due to misalignment), and at low
momenta by the multiple scattering in the middle superlayer of chambers and
energy-loss fluctuations in the calorimeter.  To be able to sign-select
muons at very high momentum (e.g., from $Z'$ decay at the highest mass,
$\sim 8$--$10\,\TeV$, that was accessible at the SSC) demanded single layer
resolutions of 75~$\mu$m and alignment between superlayers of 25~$\mu$m.  To
avoid degrading the low-momentum muons which contribute to the signal for
$H^0 \to ZZ^* \to 4\mu$, the middle muon chamber superlayer must be less
than 10\% of a radiation length in thickness.

In order to be sufficiently robust to operate at the highest luminosities
(${\cal L} = \uhl$) attainable at the SSC, the muon system was shielded by a
thick ($\ge 11\lambda$ at $\eta = 0$, increasing in the forward direction),
nearly hermetic, calorimeter. The calorimeter thickness was chosen to keep
the rate from punchthrough hadrons significantly below that from in-flight
decay muons.  In addition, a geometry for the forward calorimeter was chosen
that minimizes the background.

A feature of the muon system was the use of a 0.2~m open space between the
calorimeter and the first muon superlayer.  This clear space allowed charged
particles from electromagnetic showers initiated by high-momentum muons to
be bent out of the muon path, leading to increased reconstruction
efficiency.

\subsubsec{Calorimeter}

The general layout of the GEM calorimeter is shown in Fig. \Calorfig.  In
order to identify and measure electrons and photons, a precision
noble-liquid electromagnetic calorimeter was used, employing krypton in the
barrel and argon in the endcap.  This technology was chosen because of its
ability to achieve the required resolution, longitudinal segmentation and
pointing ability, its intrinsic radiation resistance, its ease of
calibration, and the extensive experience that has been acquired with large
liquid-argon systems. An accordion geometry was used to provide good
hermeticity and allow for faster readout than parallel-plate calorimetry
because of lower inductance and capacitance.

Behind the EM calorimeter was a hermetic hadronic calorimeter used both to
measure jet energies and, in conjunction with the forward calorimeter, to
measure $\etmiss$.  The barrel hadron calorimeter
was a hybrid system that performed the energy measurement primarily in a
noble-liquid calorimeter (in the first $\sim 6\lambda$), followed by a
relatively inexpensive copper/scintillator calorimeter which provided energy
measurement for late-developing showers and the necessary shielding for the
muon system.  In the end cap an all-liquid design was used.

\nfig\Calorfig{\SkipFig{6.5in}{gemcal.eps}}{Quarter view of the GEM
detector showing the calorimeters.}

In the forward region, additional calorimeters measured high-momentum
particles near the beam pipe. Together with the barrel and endcap
calorimeters, these forward calorimeters determined the sensitivity to
$\etmiss$. In order to detect new particles such as massive gluinos and
squarks, the forward calorimeter must cover the region $|\eta| \simle 5.5$,
be sufficiently dense to contain hadronic showers, be sufficiently fast to
cope with the high particle flux in this region, and be radiation hard. The
design had a first section consisting of a liquid-argon tube calorimeter,
followed by a liquid-scintillator-capillary and tungsten calorimeter. The
calorimeter was optimized to obtain good spatial information in the first
section and adequate shower containment in the second section.

\subsubsec{Central Tracker}

The central tracker was 1.8~m in diameter by 3.5~m long, surrounding the
interaction point. The tracker size was determined by the needs to obtain
sufficient $\pi^0$ rejection by shower shape analysis, to minimize the
calorimeter cost, to optimize the tracker resolution, and to preserve
sign-selection ability to high momenta. The layout of the tracker system is
shown in Fig. \Tracker.

\nfig\Tracker{\SkipFig{6.5in}{gemtrack.eps}}{The GEM central tracker.}

The tracker supported the goal of measuring gammas, electrons, and muons at
high $p_T$, at high luminosity, up to ${\cal L} = \uhl$.
The tracker provided good
separation of gammas and electrons by finding a charged track, and measured
the electron sign up to 600~GeV. The former is essential to the search for
$H^0 \to \gamma\gamma$ and to background rejection in $Z' \to e^+e^-$; the
latter, for the gluino search using the signature of same-sign leptons.
Another important role for the tracker was to measure the position of the
primary vertex.  Physics involving $b$, $t$, and $\tau$ decays requires full
pattern recognition capability, including secondary vertex finding and
tracking at low momenta.

The design used two technologies. For the inner section of the tracker,
silicon microstrips were used, giving the required fine segmentation and
radiation resistance. For the outer section, interpolating pad chambers
(IPC's) were chosen for their low occupancy, their ability to provide near
three-dimensional space points, their high-luminosity capability, and their
demonstrated operational resolution of $50\,\mu {\rm m}$.

\subsubsec{Electronics/Data Acquisition}

Triggering and data acquisition in GEM followed a three-level strategy to
provide a system without deadtime that provided as much information as
possible at each trigger level. It was designed for luminosities up to
$\hl$, with provision for operation at higher luminosities with modest
upgrades.  The architecture consisted of a synchronous and pipelined Level
1, an asynchronous Level 2 (possibly with special purpose hardware), and a
Level 3 processor ranch. Full granularity data was available at Levels 2 and
3. Level 1 was designed to handle up to 60 MHz input rate, with an output
rate of 10 kHz. Level 2 was designed to handle an average input rate up to
100 kHz, with an output rate of 300 Hz.  Finally, Level 3 accepted 3 kHz,
with an output rate of 100 Hz.

\newsec{MODELING THE GEM DETECTOR}

In principle, the full detector simulation programs GEANT and CALOR89
could be used to compute GEM's response to the signal and background for
any process. However, these programs are very slow for complex events at
high energy. It is impractical to use them to simulate the more than $10^6$
events often needed to determine a rare signal's background arising from a
combination of relatively likely processes. Consequently, two kinds
of simulation were done for GEM.
Detailed
simulations, based on GEANT, of each of the individual detector
systems have been
performed for single particles or for limited numbers of complete events.%
\ref\Sigem{Yu.~Fisyak, K.~McFarlane and L.~Roberts
{\it SIGEM --- Full GEANT Simulation for the
GEM Detector}, GEM--TN--92--162 (1992).}
The results of these detailed studies
have been parameterized and incorporated in |gemfast|,
the fast simulation program for GEM that was used for determining the
performance of the detector for physics processes. For example,
parameterizations were made of energy and momentum resolutions and of muon
reconstruction efficiency in the presence of other particles in the event
of interest.
Most of the results presented in this article are based on |gemfast|. Where
necessary, hybrid simulations of |gemfast| and full GEANT have been used. For
example, in the study of $\hgg$, detailed electromagnetic
shower shape studies for real photons and and jets faking photons were carried
out with GEANT. Apart from examples such as this, |gemfast| describes the
performance of GEM quite accurately.

	The key to a fast detector simulation is to use a very simple
geometry and to parameterize the response of each detector component
in a simple way. The geometry used in |gemfast| is a set of concentric
cylinders, one each for the central tracker (CT), electromagnetic
calorimeter (EC), hadronic calorimeter (HC), scintillator calorimeter
(SC), forward calorimeter (FC), and muon system (MU). The geometry is
shown in Fig.\ \FastGeom.  The radiation and absorption lengths in the
cylinders representing the calorimeters are varied with $\eta$ so as to
match the true design. Particles are tracked through each successive
volume on straight lines for neutral particles or on helixes in a
uniform magnetic field, $B = 0.8\,{\rm T}$, for charged ones. Since
the MU resolution is parameterized, particles need not be tracked
through the nonuniform field in the muon system.

	Once a particle enters a given detector system, its energy
resolution, angular resolution, and detection efficiency are
calculated using parameterizations of the single particle response.
This simple, single-particle approach is not adequate for the central
tracker reconstruction efficiency, which is sensitive to the presence
of other tracks in the same event and to pileup, and which therefore
has been investigated separately. Development of electromagnetic and
hadronic showers in transverse and longitudinal directions, including
fluctuations, is modeled and the individual tower energies are
calculated. Unstable particles are allowed to decay anywhere in the
detector using code adapted from the GEANT package. The Level\ 1
trigger response is also simulated.

\nfig\FastGeom{\dofig{50.0mm}{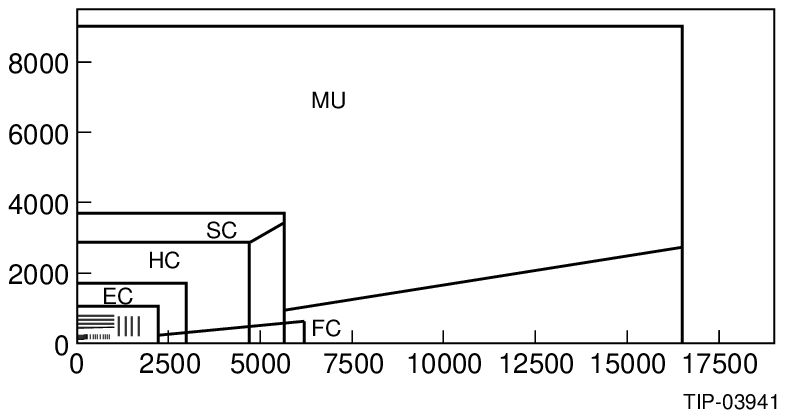}\forceright}
{Geometry used in
|gemfast|. The detector is approximated by a set of concentric
cylinders. Particles are tracked through the central tracker and
calorimeters assuming a uniform magnetic field.}

\subsec{Event Generation}

	The first step in simulating a process is to generate the events of
interest. This is handled by |gemgen|,%
\ref\isheer{I.~Sheer, {\it GEMGEN
-- A Generic Monte Carlo Generator Interface
Package}, GEM TN--93--379 (March 1993).}
which incorporates ISAJET,\cite{\isajet} PYTHIA,\cite{\pythia} and a
single particle gun. The single particle gun generates a single
particle at a given $p_T$, $\eta$, and $\phi$ or a single quark or
gluon fragmented with PYTHIA.

	The event generators are interfaced to the detector simulation
in a way that allows adding a signal event and a Poisson-distributed
number of minimum bias events in the same bunch crossing. The same or
different generators can be used for each sample. The vertex position
of each event is generated according to the expected width, $\sigma_z
= 5\,{\rm cm}$. The same approach could be used to describe pileup
from out-of-time bunches. However, this is impractical because of the
computing time required to generate minimum bias events over the tails
on the sensitive time of the detector.  Instead,
out-of-time pileup events have been generated separately, the detector
response to them has been parameterized, and the resulting noise
or inefficiency taken into account for each of the
detector systems.

\subsec{Central Tracker}

\nfig\MuReslow{\dofig{80.0mm}{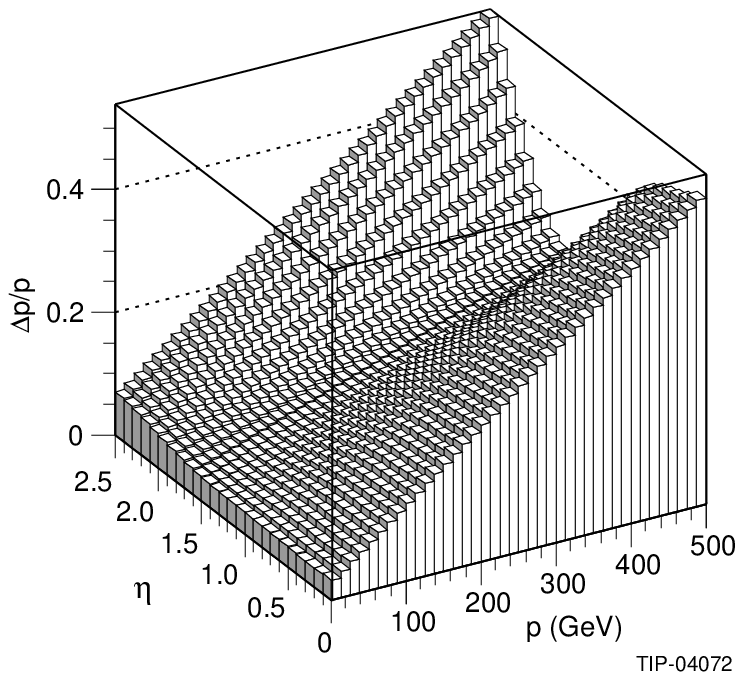}}
{Parameterization of
vertex-constrained central tracker resolution in |gemfast| for muons
or charged hadrons vs.\ $p_T$ and $\eta$ for ${\cal L} =
10^{33}\,\cmsec$. Both the silicon tracker and the interpolating pad
chambers are used.}

\nfig\MuReshi{\dofig{84.0mm}{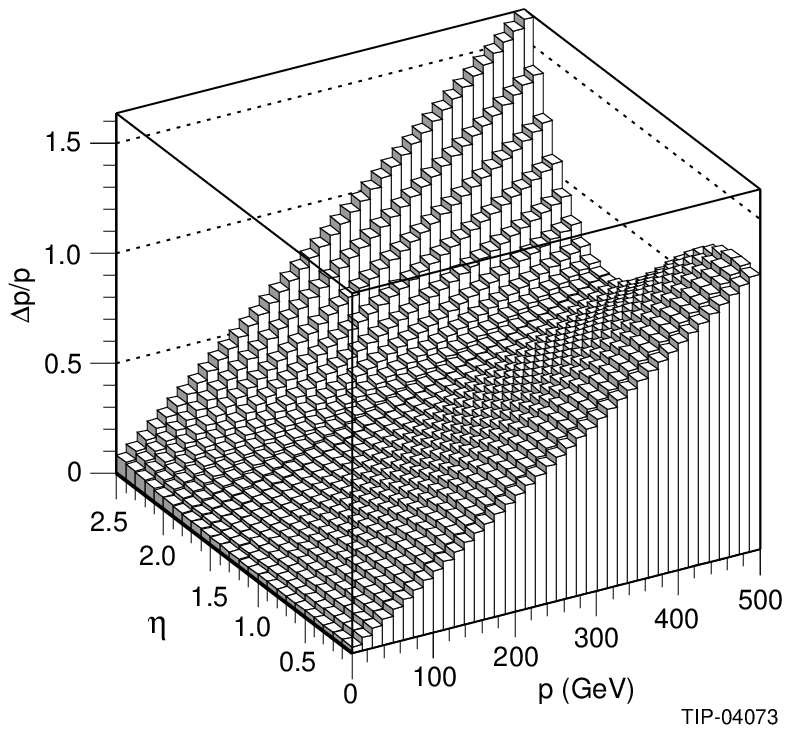}
\forceright}{Parameterization of
vertex-constrained central tracker resolution in |gemfast| for muons
or charged hadrons vs.\ $p_T$ and $\eta$ for ${\cal L} =
10^{34}\,\cmsec$. The silicon tracker is assumed to be removed.}

	The single particle momentum and vertex position resolutions
of the central tracker were calculated using a
full GEANT simulation including
the effects of the magnetic field, detector geometry, chamber positions and
resolutions, material distribution including supports and cables,
geometrical acceptance, detector efficiency, and distribution
of the interaction vertex.
The effect of
out-of-time pileup was included as an additional inefficiency due to
the detector deadtime. Both the silicon strips and the IPCs are
used at a luminosity of $10^{33}\,\cmsec$; the silicon detector
was assumed to be removed for $10^{34}\,\cmsec$. Figures\ \MuReslow\
and \MuReshi\ show the resulting parameterization of the mean muon or
pion resolution vs.\ $p_T$ and $\eta$ used in |gemfast| for
$10^{33}\,\cmsec$ and $10^{34}\,\cmsec$, respectively, assuming a
vertex constraint. The impact parameter resolution for tracks is also
parameterized, allowing modeling of vertex reconstruction on an
event-by-event basis.

	The reconstruction efficiency for isolated tracks exceeds 97\%
over the whole range of pseudorapidity $\vert\eta\vert < 2.5$.  This
is based on requiring at least ten good hits on a high-$p_T$ track. The
number of hits is calculated in |gemfast| using the positions of the
chambers and the actual origin of the track. It is difficult to
parameterize the reconstruction efficiency for non-isolated tracks, so
the simulations described here use only tracks which have $p_T >
1\,\GeV$ and which are isolated at the generator level in a
$\Delta\eta \times \Delta\phi$ region corresponding to three pads in
the IPCs.

\nfig\ERes{\dofig{74.0mm}{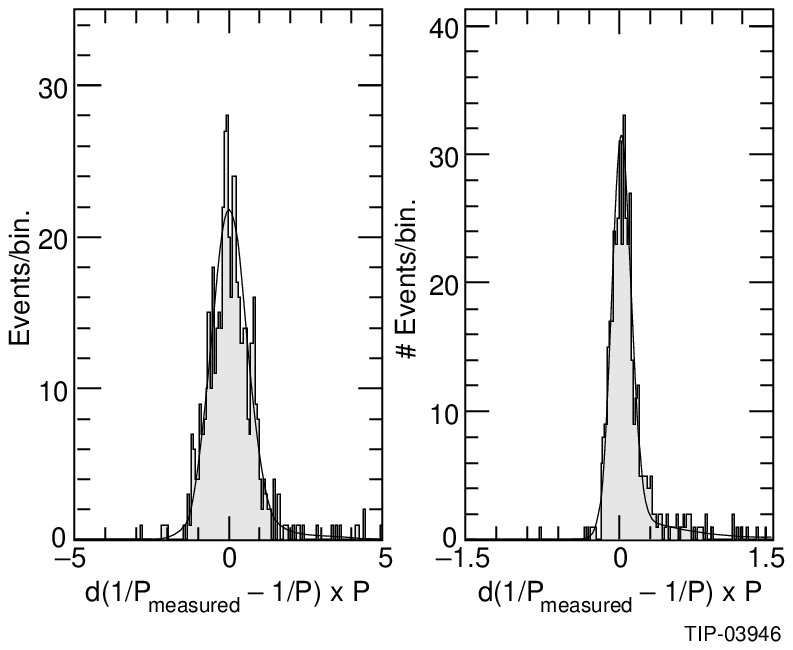}\forceright}
{Calculated
momentum resolution
function for electrons in the central tracker at $p_T = 525\,\GeV$ for
$0 < \vert\eta\vert < 0.2$ (left plot), and $p_T = 10\,\GeV$ for
$2.2 \le \eta \le 2.4$ (right plot). The histogram shows the GEANT
simulation, and the smooth curve is the parameterization used in
|gemfast|.}

	The momentum resolution for electrons has been treated
separately, taking into account the emission of bremsstrahlung photons
caused by the material in the tracker. (The photons are assumed to be
emitted nearly parallel to the electron and to hit the same calorimeter
cell, so that the calorimeter energy resolution is not degraded.)
Figure\ \ERes\ shows the electron momentum resolutions at a particular
value of $p_T$ and $\eta$ from the full GEANT simulation and from the
corresponding parameterization used in |gemfast|. The parameterization
fits the GEANT data well, including the bremsstrahlung tails.
These are important for electron sign determination, which can be done
up to $p_T \simle 600\,\GeV$.

       Material in the central tracker is parameterized as a function
of $\eta$ and $\phi$ after each layer of the silicon or IPCs, and this
is used to convert photons at appropriate space points. Secondary
$e^+e^-$ pairs are generated using code adapted from GEANT.

\subsec{Calorimeter}

\nfig\Gflashem{\dofig{71.5mm}{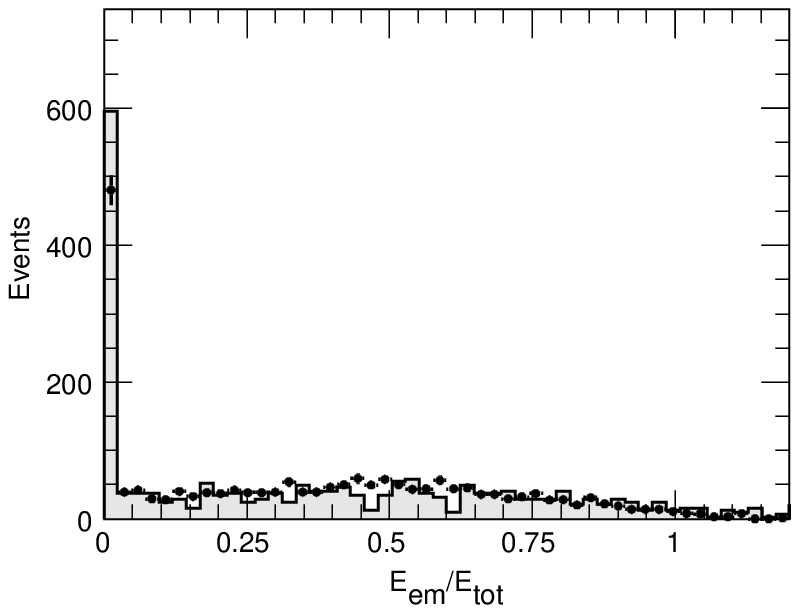}\forceleft}
{Ratio of
electromagnetic to
total energy for $100\,\gev$ pions. The histogram shows the result of a
full GEANT simulation, and the points show the GFLASH-based
parameterization used in |gemfast|.}

	The parameterized response of the central calorimeter, which
covers $\vert\eta\vert < 3$, includes shower profiles and energy
resolutions. Longitudinal and transverse electromagnetic and hadronic
shower profiles are generated using GFLASH\ 1.3, which was originally
developed to describe the H1 liquid argon calorimeter.%
\ref\Gflash{G.~Grindhammer, et al., Proc.\ of the Workshop on
Calorimetry for the Supercollider, Tuscaloosa, AL, 1989, p.151,
SLAC--PUB--5072, 1989 and Nucl.\ Inst.\ and Meth. {\bf A290}, 469
(1990).}
It has been modified to work with the |gemfast| geometry.  GFLASH
incorporates correlated fluctuations of shower profile parameters,
hadronic shower fluctuations into early $\pi^0$'s, transverse profile
variations with depth, and shower development along the true direction
of incidence as determined by tracking through the central tracker
region. It gives a good description of the shapes of both
electromagnetic and hadronic showers in uniform regions of the
calorimeter, as illustrated in Figure\ \Gflashem. GFLASH is used to
distribute shower energies among calorimeter towers. The
electromagnetic, liquid-hadronic, and scintillator tail-catcher towers
are modeled separately, but further longitudinal segmentation is
neglected for faster execution.

	The segmentation of the calorimeter is varied realistically
with $\eta$. The simulated endcap segmentation is in $\eta$-$\phi$
rather than in $x$-$y$, but this should have no effect on physics
performance. The EC and HC segmentations are $\Delta\eta \times
\Delta\phi = 0.026 \times 0.026$ and $0.08 \times 0.08$ in the middle
of the barrel.  Since the segmentation is approximately constant in
units of radiation or absorption lengths, it is about a factor of six
coarser at the small-angle edge of the endcap than in the middle of
the barrel.

	The energy resolution of the electromagnetic calorimeter has been
calculated using a GEANT simulation including an extremely detailed geometry
and very low cutoffs.  A similar simulation gave
good agreement with test beam data for the non-projective accordion.
The resolution has been parameterized in the form
\eqn\XX{
{\Delta E \over E} = {a(\eta)\over \sqrt E} \oplus b(\eta) \,,
}
with the parameters tuned to give the correct resolution as a function
of $E$ and $\eta$ for the $5 \times 5$ sum of towers used to obtain
optimal resolution for isolated electrons and photons. Typically, the
stochastic term ($a$) is about 6\% in the barrel and about 8\% in the
endcap, and the constant $b$-term is about 0.4\%. Pileup and noise are
not included in the energy resolution but are added separately.

	The pointing resolution of the electromagnetic calorimeter for
photons was calculated using the same detailed GEANT simulation.
It is parameterized as
\eqn\XX{
\Delta \theta = {a_\theta(\eta)\over \sqrt E} + b_\theta(\eta) \,,
}
for low to moderate energies, where $a_\theta$ is about $40\,{\rm
mrad}$ in the barrel and $50\,{\rm mrad}$ in the endcap and
$b_\theta(\eta)$ is about $0.5\,{\rm mrad}$. At the highest energies,
for which the shower is no longer fully contained in the EM
calorimeter, $\Delta\theta \approx 2\,{\rm mrad}$.

\nfig\Pileup{\dofig{70.5mm}{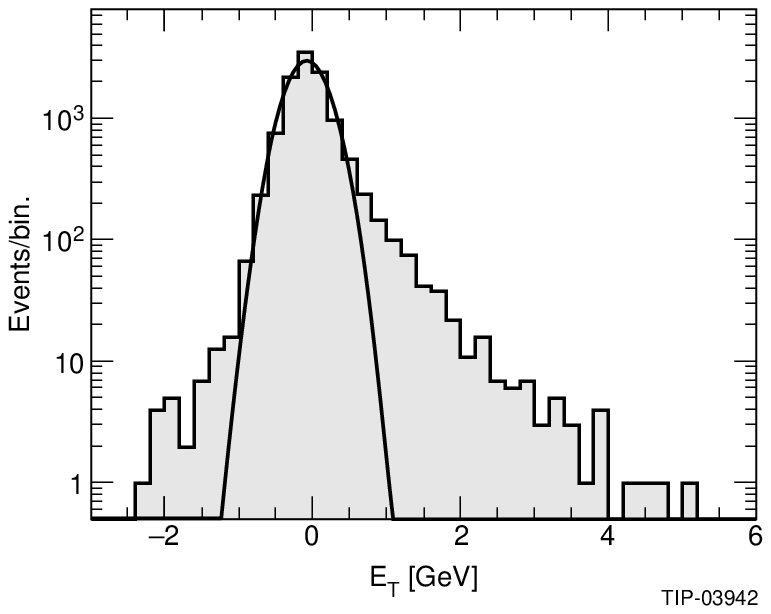}\forceleft}
{The histogram shows
the distribution of pileup, including out-of-time events, in |gemfast|
for a $0.16 \times 0.16$ electromagnetic trigger tower.}

	Thermal noise is added to each tower depending on $\eta$.
In-time pileup events are added explicitly. To determine the effects
of out-of-time pileup, a Poisson-distributed number of minimum bias
events is generated for each of the 50 preceding and 20 following
bunch crossings and simulated with |gemfast|. The calorimeter response
for each bunch crossing is weighted with a response function that
takes into account the intrinsic calorimeter response and the shaping
circuit in the electronics, using shaping times of $40\,{\rm ns}$ for
the electromagnetic calorimeter and $100\,{\rm ns}$ for the hadronic
calorimeter. The sum for each cell is calculated to provide a snapshot
of the response to pileup. This ensures that all the longitudinal and
transverse correlations among cells, caused either by individual shower
shapes or by jets, are preserved. One of these snapshots is then
superimposed on the response from the signal and in-time pileup
events. As can be seen from Fig.\ \Pileup, this approach gives larger
fluctuations and a smaller half-width than the equivalent Gaussian
noise.%
\ref\Cleland{A.~Vanyashin, GEM TN--93--375 (1993).}
For the analyses described here a sample of $10^4$ such snapshots has
typically been used.

	The GEM calorimeter design has an intrinsic $e/h \ne 1$, so
reconstructing a jet energy as a simple sum of the observed energies
in the various layers would give a large constant term in the
resolution. This effect can be reduced by using an iterative weighting
procedure.%
\nref\wgtexpt{K.\ Borras, et al., The H1 Calorimeter Group, {\it Study of
Software Compensation for Single Particles and Jets in the H1
Calorimeter}, Contributed paper to the XXV International Conference
on High Energy Physics, Singapore, 1990.}%
\nref\wgtalgo{Yu.\ Efremenko, et al., {\it Simulation Studies for GEM
Scintillating Barrel Design}, GEM TN--93--349 (1993).}%
\cite{\wgtexpt,\wgtalgo}
To save execution time this weighting is not implemented in |gemfast|.
Instead, the sampling and constant terms in the single hadron
resolution have been tuned so that the jet energy resolution from
|gemfast| matches that from the detailed GEANT simulations including
the weighting.  The resulting resolution for isolated jets is
\eqn\XX{
\left.{\Delta E \over E}\right\vert_{\rm jet} = {0.6 \over \sqrt E}
\oplus 0.04\,.
}
For many cases the effects of the clustering algorithm used to define
jets are comparable to those of energy resolution.

\nfig\PTres{\dofig{69.5mm}{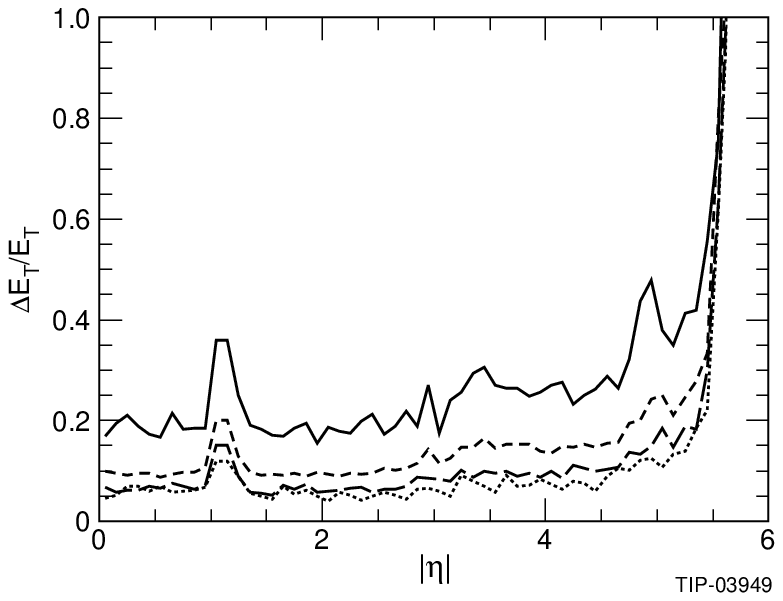}\forceright}
{Calorimetric $\et$
resolutions for hadrons vs.\ $\eta$ for $E=10\,\GeV$ (solid line),
$50\,\GeV$ (dashed line), $200\,\GeV$ (dotted line), and $500\,\GeV$
(dash-dotted line) used in |gemfast| for the calculation of $\etmiss$.
Most of the fluctuations are caused by limited statistics in the GEANT
simulation.}

	The forward calorimeter covers $3 < \vert\eta\vert < 5.5$,
with full measurement capability to $\abseta \approx 5.0$. It has been
used in the physics studies described here only to determine the
missing energy, $\etmiss$. For this, detailed simulation of the
response of individual cells is not needed. Rather, the energy and
direction of each particle hitting the forward calorimeter is smeared
according to a parameterization derived from a GEANT simulation.  The
simulation includes all the effects of dead material and shower
spreading across calorimeter boundaries. The resulting $E_T$
resolution as parameterized in |gemfast| is shown in Fig.\ \PTres.
The statistics in this simulation were not sufficient to
study potential nongaussian tails. These are modeled in |gemfast| by
adding a second Gaussian with a small amplitude and a larger width.
Test beam data for single hadrons from 50 to 100 GeV in the \Dzero\
liquid argon calorimeter show a tail composed of roughly 1\% of the
events with a standard deviation two to three times larger than the
Gaussian calorimeter resolution.%
\ref\Dzerotail{M.\ Shupe, private communication.}
This tail is slightly larger than that seen for $1\,\tev$
jets in GEANT studies of jet resolution using energy-dependent
weighting. A similar tail has been assumed for GEM.

\subsec{Muon System}

\nfig\MUres{\dofig{70.5mm}{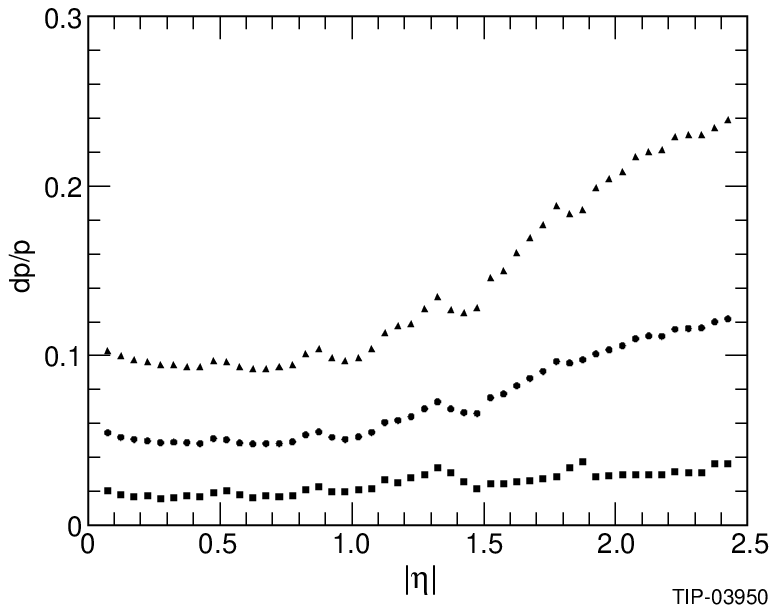}}
{Parameterization of muon
resolution vs.\ $\eta$ for $p_T=100\,\GeV$ (squares), $500\,\GeV$
(circles), and $1\,\TeV$ (triangles) in |gemfast|.}

	The muon momentum resolution has been calculated by a full
GEANT simulation including a detailed model of the detector and its
support structures. The calculation includes chamber resolutions and
alignment errors, the calculated shape of the magnetic field, the
number of CSC planes in the measurement, and multiple scattering from
the chambers and their supports. It is
described at length elsewhere.%
\ref\MuPara{T.~Wenaus, {\it A Detailed Simulation and Performance
Parameterization of the GEM Muon Detector}, GEM TN--93--297 (April 1993).}
The resulting parameterization of the resolution is shown in
Fig.\ \MUres.  The jumps in the curves come from the transitions
between various sets of chambers in the barrel and endcap.
The flattening in the resolution at $\eta \simeq 2.5$ is the effect of the
forward field shapers.

\nfig\MUacc{\dofig{70.5mm}{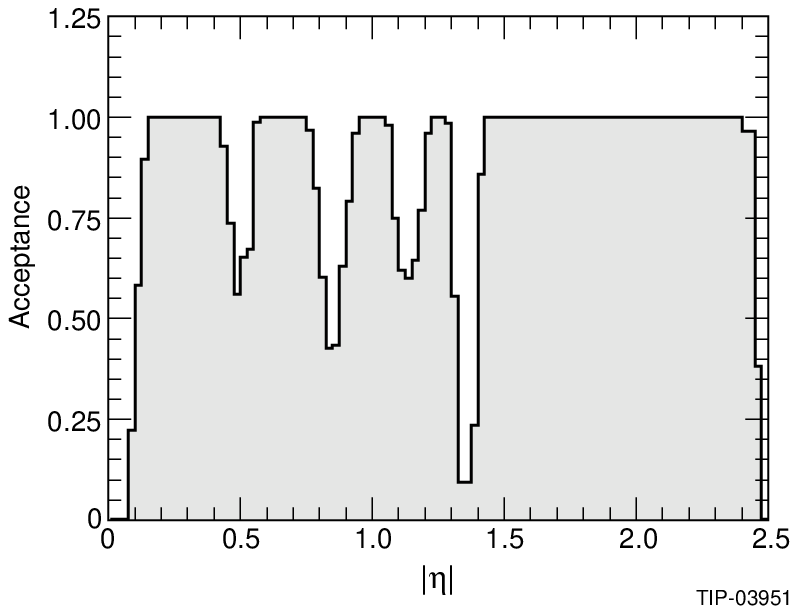}\forceright}
{Parameterization of muon
acceptance vs.\ $\eta$ in |gemfast|. Because of binning of the GEANT
data, the gaps in the barrel are wider than in reality but have
nonzero acceptance. The average acceptance is correct.}

	The geometrical acceptance for muons has been calculated
similarly, requiring that the muon pass through at least three
chambers in each superlayer. (This is overly
conservative,\cite{\MuPara} since fairly good muon resolution can be
obtained from two superlayers at low $p_T$ or from two superlayers
plus the vertex at high $p_T$.) The three-superlayer acceptance is
shown in Fig.\ \MUacc\ and is essentially independent of muon energy.
The coverage of the region $\abseta < 2.46$ is 83\%.  The losses due
to the spoiling of hits by $\delta$ rays and other electromagnetic
interactions have also been simulated by GEANT, parameterized as a
function of muon energy, and included in |gemfast|.

	The muon energy loss in the calorimeter has been calculated,
including all the processes modeled in GEANT, and has been
parameterized with a simple analytic function. The full calculation
and the parameterization are shown in Fig.\ \MUloss. The lost energy is
added to the appropriate cells of the calorimeter so that energy
losses large compared to the noise can be reconstructed in |gemfast|.
The best resolution is obtained by using the measured energy if it is
greater than a factor $k \sim 1.5$ times the most probable value, and
the truncated mean value otherwise.

\nfig\MUloss{\dofig{69.5mm}{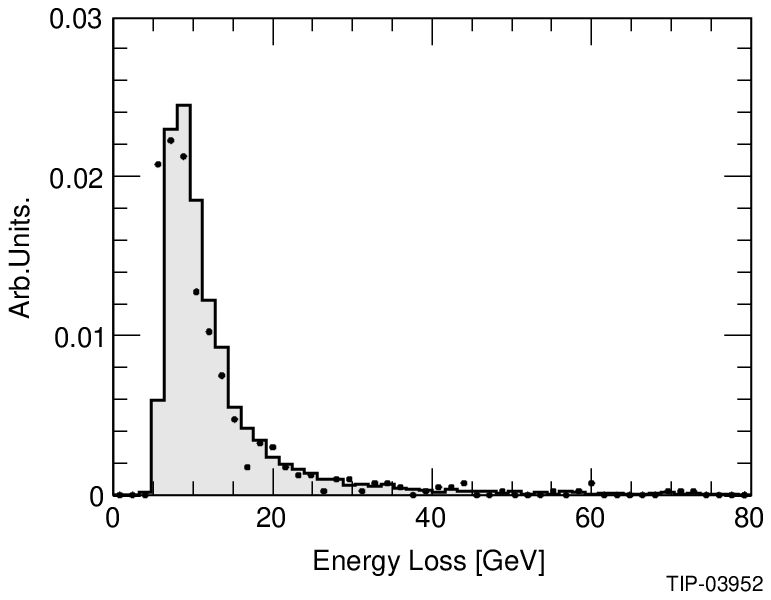}\forceleft}
{Muon energy loss in
calorimeter for $E_\mu = 500\,\GeV$ (solid curve) and {\tt gemfast}
parameterization (dots). Note that the tail at large energy loss is
fit well by the parameterization.}

\subsec{Trigger Simulation}

	The Level\ 1 trigger efficiency is simulated in |gemfast| and
included in the analyses described below. The Level\ 2 and Level\ 3
triggers are assumed to be included, to a first approximation, in the
cuts made in the analysis.

	An initial set of trigger thresholds is shown in Table~2.  The $E_x$
triggers include an isolation cut on the surrounding towers and on the
hadronic energy behind the cluster.  The $J_x$ triggers sum the
electromagnetic and hadronic sections in towers reasonably matched to the
size of a QCD jet. The $M_x$ thresholds are nominal values at which the
acceptance is 84\% and include the coarse resolution of the trigger roads.
The missing energy simulation uses the full sum of the calorimeter.  While
the highest thresholds in Table~2 have acceptable rates by themselves, the
lower thresholds can only be used either with prescaling or in combinations.

	Twelve combinations of the primitives listed in Table~2
appear to be sufficient to select all the physics processes so far
considered for GEM; they give a total Level\ 1 trigger rate less than
the design goal of $100\,{\rm kHz}$.  In particular they provide
triggers with good efficiency for low-mass Higgs bosons and the other
(relatively) low-mass processes listed in Table~3. The most
difficult trigger is that for inclusive $H \to \gamma\gamma$ at the
low end of the mass range. Some low-$p_T$ processes such as jets have
such large cross sections that they must be prescaled. No attempt has
been made to implement an efficient trigger on the inclusive-$b$ cross
section to study $b$-physics, although this might be of interest
during the low-luminosity initial period. Level\ 1 triggers on the
higher mass processes which have been considered are generally easier
than those listed in Table~3.

\widetopinsert
\GEMcaption{Table~2}{Primitives used in the GEM
trigger. The highest threshold can be used alone; the rest can be used
in conjunction with other triggers.}
\bigskip
\GEMtable
#\qquad\hfil &#\qquad\hfil &#\hfil\cr
\GEMrulerule
Name 	&Definition		&Thresholds\cr
\GEMrule
$E_x$ 	&EM cluster with $E_T > x\,\GeV$
		&$E_{8}$, $E_{16}$, $E_{50}$, $E_{80}$\cr
	&in $0.16 \times 0.16$, isolated in both\cr
	&EM and HAD calorimeters.\cr
$M_x$ 	&Muon with $p_T > x\,\GeV$.
		&$M_{10}$, $M_{20}$, $M_{30}$, $M_{40}$\cr
$J_x$ 	&Jet cluster with $E_T > x\,\GeV$
		&$J_{16}$, $J_{50}$, $J_{80}$, $J_{200}$\cr
	&in $0.48 \times 0.48$. \cr
$\Emiss_x$	&Missing energy $\etmiss > x\,\GeV$
		&$\Emiss_{50}$, $\Emiss_{100}$\cr
	&summing whole calorimeter. \cr
\GEMrulerule
\endGEMtable
\endinsert

\widetopinsert
\GEMcaption{Table~3}{Trigger efficiencies for a
variety of processes using the triggers shown in
Table~2 with standard logical notation. The
triggers listed are the principal ones among
those studied, but the efficiency is for the sum of all
combinations. The efficiency for $H \to \ell^+\ell^-\ell^+\ell^-$ at
$140\,\GeV$ could be increased about 5\% by adding an $E_{16} \wedge
M_{10}$ trigger. The last four processes are useful mainly for
calibration.}
\bigskip
\GEMtable
#\hfil &\quad\hfil # &\quad #\hfil &\quad\hfil #\cr
\GEMrulerule
Process & Mass & Trigger & Efficiency\cr
\GEMrule
$H \to \gamma\gamma$
	& $80\,\GeV$
	& $2E_{16} \vee E_{50}$
	& 78.7\%\cr
$H \to \gamma\gamma$
	& $140\,\GeV$
	& $2E_{16} \vee E_{50}$
	& 94.7\%\cr
$t \bar t H \to \gamma\gamma\ell X$
	& $80\,\GeV$
	& $2E_{16}$
	& 94.4\%\cr
$H \to \ell^+\ell^-\ell^+\ell^-$
	& $140\,\GeV$
	& $2M_{10} \vee 2E_{16}$
	& 81.8\%\cr
$H \to \ell^+\ell^-\ell^+\ell^-$
	& $400\,\GeV$
	& $2M_{10} \vee 2E_{16}$
	& 99.8\%\cr
$H \to \ell^+\ell^-jj$
	& $800\,\GeV$
	& $2M_{10} \vee 2E_{16}$
	& 99.9\%\cr
$t \bar t \to \ell\nu b X$
	& $140\,\GeV$
	& $E_{50} \vee M_{30}$
	& 75.3\%\cr
$\gluino\gluino$
	& $500\,\GeV$
	& $3J_{80} \wedge \Emiss_{100}$
	& 99.9\%\cr
$W \to e\nu$
	& ---
	& $E_{16} \wedge \Emiss_{50}$
	& 15.8\%\cr
$W \to \mu\nu$
	& ---
	& $M_{10} \wedge \Emiss_{50}$
	& 48.7\%\cr
$Z \to ee$
	& ---
	& $2E_{16}$
	& 80.3\%\cr
$Z \to \mu\mu$
	& ---
	& $2M_{10}$
	& 86.9\%\cr
\GEMrulerule
\endGEMtable
\endinsert

\newsec{STUDIES OF PHYSICS PROCESSES}

\subsec{Selection of Processes}

\nref\GEMEOI{{\it An Expression of Interest to Construct a Major SSC Detector},
SSC EOI--0020 (July 1, 1991).}%
\nref\GEMLOI{{\it GEM Letter of Intent}, The GEM Collaboration, GEM TN--92--49,
SSCL--SR--1184 (November 30, 1991).}%
\nref\response{{\it GEM Responses to the December 1991 PAC Report}, GEM
TN--92--131, (July 8, 1992).}%

	We have studied the performance of the GEM detector for a variety
of physics processes chosen both because they probe the capabilities of
important components of the detector and because of their intrinsic physics
interest:

\item{$\bullet$} Section~5.2 presents an in-depth study of the search for the
Higgs boson of the Standard Model. The signals, backgrounds and discovery
potential for $M_H = 80 - 800\,\gev$ are discussed. Depending on the Higgs
mass, the modes studied are $\hgg$; $\h (t \ol t / W) \ra \ellpm
\gamma\gamma X$;
$\h \ra ZZ^{(*)} \ra 4$ charged leptons; $\h \ra ZZ \ra \ellp \ellm \ol
\nu \nu$; and $\h \ra ZZ \ra \ellp \ellm \ts \jet \ts \jet$.

\item{$\bullet$} Flavor physics involving top-quarks is discussed in
Section~5.3. We describe the mass measurement of a heavy top-quark in the
standard decay mode $t \ra W^+ b$ using two methods: $t \ra$ isolated
$\ellp$ plus non-isolated $\mu^-$, and $t \ra 3\ts \jets$. We also discuss
the discovery of a charged Higgs boson in the nonstandard decay mode $t \ra
H^+ b$, followed by $H^+ \ra \tau^+ \nu_\tau$.

\item{$\bullet$} Jet physics is studied in Section~5.4. We discuss the
determination of the jet energy scale, using as a physics context the
search for quark substructure in high-$\et$ jets. Other jet studies
are discussed in Sections~5.2 ($\h \ra Z^0 Z^0 \ra \ellp \ellm \jet \ts
\jet$) and 5.3 ($t \ra W^+ b \ra 3 \ts \jets$).

\item{$\bullet$} Supersymmetry is discussed in Section~5.5 as an example
of missing transverse energy signatures for new physics.
The $\etmiss$ distribution is calculated for GEM, including the effects of
transition regions and dead material. The $\etmiss$ signature is studied
for a range of gluino and squark masses. In addition, the like-sign dilepton
signature for gluino production is investigated.

\item{$\bullet$} Section~5.6 is devoted to studies of high-mass-scale
physics at ultrahigh luminosity (for the SSC), $\CL \simeq \uhl$. These
physics studies include precision investigations of the properties of a
very massive $Z'$ boson in its $e^+ e^-$ and $\mu^+ \mu^-$ decay channels,
and the character of quark-lepton substructure contact interactions via the
process $\ol q q \ra \mu^+ \mu^-$.

For all these processes, the performance of the GEM design has been
determined realistically. This is an important step in optimizing the
design. Results are given below for a variety of integrated luminosities:
$10\,\fb^{-1}$ and $30\,\fb^{-1}$, corresponding to one and three years of
operation at the design luminosity of $\hl$; and $100\,\fb^{-1}$,
corresponding to one year at a peak luminosity of $\uhl$.

As noted above, the upper limit on the top-quark mass was $95\,\gev$ at
the time most of the work described here was done. Except in Section~5.3,
a top mass of $140\,\gev$ was used in the simulations. If a top mass of
$175\,\gev$ were used, this would sometimes increase signals (as in gluon
fusion of a standard Higgs boson) and/or backgrounds and sometimes decrease
them. Overall, we do not think that a significant quantitative change
would result from using the higher mass.

\gdef\Bullet{\item{$\bullet$}}
\gdef\mumu{\mu^+\mu^-}
\gdef\lqq{\hbox{\ninerm``}}
\gdef\rqq{\hbox{\ninerm''}}

\subsec{Standard Model Higgs Physics}

Here we describe searches for the minimal Standard Model
Higgs boson with the GEM detector in the mass range between 80 and 800 GeV.
This covers the interval between lower limits from LEP (which have excluded the
mass range below $\sim 60\,\GeV$,%
\ref\HRlep{See, for example, L3 Collaboration, Z.~Phys. {\bf
C57}, 355 (1993).}
and should reach 80--$90\,\GeV$ at LEP 200;%
\ref\HRlepii{{\it ECFA Aachen Workshop on LEP 200}, CERN 87--08, June
1987; and {\it LEP 200 Workshop}, CERN, September 1992.}%
and the triviality bounds discussed above.\cite{\trivial}
\nref\caltech{S.~Mrenna, et al., GEM TN--93--373.}%
The details of these Higgs boson search studies are presented in Ref.\
\caltech. Higgs bosons in extensions of the standard model such as the minimal
supersymmetric model%
\ref\HaberKane{For reviews, see H.\ E.\ Haber and G.\ L.\ Kane, Phys.\ Rept.\
{\bf 117}, 75 (1985)\semi S.~Dawson, E.~Eichten and C.~Quigg, Phys.\ Rev.\ {\bf
D31}, 1581 (1985).}
generally have similar signatures, albeit with different cross sections.

	The dominant decay modes of the standard model Higgs boson are
$W^+W^-$ and $ZZ$ for $M_H > 2M_W$ and heavy fermion pairs for $M_H <
2M_W$. The latter, and the dominant decays of $W$'s and $Z$'s into
jets, all have large backgrounds, so it is necessary to rely on rare
decays. The cleanest channel is $H\to ZZ/ZZ^* \to \ellell\ellell$
($\ell = e$, $\mu$), which has four isolated high-$p_T$ leptons in the
final state. This channel has an
inadequate rate at the lowest and highest ends of the mass range.
Hence, the search may be divided into three mass regions:

\Bullet A Higgs boson of ``intermediate'' mass ($80\,\GeV < M_H < 2 M_Z$)
will be searched for through its decays $H \to \gamma\gamma$ and $H
\to ZZ^* \to \ellell\ellell$.  Both direct $H\to\gamma\gamma$ and
lepton-associated production $(t \bar{t}/W) H \to \gamma\gamma\ell X$
will be used.

\Bullet A heavy Higgs ($2M_Z < M_H < 600\,\GeV$) will be searched for
through the channel $H \to ZZ \to \ellell\ellell$.

\Bullet A very heavy Higgs ($M_H \approx 800\,\GeV$) will also be searched
for in $\ellell\ellell$ and in the channels $H \to ZZ \to
\ellell \nu\nu$ and $H \to ZZ \to \ellell jj$.

\widetopinsert
\GEMcaption{Table 4}{Sensitivity of the GEM detector to standard
model Higgs boson signals.}
\medskip
\GEMtable
\hfil#\hfil\quad & \hfil#\hfil\qquad & \hfil#\hfil\quad & \hfil#\hfil\quad &
\hfil#\hfil\quad & \hfil#\hfil\quad & \hfil#\hfil\qquad & \hfil#\hfil\cr
\GEMrulerule
 $M_H$ & $\int {\cal L}dt$ &
 $\gamma\gamma$ & $\gamma\gamma\ell$ &
 $\ellell\ellell$ & $\ellell jj$ & $\ellell\nu\nu$ & Combined\cr
(GeV) & ($\fb^{-1}$) \cr
\GEMrule
 ~80 & 30 & 3.9$\sigma$ &  4.5$\sigma$ &&& & $6.3\sigma$ \cr
 ~90 & 30 & 4.9$\sigma$ &  4.9$\sigma$ &&& & $7.2\sigma$ \cr
 100 & 10 & 4.6$\sigma$ & 2.9$\sigma$ &&&&  $5.8\sigma$ \cr
 120 & 10 & 7.8$\sigma$ & 2.7$\sigma$ &&&&  $8.5\sigma$ \cr
 140 & 10 & 9.0$\sigma$ & 2.3$\sigma$  & 11$\sigma$  &&& $15\sigma$ \cr
 150 & 10 & 7.3$\sigma$ &              & 13$\sigma$  &&& $15\sigma$ \cr
 160 & 10 & 3.2$\sigma$ &              & 8.1$\sigma$ &&& $8.9\sigma$ \cr
 170 & 10 & &                          & 5.7$\sigma$ &&& $5.7\sigma$ \cr
 180 & 10 & &                          & 10$\sigma$  &&& $10\sigma$ \cr
 200 & 10 &&                           & 38$\sigma$ &&& $38\sigma$ \cr
 400 & 10 &&                           & 28$\sigma$ &&& $28\sigma$ \cr
 600 & 10 &&                           & 9.7$\sigma$ &&& $9.7\sigma$ \cr
 800 & 10 &&         & 4.2$\sigma^*$ & 1.0$\sigma^*$ & 4.2$\sigma^*$ &
$6.6\sigma^*$ \cr
\GEMrulerule
\multispan{8}{${}^*$ Estimated systematic errors.\hfill}\cr
\endGEMtable
\endinsert

The Monte Carlo event generator used in this study was PYTHIA 5.6 and JETSET
7.3.\cite{\pythia} The top quark mass was assumed to be 140 GeV. The study
was carried out for the SSC design luminosity ${\cal L}$ = $10^{33}\,\cmsec$.

	Table~4 summarizes the significance with which GEM could discover
the Higgs through the channels listed above.
This demonstrates the sensitivity of the GEM detector at
greater than the $5\sigma$ level for the whole range of Higgs masses
considered.  The signal can be seen with an integrated luminosity of
$10\,\fb^{-1}$, except for $M_H < 100\,\GeV$ where 20 to $30\,\fb^{-1}$ are
needed.

	The most difficult part of the Higgs spectrum is the low end (between
80 and $100\,\GeV$) and the region between the $WW$ and $ZZ$ thresholds,
170--$180\,\GeV$.  GEM is sensitive to a Higgs with $M_H$ below $100\,\GeV$
in both the two-photon and in the two-photon plus lepton channels. This
would allow the confirmation of a Higgs signal in two decay channels
necessary for a credible discovery claim in this mass region.

\bigskip

	Over the entire Higgs mass range the dominant production
mechanism at the SSC is gluon fusion, although vector-boson fusion is
also important. The Higgs can also be produced associated with a vector
boson or a top quark pair ($t \bar{t}$) which provides an isolated
lepton tag.

\nfig\TFHprod{\varfig{3.5in}{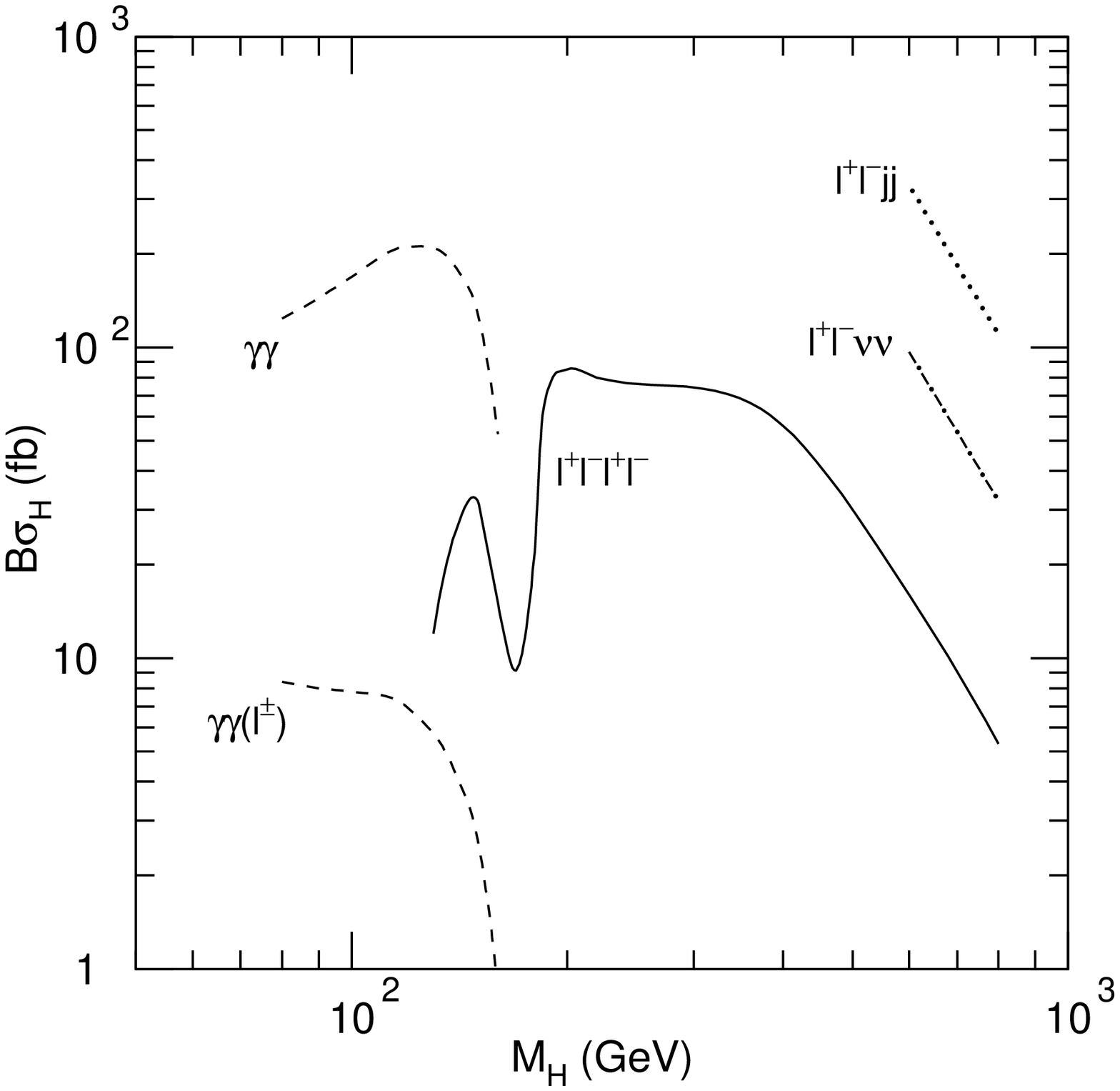}}{Lowest order Higgs
cross sections into observable modes vs.\ mass.}

	In the standard model, a Higgs boson with mass $M_H > 2M_Z$
has a width which grows like $M_H^3$, but it remains fairly narrow
for $M_H < 600\,\GeV$. Figure~\TFHprod\ shows
the lowest order Higgs production cross section multiplied by
specific decay branching ratios for the five modes studied in this
section.  These cross sections were calculated using PYTHIA 5.6
with the EHLQ-1 parton distribution functions and a top
quark mass of $140\,\GeV$. Calculations of these cross sections using
the more modern HMRS%
\ref\HRhmrs{P.~N.~Harriman, A.~D.~Martin, R.~G.~Roberts and
W.~J.~Stirling, Phys. Rev. {\bf D42}, 798 (1990) and Phys. Lett. {\bf B243},
421 (1990).}
and CTEQ%
\ref\HRcteq{J.~Botts, et al.,
MSUHEP--92--27, Fermilab--Pub--92/371, FSU--HEP--92--1225 and
ISU--NP--92--17 (1992).}
distributions differ from EHLQ-1 by less than 20\%.%
\ref\hiropdf{H.~Yamamoto, GEM TN--93--374.}
Higher order QCD corrections are not included in Fig.~\TFHprod.

\bigskip

	The physics signatures of Higgs decays all involve identification of
isolated photons, electrons and muons. Since the Higgs cross sections are
small and the potential backgrounds are large, this identification requires
particular care.  First, an isolation cut is applied, removing most of the
jet background. Then, a detailed identification algorithm is used.  The
isolation cut for selecting electron and photon candidates was done using
|gemfast| and requiring
\eqn\XX{
\sum_{R} E_T - E_T^{\gamma/e} < E_T^{\rm mean} + E_T^{\rm cut} \ts ,
}
where the sum is over calorimeter towers in a cone of radius $R$. Generally, $R
= 0.45$ was used for two-body final states and $R = 0.30 - 0.35$ for
four bodies. The transverse energy $E_T^{\gamma/e}$ of the $\gamma$ or $e$
candidate was found by summing the energy deposited in $5\times5$ cells in the
electromagnetic calorimeter; $E_T^{\rm mean}$ is the mean transverse energy
from pileup and noise; and $E_T^{\rm cut}$ is the isolation threshold imposed.

	The $E_T^{\rm cut}$ value was determined from the distribution
of the thermal and pile-up noise%
\ref\HRcleland{W.~E.~Cleland and A.~V.~Vanyashin, GEM TN--93--376,
(April, 1993).}
as described in Section~4.3. A Gaussian fit to the total noise in an
$\eta-\phi$ cone of radius $R=0.45$ gives a width of $3.4\,\GeV$. This
is reduced to $1.7\,\GeV$ if only those cells with $\vert E_T\vert >
0.5\,\GeV$ are summed. Doing this shifts the $E_T^{\rm mean}$ from
0.22 to 1.5 GeV, and reduces the signal loss from 13\% to 7.6\% for
$E_T^{\rm cut} = 4\,\GeV$, the value used here.

	For photons, there must be no reconstructed charged track in
the $5\times5$ cells. The central tracker can reject 95\% of the
electrons while keeping 96\% of the photons.
For electrons, there must be exactly one charged track in the central
$3\times3$ cells. Also, the energy and shower position measured in the EM
calorimeter must match the momentum and track position measured in the central
tracker. The background to $H \ra \gamma\gamma$ from misidentified $Z \to \ee$
is potentially large for $M_H \approx M_Z$ and is discussed below.

A more complete understanding of the capability of GEM to identify isolated
photon and electron candidates based on their shower shape requires a detailed
GEANT simulation. A sufficiently detailed geometry was available for only
three $11\times11$ cell regions of the barrel.%
\ref\HRhma{H.~Ma and M.~Leltchouk,
GEM TN--92--257 (December, 1992); GEM TN--93--356 (April, 1993).}
Therefore, the photons hitting the $5\times5$ cluster were rotated to the
nearest of these regions, and the full simulation was run. Shower shape cuts
(including lateral and longitudinal shower distributions), information from
the narrow $\theta$ strips of the first segment of the EM calorimeter, and
energy in the hadronic calorimeter are combined in a likelihood function and
used to select single electromagnetic showers. Typically this rejects 75\%
of the remaining jet background while keeping 90\% of the single photons. The
probability of a QCD jet faking an isolated electron, $\CR(e/{\rm jet})$,
was determined to be approximately $10^{-5}$ by using detailed GEANT
simulation.%
\nref\HRryhiro{R.~Y.~Zhu and H.~Yamamoto, GEM TN 92--126 (July,
1992).}%
\cite{\HRhma, \HRryhiro}
For channels needing very good resolution, such as $H \to \gamma\gamma$ and $H
\to ZZ^* \to \ee\ee$, photons and electrons were required not to be in the
region of degraded resolution between the barrel and the endcap, $1.01 <
\vert\eta\vert < 1.16$. The overall photon identification efficiency, including
this geometrical loss and other cuts, is between 80\% and 85\%. The electron
identification efficiency is 85\% -- 90\%. For signals such as $H \to ZZ \to
\ee\ee$ for $M_H > 2M_Z$, one electron was allowed to hit this
transition region.

Muons were simulated using |gemfast|. They are relatively free
of jet background and so require less stringent identification cuts. An
isolation cut was made,
\eqn\XX{
\sum_{R} E_T - \Delta E < E_T^{\rm mean} + E_T^{\rm cut} \ts ,
}
where $\Delta E$ is the energy loss of a muon in the calorimeter, calculated
as explained in Section~4.4. This eliminates muons from $b$ and $c$ decay,
secondary decay muons in jets and punchthrough.  The isolation cone radius
was taken to be 0.35 for intermediate mass Higgs searches and 0.3 for
heavier masses.  The angle and momentum of the track measured in the muon
system is then matched to that in the central tracker. The muon
identification efficiency is around 80\% for muons within $\vert\eta\vert <
2.5$, which includes factors of 85\% from the geometrical acceptance and
95\% from the muon track reconstruction and identification
efficiency.\cite{\MuPara.}

\bigskip

	Significance is used to indicate how well a Higgs signal can be
identified in the presence of background.  As in Gaussian statistics, a
probability of $1-1.35 \times 10^{-3}$ is expressed as a significance of
$3\sigma$, $1-2.85 \times 10^{-5}$ is expressed as $5\sigma$, and so on. In
this section, a $5\sigma$ significance is generally regarded as the
minimum for discovery.

\subsubsec{$H\to\gamma\gamma$ Search for $80\,{\bf GeV}< M_{\bf H} <
160\,{\bf GeV}$}

Precision electromagnetic energy resolution is essential for
the $H \ra \gamma\gamma$ search because of the small production cross section
(60 to $200\,\fb$), the narrow decay width (3 to $100\,\MeV$) of the Higgs
boson between 80 and $160\,\GeV$, and the large QCD $\gamma\gamma$ background.
There is a potentially much larger background from QCD jets fragmenting into
electromagnetic particles, which can be reduced using isolation cuts and a
shower shape analysis made possible by the fine segmentation of the
calorimeter. In PYTHIA, jets can radiate prompt photons as an option. This is
properly regarded as an approximation to part of the higher order QCD
corrections to the $\gamma\gamma$ background. Since these corrections are
explicitly known for both the Higgs signal and the $q \bar q \to \gamma\gamma$
background, the prompt photon radiation has been turned off, and the PYTHIA
signal and background cross sections have been scaled to the higher order
results.

	Events were selected using the following cuts:

\item{1.}	$\vert\eta^\gamma\vert <$ 2.5 and $p_T^\gamma >
		20\,\GeV$.

\item{2.}	Photon isolation cut with $R = 0.45$ and $E_T^{\rm
		cut} = 4\,\GeV$, and veto on $1.01 < \vert\eta\vert <
		1.16$.

\item{3.}	Photon identification based on detailed shower shape.

\item{4.}	Electron rejection, as described below.

\item{5.}	$|\cos \theta^*| < 0.7$, where $\theta^*$ is the polar
		angle of photon in the center of mass system of two
		photons.

\noindent Cut~1 simply ensures that the photons are in the overall acceptance
of the detector and are triggered on. Cuts~2 and 3 reduce the large potential
backgrounds from misidentified QCD $\gamma$-jet and jet-jet events well below
the $\gamma\gamma$ continuum. Cut~4, which is only important for $M_H \approx
M_Z$, removes the background from misidentified electrons.  Finally, cut~5
reduces the real and fake $\gamma\gamma$ backgrounds. This leaves the $H \to
\gamma\gamma$ signal as a narrow bump in the $M_{\gamma\gamma}$ distribution.

	The significance of a $H \to \gamma\gamma$ mass peak is
directly related to the $\gamma\gamma$ mass resolution $\Delta
M_{\gamma\gamma}$. This is given by
\eqn\HEmass{
{\Delta M_{\gamma\gamma} \over {M_{\gamma\gamma}}} = {1 \over 2} \Bigg[
\left({\Delta E_1 \over E_1}\right)^2 +
\left({\Delta E_2 \over E_2}\right)^2
  + \left(\Delta\theta \ts \cot{\theta\over2} \right)^2 \Bigg]^{1/2} \ts ,
}
where $E_1$ and $E_2$ are the energies of the two photons and $\theta$ is
the opening angle between them.  It is clear from this equation that
uncertainties in both energy and direction measurement would degrade
the Higgs mass resolution. The energy resolution is determined by the
EM calorimeter. The angular resolution is determined both by the
position resolution in the EM calorimeter and by the precision of the vertex
determination. GEM's calorimeter position resolution, $\Delta x = 4.4\,{\rm
mm}/\sqrt{E}$, has a negligible effect on the mass resolution.\cite{\HRryhiro}
The vertex position for a single event is well determined by the central
tracker. The only issue, therefore, is how well the correct vertex can be
selected in the presence of an average of 1.6 additional minimum bias events at
the standard SSC luminosity plus events from previous bunch crossings.

Two approaches to determining the correct vertex are possible. The first is
to use the difference in event topology resulting from the fact that Higgs
production is a harder process than most of the minimum bias events and
hence radiates more gluons. This leads to a higher multiplicity and a higher
average $p_T$. Selecting the vertex with the highest $p_T$-weighted charged
multiplicity gives the correct Higgs vertex with 95\% probability at the
standard SSC luminosity.  An independent approach is to use the pointing
provided by the longitudinal segmentation of the calorimeter; using this
pointing alone without any information from the central tracker degrades the
mass resolution by only 20\%. Using the pointing and then selecting the
closest central tracker vertex gives a vertex within $5\,{\rm mm}$ 87\% of
the time at standard luminosity. This degrades to about 65\% at $3 \times
10^{33}\,\cmsec$.  A combination of these two methods improves the
vertex-finding efficiency to 97\%.\cite{\caltech} A vertex finding
efficiency of 95\% was assumed for this analysis.

The overall $M_{\gamma\gamma}$ resolution varies slowly from $0.66\,\GeV$ at
$M_{\gamma\gamma}= 80\,\GeV$ to $0.99\,\GeV$ at $M_{\gamma\gamma}=
160\,\GeV$. Over the same range the Higgs natural width varies from
$0.003\,\GeV$ to $0.097\,\GeV$.

\nfig\HFcsts{\dofig{86.5mm}{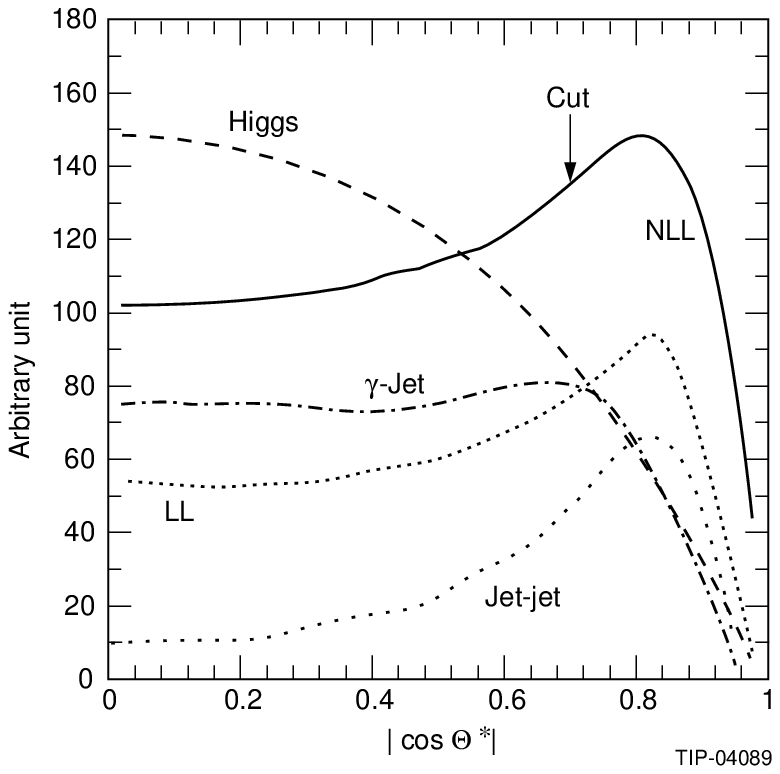}\forceright}{$\vert\cos\theta^*\vert$
distribution for an $80\,\GeV$
Higgs, NLL direct photon background, $\gamma$-jet and jet-jet background. The
cut used in this analysis is $\vert\cos\theta^*\vert <$ 0.7.}

	The $H \to \gamma\gamma$ signal and the $q \bar q \to
\gamma\gamma$ and $gg \to \gamma\gamma$ backgrounds were generated
with PYTHIA. The next-to-leading-log (NLL) corrections both to
the signal%
\ref\HRhggnll{S.~Dawson, Nucl.\ Phys.\ {\bf B359}, 283 (1991)\semi
A.~Djouadi, M.~Spira, and P.~Zerwas, Phys.\ Lett.\ {\bf 264B}, 440
(1991)\semi
S.~Dawson and R.~Kauffman, Phys.\ Rev.\ Lett.\ {\bf 68}, 2273
(1992)\semi
R.~Kauffman, Phys.\ Rev.\ {\bf D45}, 1512 (1992)\semi
C.~P.~Yuan, Phys.\ Lett.\ {\bf 283B}, 395 (1992)\semi
D.~Graudenz, M.~Spira, and P.~Zerwas, Phys.\ Rev.\ Lett.\ {\bf 70},
1372 (1993).}
and to the $q \bar q \to \gamma\gamma$ background%
\ref\HRcgg{B.~Bailey, J.~Owens and J.~Ohnemus, Phys.\ Rev.\ {\bf D46}, 2018
(1992).}
have been calculated. They enhance the signal cross section by a
factor $K \approx 1.5$ and the background by a somewhat larger factor.
The PYTHIA cross sections have been rescaled to these NLL results.

\nfig\TFHGGxsec{\varfig{3.5in}{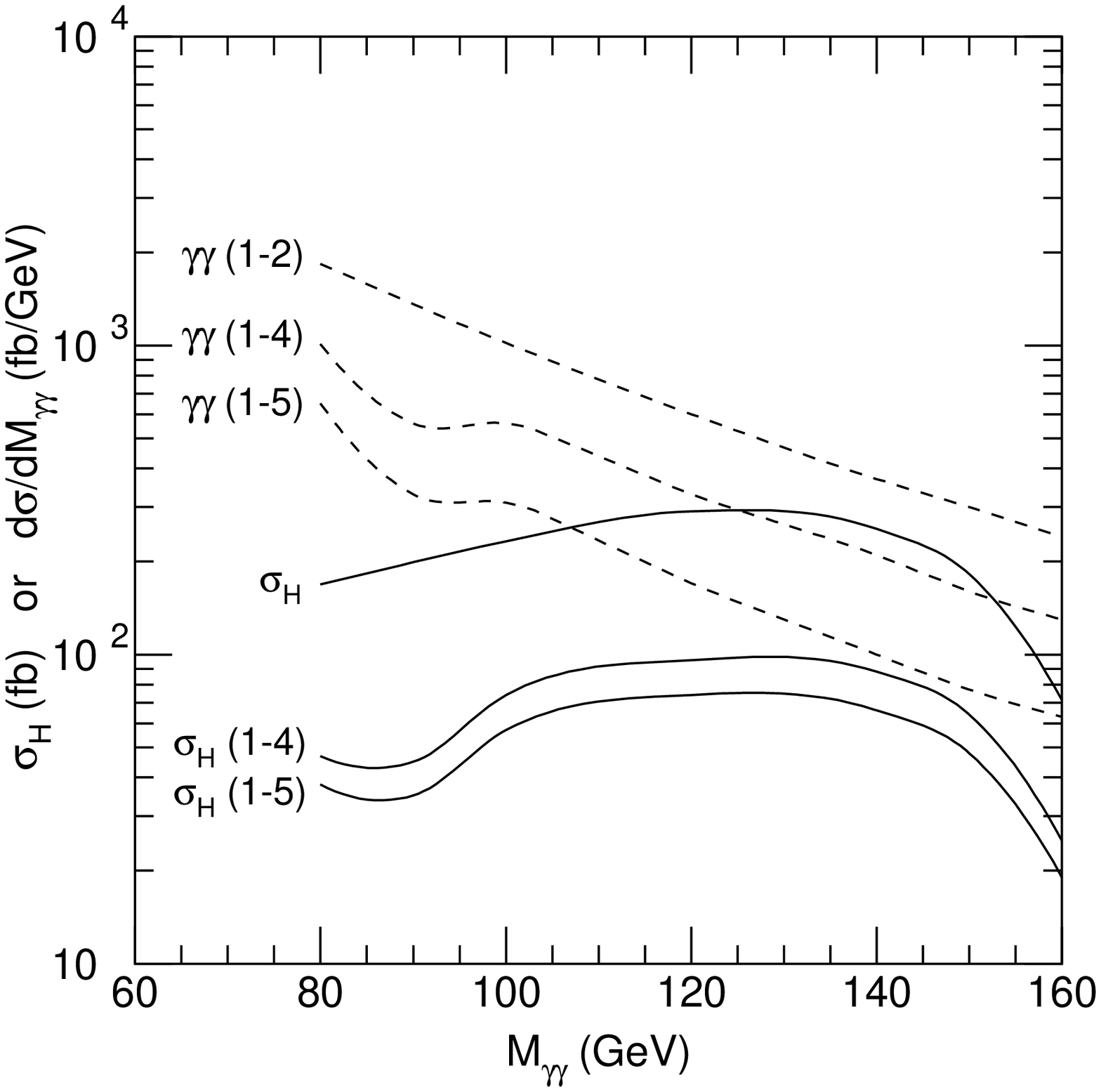}}{Cross sections for $H \to
\gamma\gamma$ signal and $\gamma\gamma$ background as functions of
$M_{\gamma\gamma}$. The mass resolution is 0.66--$1.0\,\GeV$.}

After cuts 1 to 4, the cut on $\cos\theta^*$ provides an additional
rejection of both direct photon and jet backgrounds and thus improves the
significance. Figure~\HFcsts\ shows the $\cos\theta^*$ distribution for an
$80\,\GeV$ Higgs, NLL direct photon background and the $\gamma$-jet and
jet-jet backgrounds. The improvement in significance is optimized by a cut
$\vert\cos\theta^*\vert < 0.7$, for which an improvement of 15\% is
obtained.

	Figure~\TFHGGxsec\ shows the Higgs production cross section
($\sigma_H$) and the cross sections after cuts 1--4 and 1--5. It also shows
the rate of direct photon background after cuts~1--2, 1--4, and 1--5. The
trigger efficiency for events passing all selection cuts is 98.8\%. This is
included in all signal and background rates.  The signals are small compared
to the background but still statistically significant.

\nfig\TFHGGgamq{\varfig{3in}{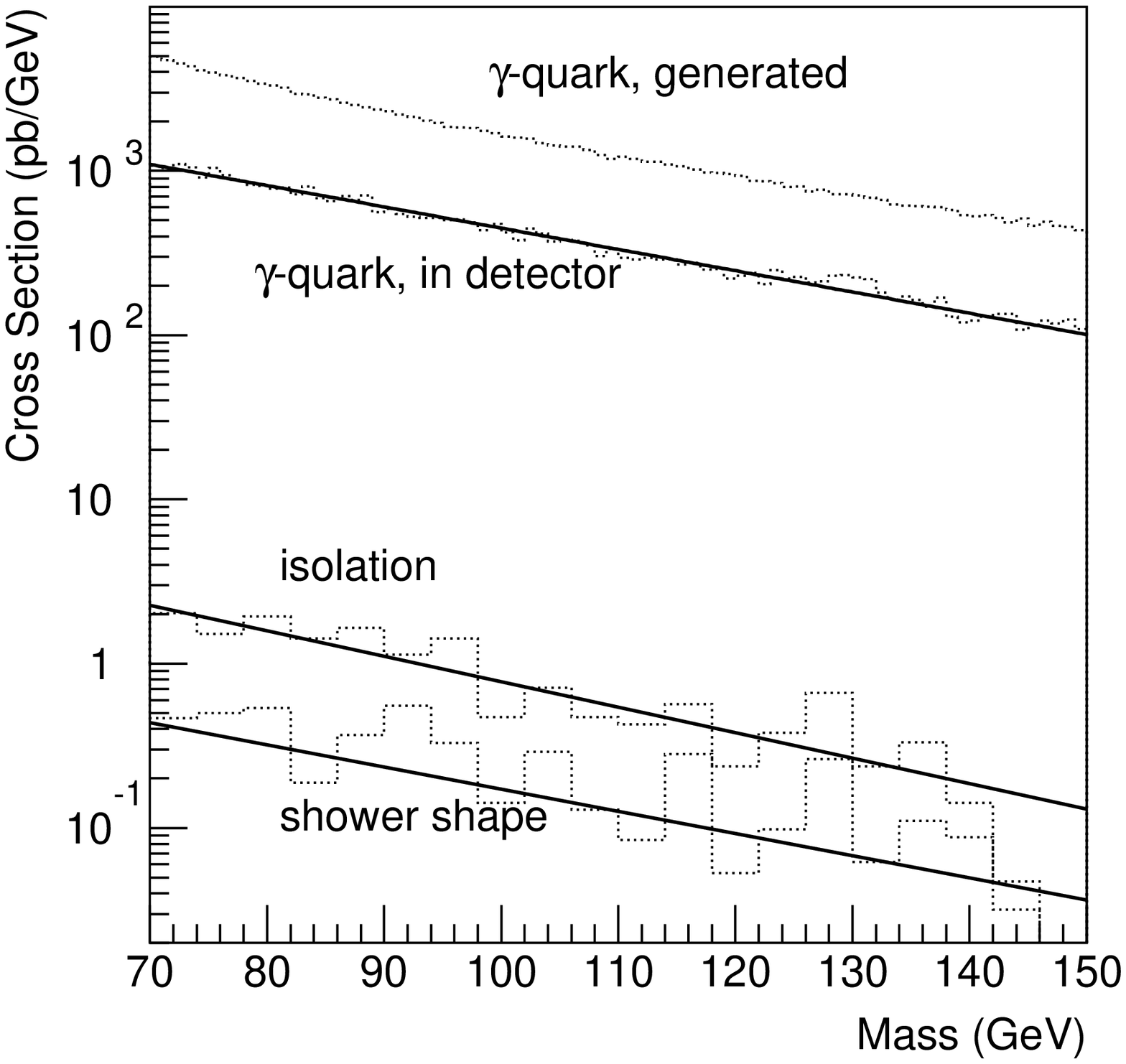}}
{Cross section for $\gamma$-quark jet background as a function of
$M_{\gamma``\gamma''}$, showing the rejection obtained by the cuts
described in the text.}

\nfig\TFHGGgamg{\varfig{3in}{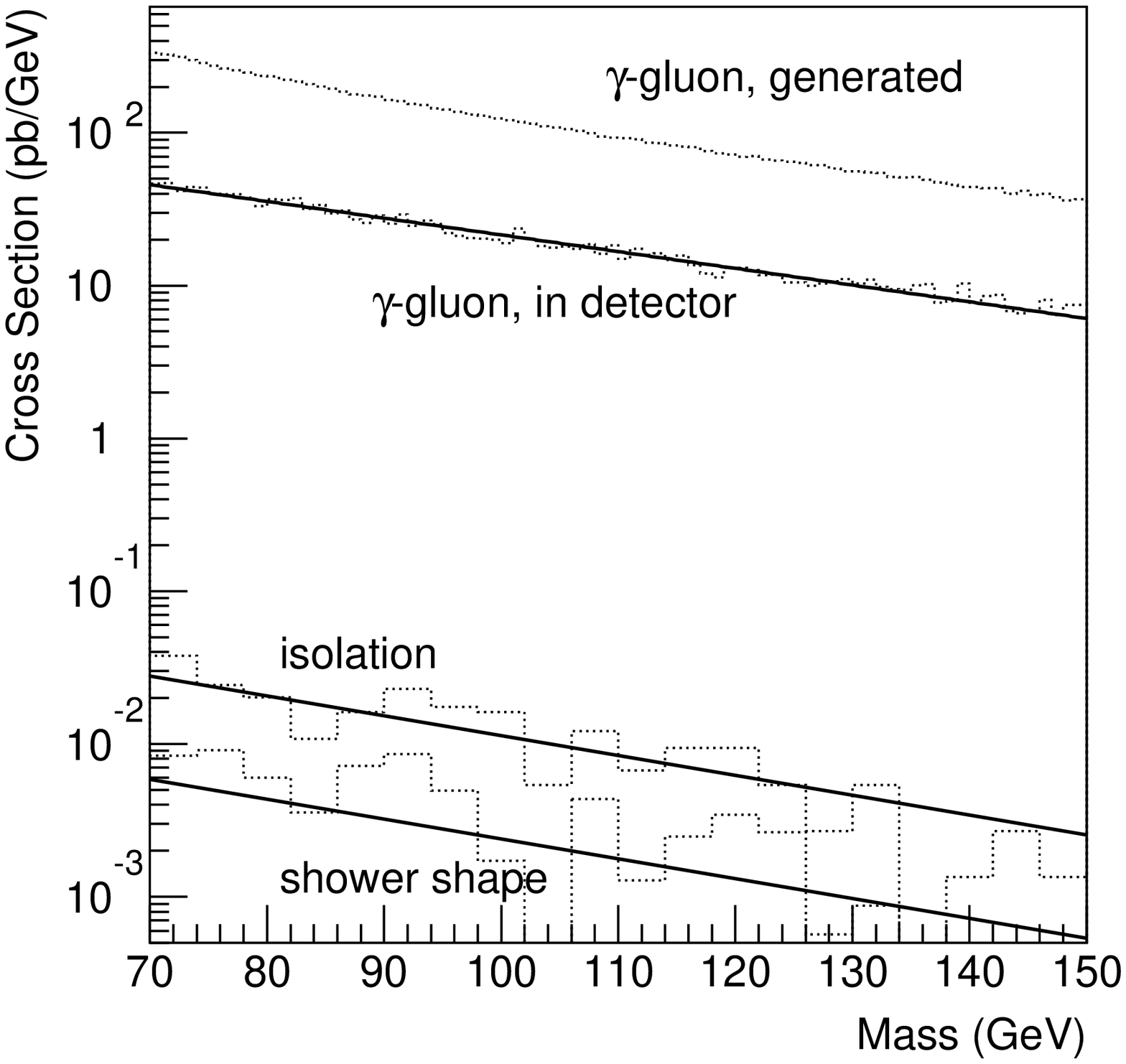}}{Same as
Fig.~\TFHGGgamq\ but for $\gamma$-gluon jet background.}

\nfig\TFHGGbkgs{\varfig{3.5in}{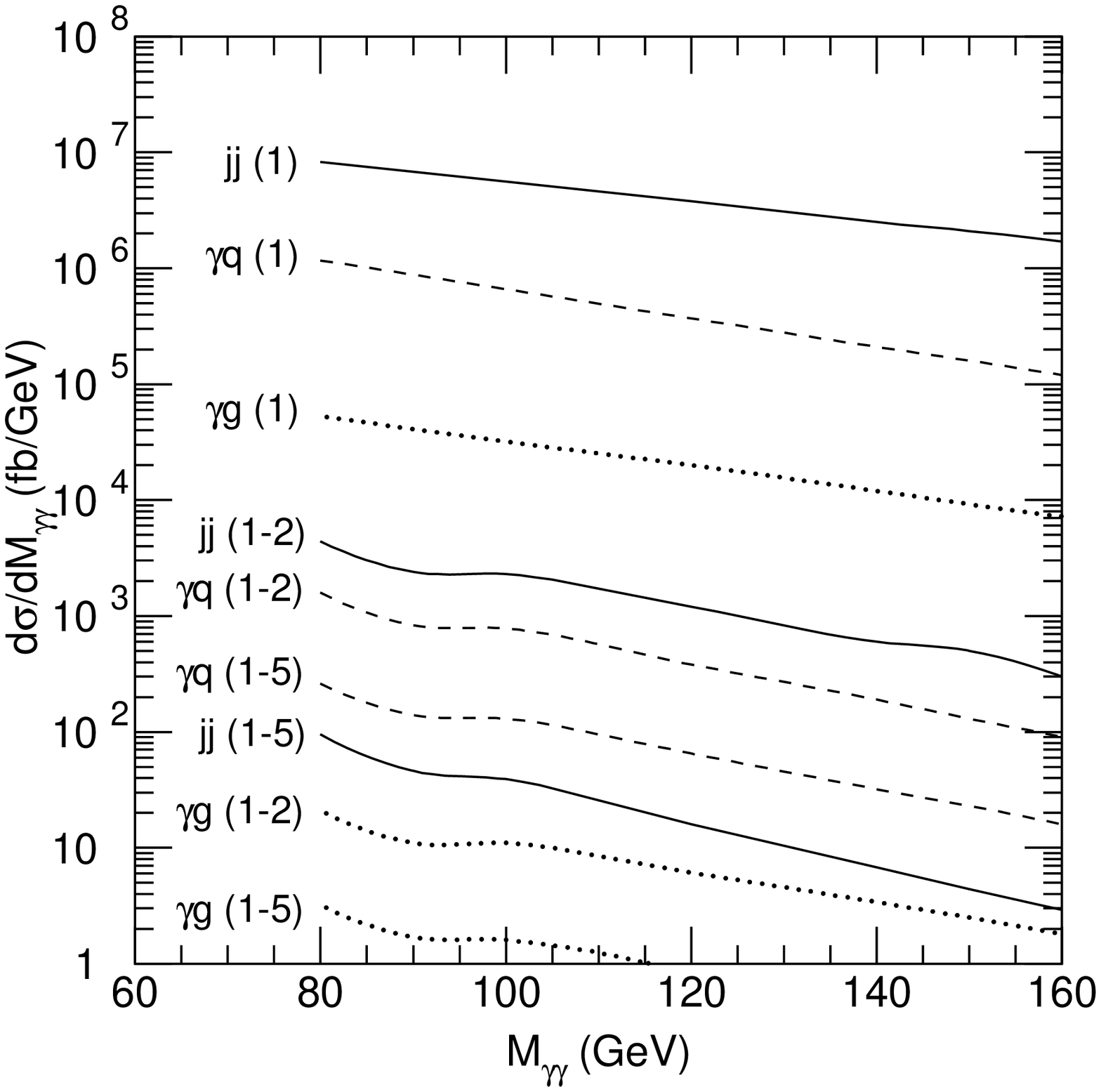}}{Jet Background Rates for
$H\to\gamma\gamma$ after various cuts defined in the text.}

\nfig\HFdy{\dofig{92.5mm}{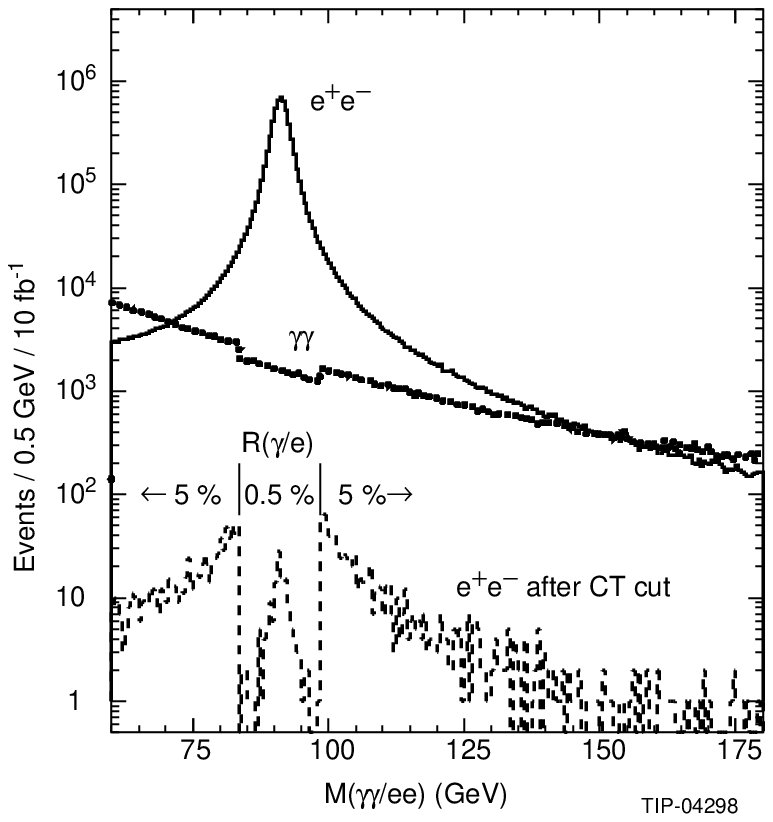}\forceright}{Drell-Yan $\ee$
cross section after event
selection cuts (dashed) and lowest order direct $\gamma\gamma$
cross section (solid). The dots show the rate of fake photon pairs
from Drell-Yan electron pairs using $\CR(\gamma/e) = 0.005$ near the $Z$
and 0.05 elsewhere.}

\bigskip

	QCD jets produce photons from $\pi^0$ and $\eta$ decays.
These sometimes appear isolated and give additional backgrounds to
$H\to\gamma\gamma$.  Both $\gamma$-jet events giving
$\gamma$``$\gamma$'' and jet-jet events producing ``$\gamma\gamma$''
must be considered. Thus, the jet background to $H\to\gamma\gamma$ is
\eqn\XX{
\sigma_{\lqq\gamma\gamma\rqq} =
\sigma_{\gamma\hbox{{\ninerm -jet}}} \CR(\gamma/{\rm jet})
+ \sigma_{\hbox{{\ninerm 2-jet}}} \CR^2(\gamma/{\rm jet}) \ts ,
}
where $\CR(\gamma/{\rm jet})$ is the probability for a jet to fake a
photon.

	The probability $\CR(\gamma/{\rm jet})$ was determined using a
combination of |gemfast| for selecting isolated events and GEANT for
simulating the shower shape cut. QED
bremsstrahlung is not included, since it is properly included in the
higher order QCD cross sections. Since $\CR(\gamma/{\rm jet})$ is
different for quark and gluon jets,\cite{\HRryhiro} separate samples
of about $10^6$ events each for $\gamma q$ and $\gamma g$ were used to
study jet rejection as a function of $M_{\gamma\gamma}$. Figures
\TFHGGgamq\ and \TFHGGgamg\ show the cross sections for the $\gamma$-jet
background for both quark and gluon jets as a function of
$M_{\gamma\lqq\gamma\rqq}$. The shower shape analysis made possible by the
fine sampling of the GEM calorimeter improves the rejection by about a
factor of 4 per photon over isolation alone. Since the jet-jet cross section
is about $10^8$ times the $\gamma\gamma$ one, it is not possible to generate
enough jet-jet events to simulate the background directly. Instead, $gg$,
$qg$ and $qq$ events were generated at the parton level, and the
probabilities of jets faking photons taken from these figures were used to
calculate the background.

\nfig\HFggmass{\dofig{93mm}{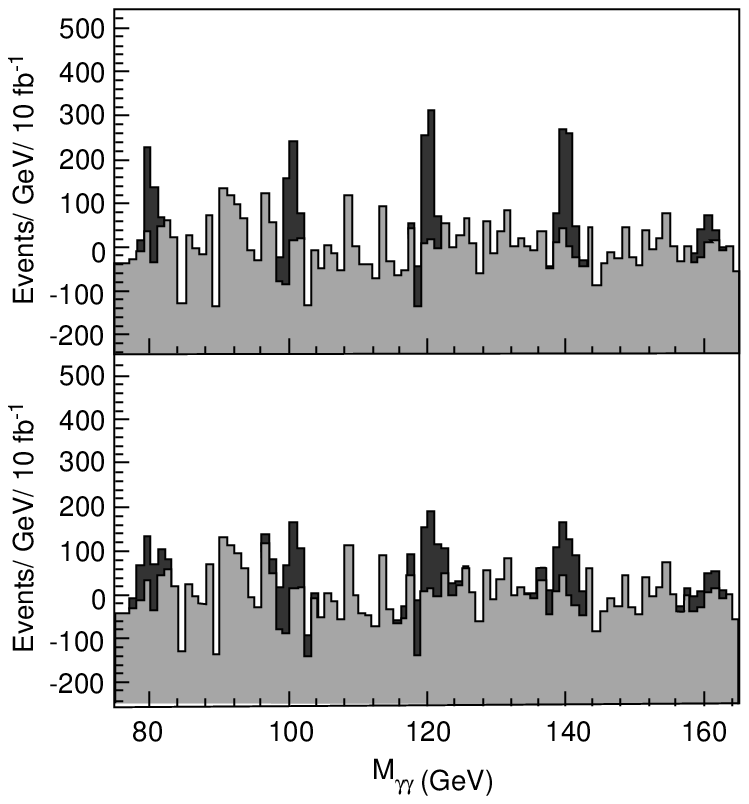}\forceleft}{Higgs mass peaks
over subtracted background,
obtained with $10\,\fb^{-1}$, for $H \rightarrow \gamma\gamma$ searches with
$M_H = 80$, 100, 120, 140 and $160\,\GeV$. Part (a) corresponds to $a=6/8.5\%$
in the barrel/endcaps and $b=0.4\%$, and part (b) to $a=14/17\%$ in the
barrel/endcaps and $b=1\%$. The GEM photon identification algorithm is used in
both cases; see Section~5.1}

\nfig\HFggmasshi{\dofig{95mm}{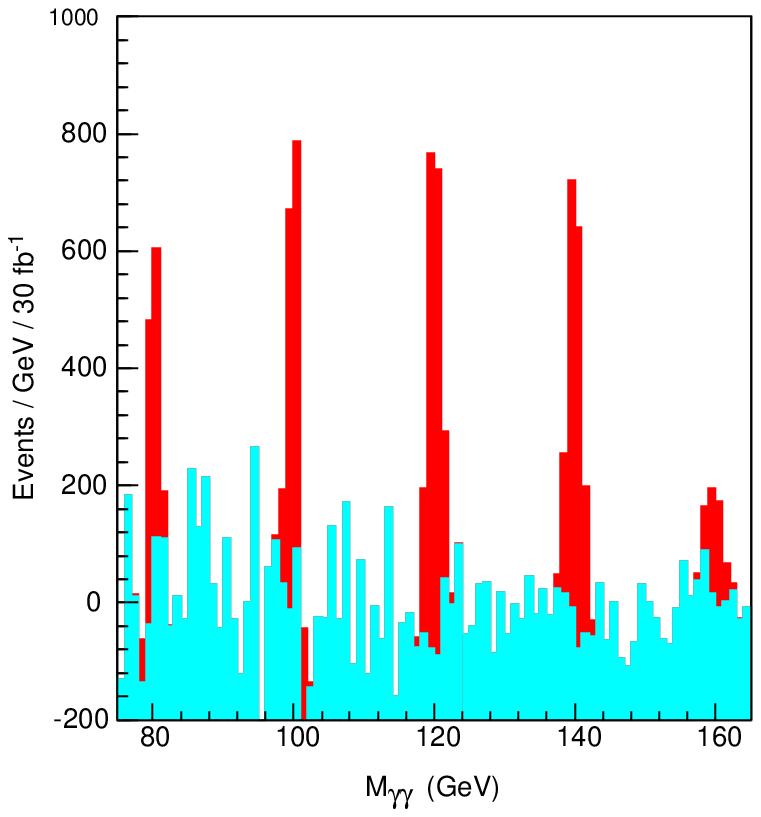}\forceright}{Higgs mass
peaks over subtracted background,
obtained with $30\,\fb^{-1}$, for $H \rightarrow \gamma\gamma$ searches
with $M_H = 80$, 100, 120, 140 and $160\,\GeV$ for the GEM detector.}

\nfig\TFHGGsig{\varfig{3.5in}{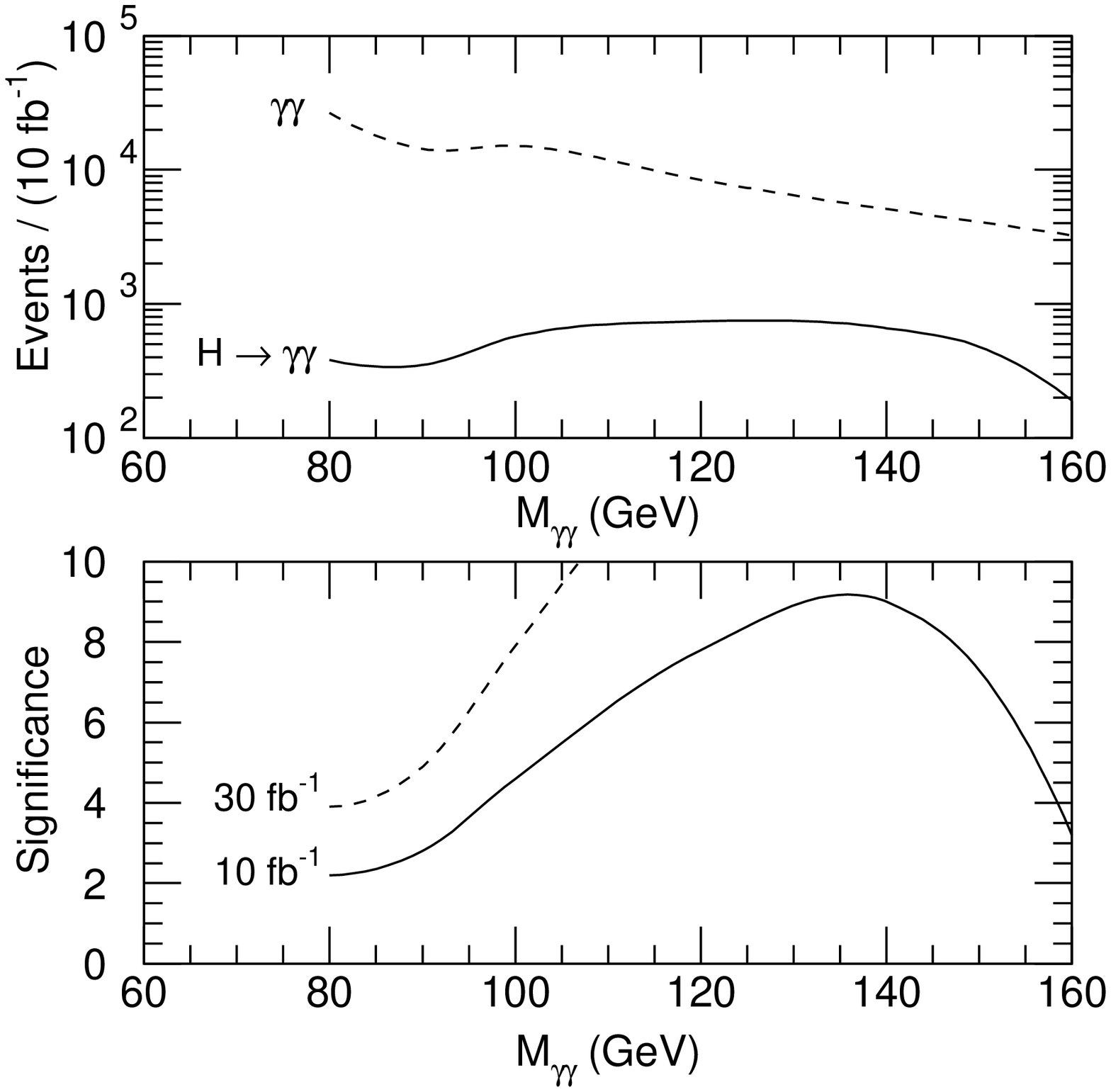}}{Number of events and
significance for $H\to\gamma\gamma$ including all acceptances.}

	Figure~\TFHGGbkgs\ shows the cross sections for the jet backgrounds
after cut~1 at the parton level (including a $K$-factor of 1.5), after cuts
1--2, 1--3 and after all cuts 1--5.  The $\gamma$-gluon background is
reduced to a negligible level compared to the NLL direct photon background.
The 2 jet background (which is dominated by gluon jets) is reduced to around
15\% of the NLL direct photon background at $80\,\gev$. The $\gamma q$
background, however, remains 40\% of the NLL direct photon background there.
Since a factor of two increase in all the jet backgrounds reduces the
significance by only 16\%, GEM's searches in the $H\to\gamma\gamma$ channel
are not very sensitive to the QCD jet rejection.

	Drell-Yan $e^+e^-$ pairs can give a background for $H \to
\gamma\gamma$ if the central tracker fails to identify both charged tracks.
Figure~\HFdy\ shows the invariant mass distribution of Drell-Yan $\ee$ after
event selection cuts and the lowest order direct photon background. The
Drell-Yan cross section is about 300 times larger than the direct photon
background at the $Z$-mass peak, but comparable elsewhere. A GEANT
simulation was performed for the central tracker to estimate the rate of
electrons faking photons, $\CR(\gamma/e)$.  If one simply looks for hits in
the central tracker forming a line, the photon acceptance is found to be
83\% for $\CR(\gamma/e) = 0.005$ and 96\% for $\CR(\gamma/e) = 0.05$.  If
one restricts the track to lie in a road defined by the pointing resolution
of the calorimeter, then the photon acceptance improves to 86\% for
$\CR(\gamma/e) = 0.002$. If one uses the vertex found to further restrict
the road, then a photon acceptance of 91\% can be achieved with the same
rejection.  In what follows, we use only the first method, since it allows
adequate rejection at all masses, as indicated in Fig.~\HFdy.

\bigskip

	Figure~\HFggmass\ shows the reconstructed $H\to\gamma\gamma$ mass
peaks after subtracting the background, for an integrated luminosity of
$10\,\fb^{-1}$, and $M_H = 80$, 100, 120, 140 and $160\,\GeV$.
Figure~\HFggmasshi\ shows the corresponding peaks after $30\,\fb^{-1}$.
Figure~\TFHGGsig\ shows the number of signal and background events after all
cuts, and the resulting signal significance using a $\pm 2\Delta M_H$ mass
bin.  With $10\,\fb^{-1}$, GEM could discover a Higgs boson, using this mode
alone, for $110 \simle M_H \simle 150\,\GeV$. With $30\,\fb^{-1}$, using this
mode alone, it could extend the discovery reach down to about $90\,\GeV$ and
up to $160\,\GeV$. While the heavier masses can also be found in the $ZZ^*$
mode, the $\gamma\gamma$ branching ratio is important to distinguish the
minimal Standard Model Higgs boson from nonminimal ones.

\subsubsec{$H (t \bar t / W) \to \gamma\gamma\ell X$ Searches for
$80\,{\bf GeV} < M_H < 140\,{\bf GeV}$}

	The processes $H t\bar t \to \gamma\gamma\ell X$ and $HW \to
\gamma\gamma\ell X$ are complementary to the inclusive $H \to \gamma\gamma$,
providing essential confirmation.%
\ref\HRhtt{W.J.~Marciano and F.~Paige, BNL Preprint, BNL--45805; \hfil\break
J.~F.~Gunion, et al., SDC report, SDC--91--00057\semi
R.~Kleiss, et al., Phys. Lett. {\bf B253}, 269 (1991)\semi
Z.~Kunszt, et al., Phys. Lett. {\bf B271}, 247 (1991).}
As shown in Fig.\ \TFHGGxsec, the signal cross section for the sum of
these two channels is of the order of few fb, but the isolated lepton
tag and photon identification cuts leads to smaller backgrounds, so
that the signal-to-background ratio is large. In these associated
production channels, most of the effects of detector resolutions on
the reconstructed mass resolution discussed previously remain
applicable. With a charged lepton in the final state, however, the
central tracker is able to determine the Higgs vertex unambiguously.

\bigskip

The main backgrounds to $H(t \bar{ t}/W) \to \gamma\gamma\ell X$ searches are:

\item{1.} $t \bar{ t}\gamma\gamma$ or $b \bar{ b}\gamma\gamma$ $\to$
$\ell\gamma\gamma$.

\item{2.} $q\bar{q}' \to W\gamma\gamma \to \ell\gamma\gamma$, where both
photons are radiated from quarks.

\item{3.} $q\bar{q}' \to W\gamma \to \ell\gamma\gamma$, where one
photon is radiated from the outgoing charged lepton.

\item{4.} $q\bar{q}/gg \to Z\gamma \to \ellell \gamma\gamma$, where
the second photon is radiated from the outgoing charged lepton.

\item{5.} $q\bar{q}/gg \to Z\gamma\to e \hbox{``$\gamma$''} \gamma$,
where the fake photon is from misidentification of the electron.

\item{6.} $t \bar{t} \to \hbox{``$\gamma\gamma$''}\ell X$, where both
photons are fakes arising from jet fragmentation.

\noindent Since all the higher order QCD corrections to the signal and
backgrounds have not been computed, leading order cross sections are used for
this entire analysis. While $t \bar t$ is initially very much larger, all
the backgrounds turn out to be comparable after the cuts described below.

Not all of these background processes are included in PYTHIA, so they
were calculated using a combination of generators. For processes 1 and
2, the initial hard scattering was generated using PAPAGENO
3.6.\cite{\papageno}
Initial and final state parton radiation, hadronization and decays
were then generated using PYTHIA. All the final state photon radiation
for processes 3, 4 and 5 was generated using PYTHIA with the QED
radiation option turned on. Since the process $gg \to Z \gamma$ is
not available in either PYTHIA or PAPAGENO, $q \bar q \to Z \gamma$
was increased by 20\%.%
\ref\HRggzz{J.~J.~van der Bij and E.~W.~N.~Glover, Phys. Lett. {\bf B206},
701 (1988) \semi
E.~W.~N.~Glover and J.~J.~van der Bij, Nucl. Phys. {\bf B321}, 561 (1989).}
The $t \bar{t}$ background (process 6) was generated with PYTHIA with
QED radiation turned on. The study was carried out with $2.6\times 10^6$
events. Only 3 events survived all cuts; none of them contained two radiated
photons. The $b \bar{ b}\gamma\gamma$ background was also simulated, and it is
small compared to the $t \bar t$ background after isolation cuts.

	The following cuts were made to reject the backgrounds:

\item{1.} $|\eta_\ell|<$ 2.5 and $p_T^\ell >$ 20 GeV.

\item{2.} $|\eta_\gamma|<$ 2.5 and $p_T^\gamma >$ 20 GeV.

\item{3.} Photon and lepton isolation with $E_T^{\rm cut}$ = 5 GeV and $R
		= 0.45$ and 0.3 for photon and lepton respectively.

\item{4.} Photon and lepton identification by shower shape and track
matching.

\item{5.} $p_T^{\gamma\gamma} > 40\,\GeV$.

\noindent The first two cuts ensured that the lepton and photons could be
detected. Cuts~3 and 4 identify the photons and lepton and reject jet
backgrounds. Finally, cut~5 was found to reject the backgrounds, especially
$t \ol t$ listed above. It reduces the background by a factor of 2 to 3
while losing only 20\% of the signal. The trigger efficiency for events
passing these selection cuts is 99\%.

\nfig\TFHGGLxsec{\varfig{3.5in}{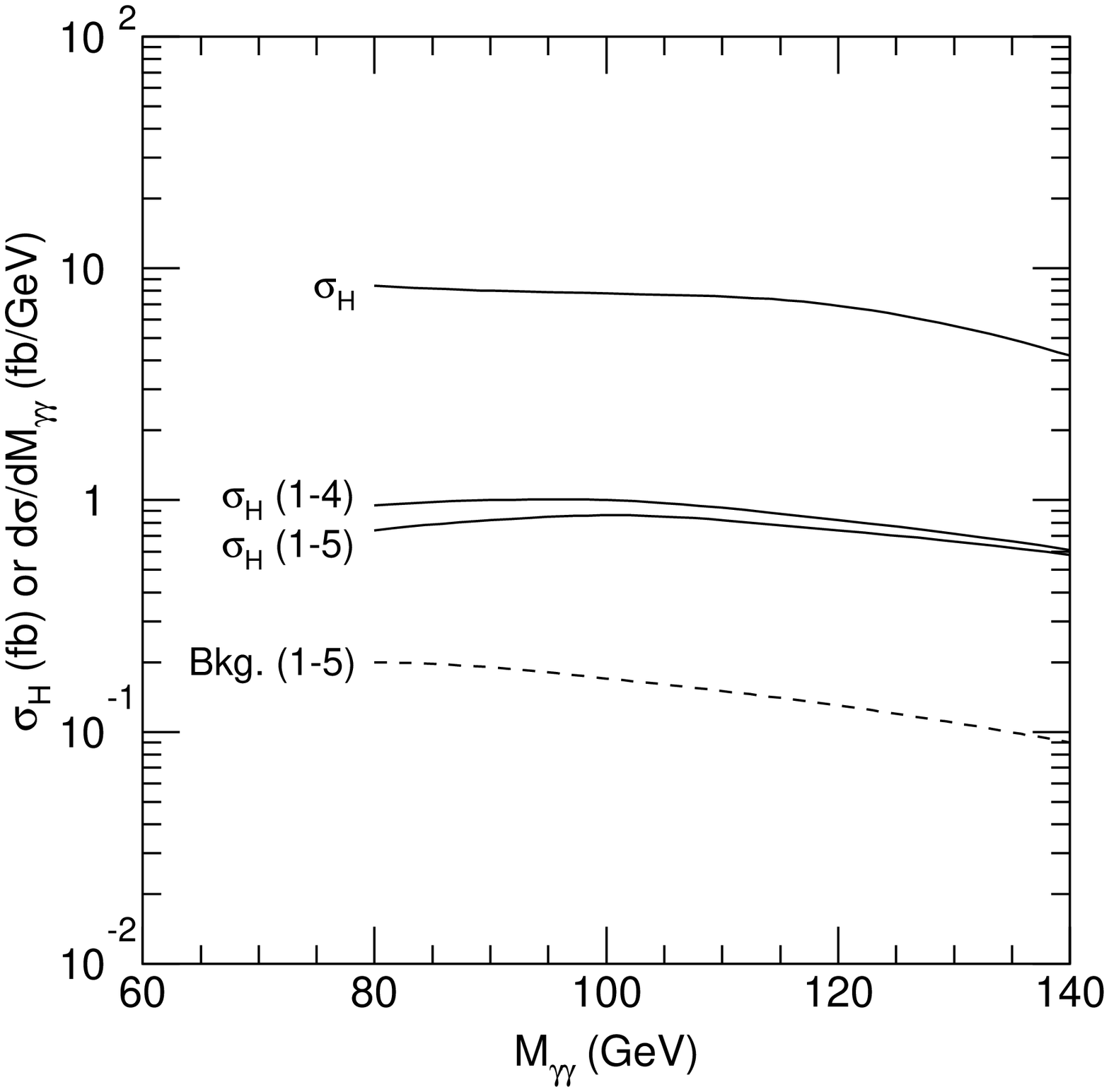}}{Signal and background
for $H(t \bar{t}/W) \to \gamma\gamma\ell X$ after the cuts described
in the text.}

\nfig\HFtt{\varfig{3.5in}{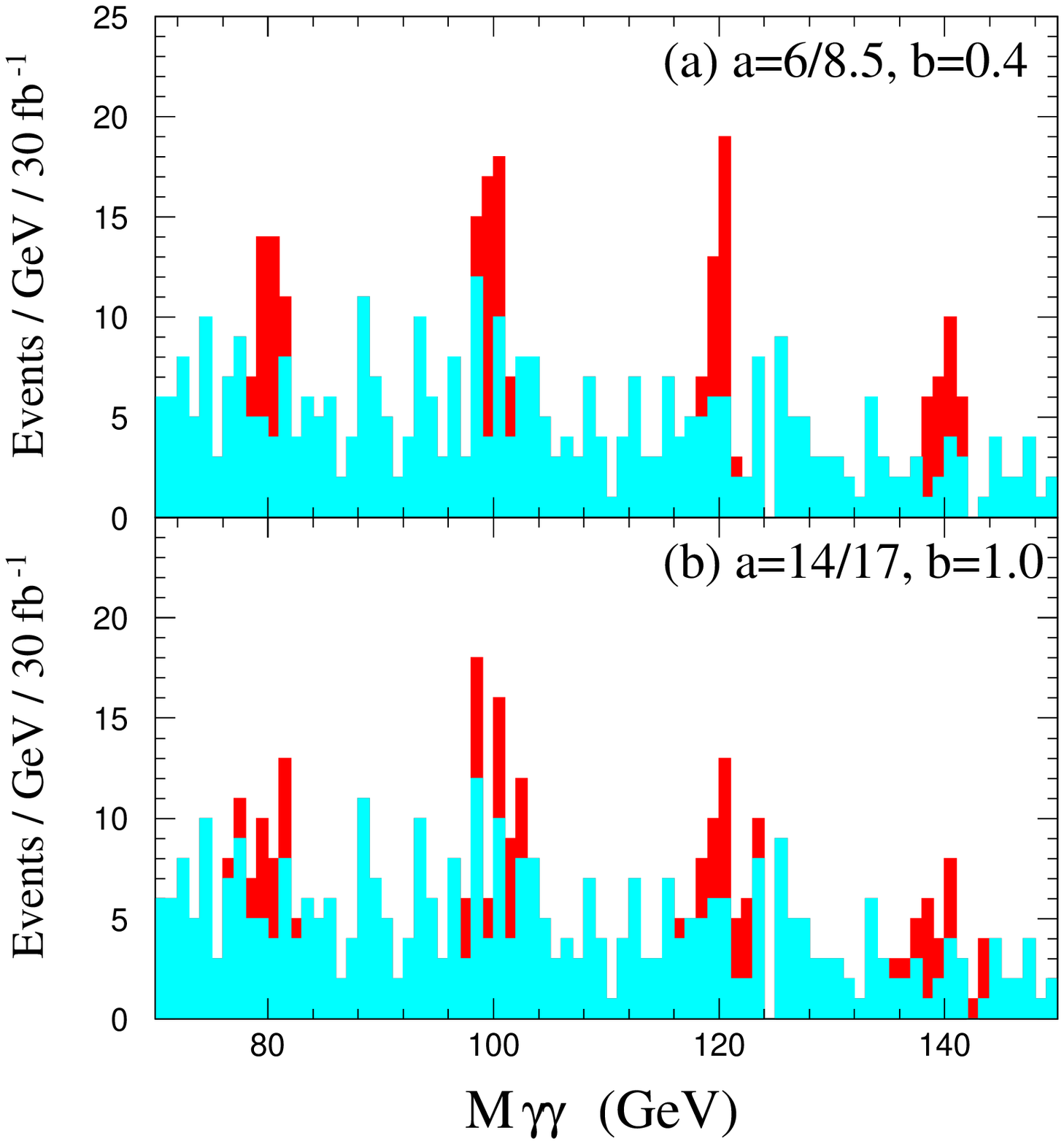}\forceright}{Two-photon mass peaks
for $H(t \bar{
t}/W)\to\ell\gamma\gamma X$ (black area) over sum of all backgrounds
(gray area), for an integrated luminosity of 30 fb$^{-1}$ and
$M_H$ = 80, 100, 120 and 140 GeV.}

\nfig\TFHGGLsig{\varfig{3.5in}{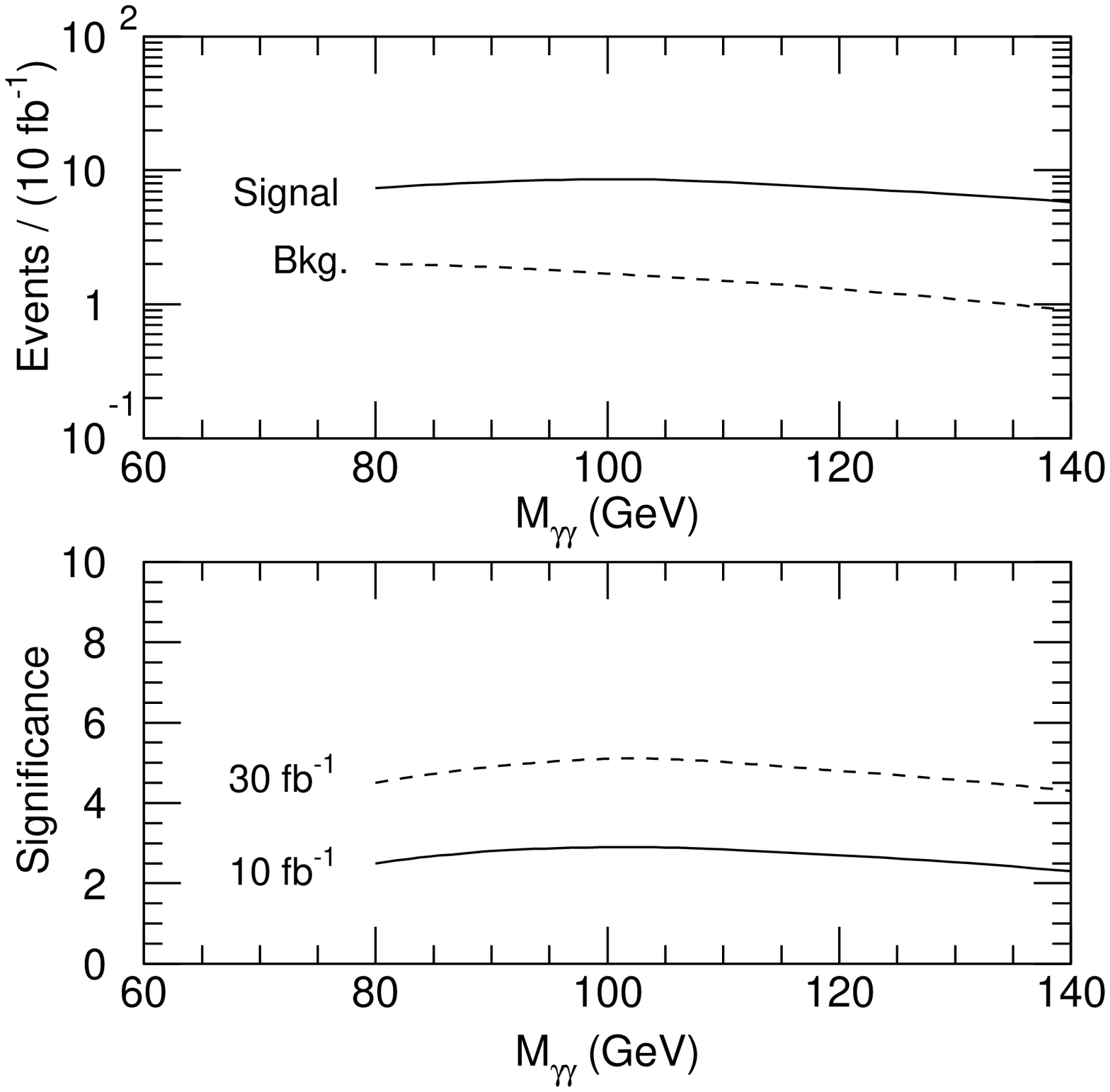}}{Number of events and
significance for $H(t\bar{t}/W) \to \ell\gamma\gamma X$ for various
masses and integrated luminosities.  Signal and background are
integrated over $\pm 2\Delta M_H$.}

	Figure \TFHGGLxsec\ shows the Higgs production cross section
($\sigma_H$) and the cross sections after event selection cuts 1--4
and 1--5 together with the total background rate after all cuts as a
function of mass. The largest backgrounds are the fake $\gamma\gamma$
pairs from $t \bar{t}$ production and the QED radiation $Z \gamma \to
\ell^+\ell^-\gamma\gamma$. A $Z$-mass veto improves the rejection of
the latter background, but it does not improve the significance.

	Figure~\HFtt\ shows the $\gamma\gamma$ invariant mass spectrum
collected in $30\,\fb^{-1}$ with Higgs signals of 80, 100, 120 and
$140\,\GeV$ superimposed on the sum of all backgrounds.  Figure
\TFHGGLsig\ shows the significance for the $H(t \bar{t}/W) \to
\gamma\gamma\ell X$ search with integrated luminosity of 10 and
$30\,\fb^{-1}$. Although the numbers of signal and background events are
small, this process provides higher significance than the inclusive $H \to
\gamma\gamma$ at very low masses. The combination of this channel and the
inclusive $H \to \gamma\gamma$ provides a more robust search for low
mass Higgs bosons than either one alone.

\subsubsec{$H\to ZZ^* \to\ellell\ellell$~Searches for $140\,\GeV < M_H <
2M_Z$}

	The $H\to ZZ^{(*)} \to \ellell\ellell$ decay provides the cleanest
Higgs signal. Because of the four isolated leptons in the final state, most
of the QCD background can be rejected by an isolation cut.  For a Higgs mass
larger than $2M_Z$, both $\ellell$ pairs have an invariant mass of $M_Z$, so
a mass constraint can also be used to reject background. If the Higgs mass
is less than $2M_Z$, one (or both) $Z$-bosons are off the mass shell. In
either case, good mass resolution is important for rejecting backgrounds.

The background processes considered are:

\item{1.} $ZZ/ZZ^*\to \ellell\ellell$.

\item{2.} $Q \bar{Q} Z \to \ellell\ellell + X$, where $Q = b$ or $t$.

\item{3.} $t \bar{t}\to W^+ b W^- \bar{b}$, in which the two $W$-bosons decay
semileptonically and the $b$-jets give isolated leptons.

\noindent The $ZZ/ZZ^* \to \ellell\ellell$ background is irreducible.
Since the cross section of $gg \to ZZ^*$ is not yet available in either
PYTHIA or ISAJET, its contribution was accounted for by multiplying the
contribution of $q\bar{q} \to ZZ^*$ by 1.65.\cite{\HRggzz}

The following cuts were used to reject the other backgrounds:

\item{1.} $|\eta^\ell|~< 2.5$ and $p_T^\ell > 10\,\GeV$. For electrons, the
region $1.01 < \abseta < 1.16$ was excluded.

\item{2.} Lepton isolation with $R = 0.35$ and $E_T^{\rm cut}$ = 5
GeV; see Section~5.1.

\item{3.} Lepton identification and track matching.

\item{4.} $10\,\GeV \le M_{\ell\ell}^{(l)} \le 100\,\GeV$ and
$70\,\GeV \le M_{\ell\ell}^{(h)} \le 100\,\GeV$ to suppress the
continuum background, where $M_{\ell\ell}^{(l)}$ and $M_{\ell\ell}^{(h)}$
are the low and high invariant masses of two $\ellell$~pairs.

\noindent The isolation, $p_T$, and mass cuts help reject the heavy flavor
background. The trigger efficiency for events passing these selection cuts is
higher than 98\% for the four-electron mode and 99\% for the other two modes.

\widetopinsert
\GEMcaption{Table 5}{Signal and Background for $H \to ZZ^* \to
\mumu\mumu$ and $H\to ZZ ^* \to \ellell\ellell$ for an integrated
luminosity of $\sscy$. The signal cross section after cuts is
$\sigma_{\rm accep}$.}
\medskip
\GEMtable
\hfil#\hfil\qquad & \hfil#\hfil\quad & \hfil#\hfil\quad &
\hfil#\hfil\quad & \hfil#\hfil\quad & \hfil#\hfil\cr
\GEMrulerule
$M_H$ (GeV)      & 140 & 150 & 160 & 170 & 180\cr
\GEMrulerule
\multispan{6}{\hfill $H\to \ee\ee$ \hfill}\cr
\GEMrule
$\Delta M_H$ (GeV)& 1.05 & 1.06  & 1.13  & 1.23 & 1.33\cr
$\sigma_{\rm accep}$ & 1.2  & 1.7  & .86  & .62  & 1.6\cr
Background (fb/GeV)   & .025    & .025    & .025    & .025    & .040\cr
\GEMrulerule
\multispan{6}{\hfill $H\to \mumu\mumu$ \hfill}\cr
\GEMrule
$\Delta M_H$ (GeV) & 1.59 & 1.62  & 1.73  & 1.84  &2.22\cr
$\sigma_{\rm accep}$ (fb)& .81  & 1.1   & .56  & .36   & .92\cr
Background (fb/GeV)   & .016 & .016  & .016 & .016  & .026\cr
\GEMrulerule
\multispan{6}{\hfill $H\to \ee\mumu$ \hfill}\cr
\GEMrule
$\Delta M_H$       & 1.36 & 1.46 & 1.56  & 1.71  &1.77\cr
$\sigma_{\rm accep}$ (fb)& 1.9 & 2.6  & 1.4 & 0.89   & 2.4\cr
Background (fb/GeV)   & .038 & .038 & .038  & .038  & .062\cr
\GEMrulerule
\multispan{6}{\hfill $H\to\ellell\ellell$ \hfill}\cr
\GEMrule
Significance       & 11  & 13   & 8.1  & 5.7   & 10\cr
\GEMrulerule
\endGEMtable
\endinsert

\nfig\HFeeee{\dofig{104.5mm}{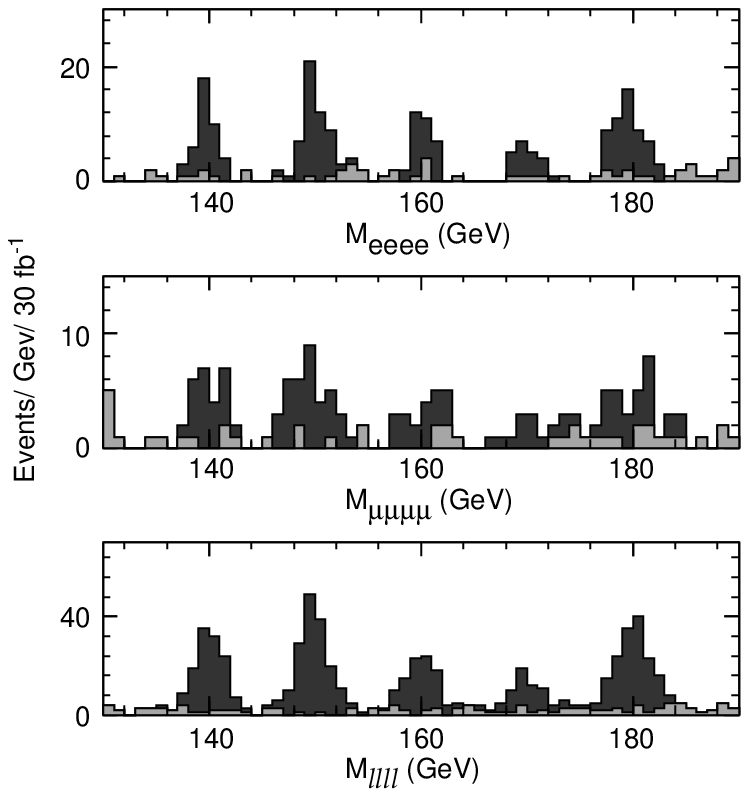}\forceright}{Higgs mass
peaks for $H \to ZZ^* \to \ellell
\ellell$ over the sum of all backgrounds plotted for an integrated
luminosity of $30\,\fb^{-1}$, and $M_H = 140$, 150, 160 and $170\,\GeV$.
(a) Four electrons. (b) Four muons. (c) All leptons.}

Table~5 shows the cross section after event selections ($\sigma_{\rm
accep}$) and the corresponding background rate for the various $H \to
\ellell\ellell$ channels. The integrated luminosity is $\sscy$. While the
mass resolution is excellent in the four-electron channel, it is not as
good in the muon channels because of multiple scattering of the relatively
low-energy muons in the second muon chamber superlayer.  (This is an
instance in which simulations directly influenced the detector. Changes
were made to the muon system design following the TDR work to increase its
acceptance and resolution for this four-muon process.) The significance
for all channels combined is also listed in Table~5. Using all the
four-lepton channels, GEM can discover a Higgs boson in this mass range
with an integrated luminosity of $\sscy$. The most difficult mass is
$170\,\GeV$ where the signal is only $5.7\sigma$. Figure~\HFeeee\ shows the
$\ee\ee$, $\mumu\mumu$, and $\ellell\ellell$ invariant mass spectra
collected in $30\,\fb^{-1}$ for $M_H = 140$, 150, 160, 170 and $180\,\GeV$
superimposed on the sum of all backgrounds. The increased integrated
luminosity makes even the peak at $170\,\GeV$ unambiguous.

\subsubsec{$H\to ZZ \to \ellell\ellell$ Searches for $2M_Z < M_H <
800\,\GeV$}

\widetopinsert
\GEMcaption{Table 6}{Signal and background for $H\to ZZ \to
\ellell\ellell$ for an integrated luminosity of $\sscy$.}
\medskip
\GEMtable
\hfil#\hfil\qquad & \hfil#\hfil\quad & \hfil#\hfil\quad &
\hfil#\hfil\quad & \hfil#\hfil\cr
\GEMrulerule
$M_{ H}$ (GeV)   & 200 & 400 & 600 & 800\cr
\GEMrulerule
\multispan{5}{\hfill Signal (fb) \hfill}\cr
\GEMrule
$\sigma_H$  & 85  & 56  & 16  & 5.3\cr
Mass Bin (GeV)   & $\pm$4.7 &350--450 & 500--800 & 600--1200\cr
$\sigma_{\rm accep}$ & 21& 14 & 4.3  & 1.5\cr
\GEMrulerule
\multispan{5}{\hfill Background (fb) \hfill}\cr
\GEMrule
$ZZ$	         & 3.0 & 2.3 & 1.0 & .6\cr
\GEMrulerule
\multispan{5}{\hfill Significance \hfill}\cr
\GEMrule
Significance     &  38   & 28   & 9.7 & 4.7\cr
\GEMrulerule
\endGEMtable
\endinsert

\nfig\HFllll{\dofig{88mm}{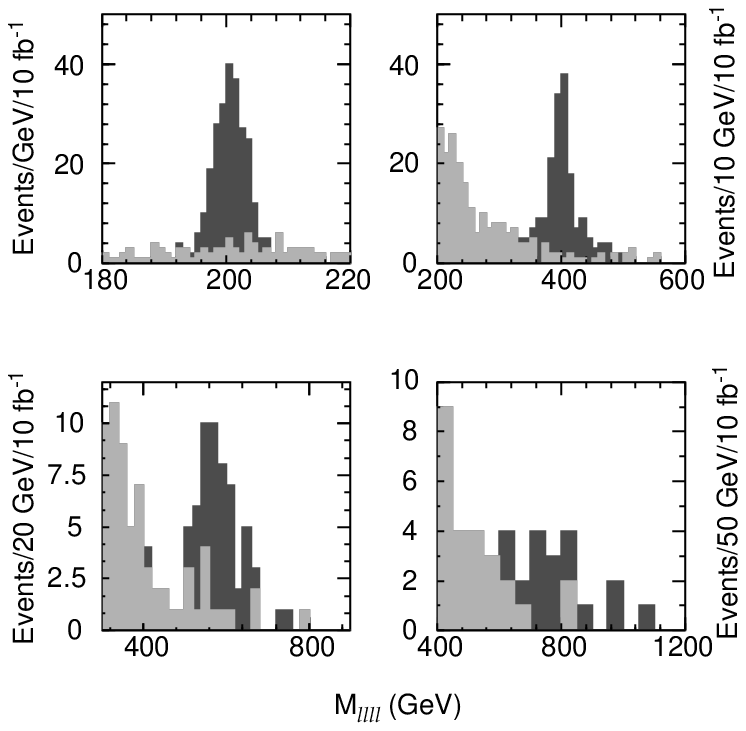}\forceleft}{Higgs mass peaks and
background for $H\to ZZ
\to \ellell\ellell$ for an integrated luminosity of $10\,\fb^{-1}$.
(a) $M_H = 200\,\GeV$, (b) $400\,\GeV$. (c) $600\,\GeV$. (d)
$800\,\GeV$.}

For a heavy Higgs decaying into $\ellell\ellell$, PYTHIA was used to
generate the signal and the $ZZ$ background. All other backgrounds are
negligible. The event selection cuts were taken to be:

\item{1.} $|\eta^\ell| < 2.5$ and $p_T^\ell > 10\,\GeV$.

\item{2.} Lepton isolation with $R = 0.3$ and $E_T^{\rm cut} = 5\,\gev$.

\item{3.} Lepton identification and track matching.

\item{4.} At least one $Z$ with $p_T >
\textstyle{{1 \over{4}}} \ts \sqrt{M_{ZZ}^2 - 4M_Z^2}$.

\item{5.} $|M_{\ell\ell} - M_Z| < 10\,\GeV$ for both lepton pairs.

\noindent Cuts~1, 2, 3, and 5 are self-explanatory. If the transverse
momentum of the $ZZ$ system is neglected, then cut~4 requires
$\sin\theta^* > 0.5$ and so reduces the more peripheral $q \bar q \to
ZZ$ background. The trigger efficiency for events passing these
selection cuts is higher than 99\%.

Table~6 shows the production cross section ($\sigma_H$) and the cross
section after event selection cuts ($\sigma_{\rm accep}$) for $M_H = 200$,
400, 600 and 800 GeV.  (A Higgs mass of $800\,\gev$ is somewhat greater than
the ``triviality'' bound,\cite{\trivial} but we consider it because it is the
greatest mass at which the Higgs still appears to give a distinct
resonance. Such a mass is possible, but it will be accompanied by
unspecified new physics at or below a few~TeV.)
The rate of the $ZZ$ background and the combined significance are also
listed for an integrated luminosity of $\sscy$. For such a data sample,
discovery of heavy Higgs bosons in the $\ellell\ellell$ modes will be
possible for all masses below about $700\,\gev$.  For the heaviest masses
one would seek confirmation of these four-lepton signals in other channels,
as we discuss next. We also expect that GEM's capabilities at ultrahigh
luminosity (see Section~5.6) will permit the discovery of a very heavy Higgs
in the four-lepton channel in about one year with a data sample of order
$\sscd$.

Figure~\HFllll\ shows the $\ellell\ellell$ invariant mass spectra collected
in $10\,\fb^{-1}$ for $M_H = 200$, 400, 600 and $800\,\GeV$ superimposed on
the background. While the $800\,\GeV$ peak looks ragged, the calculated
statistical significance is $4.7\sigma$ in $10\,\fb^{-1}$. There is also a
systematic uncertainty caused by lack of knowledge of the shape of the
background. This is discussed at the end of the following subsection.

\subsubsec{$H \to ZZ \to \ellell\nu\bar\nu$ Searches for $M_{\bf H} =
800\,{\bf GeV}$}

The branching ratio for $H \to ZZ \to \ellell\nu\bar\nu$ is six times that
for $H \to ZZ\to \ellell\ellell$.  Since one $Z$ decays to $\nu\bar\nu$, the
signature is a high-$p_T$ $Z \ra \ellell$ plus missing energy, and the mass
of the Higgs boson cannot be directly reconstructed.  The total signal cross
section is $32\,\fb$.

The following backgrounds were considered:

\item{1.} $q\bar{ q}/gg\to ZZ$; the cross section is $73\,\fb$ for $p_T^Z >
150\,\GeV$.

\item{2.} $q\bar{ q}\to Zg$ and $qg\to Zq$, with $Z\to\ellell$; the cross
section is $66\,\pb$ for $p_T^Z > 150\,\GeV$ before any $\etmiss$ cut.

\item{3.} $q\bar{q}, \ts gg \to t \bar{t}$, with $t \to b \ell\nu$; the cross
section is $380\,\pb$ before cuts.

\noindent To separate the signal from these backgrounds, events with an
$\ee$ or $\mumu$ pair were selected as follows:

\item{1.} $|\eta^\ell| < 2.5$ and $p_T^\ell > 20\,\GeV$ for each
lepton.

\item{2.} Lepton isolation with $R = 0.3$ and $E_T^{\rm cut} =
5\,\GeV$.

\item{3.} Lepton identification and track matching.

\item{4.} $| { M}_{\ell\ell} - { M}_Z | < 10\,\GeV$.

\item{5.} $E_T^Z > 250\,\GeV$.

\item{6.} $\etmiss > 250\,\GeV$.

\noindent Cuts 5 and 6 are primarily intended to reject the large $Z + {\rm
jets}$ background; see Fig.~\HFptz.  The trigger efficiency for events
passing all selection cuts is higher than 99\%. Table~7 lists the cross
sections after cuts 1--4, 1--5, and 1--6, for the signal and background in
the $\ee \nu\bar\nu$ and $\mumu\nu\bar\nu$ channels.

Figure~\HFmt\ shows the reconstructed transverse mass, $M_T^2 = 2 E_T^Z
\etmiss \ts (1 - \cos \Delta \phi)$, where $E_T^Z$ is the transverse energy
of the $Z$ and $\Delta\phi$ is the azimuthal angle between the direction of
the $Z$-boson and the $\etmiss$-vector. This distribution is not
sensitive to the degradation of the $\etmiss$ resolution that results from
adding to the calorimeter response a 1\% nongaussian tail with twice the
normal width (see Section~4.3). There are 105 signal events over a total
background of 91 for an integrated luminosity of $\sscy$. Since the signal
and background distributions are similarly shaped and the signal to
background ratio is only 1.15:1, this cannot be regarded as a convincing
discovery channel. A knowledge of the $ZZ$ continuum background to $\sim
25$\% would serve to give a $5\sigma$ systematic-limited significance.
This knowledge of the $ZZ$ background should be achievable by comparison
with $WZ$ and $Z +  \ts \jets$ production. The statistical $4.7\sigma$
significance of the four-lepton signal is reduced to $4.2\sigma$ by the
same background uncertainty. Combining the $\ell\ell\ell\ell$ and $\ell\ell
\ol \nu \nu$ channels with Gaussian statistics gives a $6.6\sigma$ signal
for $\sscy$.

\widetopinsert
\GEMcaption{Table 7}{Signal and Background (fb) for $H \to
\ellell \nu\bar{\nu}$}
\medskip
\GEMtable
\hfil#\hfil\qquad & \hfil#\hfil\quad & \hfil#\hfil\quad &
\hfil#\hfil\quad & \hfil#\hfil\cr
\GEMrulerule
\multispan{5}{\hfill $\ee \nu\bar{\nu}$ Channel \hfill}\cr
\GEMrulerule
              &   Signal &  $ZZ$      &  $Z + jets$     &  $t\bar{t}$\cr
\GEMrule
$\sigma_{H}$ &  16.0  &  36.7    &  3.3$\times 10^4$ & 2.0$\times 10^5$\cr
\GEMrule
After cuts 1--4&  9.5   &  13.8   &  1.2$\times 10^4$ & 53.5\cr
\GEMrule
After cuts 1--5&  7.4   &  4.2    &  3.1$\times 10^3$ & 0.36\cr
\GEMrule
After cuts 1--6&  6.2   &  3.2    &  2.2              & 0.0\cr
\GEMrulerule
\multispan{5}{\hfill $\mumu\nu\bar{\nu}$ Channel \hfill}\cr
\GEMrulerule
              &   Signal &  $ZZ$      & $Z + jets$     &  $t\bar{t}$\cr
\GEMrule
After cuts 1--4 &  6.5   &  9.4    &  8.1$\times 10^3$ & 36.2\cr
\GEMrule
After cuts 1--5 &  5.0   &  2.9     &  2.1$\times 10^3$ & 0.24\cr
\GEMrule
After cuts 1--6 &  4.2   &  2.2     &  1.5              & 0.0\cr
\GEMrulerule
\endGEMtable
\endinsert

\subsubsec{$H \to ZZ \to \ellell jj$ Searches for $M_{\bf H} = 800\,{\bf
GeV}$}

The branching ratio for $H \to ZZ \to \ellell jj$ for $\ell = e$ and $\mu$ is
approximately 20 times higher than that into all the $\ellell\ellell$ modes.
For $M_H = 800\,\gev$, the cross section is $110\,\fb$. The
signal-to-background ratio, however, is much worse than in the four-lepton
channel because of the large background from $Z+ \ts \jets$ production. In
addition, since the width of an $800\,\GeV$ Higgs is $270\,\GeV$, the signal
will be seen only as a broad excess of $ZZ$ pairs over the background.

The following backgrounds were studied in this analysis:

\item{1.} $ZW$ or $ZZ$ with one $Z\to\ee$ or $\mumu$.

\item{2.} $Z+ \ts \jets$ with $Z \to \ee$ or $\mumu$.

\item{3.} $t\bar{t}$ with $t\to bW$ and $W\to e\nu$ or $\mu\nu$.

\noindent The irreducible $ZZ$ background has a production cross section of
$130\,\fb$, for $p_T^Z > 200\,\GeV$. The $Z+ \ts \jets$ background has a
much larger rate of $27\,\pb$ for $p_T^Z > 200\,\GeV$. The $Z$-mass
constraint and high $p_T$ of leptons and jets were used to reduce this
background. (The possibility of tagging forward jets and vetoing on central
jets to enhance the signal-to-background ratio is discussed below.) The
cross section for the $t \bar{t}$ background with semileptonic decays to
$\ee + X$ and $\mumu + X$ is $380\,\pb$ before the $Z$-mass constraint is
imposed. A sample of $1.5\times10^6$ events was generated for this
background. It was largely eliminated by $\pt$ and isolation cuts and the
$Z$-mass constraint.

\nfig\HFptz{\dofig{90mm}{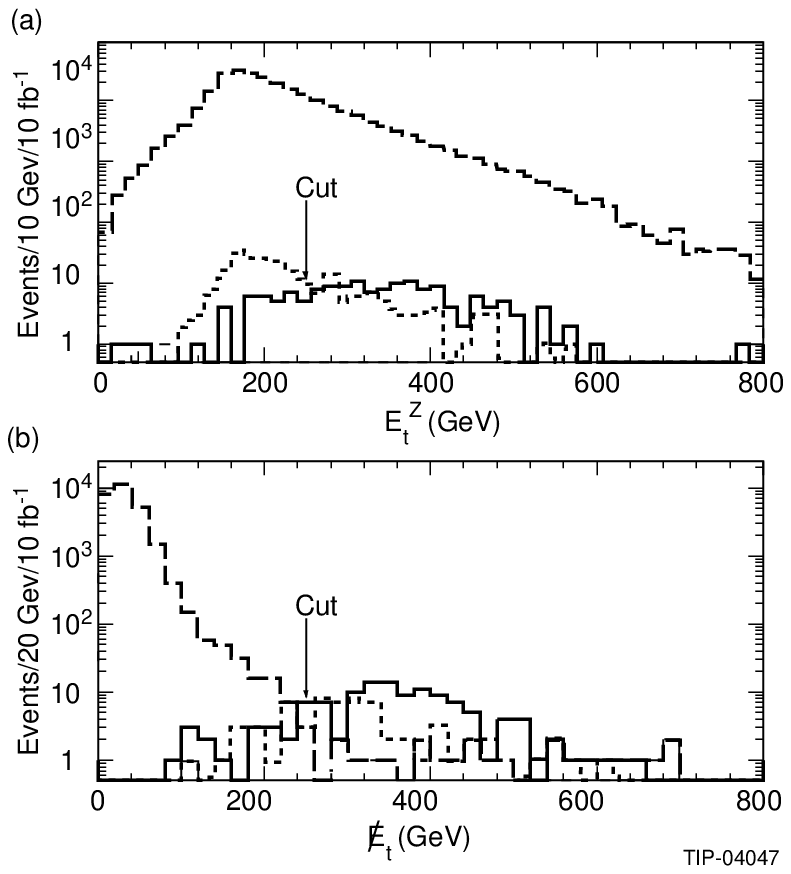}}{Distributions for an
$800\,\GeV$ $H \to \ellell \nu \bar\nu$ signal (solid), $ZZ$
background (dots), and $Z+ \ts \jets$ background (dash) for
$10\,\fb^{-1}$. The $t\bar{t}$ background is small. (a) Transverse
energy of the reconstructed $Z$.  (b) Missing transverse energy
($\slashchar{E}_T$).}

\nfig\HFmt{\dofig{88.5mm}{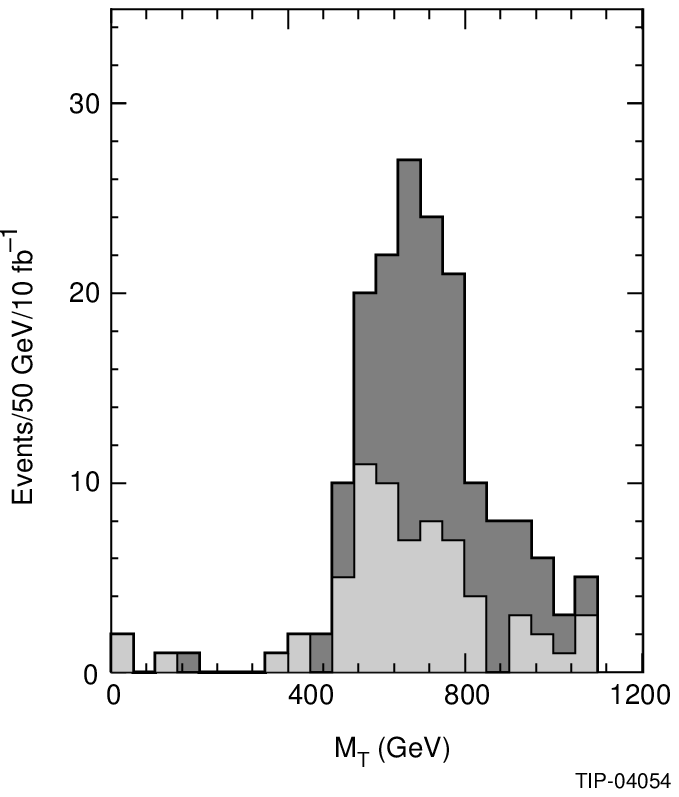}\forceright}{Transverse mass
$M_T$ of an
$800\,\GeV$ Higgs
signal for $H \to ZZ \to \ell^+\ell^-\nu\nu$ (black area) and sum of
all backgrounds (gray area), obtained for an integrated luminosity of
10 fb$^{-1}$.}

Leptons were selected as follows:

\item{1.} $|\eta^\ell|~< 2.5$ and $p_T^\ell > 70\,\GeV$
for each lepton.

\item{2.} Lepton isolation with $R = 0.3$ and $E_T^{\rm cut} =
5\,\GeV$.

\item{3.} Lepton identification and track matching.

\item{4.} $|M_{\ell\ell} - M_Z| < 10\,\GeV$.

\item{5.} $p_T^{\ell\ell} > 230\,\GeV$.

\noindent The hadronically-decaying $Z$-boson was reconstructed with the
following algorithm:

\item{1.} Find all jets at $\abseta < 2.5$ using a large clustering cone,
$R = 0.9$.

\item{2.} Find all jets with a small cone, $R = 0.3$, and match
these narrow jets with those found using the larger cone.

\item{3.} Require the highest $p_T$ jet found with $R = 0.9$ to have $p_T
> 250\,\GeV$ and to be composed of two jets found with $R=0.3$, each having
$p_T~> 80\,\GeV$.

\item{4.} Reconstruct the mass $M_{jj}$ of the highest-$p_T$ dijet with $R =
0.9$, and require that $|M_{jj} - M_Z| < 15\,\GeV$.

\noindent The trigger efficiency for events passing all these selection
cuts is higher than 99\%.

\nfig\HFlljj{\dofig{87mm}{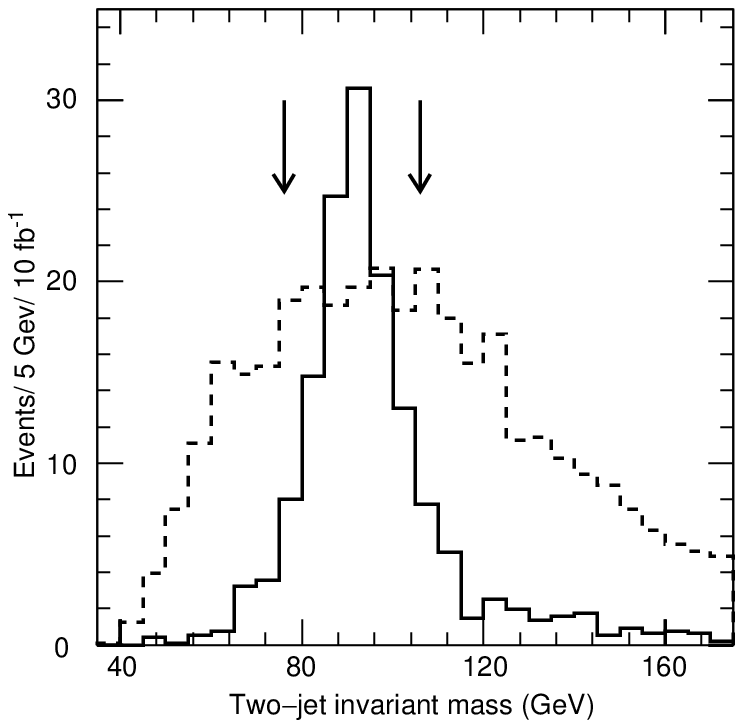}}{Dijet invariant mass distribution
for $H \to
ZZ \to \ell^+\ell^-jj$ for $M_H=800\,\GeV$ (solid) and sum of all
backgrounds (dashed), obtained for an integrated luminosity of
$10\,\fb^{-1}$. The background  is multiplied by a factor of 0.2. to facilitate
comparison of the shapes.}

\widetopinsert
\GEMcaption{Table 8}{Signal and background cross sections (fb) for $H \to
ZZ \to \ellell jj$ for $\sscy$. The invariant mass $M_{\ell\ell jj}$
signal region is 600 to $1000\,\gev$.}
\medskip
\GEMtable
\hfil#\hfil\quad & \hfil#\hfil\quad & \hfil#\hfil\quad &
\hfil#\hfil\quad & \hfil#\hfil\cr
\GEMrulerule
Cut   & Higgs & $t\bar {t}$ & $Z+$jets & $ZW/ZZ$\cr
\GEMrule
After lepton cuts & 41  & 130 & 3240 & 20\cr
After jet cuts & 12  & 0.5 & 65   & 1.5\cr
Within $600<M_{\ell\ell jj}<1000$ & 11  & 0.3 & 42   & 0.8\cr
\GEMrulerule
\endGEMtable
\endinsert

	Figure~\HFlljj\ shows the reconstructed dijet mass $M_{jj}$ for the
signal and backgrounds, and the dijet mass cut is indicated.  Note that the
background has been multiplied by 0.2 in this plot. The mass
resolution is about $9\,\GeV$.  Table~8 shows the cross sections for signal
and backgrounds after the lepton and jet cuts for $600 < M_{\ell\ell jj} <
1000\,\GeV$.

\nfig\HFlljjM{\dofig{87mm}{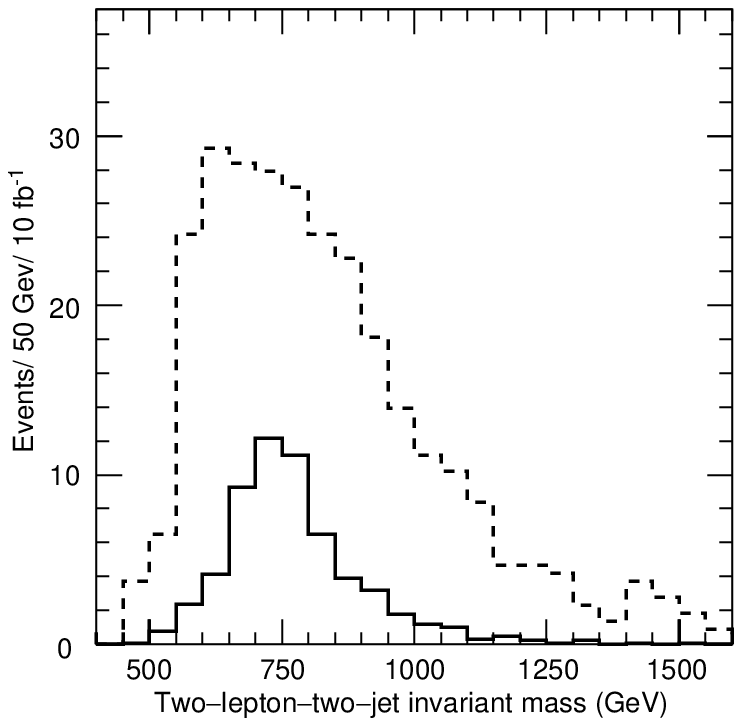}\forceright}{Invariant mass
distribution
$M_{\ell^+\ell^- jj}$ for 800 GeV $H\to ZZ \to \ell^+\ell^- jj$ signal
(solid) and sum of all backgrounds (dashed), obtained for an
integrated luminosity of $10\,\fb^{-1}$.}

	The invariant mass $M_{\ell\ell jj}$ of the $Z$-jet system is shown
in Fig.~\HFlljjM. The signal and background events passing all these cuts
are 110 and 430 respectively for $10\,\fb^{-1}$. The signal and background
have a very similar shape, with no recognizable peak in the mass spectrum.
Since the signal is 25\% of the background, the latter must be known to
better than 5\% to achieve a $5\sigma$ significance. Thus, this mode cannot
be regarded as a discovery channel for the Higgs.
\nref\sergey{S.~Schevchenko, Private communication.}

The possibility of tagging forward jets to enhance the signal-to-background
was studied.\cite{\sergey,\caltech} For this, a segmentation of the forward
calorimeter of $\Delta \eta \times \Delta \phi = 0.3 \times 0.3$ was
assumed; similar results were obtained with $0.2\times 0.2$. A resolution
of $100\%/\sqrt{E} \oplus 5\%$ was used to smear the hadronic energies. One
forward jet was required in $2.5 < |\eta| < 4.5$ with $E > 2\,\tev$ and
$\pt > 50\,\gev$. After the $\ell\ell\ts\jet\ts\jet$ cuts described above,
the signal was reduced to 24 events on a background of~32. Again, the shape
of signal and background were similar, so there is no significant advantage
in tagging a single forward jet. Requiring two tagged forward jets is
expected to reduce the signal and backgrounds to a level comparable with
the four charged lepton
channel and, so, will not be of help. Finally, a central jet veto
was found not to
be useful after the high-$\pt$ cuts on the jets from $Z$ decay were
imposed.

\subsec{Heavy Flavor Physics}

\gdef\hp{H^+}

\gdef\hpm{H^\pm}
\gdef\hcs{H^+ \ra c \ol s}
\gdef\htn{H^+ \ra \tau^+ \nu_\tau}

As we have emphasized, while the characteristic energy of electroweak
symmetry breaking lies within reach of multi-TeV hadron colliders such as
the SSC and LHC, the scales of flavor symmetry and the interactions that
break it are unknown; they may be near 1~TeV or much higher. Nevertheless,
searching for the origin of flavor symmetry and its breakdown at hadron
colliders is just as important as it is for electroweak symmetry breaking.
Given how little is known of this physics, it is essential that detectors
be designed to allow for the broadest possible investigations. This can be
done only by testing designs against a wide variety of scenarios for flavor
physics.

Top-quark physics, in particular, will be an important part of a
supercollider's physics program for many reasons. Even if the top quark has
been found at the Tevatron Collider,\cite{\topmass} precise measurements of
its mass and and other properties may not be possible there because of
limitations in the detectors and the machine itself. The top quark's
apparently very large mass of 175~GeV is our most dramatic example of
flavor symmetry breaking. Thus, if other manifestations of flavor physics
are accessible at supercollider energies, the top quark will be an
important key to finding them. If charged scalar bosons, $\hpm$, exist,
then either $t \ra \hp b$ or $\hp \ra t \ol b$ is expected to be an
important decay mode. Also, many technicolor models contain a color-octet,
spin-zero ``technieta'' boson, $\eta_T$, with mass in the range $300 -
500\,\gev$. The $\eta_T$ would be produced via gluon fusion and
is expected to decay into $\ol t t$ pairs.%
\nref\etaT{S.~Dimopoulos, Nucl.\ Phys.\ {\bf B168}, 69 (1980)\semi
T.~Appelquist and G.~Triantaphyllou, Phys.\. Rev.\ Lett.\ {\bf 69}, 2750
(1992).}%
\nref\eekleta{E.~Eichten and K.~Lane, Phys.~Lett.~{\bf 327B}, 129 (1994).}%
\cite{\ehlq,\etaT,\eekleta}
Its cross section at the LHC may be enormous.
At the same time, $t$-quark production processes can be serious backgrounds
to new physics. An example of this was encountered in Section\ 5.2,
where it was seen that isolated-photon backgrounds arising in $\ol tt$
production were an important background to the $\ol t t \h \ra \ellpm
\gamma\gamma$ signal for production of an intermediate mass Higgs boson.
Thus, it is essential to know as much as possible about the rate and other
characteristics of $t$-quark production. Finally, the fact that
$t$-production gives a pure sample of $W$-bosons, with $M_W$ and $m_t$
accurately known, may be useful for calibrating the calorimeters for jet
energy measurements.

The work reported in this section was done over a year before the
announcement by the CDF Collaboration of its evidence for the top-quark.
The top-quark masses considered in this section are 140~GeV (as generally
used in this report) and approximately 250~GeV. These bracket fairly well
the 175~GeV value reported by CDF. Therefore, we have not felt it necessary
to repeat the analyses for the CDF central value. This section discusses
the following examples of top physics in the context of the GEM detector:

\item{1.} Discovery and measurement of the mass of heavy top quarks ($m_t
\simeq 250\,\gev$) pair-produced by $gg, \ts \ol q q \ra t
\ol t$ and decaying via the standard mode $t \ra W^+ b$. In this case, one
$t$-quark gives an isolated electron or muon from $W$-decay plus a
non-isolated muon from the decay $b \ra c \mu \ol \nu_\mu$, and the top
mass is determined from the invariant mass distribution, $\Mlm$, of the
isolated lepton and the non-isolated muon.%
\nref\CYtopll{C.~Yanagisawa, {\it Top-quark Detection in the Multileptons Mode
with the GEM Detector}, GEM TN---93--371 (1993).}%
\cite{\response,\CYtopll} This study may be viewed as a model of using
multi-lepton modes to search for a very heavy quark decaying into a light
one plus a $W$-boson.

\item{2.} Discovery and mass measurement of top quarks in $\ol t t$
production, both with $m_t = 140\,\gev$ and with $m_t = 250\,\gev$. Again,
the standard $t \ra W^+ b$ decay mode is assumed. Events are tagged by one
isolated high-$\pt$ lepton ($e$ or $\mu$) from decay of one of the
$W$-bosons and several high-$\pt$ jets. The other $W$-boson and its
parent $t$-quark are observed in their hadronic decay modes.%
\ref\CYtopjjj{C.~Yanagisawa, {\it Top-quark Detection in the Multijets
Mode with the GEM Detector}, GEM TN--93--372 (1993).}

\item{3.} Discovery and study of a charged scalar $\hp$ produced in
the decay of a heavy top quark, $t \ra \hp b$. The scalar will be assumed
to decay as $\htn \ra$ one or three prongs.%
\ref\MMhtn{M.~Mohammadi and D.~Skrzyniarz, GEM TN--93--363 (1993).}
For this search, one $t$-quark is tagged using the standard mode $t \ra
W^+ b \ra$ isolated lepton.

\subsubsec{Heavy Top-Quark Detection and Mass Measurement via the $\Mlm$
Distribution}

This analysis\cite{\CYtopll} uses $\ol t t$ events with two leptonic top
decays and an additional non-isolated muon from $b \to \mu X$. The top mass
is determined from the mass $M_{\ell\mu}$ of the non-isolated muon and the
isolated lepton of the opposite sign. While the $M_{\ell\mu}$ distribution is
broad, this analysis avoids hadronic energy measurements and minimizes the
sensitivity to the top production dynamics.

ISAJET 6.36 with the EHLQ-1 distribution functions was used to generate 60K
$\ol t t$ events for each of the masses $m_t = 200$, 230, 250, 270 and
$300\,\gev$. (The $t$-quarks were generated with $50\,\gev < \pt <
1000\,\gev$, and the appropriate decays were forced.)  The signal events are
those with one isolated electron and one isolated muon of opposite sign,
together with one non-isolated muon of either sign. Choosing isolated
leptons of different flavors eliminates backgrounds from $Z^0 \ra
e^+e^-, \ts \mu^+ \mu^-$.  The signal cross sections (assuming $B(W \ra \ell
\nu) = B(b \ra \mu^- + X) = 1/9$) for $m_t = 200$, 250 and $300\,\gev$ were
found to be $19.9\,\pb$, $8.2\,\pb$ and $4.0\,\pb$, respectively.

The signal cross sections are much larger than all backgrounds, as discussed
below. To optimize the separation between isolated leptons from $W$-decay
and non-isolated ones from $b$-decay, the following cuts were imposed to
select isolated electrons:
\eqn\isole{\eqalign{
& \vert\eta_e\vert < 2.4  \cr
&\sum_{R = 0.2} \et({\rm EC}) > 40\,\gev \cr
&\sum_{R = 0.4} \et - \sum_{R = 0.2} \et({\rm EC}) < 10\,\gev \cr
& 0.8 < \et(R=0.2)/\pt({\rm CT}) < 1.2 \,,\cr}}
\noindent where $\et$ is the transverse energy in a cell of the full
calorimeter; $\et({\rm EC})$ is the transverse energy in EM calorimeter
cells; $\pt({\rm CT})$ is the electron momentum as measured in the tracker.
As usual, the sums are over $\eta-\phi$ cones of radius $R$.
The cuts for isolated muons were:
\eqn\isolmu{\eqalign{
& \vert\eta_\mu\vert < 2.4 \cr
&\pt > 40\,\gev \cr
&\sum_{R = 0.4} \et - \sum_{R = 0.1} \et < 2\,\gev\,. \cr}}
\noindent Non-isolated muons were required to satisfy:
\eqn\inclmu{\eqalign{
& \vert\eta_\mu\vert< 2.4 \cr
&\pt > 20\,\gev \cr
&\sum_{R = 0.4} \et - \sum_{R = 0.1} \et > 10\,\gev\,. \cr}}

The choice $R = 0.2$ for the cone defining the electron energy is
not critical because the electron energy measurement does not need to be
very precise here. The $\Mem$ and $\Mmm$ distributions turn out to be very
similar for this choice of isolation parameters and not very sensitive to
moderate changes. For the relatively low-$\pt$ muons in the top-quark
decays, the muon energy may be corrected simply by adding
the average energy loss in the calorimeter for a given momentum and rapidity.
For a top mass of $250\,\gev$, the isolation cut \isolmu\ retained 78\% of
the muons from $W$-decay while rejecting 99\% of the muons from $b$-decay.
The non-isolation cut in \inclmu\ accepted 89\% of $b \ra \mu + X$ and
rejected 91\% of $W \ra \mu \nu_\mu$.

There was a 30\% contamination of the $t \ol t \ra e^\pm_{\rm isol} \ts
\mu^\mp_{\rm isol} + \mu_{\rm non-isol}$ signal due to non-isolated
muons in the decay of $c$-quarks which, in turn, came from the decay of
the ``wrong'' $t$-quark or from $g \ra c \ol c$. This background was
reduced to the 15\% level by selecting events in which the isolated lepton
and non-isolated muon of opposite sign are close to each other and, hence,
more likely to be from the decay of the same $t$-quark.
Events were required to satisfy
$$
\delta \phi(\ell \mu) < 90^\circ\,.
$$
The acceptance of this cut was found to be 67\%, 56\% and 49\% for $m_t =
200$, 250 and $300\,\gev$.  A detailed simulation showed that the
reconstruction efficiency was greater than 95\% even for non-isolated muons
from $t \to b \to \mu$ decays.  The overall acceptance ranged from 1.1\% for
$m_t = 200\,\gev$ to 1.8\% for $m_t = 300\,\gev$ both for the $t, \ts \ol t
\ra e^\pm_{\rm isol} \ts \mu^\mp_{ \rm non-isol}$ events and for the
$\mu^\pm_{\rm isol} \ts \mu^\mp_{\rm non-isol}$ events. These efficiencies
are low because the analysis was designed to obtain a very clean sample. The
total number of $\ell^\pm_{\rm isol} \ts \mu^\mp_{\rm non-isol}$ events
expected with an integrated luminosity of $\sscy$ is 4500, 2400 and 1400 for
$m_t = 200$, 250 and $300\,\gev$.

Backgrounds to the $ t \ol t \ra e^\pm_{\rm isol} \ts \mu^\mp_{\rm isol}
\ts \mu_{\rm non-isol}$ signal from production of $W^\pm + \ts \jets$ and
$Z^0 + \ts \jets$ were considered. The most important contributions come
from $b$-quark jets and $Z^0 \ra \tau^+ \tau^- \ra e^\pm + \mu^\mp +
\etmiss$. After the selections described above, the $W^\pm + \ts \jets$
background was found to be less than 9\% of the signal for the $250\,\gev$
top quark; the $Z^0 + \ts \jets$ background was negligible.\cite{\CYtopll}

\nfig\TMlm{\dofig{88mm}{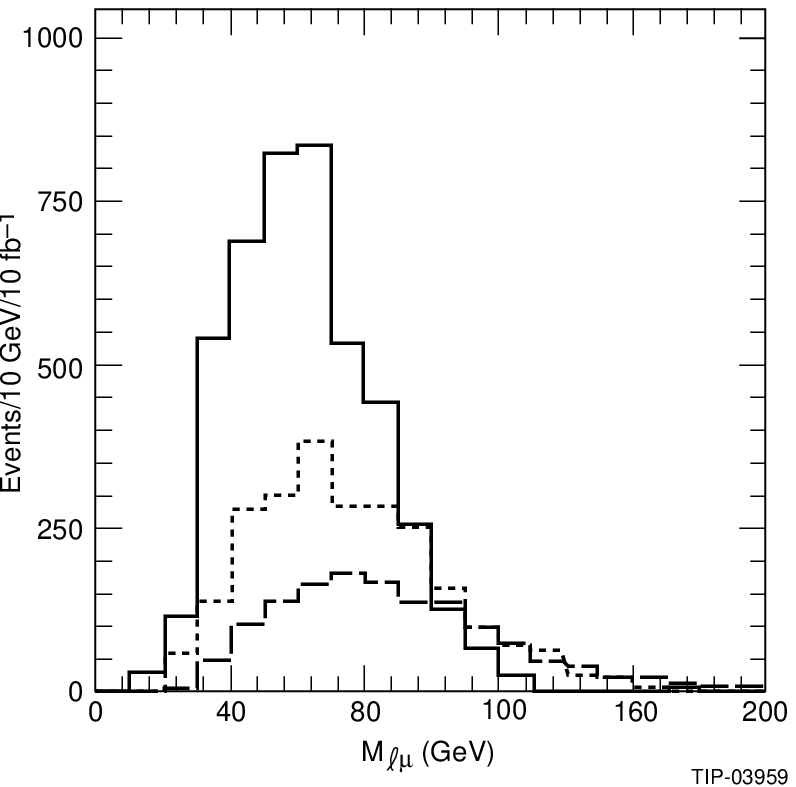}}{$\Mlm$ distribution for an isolated
lepton and a nonisolated muon of opposite sign for $m_t = 200$ (solid), 250
(dotted), and $300\,\gev$ (dashed).}

\nfig\TMMlm{\dofig{87.5mm}{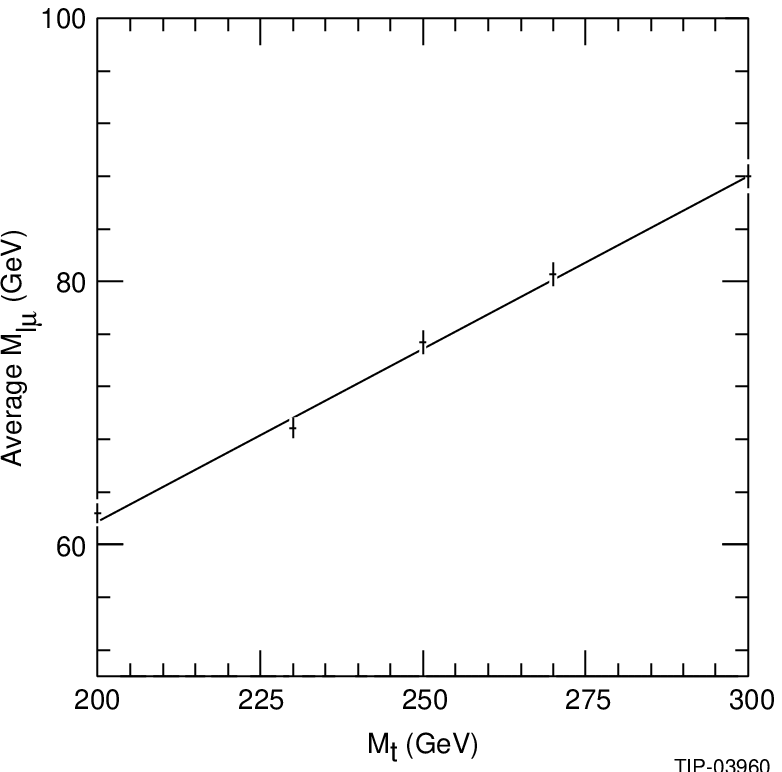}\forceright}{Mean value of
$\Mlm$ vs.\ $m_t$.}

\nref\kltop{K.~Lane, {\it Top-Quark Production and Flavor
Physics --- The Talk}, to appear in the Proceedings of the 27th International
Conference on High Energy Physics, Glasgow, 20-27th~July 1994; Boston
University Preprint BUHEP--94--25 (1994).}

The shapes of the $\Mem$ and $\Mmm$ distributions are nearly identical.
Thus, adding them together to form a $\Mlm$ distribution adds statistical
weight to the $m_t$ determination without introducing significant
systematic error. The $\Mlm$ distributions are plotted in Fig.\ \TMlm\ for
several values of $m_t$. The mean value of $\Mlm$ should vary linearly with
$m_t$ and does, as can be seen from Fig.\ \TMMlm (also see
Ref.~\kltop). A statistical error of
$2.3\,\gev$ on the mass of a $250\,\gev$ top quark is expected for $\sscy$
of data.

The systematic error in this method of determining $m_t$ arises from
imperfect knowledge of the $b$-quark fragmentation function and of the $\pt$
distribution of the $t$-quarks. These affect the momentum distributions of
the non-isolated muon and the isolated lepton, respectively. The effect of
heavy-quark fragmentation was studied by varying the parameter $\epsilon$ in
the Peterson fragmentation function from its nominal value of $\epsilon =
0.006$.\cite{\PDG} The range considered, $1 \sigma = 0.002$, led to a
variation of $2.3\,\gev$ in the mass of a $250\,\gev$ $t$-quark. A measure
of the systematic error due to the $\pt(t)$-distribution was obtained by
varying the amount of initial state radiation in $\ol t t$ production. This
resulted in a $3.4\,\gev$ change in the mass determined for the $250\,\gev$
$t$-quark. Finally, next-to-leading-order QCD corrections to the $\ol t t$
cross section increase the magnitude of the cross section by about 60\%, but
do not significantly change the shape.%
\nref\qcdcorr{P.~Nason, S.~Dawson and R.~K.~Ellis, Nucl.\ Phys.\ {\bf
B303}, 607 (1988)\semi
W.~Beenakker, H.~Kuijf, W.~L.~van~Neerven and J.~Smith, Phys.\ Rev. {\bf
D40}, 54 (1989)\semi
E.~Laenen, J.~Smith and W.~L.~van~Neerven, Nucl.\ Phys.\ {\bf B369}, 543
(1992);\ FERMILAB--Pub--93/270--T\semi
K.~Ellis, {\it Top-Quark Production Rates in the Standard
Model}, invited talk in Session Pa--18 given of the 27th International
Conference on High Energy Physics, Glasgow, 20--27th July 1994.}
\cite{\qcdcorr} Still higher-order corrections are expected to amount to
$\pm 15\%$. We conclude that, with an integrated luminosity of $\sscy$, it
would be possible to use the $\Mlm$ method to determine the mass of a
$250\,\gev$ top quark to within an error of $\pm 2.3\,\gev \ts {\rm
(statistical)}$ and $\pm 4.1\,\gev \ts {\rm (systematic)}$.

\subsubsec{Top-Quark Detection and Mass Measurement via the $M_{{\bf jjj}}$
Distribution}

The most direct measurement of the top-quark mass comes from the nonleptonic
decay $t \to W b \to \bar qq' b$, giving three jets. This is also the most
precise measurement if systematic uncertainties associated with jet
definition and energy measurement are under control. Furthermore, it is
important for flavor physics spectroscopy to observe the top quark in
nonleptonic decay modes.

ISAJET was used to generate 300K $\ol t t$ events for $m_t = 140$ and for
$250\,\gev$.\cite{\CYtopjjj} As above, the top quarks were generated with
$50\,\gev < \pt < 1000\,\gev$. Events were selected in which one $W$ decayed
to an electron or muon while the other $W$ decayed nonleptonically.  The
physics background to $\ol t t \ra \ellpm_{\rm isol} + \ts \jets$ comes
mainly from production of $W + \ts \jets$. Since both the signal cross
section and the signal/background ratio are much higher than at the
Tevatron,\cite{\topmass} it is possible to make cuts that render this
background unimportant. Thus, a non-isolated muon tag was not required. The
ISAJET cross sections for these events, including $W$-boson branching
ratios, are $4.1\,\nb$ and $0.44\,\nb$ for
the two values of $m_t$. The isolated leptons were required to satisfy the
cuts in Eqs.\isole\ and \isolmu.

To eliminate combinatorial backgrounds in the multijet mass distributions,
the $t$-quark and individual jets were forced to be at high $\pt$ and to be
in the hemisphere opposite the isolated lepton. For such events, the three
jets from the decays of the $t$ tend to be close, and a cone of small radius
$R=0.4$ was therefore used. Individual jets were required to have $\pt >
30\,\gev$ ($50\,\gev$) for $m_t = 140\,\gev$ ($250\,\gev$). In both cases,
jets were required to satisfy $\delta \phi(\ell \ts \jet) > 90^\circ$.
Finally, the three highest-$\pt$ jets satisfying these constraints were
required to have $\vert \vec \pt(3 \ts \jets) \vert > 200\,\gev$
($300\,\gev$) for $m_t = 140\,\gev$ ($250\,\gev$). The combined geometrical
acceptance and efficiency of these cuts was found to be 0.70\% and 1.0\% for
$m_t = 140$ and $250\,\gev$. The number of events obtained per $\sscy$,
taking into account trigger and reconstruction efficiencies and these
acceptances, were 240K and 40K respectively. The $W + \ts \jets$ background
was found to be negligibly small for the $140\,\gev$ case. It contributed
15\% to the three-jet mass spectrum for $250\,\gev$, but this had no important
effect on the determination of $m_t$.

\nfig\TjjjMw{\dofig{156mm}{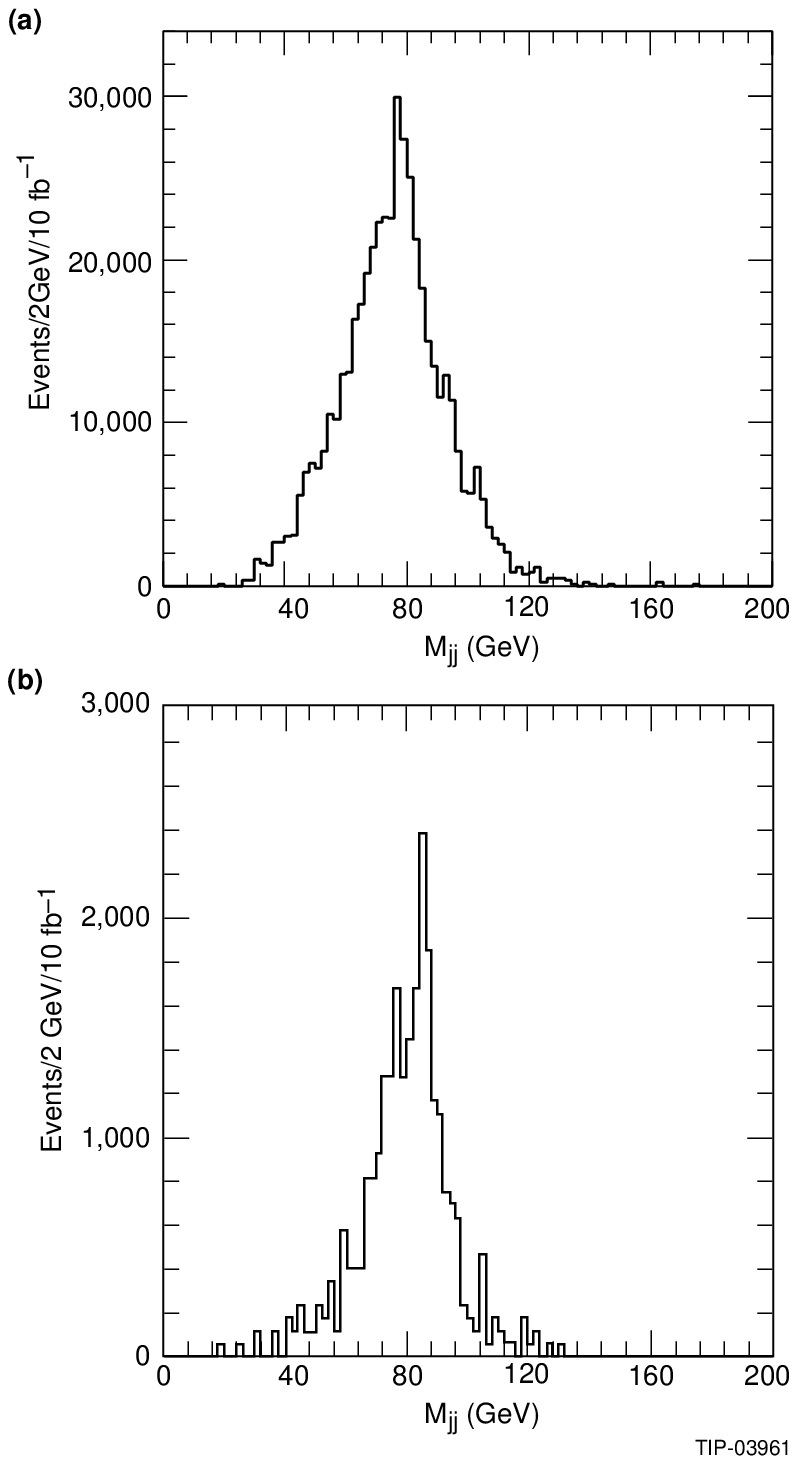}\forceleft}{Dijet mass distribution
in $t \ol t$ events with $m_t = 140\,\GeV$. (a) No $b$-tagging was assumed.
(b) $b$-tagging was assumed with 20\% efficiency.}

Consider first the case of $m_t = 140\,\gev$. The $W \ra 2 \ts \jets$ and $t
\ra W b \ra 3 \ts \jets$ mass distributions, $M_{jj}$ and $M_{jjj}$, are
shown in Figs.\ \TjjjMw\ and \TjjjMt. Here, the trijet search
region was $110\,\gev < M_{jjj} < 170\,\gev$, and at least one dijet pair was
required to satisfy $\vert M_{jj} - M_W \vert < 20\,\gev$. All dijet pairs
passing the cuts appear in Fig.\ \TjjjMw\ (a).  The fitted $W$-mass peak is
at $75.8\,\gev$ in this figure, and the resolution is $13.9\,\gev$.
The large combinatorial background is a consequence of the
kinematics: for $m_t = 140\,\gev$ and $M_W = 80\,\gev$, the
three jets tend to be roughly equidistant in $\eta-\phi$ space. The
background comes from picking up the $b$-jet from the same or, less
frequently, the other $t$-quark decay. The signal-to-noise ratio in the
$W$-peak region is about 2:1.

This signal-to-noise can be improved considerably by requiring that one of
the jets passing the above cuts be tagged as a $b$ jet by the tracker.
Figure\ \TjjjMw\ (b) shows the $M_{jj}$ distribution for events with an
identified $b$ jet which is then excluded from the dijet mass. The $b$-tag
efficiency for this plot was assumed to be 20\%. The $W$ signal-to-noise
ratio is improved to 3.9:1. The fitted $W$ mass in this plot is at
$76.9\,\gev$ with a resolution of $8.4\,\gev$. We emphasize that this
$b$ tag is not required for determining $m_t$. This is seen in Fig.\
\TjjjMt, where the top-quark peak appears clearly above little combinatorial
background. The mass was determined to be $138.2\,\gev$ with a resolution of
$8.1\,\gev$.

\nfig\TjjjMt{\dofig{79.5mm}{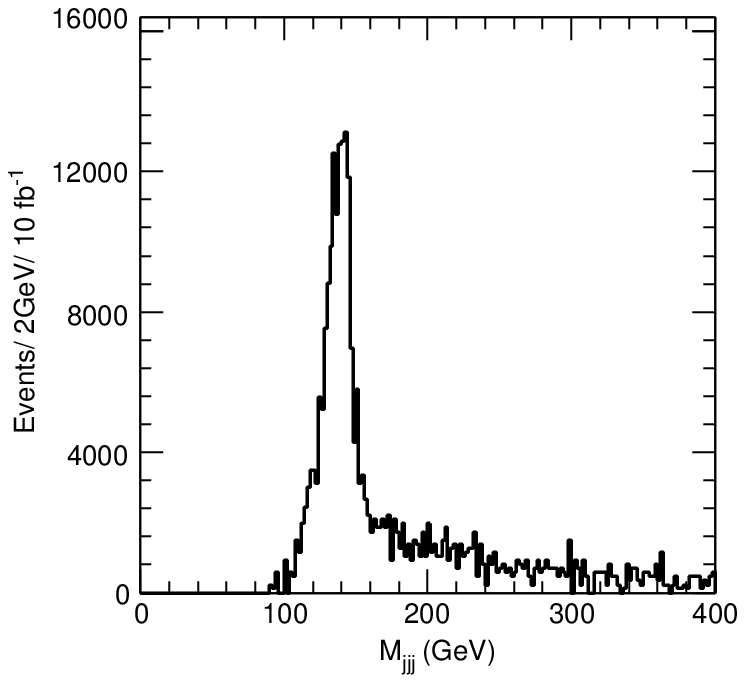}\forceright}{Trijet mass
distribution in $t \ol t$ events with
$m_t = 140\,\GeV$, without $b$-tagging.}

For $m_t = 250\,\gev$, the dijet and trijet mass distributions are shown in
Figs.\ \TheavyjjjMw\ and \TheavyjjjMt. The search region was $200\,\gev <
M_{jjj} < 300\,\gev$. As above, at least one dijet was required to have
invariant mass within $20\,\gev$ of $M_W$ and the invariant mass of the two
closest jets was plotted. There was no need for a $b$-tag to sharpen the
dijet mass distribution, since for such a heavy top-quark the two jets from
the $W$ are closer to each other than either is to the $b$ jet.  Thus, the
combinatorial background under the $W$-peak is much smaller if one selects
the closest two jets. The fitted $W$ mass was found to be $79.8\,\gev$, with
a resolution of $7.1\,\gev$. The top-quark mass was determined to be
$247.4\,\gev$ with a resolution of $14.7\,\gev$.

\nfig\TheavyjjjMw{\dofig{79.5mm}{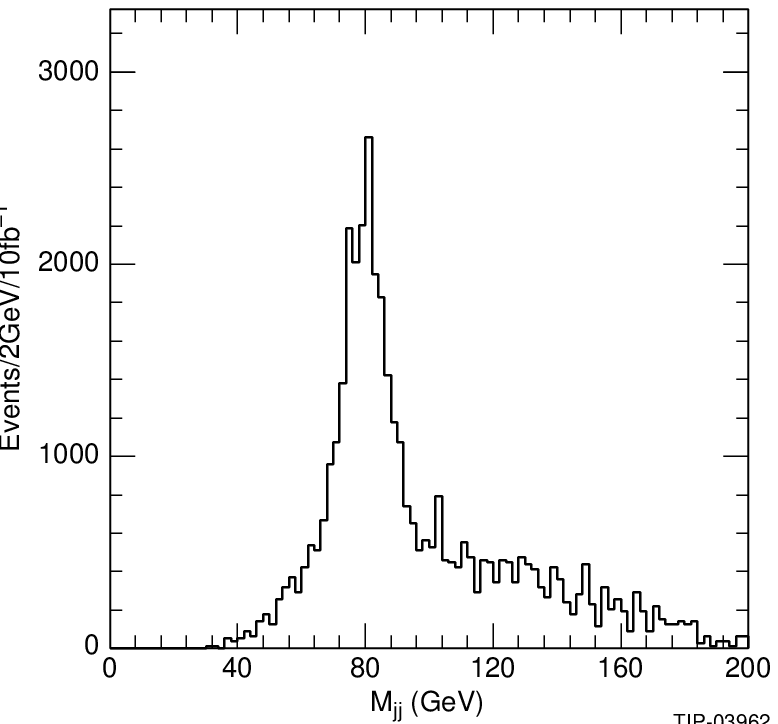}}{Dijet mass
distribution in $t \ol t$ events with
$m_t = 250\,\GeV$.}

\nfig\TheavyjjjMt{\dofig{79.7mm}{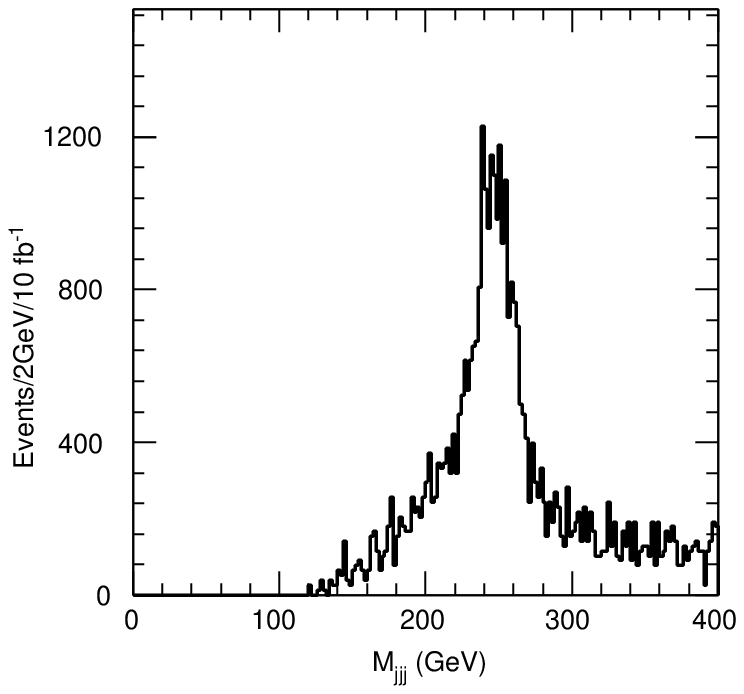}\forceright}{Trijet mass
distribution in $t \ol t$ events
with $m_t = 250\,\GeV$.}

The statistical uncertainty on the top-quark mass for $\sscy$ of data is
approximately $0.03\,\gev$ ($0.11\,\gev$) for $m_t = 140\,\gev$
($250\,\gev$). Systematics dominate the error in this $m_t$ measurement.
The largest effect is the uncertainty in the jet energy measurement. More
detailed studies of the calorimeter --- beam tests as well as
simulations --- will be required to yield adequate correction functions. It
will be helpful that the $t \ra W b$ process is self-calibrating: The
position and width of the $W$ peak will help calibrate the energies of jets
in $t$-decay and determine the systematic uncertainty in $m_t$.

The top quark will be a signal of flavor physics. One example is the
$\eta_T$, a spin-zero color-octet boson occurring in many technicolor
models.\cite{\ehlq,\etaT} In modern models,\cite{\wtc} the
$\eta_T$ is expected to have a mass in the range $300 - 500\,\gev$.%
\nref\wtcsig{K.~Lane and M.~V.~Ramana, Phys.\ Rev.\ {\bf D44}, 2678 (1991).}
\cite{\wtcsig,\eekleta} It is produced in $pp$ collisions via gluon
fusion with a cross section of order 1--50~nb, depending on its mass,
the number of technicolors, and other model-dependent factors.
It is expected to decay predominantly to $t \ol t$ with a width of order
10--100~GeV, again model-dependent.

In a search for the $\eta_T$, one would tag $t \ol t$ production in the
$\ellpm_{\rm isol} + \ts \jets$ mode and look for an enhancement in the $t
\ol t$ invariant mass distribution. This invariant mass can be determined
up to a quadratic ambiguity by assuming that the missing-$\et$ is due to $W
^\pm \ra \ellpm \nu_\ell$. The resolution on this invariant mass for $M_{t
\ol t} \simeq 400\,\gev$ is about $15\,\gev$ {\it plus} the contribution
from the $\etmiss$ resolution. For a $20\,\gev$ wide $\eta_T$, and assuming
that the $M_{t \ol t}$ resolution is $30\,\gev$, the $\eta_T$ appears as an
enhancement $75\,\gev$ wide. The underlying $t \ol t$ cross section in such
a region is about $3\,\nb$ for $m_t = 140\,\gev$ and $M_{\eta_T} =
300$--500~GeV. There should be no difficulty discovering the $\eta_T$ over
such a background.

\subsubsec{Discovery of a Charged Scalar in the Decay of a Heavy Top Quark}

Charged color-singlet scalar bosons, $\hpm$, occur in  multi-Higgs-doublet
models, in all supersymmetric extensions of the Standard Model and,
typically, as technipions in models of dynamical electroweak symmetry
breaking. Generally, their couplings are Higgs-like. That is, $\hpm$ tend
to decay to the heaviest fermion pairs kinematically accessible and they
often must be in the same generation. Since $\hpm$ are color-singlets, they
are copiously produced only if there is a heavier quark which can decay
into them.

We study discovery of an $\hpm$ in the decay products of a heavy top quark.
The masses $m_t = 250\,\gev$ and $M_{H^+} = 150\,\gev$ were used for the
discussion that follows. At the end, results for other mass combinations
will be described. It was assumed that both $t \ra W^+ b$ and $t \ra \hp b$
decays are allowed. The charged scalar was assumed to decay into $c \ol s$
and $\tau^+ \nu_\tau$.  Only the decay mode $\htn$ is considered
here,\cite{\MMhtn} since for $\hcs$, it should be straightforward to
discover the $\hp$ and measure its mass using the same technique as
described for the study of $t \ra W^+ b \ra 3\ts \jets$ in the previous
section.

The various branching ratios for $t$ and
$\hp$ decay are model-dependent. In the absence of experimental support for
any particular model, it is appropriate to assume only that $B(t \ra W^+ b)
+ B(t \ra \hp b) = 1$ and $B(\hcs) + B(\htn) = 1$, and to study the reach
of the detector for $\hpm$ as a function of the branching ratios. For
comparison with other simulations,%
\nref\SDC{Solenoidal Detector Collaboration Technical Design Report,
SDC--92--201, SSCL--SR--1215 (1992).}%
\nref\barnett{R.~M.~Barnett, R.~Cruz, J.~F.~Gunion and B.~Hubbard, Phys.\
Rev.\ {\bf D47}, 1048 (1993).}%
\cite{\SDC,\barnett}
we also present results in terms of the two-Higgs-doublet model occurring in
the minimal supersymmetric standard model (MSSM).\cite{\HaberKane}

The presence of $t \ra \hp b$, followed by $\htn$, is signalled by a
breakdown of lepton universality expected if only $t \ra W^+ b$ were
allowed (also see Refs. \SDC\ and \barnett). In addition to the $\pi^+
\nu_\tau$ decay mode used as a $\tau$-tag in those references,
all hadronic $\tau$-decay modes were used here.


ISAJET 6.50 was used to generate 15K each of $t \ol t \ra W^+W^-b \ol b$ and
$t \ol t \ra W \hpm b \ol b$ events. The ISAJET decay table was updated to
include all major $\tau$ decay modes. Hadronic modes account for 64.5\% of
all decays while the $\pi^+$ mode is only 12\%. Signal events were selected
by requiring an isolated electron or muon and a tau candidate.  The
criteria for an isolated $\mu^\pm$ were taken to be:
\eqn\htnisolmu{\eqalign{
& \vert\eta_\mu\vert < 2.4 \cr
&\pt > 20\,\gev \cr
&\sum_{R = 0.4} \et - \sum_{R = 0.1} \et < 2\,\gev \cr}}
\noindent Muons were required to be fully reconstructed according to the
efficiency parameterized in |gemfast|. For an isolated $e^\pm$ the
criteria used were
\eqn\htnisole{\eqalign{
& \vert\eta_e\vert < 2.4  \cr
&\sum_{R = 0.3}\et > 50\,\gev \cr
&\sum_{R = 0.2} \et({\rm EC}) > 40\,\gev \cr
&\sum_{R = 0.4} \et - \sum_{R = 0.2} \et({\rm EC}) < 10\,\gev \cr
& 0.8 < \et(R=0.2)/\pt({\rm CT}) < 2.0 \ts .\cr}}

For analysis of the $\tau$-lepton's two-body decays, it is important
to account properly for the polarization it has in  $W^\pm$ and $\hpm$
decays. The decay $W^\pm \ra \tau^\pm \nu_\tau$ conserves chirality (the
same as helicity for a high-energy $\tau$), while $\hpm \ra \tau^\pm
\nu_\tau$ maximally violates it. Then, for example, the $\pi^+$ occurring
in $\tau^+ \ra \pi^+ \ol \nu_\tau$ tends to follow the $\hp$ direction of
motion and so has higher $\pt$ than $\pi^+$ from $W^+ \ra \tau^+ \nu_\tau$
decays. This effect is enhanced here because $M_{H^+}$ is larger than
$M_W$. The polarization correlations were implemented in ISAJET for
two-body $\tau$ decays. The $\tau$-polarization effects are small for the
three-prong decays and require no special simulation.%
\ref\MMthesis{M.~Mohammadi, Ph.~D.~Thesis, University of Wisconsin
(1987), unpublished.}

Isolated $\tau$ candidates were selected by requiring:
\eqn\htnisoltau{\eqalign{
& \vert\eta_\tau\vert < 2.4  \cr
& N_{\rm ch} = 1 \ts\ts {\rm or} \ts\ts 3\cr
&\sum_{R = 0.3}\et > 50\,\gev \cr
& \delta \phi(\ell \ts \tau) > 100^\circ \cr
&\sum_{R = 0.4} \et - \sum_{R = 0.2} \et < 10\,\gev \cr}}
\noindent For the charge multiplicity ($N_{\rm ch}$) cut, tracks were required
to have $\pt > 1\,\gev$ and to lie within a cone of $R=0.1$ around the
calorimeter jet axis. The leading track in this cone was required to have
$\pt > 15\,\gev$. Electrons from $\tau$ decays were not included in the
$\tau$ sample if they passed the isolated $e$ criteria in Eq.\htnisole.
Muons from $\tau$-decays are predominantly isolated and are thus rejected by
the absence of significant activity in the calorimeter.  The central
tracker was not used to tag $\tau$ leptons by their displaced vertices. If
it were, it would further enhance the significance of the nonuniversality
signal.

In addition, we required a $b$ jet tagged by the central tracker. The
$b$-jet criteria were $\vert \eta_b \vert < 2.4$, $\delta \phi(\ell \ts b) <
100^\circ$, and scalar $\et > 30\,\gev$. A tagging efficiency
$\epsilon(b\hbox{-}{\rm tag}) = 20\%$ was assumed in this study.
Alternatively, $b$ jets could be tagged by a non-isolated muon, but this
would decrease the signal significances somewhat.\cite{\MMhtn}

The number of standard deviations $N_\sigma$ by which the number of
$\tau$ events exceeds the expectation from lepton universality is
\eqn\SDtau{
N_\sigma = \ {N(W \hpm \ra \ell_{\rm isol} \tau) \over
{\sqrt{N(WW \ra \ell_{\rm isol} \tau) + N(W \hpm \ra \ell_{\rm isol}
\tau) }}} \ts .}
Here, the event numbers are given by
\eqn\Ntau{\eqalign{
N(W W \ra \ell_{\rm isol} \tau) &= 2 N_{\ol t t} \ts B^2(t\ra W^+ b)
\cr
&\quad \times B(W \ra \ell \nu_\ell) \ts B(W \ra \tau \nu_\tau)
       \ts B(\tau \ra {\rm hadrons}) \cr
&\quad \times \epsilon(W \ra \ell) \ts \epsilon(W \ra \tau) \ts
\epsilon(b\hbox{-}{\rm tag})
   \epsilon(t\hbox{-}{\rm tag}) \ts \epsilon(t\hbox{-}{\rm trigger})
\ts; \cr\cr
N(W \hpm \ra \ell_{\rm isol} \tau) &= 2 N_{\ol t t} \ts B(t\ra W^+ b)
       \ts B(t \ra \hp b) \cr
&\quad \times B(W \ra \ell \nu_\ell) \ts B(\hp \ra \tau \nu_\tau) \ts B(\tau
\ra
       {\rm hadrons}) \cr
&\quad \times \epsilon(W \ra \ell) \ts \epsilon(\hp \ra \tau)
       \epsilon(b\hbox{-}{\rm tag)} \ts \epsilon(t\hbox{-}{\rm tag})
       \ts \epsilon(t\hbox{-}{\rm trigger}) \ts. \cr }
}
For a $250\,\gev$ top quark, the number of events produced with an
integrated luminosity of $\sscy$ is $N_{\ol t t} = 1.5 \times 10^7$.  The
efficiencies $\epsilon(W \ra \tau)$ and $\epsilon(\hp \ra \tau)$ are the
ratios of the numbers of true $\tau$-leptons passing the above cuts to the
number generated, including $\tau \ra e$ events
misidentified as hadronic $\tau$-decays. For $m_t = 250\,\gev$ and
$M_{H^+} = 150\,\gev$, these efficiencies were found to be $\epsilon(W
\ra \tau) = 9.8\%$ and $\epsilon(\hp \ra \tau) = 14.7\%$. The efficiency for
finding $W \ra \ell\nu_\ell$ events was 36\% for electrons and
45\% for muons. The average value $\epsilon(W \ra \ell) = 40\%$ was used.
The top-tagging efficiency, $\epsilon(t\hbox{-}{\rm tag})=69\%$,
is the fraction of top events remaining after the tau-selections.
The efficiency for triggering on top quarks
at Level\ 1 was found to be 93\%.\cite{\MMhtn}

The backgrounds to the $\ol t t \ra W^\mp \hpm \ra \ell^\mp + \tau^\pm + X$
signal come from (1) $t \ol t \ra WW b \ol b$ and $W\hpm b \ol b$ events in
which $W^\pm$ and $\hpm$ decay to jets which fake a $\tau$; (2) $W +
\ol Q Q$ events with $Q = c,b$; and (3) $\ol b b$ production. To study the
first background, 40K two-jet events were generated with $\pt$ in the range
$50 -1000\,\gev$. All jets found by |gemfast| with $\et > 25\,\gev$ were
subjected to the $\tau$-selection criteria above. A rejection
factor of ${\cal R}(\tau/\jet) = 0.0027$ was found. Since there are two to
three jets in the hemisphere opposite the isolated lepton, this background
amounts to 8.4\% of the $\ol t t \ra WW \ra \ell_{\rm isol} + \tau$ signal
and $1.4\% \times (1 - B(\hp \ra \tau \nu_\tau))/B(\hp \ra \tau \nu_\tau)$ of
the $\ol t t \ra WH \ra \ell_{\rm isol} + \tau$ signal. Including these
backgrounds reduces $N_\sigma$ by 3\%. The other backgrounds have been
shown to reduce $N_\sigma$ by less than 3\%.\cite{\SDC,\barnett,\MMhtn}

\nfig\THplus{\dofig{3in}{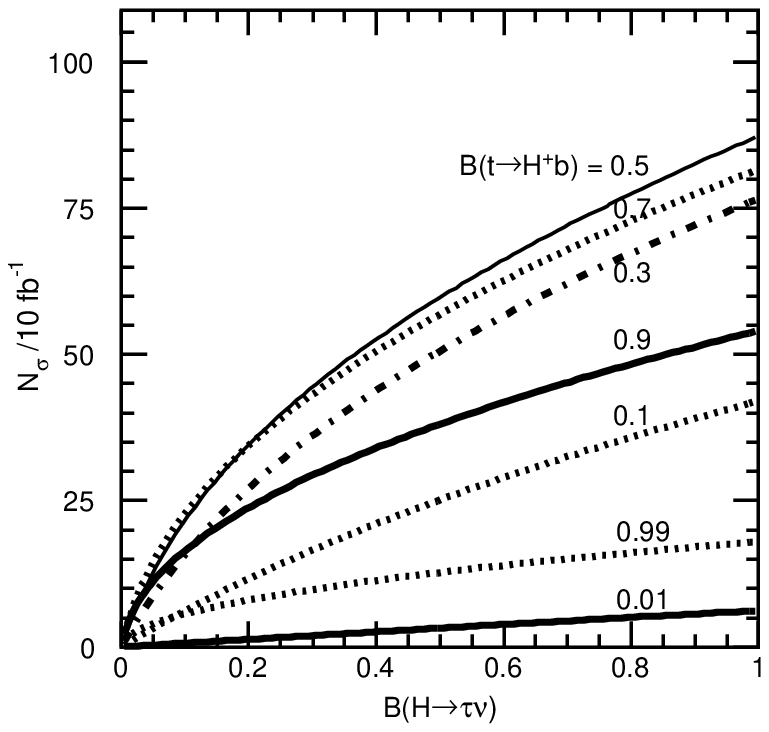}\forceleft}{Significance of
the $t \ra \hp b$, $\hp \ra \tau^+
\nu_\tau$ signal vs.\ $B(\hp \to \tau^+ \nu_\tau)$ for various $B(t \ra \hp
b)$. Here, $m_t = 250\,\gev$ and $M_{H^+} = 150\,\gev$.}

\nfig\THplusmssm{\dofig{3in}{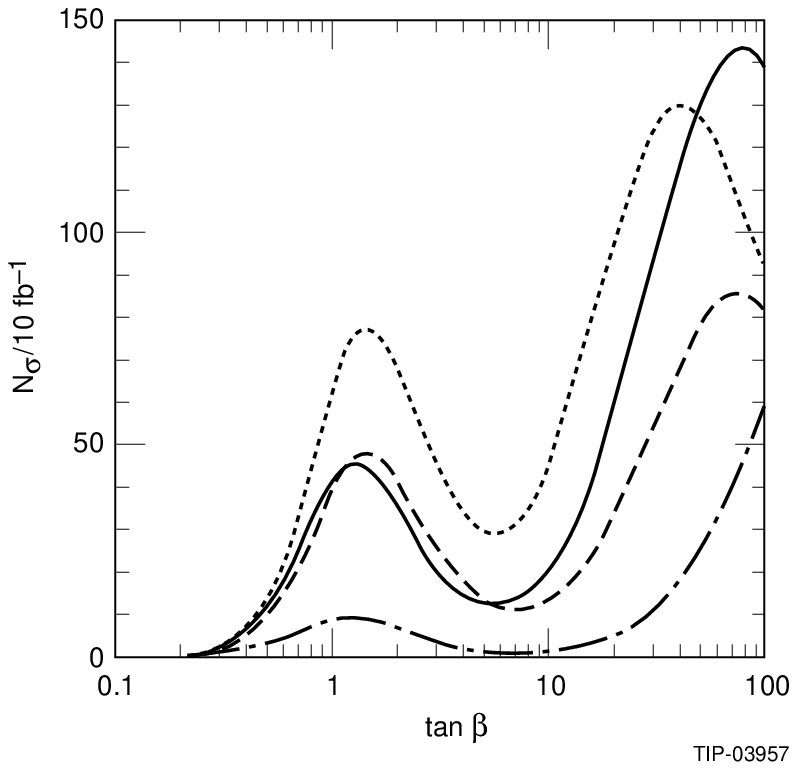}}{Significance
of the $t \ra \hp b$, $\hp \ra
\tau^+ \nu_\tau$ signal vs.\ $\tan \beta$ for various $m_t = 250\,\gev$ and
$M_{H^+} = 225\,\gev$ (dash-dotted); 250 and $150\,\gev$ (dotted); 150 and
$125\,\gev$ (solid); 150 and $100\,\gev$ (dashed).}

The significance for $\sscy$ is plotted against $B(\htn)$ in Fig.\
\THplus\ for $B(t \ra \hp b) = 0.01 - 0.99$. Backgrounds are not
included in these plots because they are so small. A signal ($N_\sigma
> 5$) for the $t \ra \hp \ra \tau^+$ decay channel can be discovered if
$B(t \ra \hp b) \simge 0.02$ and $B(\hp \ra \tau^+\nu_\tau) \simge
0.05$.

We have also computed $N_\sigma$ for the two-Higgs-doublet model used in
the MSSM. In this model, $B(t \ra \hp b)$ and $B(\htn)$
are determined by a single parameter, $\tan \beta$, the ratio of the
vacuum expectation values of the two Higgs doublets. The significances
were determined for four cases: $(m_t, M_{H^+}) = (250,\ts 225)$, (250,
150), (150, 125), and $(150, \ts 100)$ to facilitate comparison with
Ref.\ \SDC. The significances expected in GEM are shown in Fig.\
\THplusmssm. For all but the heaviest mass combination, GEM is able to
detect a $5\sigma$ enhancement in $\tau$ production for $\tan \beta >
0.4$.  For $(m_t, M_{H^+}) = (250,\ts 225)$, the discovery level covers
the range $0.7 < \tan \beta < 2.0$ and $\tan \beta > 20$. By tagging
the $\tau$ lepton in all its hadronic modes instead of just $\tau \ra
\pi \nu_\tau$, the significances have been increased by a factor of
$2.3 - 5$.

\subsec{Jet Physics at Large Transverse Momentum}

\nref\TNjet{R.~Carey, {\it Jet Studies with the GEM Detector},
GEM--TN--93--377 (1993).}

Although GEM emphasized precision measurement of photons and leptons up to
the highest SSC luminosities, the accurate identification and measurement
of jets is also important.  The search for a high mass Higgs boson
(Section\ 5.2) and heavy flavor physics (Section\ 5.3) are two examples of
new physics requiring accurate understanding and measurement of jets.
Hadronic jets can also be important backgrounds to new physics; the search
for $\hgg$ is an example of this. The errors in jet measurements introduced
by analysis effects, such as jet definition and clustering algorithms, and
by instrumental effects, such as detector resolution, $e/h$ for the
calorimeters, and cracks and dead spaces, must be carefully studied and
kept under control. This section discusses the main issues involved in
measuring the high-$\pt$ jet cross section accurately. The search for signs
of quark substructure in deviations from QCD of the jet cross section forms
the context for this discussion.

If quarks and leptons are composite, with structure at the scale
$\Lambda$, the most visible manifestation at subprocess energies
$\rshat \ll \Lambda$ is the presence of four-fermion contact interactions,
$\CL_{\Lambda}$, involving the composite quarks and leptons.%
\ref\elp{E.~J.~Eichten, K.~D.~Lane and M.~E.~Peskin, Phys.\ Rev.\ Lett.\
{\bf 50}, 811 (1983).}
These interactions induce terms in the cross section for dijet and dilepton
production that are of order $\pi \shat / \Lambda^4$, leading to
significant excesses at ``low'' $\shat$. It is known from experiments at
$e^+ e^-$ and hadron colliders that $\Lambda \simge 1 - 2\,\tev$, above the
scale of electroweak symmetry breaking.%
\nref\CDFlim{F.~Abe et al., The CDF Collaboration, Phys.\ Rev.\ Lett.\
{\bf 68}, 1104 (1992); Fermilab--PUB--91/231--E.}%
\cite{\PDG,\CDFlim}
Thus, $\CL_{\Lambda}$ must be $SU(3) \otimes SU(2) \otimes U(1)$ invariant,
and the composite quark and lepton fields appearing in it are electroweak,
not mass, eigenstates. This raises the possibility that unacceptably large
flavor-changing neutral currents will appear in the contact interactions.
The most stringent limit on such interactions comes from the allowed
magnitude of new effects in the neutral kaon system. If $\vert \Delta S
\vert = 2$ contact interactions exist, they must have $\Lambda \simge
500\,\tev$, well above the reach of the SSC. For the discussion of contact
interactions accessible at the SSC, therefore, we shall assume that
$\CL_{\Lambda}$ is symmetric under interchanges of the three generations of
quarks and leptons.

Quark substructure shows up directly as an excess of jets at high $\pt$ or
$\rshat$. In this section the contact interaction
\eqn\Lqq{
\CL_{\rm qq} = - {4 \pi \over {2 \Lambda^2}} \ts \ol Q_{La} \gamma^{\mu}
Q_{La} \ts \ol Q_{Lb} \gamma_{\mu} Q_{Lb}}
is used as a model to modify QCD jet production. Here, $Q_{La} = (u_a,
d_a)_L$ are left-handed quark fields and $a,b = 1,2,3$ label the
generations. This model for the four-quark contact interaction is
essentially the one discussed in Ref.~\ehlq\ except that all quarks
are taken to be composite. PYTHIA~5.6 was used to generate the jet
events for QCD and the quark substructure signal.%
\ref\torsjo{We are greatly indebted to T.~Sjostrand for embedding the quark
and lepton compositeness routines in PYTHIA and for continued help on the
event simulations.}
Several different choices of parton distribution functions (PDFs) were used
and results compared: EHLQ Set\ 1,\cite{\ehlq} the CTEQ Set\
1L,\cite{\HRcteq} and Morfin-Tung Set 2.%
\ref\MT{J.~Morfin and W.~-K.~Tung, Fermilab--Pub--90/74, (1990).}
The signal region lies above $\et = 4\,\tev$ for the jets.
The study described here was made for an integrated luminosity
of $\sscy$. We shall show that GEM could reach a substructure scale of about
$25\,\tev$ with this data sample.

The most important issue in searching for quark substructure is to be
certain that an observed excess of high-$\et$ jets is not an artifact of
the detector nor of the analysis. The jet cross section must be
normalized to the QCD expectation at lower $\et$
to reduce uncertainties due to luminosity
(approximately 3\% at the  Fermilab Tevatron\cite{\topmass}).
Since the cross section for jets at the lowest
transverse energies is not well-known theoretically, the normalization
region used is the middle of the jet $\et$ spectrum. Finally, a scheme must
be developed that corrects for the calorimeter's lack of compensation.

For this study signal events were generated with total scalar
transverse energy, $\sum \et \simge 9\,\tev$ and the normalization
sample was generated with $\sum \et > 4.8\,\tev$.  The only kinematic
cut imposed on jets is $\abseta < 1.1$. This enhances the roughly
isotropic signal relative to the forward-backward peaked QCD
background. More than one jet can be taken from a single event.  For
these central high-$\et$ jets, the triggering efficiency is close to
100\%. The discovery criterion adopted for this analysis is an excess
of 100 events in an $\et$-region in which the observed cross section
is twice as large as the QCD expectation.

Jet clustering was done with a fixed-cone algorithm using a large
cone, $R = 0.9$.  The large clustering
radius was chosen to reduce energy loss out of the clustering cone.
This cone algorithm was found to be efficient and insensitive to
detector variations and to have good angular resolution.\cite{\TNjet}
If the jet-cone radius was decreased to 0.7, the main effect was to
shift the energy scale of the jet-$\et$ spectrum downward by 1.2\%.
This shift should be calculable in perturbation theory.

A full experimental analysis would have to include the development of
a jet energy correction function and an unsmearing procedure for the
inclusive jet-$\et$ spectrum. The correction function must include the
effects of the underlying event, showers spreading outside the clustering
cone, and detector noise, inefficiencies and nonlinearities. Most of
these corrections are reasonably well-understood and tend not to be
important for very high-$\et$ jets. The most important problem is to
determine the jet energy scale. Jets whose $\et$ is mismeasured upward
easily produce a false compositeness signal. The jet energy scale
for the calorimeter would be established using test beam and
collider data and the energy scale correction would depend on, among
other parameters, the jet rapidity and the fraction of electromagnetic
energy observed in a jet.

\nfig\Jcomp{\dofig{84.5mm}{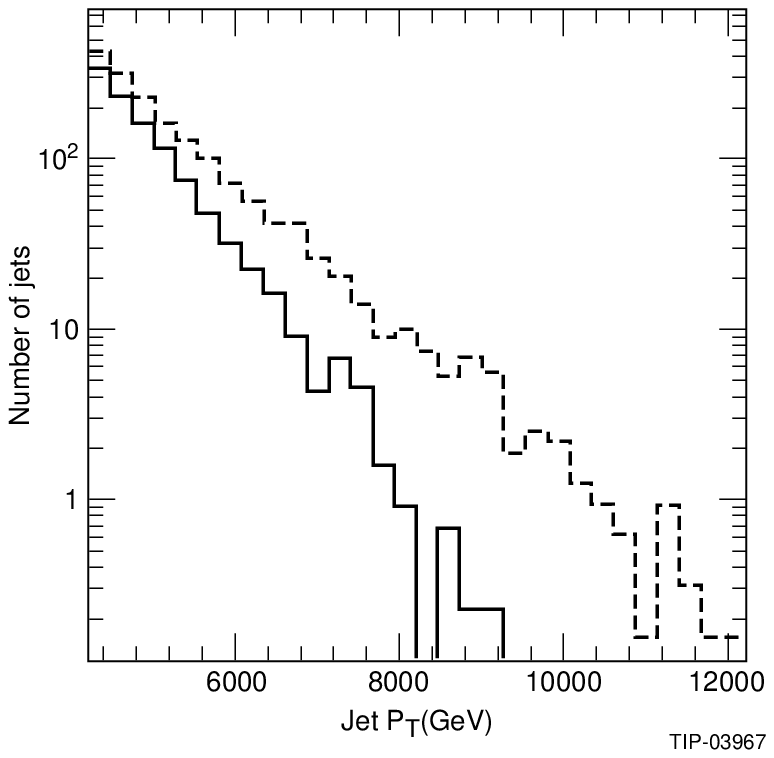}}{Inclusive jet $\et$
spectrum for substructure
at the scale $\Lambda = 25\,\tev$ (dashed) and for QCD (solid).}

\nfig\JTpdfs{\varfig{3.5in}{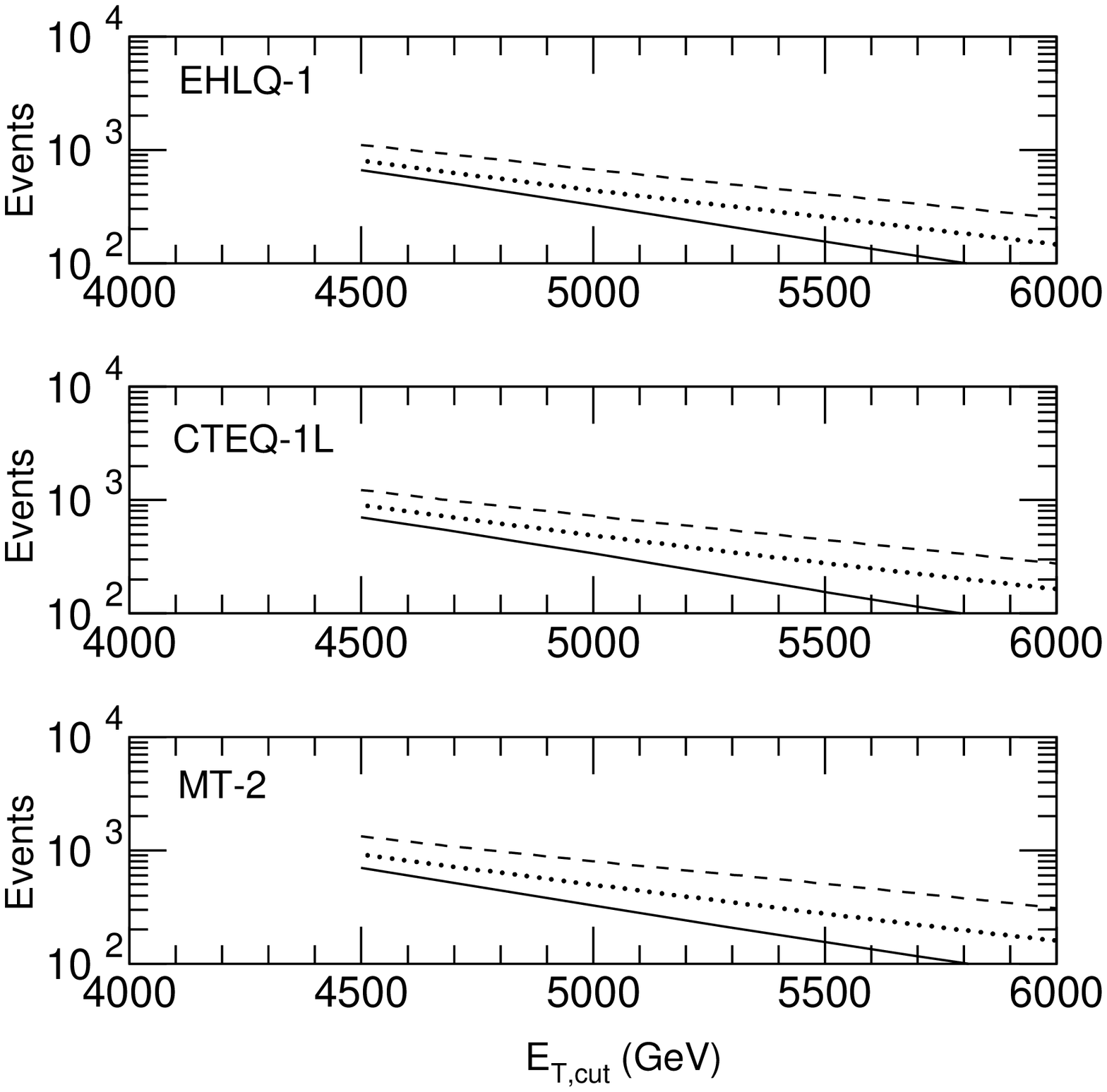}}
{Number of jets expected with $\sscy$ with $\et > \et^{\rm cut}$,
for various parton distributions. The solid line is QCD (no substructure),
dotted line is $\Lambda=30\,\tev$ and dashed line is $\Lambda=25\,\tev$.}

The reach in $\Lambda$ is limited by incomplete knowledge of the
parton distributions and of the jet energy resolution and scale. A
number of jet energy reconstruction schemes, particularly weighting
algorithms which attempt to boost the hadronic part of a shower, have
been used to improve jet energy resolution and
linearity.\cite{\wgtexpt,\wgtalgo} We have estimated the systematic
effects due to nongaussian tails in the energy response and to
nonlinearity arising from lack of compensation in the
calorimetry. Nongaussian tails were modeled in |gemfast| by adding a
second Gaussian of 1\% the amplitude and three times the width of the
normal Gaussian energy resolution, as described in Section~4.3. To
estimate the uncertainty due to parton distributions we compared the
rates of the CTEQ-1L, EHLQ-1 and MT-2 distribution functions.

The inclusive cross sections for jets with $\abseta < 1.1$, using the
CTEQ-1L distribution functions, are shown in Fig.\ \Jcomp\ for
$\Lambda = 25$ and QCD ($\Lambda=\infty$).  Fig. \JTpdfs\ shows the
number of jets expected in $\sscy$ with $\et > \et^{\rm cut}$. It is
clear that a quark substructure signal at the scale $\Lambda =
25\,\tev$ could be discovered easily in this integrated luminosity
with a detector whose calorimeter is as linear as the one modeled
here. The variation in jet rates obtained using the CTEQ-1L, EHLQ-1
and MT-2 distributions is indicated in Fig.\ \JTpdfs. The differences are
small compared to the signal and produce a $10 - 20\%$ variation in
the observable $\Lambda$ according to our discovery criterion.

Nonlinearity in the charged pion response was modeled using a GEANT
simulation of the GEM detector incorporating the expected $e/h$ values
for each calorimeter. Figure\ \Jlinear\ shows the charged pion
response, averaged over $\abseta < 1.1$, as a function of energy. The
relative response has been set to 1.0 at $200\,\gev$. The solid curve
is the average response obtained using energy-independent weights for
each calorimeter layer; the dashed curve shows the response with
energy-dependent weighting.  The extrapolation to high energies shown
here gives noticeable errors in both cases, but mimics the
extrapolation from test beam energies that will have to be carried out
when calibrating the real detector.

The effect of the nonlinearity in the two weighting schemes is illustrated
in Table~10. The CTEQ-1L distribution functions were used here. As expected
from Fig.\ \Jlinear, the energy-independent weighting scheme tends to
increase the measured $\et$ of the jets, while the energy-dependent scheme
reduces jet-$\et$. The effects are small, of order 5\% and the scale
$\Lambda = 25\,\tev$ is still easily within reach for a data sample of
$10\,\fb^{-1}$.

\nfig\Jlinear{\dofig{75.5mm}{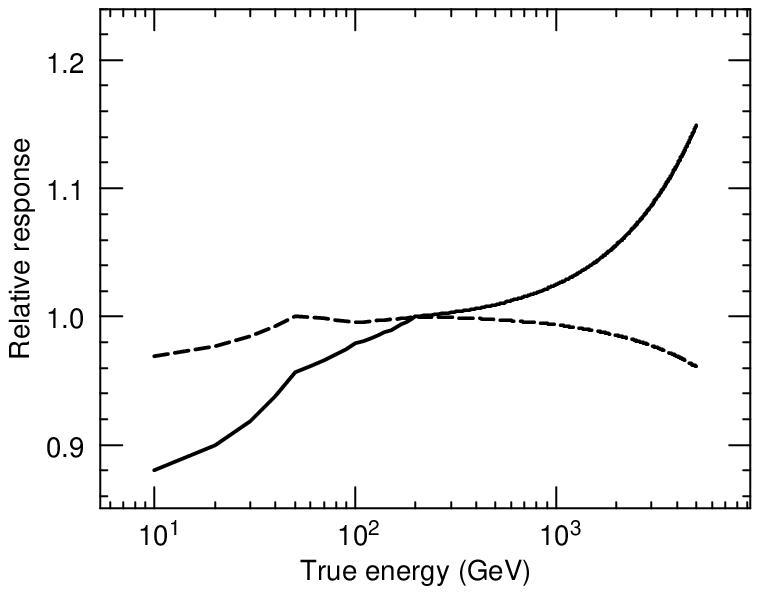}\forceleft}{Calorimeter
response to single pions relative to that at $200\,\gev$,
using energy-independent weights (solid curve)
and energy-dependent weights (dotted curve).}

\widetopinsert
\GEMcaption{Table 10}{Number of jets expected with $\sscy$ using
different energy-weighting schemes as described in the text, and the
CTEQ-1L parton distributions.}
\medskip
\GEMtable
#\hfil & \qquad\hfil# & \qquad\hfil# & \qquad\hfil# & \qquad\hfil#\cr
\GEMrulerule
$\et^{\rm cut}(\gev)$ & 4500 & 5000 & 5500 & 6000\cr
\GEMrulerule
\multispan5 \hfil Perfect Calorimeter \hfil\cr
\GEMrule
 $\Lambda = \infty$ & 704  &  339 & 155  &  74\cr
  $25\,\tev$        & 1226 &  724 & 447  & 276\cr
\GEMrulerule
\multispan5 \hfil{Energy-independent Weights}\hfil\cr
\GEMrule
 $\Lambda = \infty$ & 718  &  349 & 167  &  79\cr
  $25\,\tev$        & 1257 &  739 & 468  & 286\cr
\GEMrulerule
\multispan5 \hfil{Energy-dependent Weights}\hfil\cr
\GEMrule
 $\Lambda = \infty$ & 676  &  321 & 148  &  70\cr
  $25\,\tev$        & 1189 &  701 & 435  & 267\cr
\GEMrulerule
\endGEMtable
\endinsert

Finally, we can estimate the reach in the quark compositeness scale
$\Lambda$ attainable with a data sample of $\sscd$. We anticipate no
special difficulties for high-$\et$ jet measurement associated with
operations at $\CL \simeq \uhl$. Thus, the reach in $\Lambda$ can be
determined from the fact that the subprocess cross section goes as
$\shat/\Lambda^4$. This yields $\Lambda \simeq 45\,\tev$, a factor of
40 greater than the limit set by existing hadron collider data.

\gdef\lsp{\widetilde\chi_1^0}

\subsec{Supersymmetry}

	Supersymmetry (SUSY) is theoretically attractive because it
eliminates the quadratic divergences in the Higgs sector and so allows light
elementary Higgs bosons to occur naturally. Its study also provides a good
testing ground for many aspects of GEM, including missing energy, jets and
leptons. The minimal supersymmetric standard model\cite{\HaberKane}
has two Higgs doublets and superpartners (denoted by a tilde) for all normal
particles. In particular there are four neutralinos, $\widetilde\chi_i^0$,
which are linear combinations of the partners of the photon, $Z$, and neutral
Higgs bosons, and two pairs of charginos $\widetilde\chi_i^\pm$.  There is a
conserved $R$ parity carried by all superparticles, so they must always be
produced in pairs and decay to the lightest supersymmetric particle, which
is absolutely stable. We shall assume that this lightest particle is the
lightest neutralino $\lsp$. If SUSY is broken at the electroweak scale, the
masses of all of these particles should be less than about $1\,\TeV$.

	In this section we shall discuss only searches for gluino and squark
production.  Neutralino, chargino, and slepton production are all important
parts of the search for supersymmetry but have smaller cross sections. While
MSSM Higgs bosons have rather different production cross sections and decay
modes than Standard Model ones, the methods of searching for them are
generally similar.  Figure\ \SSzhu\ shows the regions of MSSM parameter space
in which GEM would be able to detect one or more of the neutral Higgs bosons
using the $\gamma\gamma$ and $\ell^+\ell^-\ell^+\ell^-$ decay channels.  The
$\tau^+\tau^-$ decay modes are not shown but appear to offer potential.%
\ref\Htotautau{S.~Mrenna, GEM--TN--501 (1993)}

\nfig\SSzhu{\varfig{3.25in}{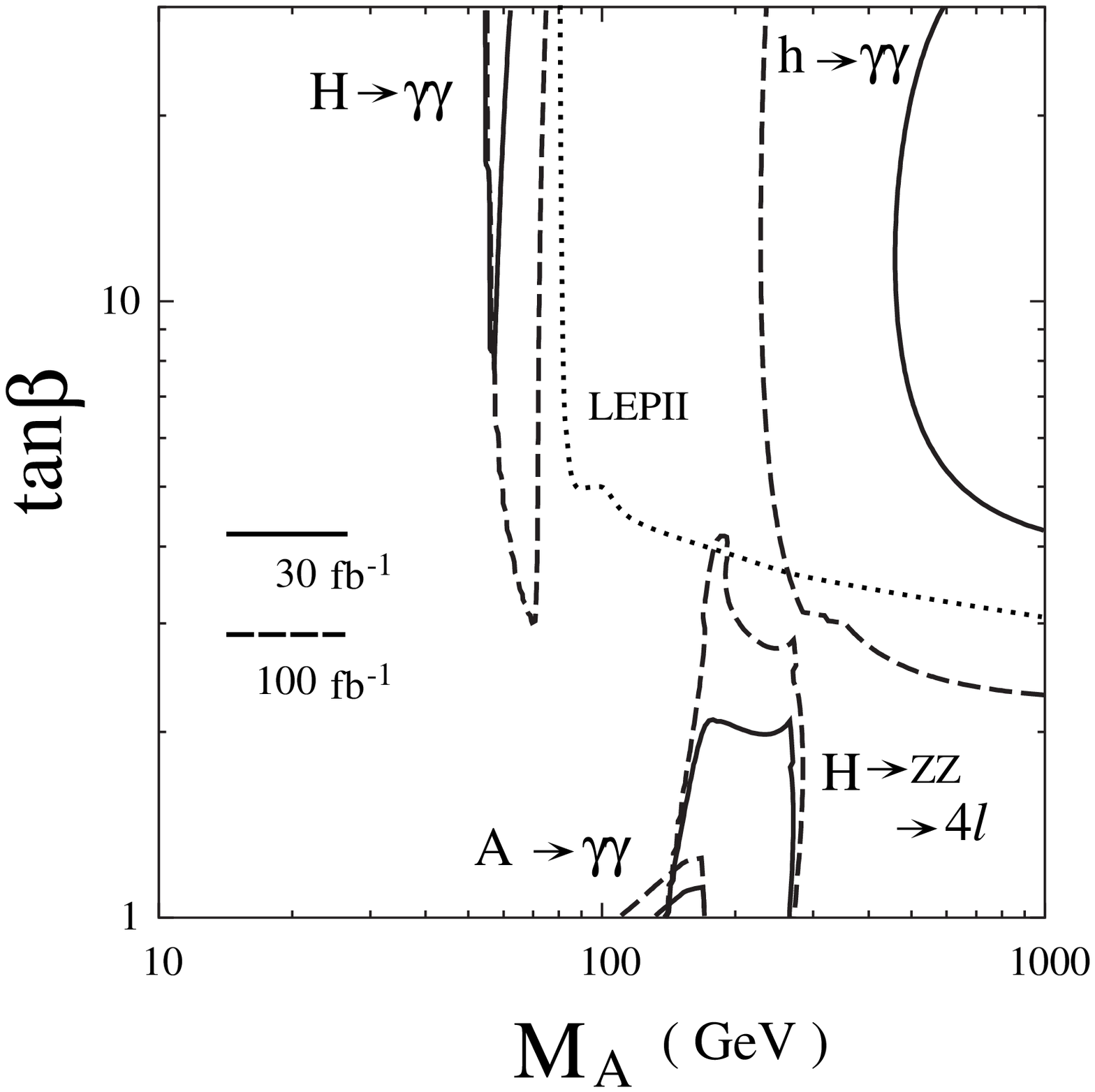}}
{Regions of the $(M_A,\tan\beta)$ plane in which GEM could discover the
neutral Higgs bosons of the MSSM at the $5\sigma$ significance level,
in the $\gamma\gamma$ and $\ell^+\ell^-\ell^+\ell^-$ decay channels,
with integrated luminosities of $30\,{\rm fb}^{-1}$ and $100\,{\rm
fb}^{-1}$. Also shown is the region covered by LEP II at $\sqrt{s}=190\,\GeV$
with an integrated luminosity of $500\,{\rm fb}^{-1}$.}

\subsubsec{$\etmiss$ Signature for Gluinos and Squarks}

	Since the lightest supersymmetric particle $\lsp$ is neutral
and interacts weakly with matter, it escapes from the detector. Thus,
one of the basic signatures for SUSY is missing transverse energy,
$\etmiss$, plus multiple jets. A stringent test for
GEM's missing energy resolution is to be able to detect, in this mode,
gluinos and squarks with masses as light as $300\,\GeV$. This is near
the limit expected from the Tevatron and is also the mass range
expected in some SUSY grand unified models.%
\nref\Arnowitt{R.~Arnowitt and P.~Nath, Phys.\ Rev.\ Lett.\ {\bf 69},
725 (1992).}%
\nref\RossRoberts{R.~Ross and R.~Roberts, Nucl.\ Phys.\ {\bf B377},
571, 1992.}%
\cite{\Arnowitt,\RossRoberts}
The MSSM typically produces cascade decays from one supersymmetric
particle to another. The events can have many jets and leptons, and
the missing energy from the final lightest supersymmetric particle
$\lsp$ can be small compared to the parent mass. A typical decay
sequence for a relatively light squark and gluino with $M_\squark >
M_\gluino$ might be
\eqn\XX{
\eqalign
{
\widetilde u_L &\to \widetilde g u\,, \cr
\widetilde g &\to \widetilde\chi_1^+ \bar u d\,, \cr
\widetilde \chi_1^+ &\to \widetilde\chi_1^0 e^+ \nu\,. \cr
}
}
Decay chains can be even more complex for heavier masses. All of these
possible decays are included in the version of ISAJET%
\ref\Isasusy{H.~Baer, F.~E.~Paige, S.~D.~Protopopescu and X.~Tata,
FSU--HEP--930329. Production and decay of $\widetilde t$ is not yet included.}
used for this analysis.%
\ref\Sasha{F.~E.~Paige and A.~Vanyashin, GEM TN--93--376 (1993).}

\widetopinsert
\GEMcaption{Table~9}{Choices of MSSM parameters
for the three cases considered. These were chosen to have different
event topologies and to span the whole mass range. All the squarks and
all the sleptons are taken to be degenerate for simplicity. All masses
are in GeV. See Ref.~\HaberKane\ for the notation.}
\bigskip
\GEMtable
\hfil#\hfil\quad&\hfil#\quad&\hfil#\quad&\hfil#\cr
\GEMrulerule
Parameter 		& Case~I 	& Case~II 	& Case~III\cr
\GEMrule
$M_\gluino$		& 300		& 350		& 2000\cr
$M_\squark$		& 600		& 325		& 2500\cr
$M_{\widetilde\ell}$	& 500		& 200		& 1500\cr
$M_A$			& 300		& 300		& 300\cr
$\mu$			& $-300$	& $-300$	& $-1000$\cr
$\tan\beta$		& 2		& 2		& 2\cr
\GEMrulerule
\endGEMtable
\endinsert

	There are a number of other parameters in the MSSM besides the
gluino and squark masses, and it is beyond the scope of this study to
explore the MSSM parameter space completely. Instead, the
representative choices listed in Table~9 have
been considered. Case~I has a light gluino and a heavier squark; it is
generally similar to the models of Ref.~\Arnowitt\ and to the case
considered in previous GEM studies.\cite{\response} Case~II has a
squark slightly lighter than the gluino and is generally similar to
the models of Ref.~\RossRoberts. Since $\gluino \to \squark q$
dominates for $M_\gluino > M_\squark$, the signatures in this case are
similar to those for squark pair production. One might think that this
case would be more difficult to detect because the events contain just
two hard jets from $\squark \to \lsp q$. It is actually easier,
because the branching ratios for $\squark_L \to \widetilde\chi_1^\pm
q$ and $\squark_L \to \widetilde\chi_2^0 q$ turn out to be large and
to provide multijet signatures, and the dominant decay $\widetilde q_R
\to \lsp q$ gives a harder $\etmiss$ distribution. Finally, Case~III
has all the masses pushed close to the upper limit for SUSY to be
related to the electroweak scale.  It tests the ability of GEM to
cover the top of the plausible mass range for weak-scale supersymmetry.

	For the three cases, samples of 70K, 25K, and 35K,
respectively, of gluino and squark signal events were generated with a
version of ISAJET containing all the MSSM decay modes.\cite{\Isasusy}
The total production cross sections for all combinations of gluinos
and squarks are
\eqn\XX{
\eqalign
{
\hbox{\rm Case\ I:} 	& \qquad \sigma = 8.27\,\nb\,, \cr
\hbox{\rm Case\ II:}	& \qquad \sigma = 7.60\,\nb\,, \cr
\hbox{\rm Case\ III:}	& \qquad \sigma = 0.81\,\pb\,. \cr
}
}
The Monte Carlo statistics are therefore small compared to those
obtained in $10\,\fb^{-1}$ for the first two cases but comparable in
the third. This is reflected in the error bars on the plots shown below.

	The signal events are characterized by multiple jets and large
missing energy. For the lower masses in Cases~I and II the dominant
Standard Model physics background comes from heavy flavors decaying
into neutrinos, and the dominant detector-induced background comes from
mismeasuring QCD jets. A total of 1.5M QCD jets of all types in ten
$p_T$ ranges covering $50 < p_T < 3200\,\GeV$ was generated with
ISAJET to determine both kinds of backgrounds. For the high masses in
Case~III, the backgrounds from $W \to \ell\nu$ and $Z \to \nu\bar\nu$
are also significant. A total of 40K $W \to \ell\nu$ and 80K $Z \to
\nu\bar\nu$ events was generated covering the same $p_T$ range.

	The detector response to all events was simulated with
|gemfast|. The missing energy was calculated using the single-particle
$p_T$ resolution of the forward calorimeter determined from GEANT plus
an additional 1\% nongaussian tail three times as wide as the main
peak, as described in Section~4.3. The effect of this nongaussian
tail is small compared to the effects of angular resolution in the
forward calorimeter and of the hole for the beam pipe, so its exact
parameterization is not crucial.

\nfig\SSi{\dofig{84.5mm}{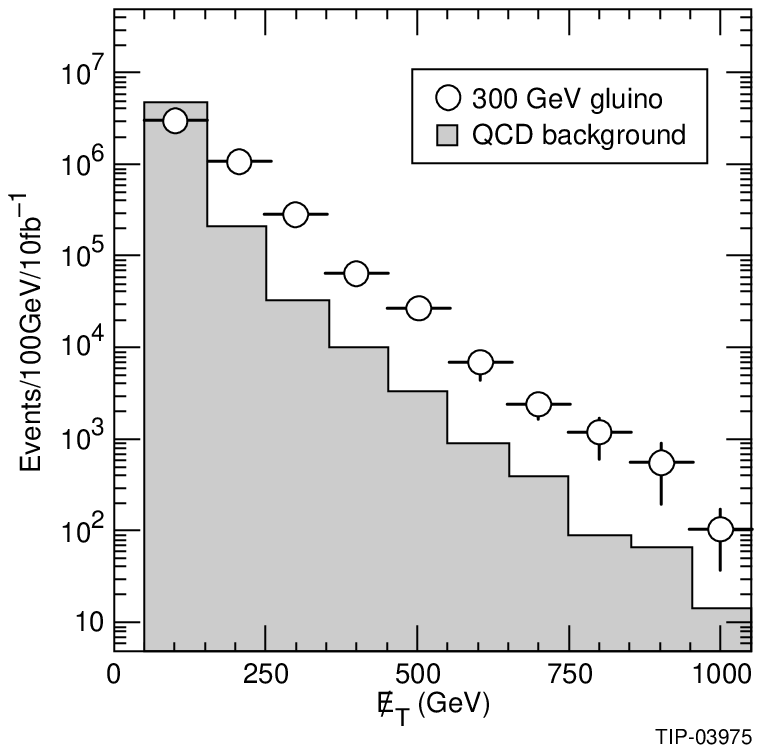}}{$\etmiss$ signal for Case~I
MSSM parameters defined in Table~9 (open circles)
and for QCD background (histogram) after requiring at least 5 jets
with $p_T > 75\,\GeV$ and the sphericity and lepton veto cuts
described in the text.}

\nfig\SSiSB{\dofig{86mm}{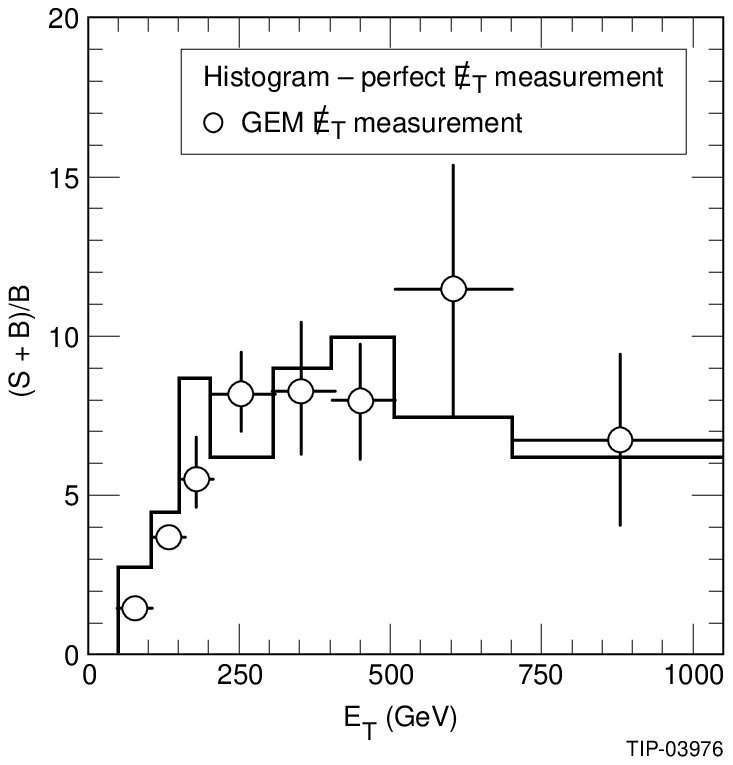}\forceright}{Open circles: Ratio of
signal and background curves from Fig.~\SSi. Histogram: The same ratio
for a perfect $\etmiss$ measurement.}

	In the inclusive $\etmiss$ cross section, the Standard Model
physics background is larger than the signal. Furthermore, the
detector-induced background from mismeasured jets in the forward
region is several times larger than the real neutrino background, even
for an ideal calorimeter covering $\eta < 5.5$.%
\ref\Sasha{F.~E.~Paige and A.~Vanyashin, GEM TN--92--70 (1992).}
First consider the lower-mass Cases~I and II. Since gluinos and squarks are
centrally produced with $\pt \sim M_\gluino$, they give multiple jets and
``round'' events in addition to $\etmiss$. Jets with $p_T > 75\,\GeV$ were
found using the |gemfast| fixed cone algorithm with $R = 0.7$. The minimum
number of jets, $N_{\rm jet}$, was varied between two and five. To identify
round events, the sphericity in the transverse momentum plane, $S_T$, was
calculated by summing all calorimeter cells with $E_T > 0.5\,\GeV$ and
$\vert\eta\vert < 3$. A cut on $S_T > 0.2$ provided good separation of
signal and background.  After these cuts the signal to background ratio
$S/B$ for $\etmiss \sim 250\,\GeV$ was about 3 for Case~I and about 5 for
Case~II. The larger $S/B$ for Case~II reflects the harder $\etmiss$ spectrum
from $\widetilde q_R$ decays mentioned earlier.

        Semileptonic decays of gluinos and squarks are important; see
Section~5.3.2 below. However, a lepton veto further improves the
$S/B$ for the $\etmiss$ distribution by rejecting $t \bar t$ and other
Standard Model backgrounds. Events were vetoed if they contained a
muon or an isolated electron. An electron was identified as an
isolated electromagnetic cluster in the calorimeter with $p_T >
20\,\GeV$ and $\vert\eta\vert < 2.5$, matched to a single track in the
central tracker with a loose matching constraint,
\eqn\XX{
\left\vert E/p - 1 \right\vert < \max(0.5,3{\Delta p \over p})\,.
}
Isolated muons with $p_T > 20\,\GeV$ and $\eta < 2.5$ were identified
using the standard |gemfast| muon reconstruction. The efficiency of the
lepton identification is not crucial for this analysis; even if it
were perfect, there still would be background from $\tau$-decays of
$b$ and $t$ quarks.

\nfig\SSii{\dofig{84mm}{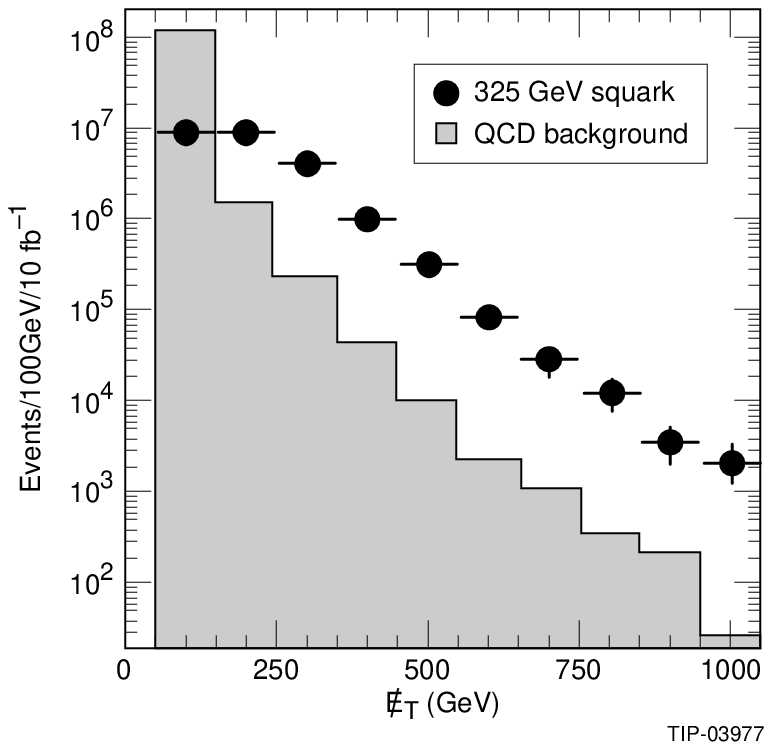}\forceright}{$\etmiss$ signal for
Case~II MSSM parameters defined in Table~9 (solid
circles) and for QCD background (histogram) after requiring at least 2
jets with $p_T > 75\,\GeV$ and making the sphericity and lepton veto
cuts described in the text.}

	The signal and background $\etmiss$ distributions for Case~I
with at least five jets and the sphericity and lepton veto cuts
described above are shown in Fig.~\SSi. The $(S+B)/B$ ratio, shown in
Fig.~\SSiSB, reaches about 8 for $\etmiss = 250\,\GeV$. Figure~\SSiSB\
also shows the $(S+B)/B$ ratio obtained using $\etmiss$ calculated
from the the missing $\nu$ and $\lsp$ momenta, with the rest of the
analysis unchanged. While the GEM calorimeter performance increases the
background at low $\etmiss$, it is about as good as that of a perfect
detector in the region for which the ratio is large.

\nfig\SSnjet{\dofig{85mm}{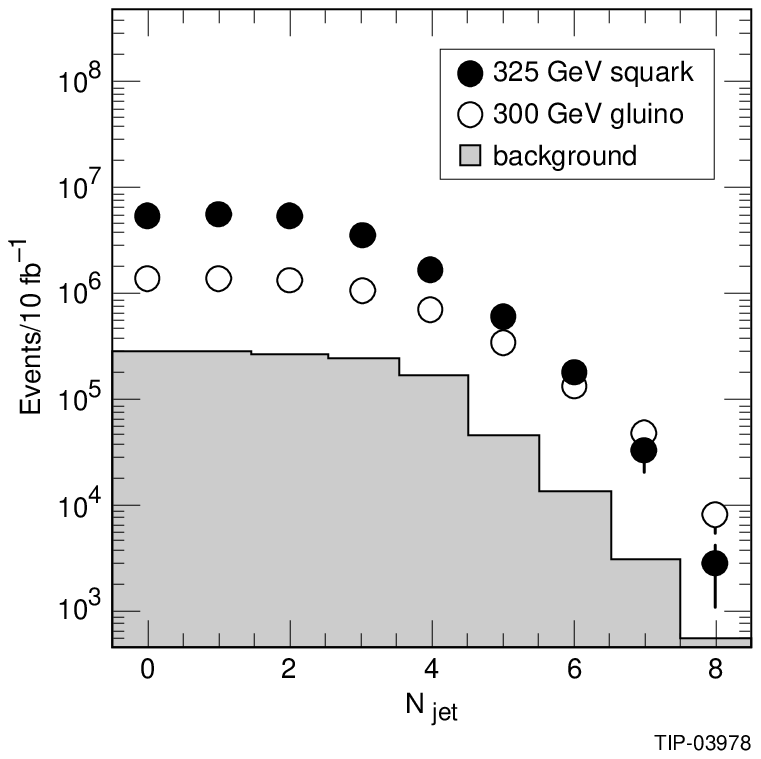}\forceleft}{Event numbers with
$\etmiss > 250\,\GeV$ and the sphericity and lepton veto cuts
described in the text vs.\ the minimum number $N_{\rm jet}$ of jets
with $p_T > 75\,\GeV$. Open circles: Case~I signal. Solid circles:
Case~II signal. Histogram: QCD background. Case~II has more events
with low jet multiplicity because $\widetilde q_R \to \lsp q$
dominates.}

	Figure~\SSii\ shows the signal and background for Case~II,
requiring at least two jets with $p_T > 75\,\GeV$ and the same
sphericity and lepton veto cuts. For this case the direct decay
$\squark_R \to \lsp q$ dominates and leads to a significantly harder
$\etmiss$ spectrum and to lower jet multiplicity. The $(S+B)/B$ ratio
is even larger in this case. Figure~\SSnjet\ plots the signals for
Cases~I and II and the Standard Model background for $\etmiss >
250\,\GeV$ and $S_T> 0.2$ vs.\ the minimum number, $N_{\rm jet}$, of
jets with $p_T > 75\,\GeV$. Both signals and backgrounds are constant
for $N_{\rm jet} \le 2$ because it is impossible to have a large
sphericity with only one jet. The signal falls off faster with
increasing $N_{\rm jet}$ for Case~II than for Case~I because
$\widetilde q_R \to \lsp q$ is dominant and gives a large rate for two
jets plus $\etmiss$. Thus, the $N_{\rm jet}$ dependence provides a
handle to distinguish among models.

	Given the large number of signal events, the statistical
significance of the signals is not an issue. The $t$, $W$ and $Z$
backgrounds can be checked using isolated lepton samples; the $b$ and
$c$ backgrounds can be checked using muons in jets. The $\etmiss$
resolution of the detector can be studied using inclusive data on QCD
jets and on $\gamma + \jets$ events. Given all these constraints, the
background should be reliably known, so observation of a signal 5--10
times that expected from the Standard Model should be very convincing.
The difficult problem of extracting the masses and other model
parameters is briefly discussed in Section~5.3.3.

\nfig\SSiii{\dofig{85mm}{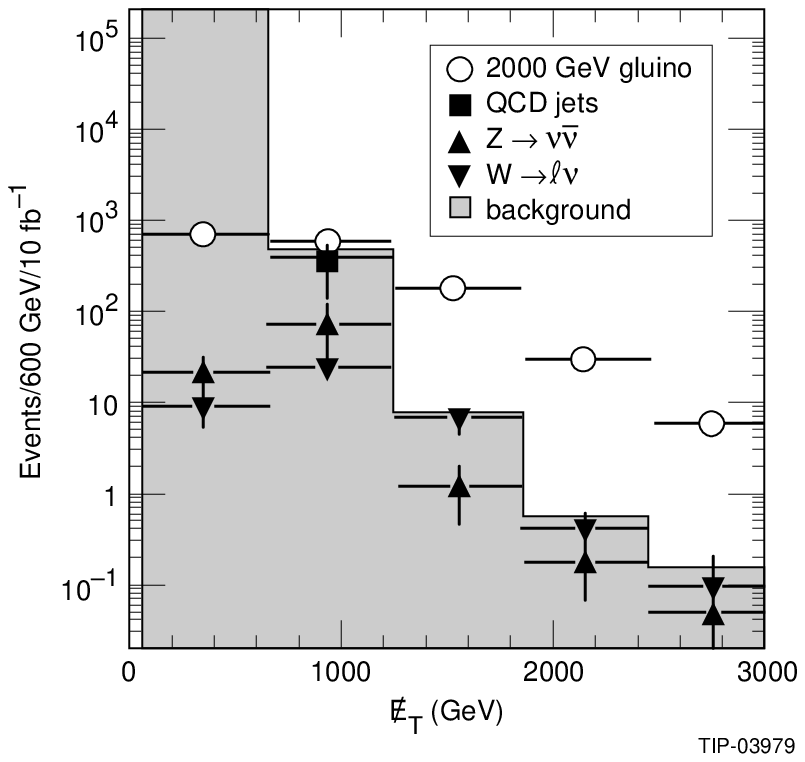}\forceright}{$\etmiss$ signal for
Case~III MSSM parameters defined in Table~9 after
requiring at least 4 jets with $p_T > 300\,\GeV$ and making the
sphericity and lepton veto cuts described in the text. Open circles:
Signal. Solid squares: QCD background. Triangles: $W \to \ell\nu$ and
$Z \to \nu\bar\nu$ backgrounds. Histogram: Sum of all backgrounds.}

	For heavy gluino and squark masses such as those in Case~III,
$\etmiss$ is so large that the $\etmiss$ resolution is not important.
Figure~\SSiii\ shows the signal and background $\etmiss$ distributions
for Case~III with at least four jets having $p_T > 300\,\GeV$ and with
the sphericity and lepton veto cuts identical to those for lighter
masses. Heavy flavor backgrounds, detector-induced backgrounds from
mismeasured QCD jets, and $W$ and $Z$ backgrounds are included. The
QCD background dominates for low $\etmiss$ but falls more rapidly than
the $W$ and $Z$ backgrounds, and so all must be considered. Since
several hundred signal events survive these cuts with large $S/B$, it
is evident that with an integrated luminosity of $\sscy$ GEM could
discover SUSY in this channel up to masses of order $2\,\TeV$, about
the upper limit if SUSY is related to electroweak symmetry breaking.

\subsubsec{Leptonic Signatures}

	In addition to the $\etmiss$ plus multijet signatures
described above, there are many other signatures for supersymmetry,
including a number involving two or more leptons.%
\ref\HBaer{H.~Baer, X.~Tata and J.~Woodside, Phys.\ Rev.\ {\bf D45},
142 (1992).}
In particular, since the gluino is a self-conjugate
Majorana fermion, $\gluino\gluino$ and $\gluino\squark$ pairs can give
isolated $\ell^\pm\ell^\pm$ pairs. Observing such like-sign pairs is
essential for establishing the Majorana nature of any gluino signal.
It also helps in separating gluinos and squarks. The dominant standard
model $\ell^\pm\ell^\pm$ background is expected to be from $t \bar t$
events in which either a $b \to \ell X$ lepton appears isolated or an
isolated lepton sign is wrongly determined.  These backgrounds,
calculated previously in Ref.~\HBaer, are found to be negligible.  For
light gluinos, such as those in Cases~I and II, the cross sections are
so large that it is sufficient to use only the $\mu^\pm\mu^\pm$
channel, for which the lepton signs are very well determined in GEM.
Therefore, only the issue of measuring the signs of electrons from
Case~III is addressed here.

	The same sample of Case~III signal events described in the
previous subsection was used for this analysis. While it is possible
to enhance the leptonic sample by forcing a particular decay chain,
e.g. $\gluino \to \widetilde\chi_1^\pm q \bar q'$,
$\widetilde\chi_1^\pm \to \lsp \ell^\pm \nu$, there are many such
chains possible, no one of which obviously dominates. It was therefore
decided to use the inclusive sample. For the background, only $t \bar
t$ events, which are expected to dominate, were considered. A total
sample of 30K $t \bar t$ events in ten bins with $50 < p_T <
3200\,\GeV$ were generated, forcing the decays $t \to e^+
\nu_e b$ and $\bar t \to \mu^- \bar\nu_\mu \bar b$. This sample was used
to determine the principal detector-induced background, that from
misidentification of $e^\pm$ signs in the central tracker. From this,
the total $\ell^\pm\ell^\pm$ background was determined.

	Electrons and muons with $p_T > 50\,\GeV$ and $\vert\eta\vert
< 2.5$ were identified using the relatively loose cuts described in
the previous subsection. These cuts, optimized for background
rejection rather than for signal detection, appear to be adequate to
identify this signal. At least two such leptons were required
satisfying the isolation criterion
\eqn\XX{
\mathop{{\sum_{R=0.2}\!}'} E_T < 0.1p_{T\,\ell} + 5\,\GeV\,.
}
Here, the prime on the sum indicates that the lepton itself is not
included. This cut effectively rejects\cite{\HBaer} the background
from $t \to \ell^+ \nu b$ and $\bar t \to \bar b X$, $\bar b \to
\ell^+ X$. In addition a missing energy $\etmiss > 500\,\GeV$ and a
transverse sphericity $S_T > 0.2$ were required. After these cuts, the
total dilepton rates for the signal and for the $t \bar t$ background
were comparable, so a very large rejection of unlike-sign pairs is not
needed.

\nfig\SSleps{\dofig{85mm}{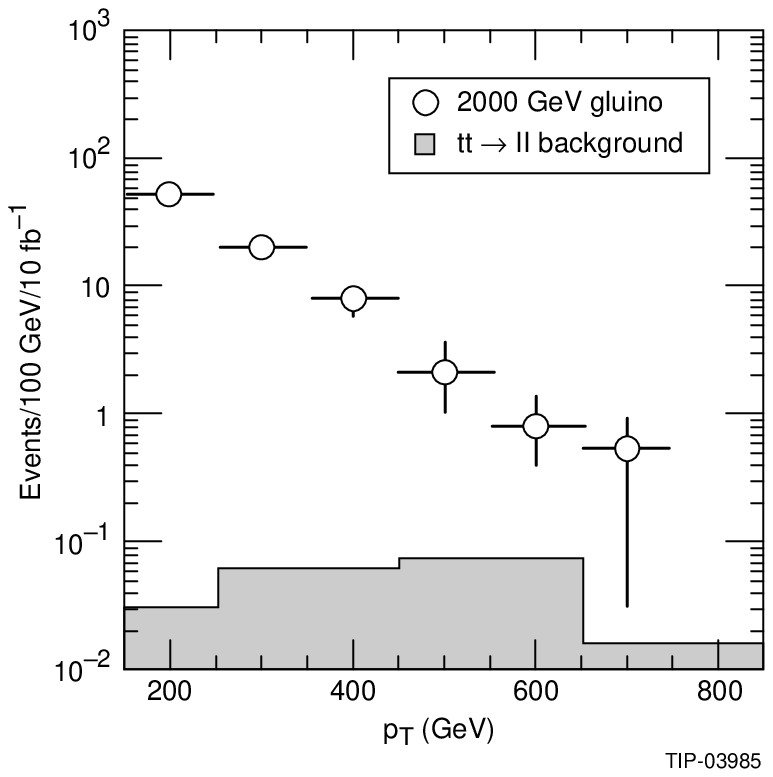}}{$p_T$ distributions for
the highest-$p_T$ isolated lepton in dilepton events containing two
isolated like-sign leptons. Open circles: Signal events generated with
Case~III MSSM parameters. Histogram: $t \bar t$ background from
mismeasured electrons in the GEM central tracker.}

	Figure~\SSleps\ shows the transverse momentum distribution of
the highest $p_T$ lepton in like-sign dilepton events satisfying the
above cuts. The lepton spectrum is soft relative to
the gluino mass because the leptons arise
from cascade decays.  Figure~\SSleps\ also shows the $t \bar t \to
\ell^+ \nu b\ell^- \bar\nu \bar b$ background in which an electron
sign is mismeasured by the GEM central tracker. The probability of
mismeasurement was determined using the |gemfast| parameterization of
the central tracker electron resolution, including the tail from
bremsstrahlung, as described in Section~4.2. Muon signs are assumed
perfectly determined, an excellent approximation at these momenta.
Since the cuts not dependent on the electron sign reduce the
background to the order of the signal, and since most of the signal
leptons have $p_T < 600\,\GeV$, for which the central tracker
determines signs with 95\% reliability, it is not surprising that the
background is small compared to the signal. The signal comprises
several tens of events per $10\,{\rm fb^{-1}}$ when both electrons and
muons are combined. It would be uncomfortably small if one had to rely
on only the $\mu^\pm\mu^\pm$ signal, which is a factor of four
smaller.

\subsubsec{SUSY Parameter Determination}

	In the MSSM there are at least eight mass parameters
($M_\gluino$, $M_\squark$, $M_{\widetilde t_L}$, $M_{\widetilde t_R}$,
$M_{\widetilde\ell_L}$, $M_{\widetilde\ell_R}$, $M_{\widetilde\nu_L}$,
and $M_A$), two additional parameters related to the Higgs sector
($\mu$ and $\tan\beta$), and still more parameters related to
$\widetilde t$ decay. Nonminimal SUSY models have even more
parameters. Since all supersymmetric particles in the MSSM ultimately
decay into the $\lsp$, which is invisible, it is not possible to
reconstruct any masses directly.

	The approximate mass scale can be inferred from the $\etmiss$
scale at which the signal deviates from the standard model background,
as can be seen by comparing Figs.~\SSi, \SSii, and \SSiii. The mean
$\etmiss$ for the distribution of the excess of events can be
calculated very accurately for low masses because of the high
statistics. However, the relationship of this mean to the masses is
model dependent. For example, the missing energy is lower and the jet
multiplicity is higher if $M_\squark > M_\gluino$ than if
$M_\gluino > M_\squark$.

	A large number of possible signatures may be used
to determine the parameters. These include the
$\etmiss$ cross section with multiple leptons,\cite{\HBaer}
multilepton cross sections arising from production of
$\widetilde\chi_1^\pm \widetilde\chi_2^0 \to 3\ell$ and similar
channels, and the observed cross sections or limits for $h,H \to
\gamma\gamma$; $h,H \to 4\ell$; and $t \to H^+b$. The tools to
simulate these signatures have recently been developed,\cite{\Isasusy}
and the methods to determine all of the MSSM parameters from these
signatures are being studied. However, it is clear from the previous
two subsections that a detector such as GEM, with reasonable calorimetry
out to $\eta \simeq 5.5$ and tracking capabilities for charged leptons,
is capable of observing clean samples of events in the relevant channels.


\gdef\wppm{W^{'\pm}}

\subsec{High Mass Physics at Ultrahigh Luminosity}

While the studies presented above show that, at the SSC design energy of
40~TeV, the standard integrated luminosity of $\sscy$ is sufficient for a
wide range of interesting physics, there remain many potential new physics
signals which require an integrated luminosity of order $\sscd$.
Practically speaking, this requires extended running at $\CL \simeq \uhl$.
These new physics opportunities are almost all at very high mass scales,
with effective cross sections of order $1-10\,\fb$.  Examples include
multi-TeV, weakly coupled $Z'$ and $W'$ bosons, quark-lepton substructure,
and very massive technirho vector bosons. In these and most other examples,
the new physics is signalled by an excess of well-isolated leptons at high
$\pt$. Robust and accurate measurement of such leptons was a primary design
goal of GEM, as it is for the LHC detectors.

This section demonstrates the capacity of GEM at ultrahigh luminosity for
precision measurements of the masses, widths and chiral couplings of
$4\,\tev$ $Z'$ bosons via $\ee$ and $\mup\mum$ final states and for the
discovery and study of quark-lepton substructure at the scale $\Lambda =
25\,\tev$ in Drell-Yan production of high-mass dimuons. For both, we show
that the GEM design permits distinguishing different models with a data
sample of $\sscd$. We also find that the reach of the GEM detector is $\mzp
\simeq 8\,\tev$ for new gauge bosons and $\Lambda = 30 - 35\,\tev$ for
quark-lepton substructure. Finally, we briefly describe approaches to
studying high-mass $\wppm \ra \ellpm \nu_\ell$ and quark-lepton
substructure in $\ol q q' \ra \mu^\pm \nu_\mu$. We believe these are
interesting because they effectively allow high-mass studies even when
there is a neutrino and even when there is no well defined subsystem
invariant mass.%
\ref\footrc{QCD radiative corrections have not been taken into account in
these studies. They are likely to increase the Drell-Yan signal rates.
Since all the physics backgrounds are very small after event selections are
made, higher-order corrections to them are not expected to change our
conclusions. The EHLQ-1 distribution functions were used for all simulations
in this section.}

The isolated high-energy lepton signals of these and most other new physics
processes are relatively free of physics backgrounds.  For example, a
rejection factor of $\CR(e/\jet)$ of $\CO (10^{-4})$ reduces the jet
backgrounds to $Z' \ra e^+ e^-$ to the level of a few percent.
Straightforward isolation cuts on muons completely remove the main physics
backgrounds to $\ol q q \ra \mup \mum$.  The most stringent requirements
will be reliable operation of the trigger and detector and pattern
recognition at such high luminosities.

\subsubsec{Precision Studies of New Heavy $Z'$ Gauge Bosons}

Extensions of the standard electroweak gauge group, $SU(2) \otimes U(1)$,
involve additional neutral gauge bosons $Z'$ and sometimes charged $\wppm$
vector bosons. In the models of interest, the new weak bosons couple to
quarks and leptons with strength of $\CO (e)$. In one year of running at
$\CL = \hl$, one could observe $\mzp \simeq 4-6\,\tev$, with 10 $Z' \ra
\ee$ events in a narrow $\mee$ range where none are expected.\cite{\ehlq}
If such a boson were found, high-statistics studies would be needed to
determine its nature.  This subsection considers a $4\,\tev$ $Z'$ boson.
For this mass, hundreds of events would be detected per year at ultrahigh
luminosity.  Since the couplings of such $Z'$ bosons must be
flavor-conserving, they may be detected in both $Z' \ra \ee$ and $Z' \to
\mup\mum$ modes. Very precise measurements of the $Z'$ mass and width can
be made via the $\ee$ decay. As we shall describe below, the chiral nature
of the $Z'$ couplings to quarks and leptons can be investigated in the
$\mup\mum$ mode by measuring the distribution in the angle $\theta$ between
the outgoing $\mum$ and the incoming quark.

The goal of the studies presented here is to determine GEM's ability to
distinguish two different $Z'$ models by the bosons' widths and angular
distributions. The two models considered were:

\item{1.} A left-right model, in which $SU(2)_L \otimes SU(2)_R \otimes
U(1)$ breaks down to $SU(2)_L \otimes U(1)$. The $SU(2)_R$ coupling
constant was taken to be the same as the $SU(2)_L$ one. The extra $Z$-boson
of this model is called $Z_1$ below.

\item{2.} A model in which the grand unification group $SO(10)$ breaks
down to $SU(5) \otimes U(1)$, then to $SU(2)_L \otimes U(1)$. Such a
model may have an extra $Z' \equiv Z_2$ well below the unification scale.

\noindent The left-and right-handed couplings to quarks and leptons in the
two models are given by
\eqn\zprime{\eqalign{
 g_{u L} = \beta \ts &, \quad g_{u  R} = \beta + \gamma \ts;\cr
 g_{d L} =  \beta \ts &, \quad g_{d R} = \beta - \gamma \ts;\cr
 g_{\nu L} = -3 \beta \ts &, \quad g_{\nu R} = -3\beta + \gamma \ts;\cr
 g_{\ell L} = -3 \beta \ts &, \quad g_{\ell R} = -3\beta - \gamma \ts.\cr}}
The parameters $\beta$ and $\gamma$ are
\eqn\betgam{\eqalign{
& \beta = 0.0528 \ts, \quad \gamma = - 0.3630 \qquad {\rm (Model \ts\ts 1)}\cr
& \beta = 0.0979 \ts, \quad \gamma = - 0.1958 \qquad {\rm (Model \ts\ts 2)}
\ts .\cr}}

In calculating the $Z'$ decay widths, it is assumed that there are three
generations of quarks and leptons and that right-handed neutrinos exist and
are much lighter than $\mzp$. If there are no other significant decay modes,
the widths are given by
\eqn\zwidth{
\Gamma(Z' \ra f_i \ol f_i) = {2 \alpha \mzp C_i \over {3 \sin^2 2 \theta_W}}
\ts \left(g_{i L}^2 + g_{i R}^2 \right) \ts,}
where $C_i =$ 3 for quarks and 1 for leptons. The full widths in the two
models are $\Gamma(Z_1) = 105\,\gev$ and $\Gamma(Z_2) = 67.7\,\gev$.

\subsubsubsec{$Z' \ra e^+ e^-$}

PYTHIA 5.6 was used to generate 3000 events for each of the two $Z'$
models, including full $\gamma/Z/Z'$ interference.%
\ref\zpsmehs{S.~McKee and E.~H.~Simmons, {\it $Z'\ra \ee$ studies at $\uhl$
with the GEM Detector}, GEM PN--93--6 (1993).}
Pileup at $\CL = \uhl$ was simulated by merging with the hard scattering
process a set of minimum bias events. The latter were Poisson-distributed,
with a mean number of 16~events per beam crossing. The total signal cross
sections for $3.75\,\tev <\mee < 4.25\,\tev$ are $5.44\,\fb$ for Model\ 1
and $4.54\,\fb$ for Model\ 2. No rapidity cut was imposed on the generated
events.

All tracks found by the central tracker (which, at $\uhl$, has its inner
silicon layers removed, leaving 8~planes of tracking) were processed and a
most likely $Z'$~vertex was found using momentum-weighted $z$~locations.
Then the $5\times 5$ towers in the EM calorimeter were examined. Electron
candidate towers were required to satisfy the following criteria:

\item{1.} $\abseta < 2.46$ so as to be within the tracker's coverage.
Electrons falling in the transition region between barrel and endcap ($1.01
< \abseta < 1.16$) were excluded because their energies are not as
precisely measured. The geometrical acceptance of the detector for the $Z'
\ra \ee$ events is 78\%.

\item{2.} $\et > 250\,\gev$ in a $5 \times 5$ tower in the EC; 99.4\%
of the events with $\abseta < 2.46$ passed this cut.

\item{3.} Transverse energy leakage into the $3\times 3$ HC tower behind
the $5 \times 5$ EC tower less than 5\% of the energy in the EC tower;
99.2\% of the remaining events passed this selection. If a high energy EC
cluster was isolated at this level, any energy in the first layer of the HC
behind the cluster was added to the EC energy to improve the resolution.

\item{4.} A charged track candidate with $\pt > 1\,\gev$ must be within $R
= 0.0632$ ($\Delta \eta = 0.06 \times \Delta \phi = 0.02$) of the
centroid of the EC tower. Essentially all electron candidates passed this
cut. Much tighter track restrictions will be imposed below to reduce jet
backgrounds.

\item{5.} A calorimetric isolation cut of
\eqn\eisol{\sum_{R = 0.55} \et < 1.15 \sum_{5 \times 5} \et(\rm EC) \ts.}
was applied. This accepted 99.6\% of the signal events.

\noindent The trigger efficiency for these events is close to 100\%. Thus,
GEM would collect 415 $Z_1 \ra \ee$ and 350 $Z_2 \ra \ee$ events per
$\sscd$ run.

Backgrounds to the $Z' \ra \ee$ signal come from misidentifying jets as
electrons in QCD jet production and in $pp \ra W + \ts \jets$ with $W \ra e
\nu$. They may also come from isolated, high-mass $\ee$ pairs from $\ol t t$
production. The cross section for QCD jets with $\abseta < 2.46$ and
invariant mass $M_{jj} = 4000 \pm 250\,\GeV$ is $704\,\pb$. The
corresponding cross section for $W + \ts \jets \ra e + \ts \jets$ is about
$25\,\fb$. The rate for $\ol t t \ra \ee +X$ with generated $M_{\ol t t} >
3.5\,\tev$ and $\pt(t,\ol t) > 100\,\gev$ is $20.4\,\fb$, for $m_t =
140\,\gev$.

To be considered as background electron candidates, jets had to have $\et >
200\,\gev$, one or two charged tracks in a cone of $\Delta \eta \times
\Delta \phi = 0.09 \times 0.03$ about the jet direction, and at least one
charged track with $\pt > 50\,\gev$. This reduced a generated sample of
120K jets to 9360 (in 8890 events). These were then subjected to the cuts
used to select electron candidates and 26~jets passed. This implies a jet
rejection factor of $\CR(e/\jet) \simeq 2.9\times 10^{-4}$ at the 95\%
confidence level. The surviving dijet rate is $0.06\,\fb$, less than 2\% of
the accepted signal rate.  The $W+\jet$ background is negligible. After the
isolation cut, the $\ol t t \ra \ee +X$ background is completely eliminated
by requiring $\mee > 3.5\,\tev$. (See the discussion below of the $\ol t t
\ra \mup\mum +X$ background to $Z' \ra \mup\mum$.)
When the requirement of exactly one {\it reconstructed} track with $\pt >
50\,\gev$ was applied to the $Z'$ candidates, 83\% of the events passed,
for a net acceptance of 64\%.

The $\ee$ mass resolution depends on the resolutions in both the electron
energy and angle (see Eq.~\HEmass). In the EC energy resolution, only the
constant term matters for a $4\,\tev$ $Z'$. For the GEM baseline, this is
0.4\%. (Any bremsstrahlung from the passage of the very high energy
electrons through the CT material enters the same $5 \times 5$ EC tower
which defines the electron shower and so does not affect the resolution.)
For the high energies of interest here, the $e^+$ and $e^-$ directions can
be determined to an accuracy of about $2.4\,{\rm mrad}$ using the
longitudinal segmentation in the EC. This gives $\Delta \theta = 3.4\,{\rm
mrad}$. The angular resolution can be significantly improved by using the
tracker. The vertex location found using the two highest-$\et$ electrons
has a resolution of 1.5~mm, while the EC hit location has a resolution of
0.25~mm for such high energy showers. The resulting angular resolution is
$\Delta \theta = 1.5\,{\rm mrad}$. Then, since the typical opening angle
between the electrons is greater than $60^\circ$, $\Delta \mee \simeq 0.003
\times \mzp$, dominated by the EC energy resolution. This is appreciably
less than $\gzp/2.35$ for the two models under consideration. Indeed,
models in which the the $Z'$~width was artificially made as small as
$25\,\gev$ were considered and the width was shown to be
resolvable.\cite{\zpsmehs}

Several luminosity-dependent effects were considered that might affect the
$\mee$ resolution.\cite{\zpsmehs} Pileup noise in the calorimeters is
unimportant. The effect of pileup tracks
on the performance of the tracker for the purpose of vertex determination was
included.

Figures\ \LZone\ and \LZtwo\ show the $\mee$ distributions for the two $Z'$
models as generated by PYTHIA and as reconstructed by |gemfast|. The data
in these figures correspond to about $200\,\fb^{-1}$. These
distributions are fit very well by a Lorentzian. The reconstructed masses
determined from this fit are within $2\,\gev$ of the generated $4\,\tev$.
The reconstructed widths are $\Gamma_{Z_1} = 123.0 \pm 5.2\,\gev$ and
$\Gamma_{Z_2} = 88.5 \pm 4.3\,\gev$. These are to be compared with the
generated widths of $116.4 \pm 5.0\,\gev$ for $Z_1$ and $80.9 \pm
3.9\,\gev$ for $Z_2$.%
\ref\footi{The discrepancy between the theoretical widths determined from
Eq.\zwidth and those from fitting the PYTHIA distributions corresponds to
about a 0.5\% Gaussian noise term that we have not yet understood. Since
PYTHIA and |gemfast| widths agree within the expected detector resolution,
this is not a matter of immediate concern.}
Within errors, the generated and reconstructed widths differ by an amount
that corresponds to the detector resolution estimated above. This was
verified for the artificially narrow $Z'$ models. While the error in $\gzp$
will be dominated by statistics, the main error in $\mzp$ will be
systematic, arising from possible nonlinearity in the EM energy scale in
the TeV region.

\nfig\LZone{\dofig{85.5mm}{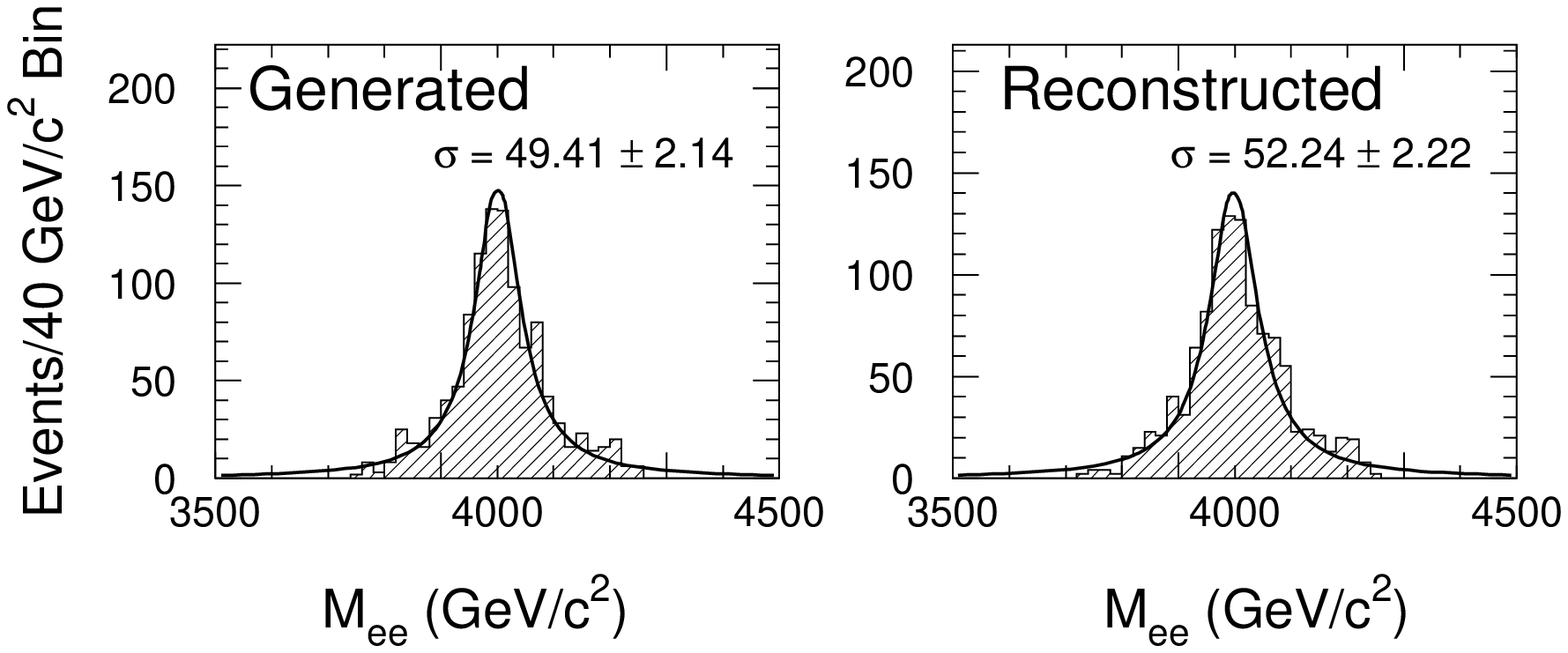}\forceleft}{The $\mee$ distribution
for a $4\,\TeV$ mass $Z'$  of Model~1 as described in the text. Shown are
the generated mass distribution (all events within acceptance) and the
reconstructed distribution (passing geometric and calorimetric cuts 1--5).
The number of events corresponds to $200\,\fb^{-1}$.}

\nfig\LZtwo{\dofig{85.5mm}{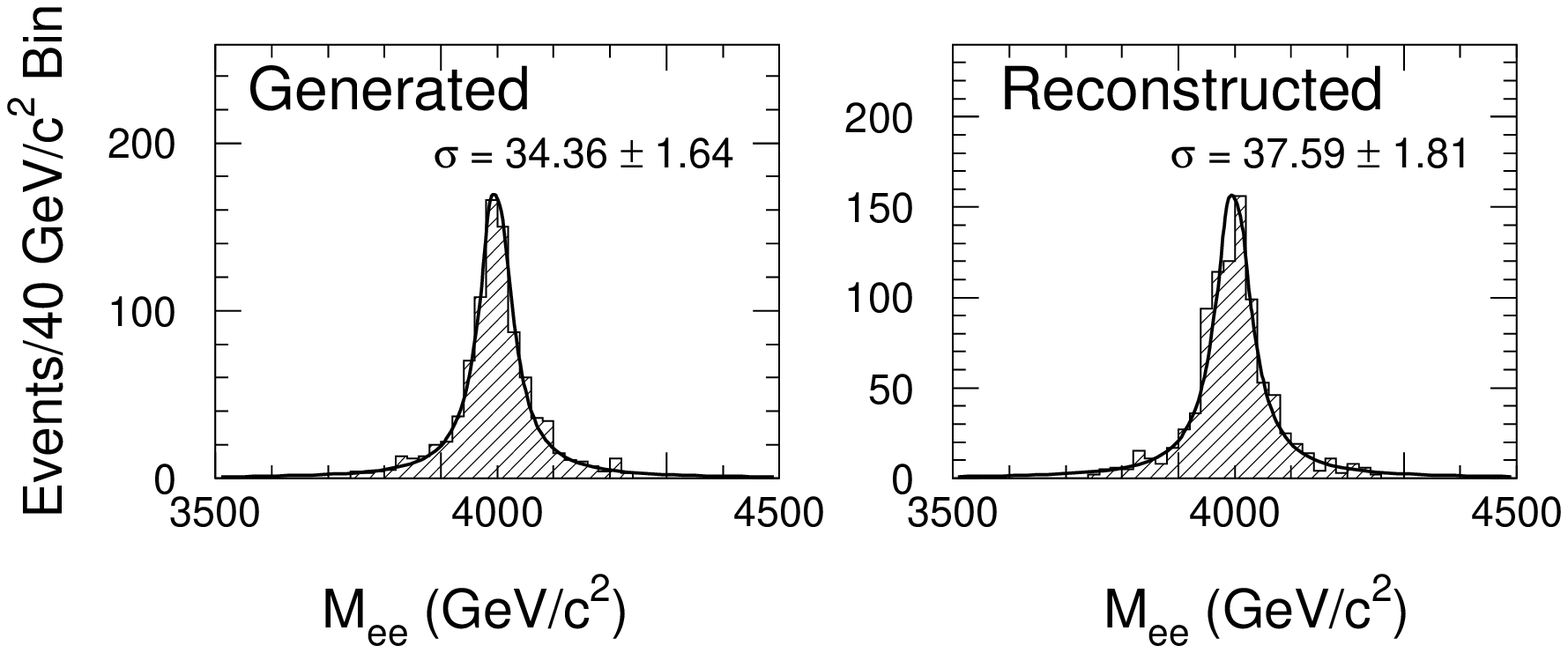}\forceright}{The $\mee$
distribution for a $4\,\TeV$ mass $Z'$  of Model~2 as described in the
text. Shown are the generated mass distribution (all events within
acceptance) and the reconstructed distribution (passing geometric and
calorimetric cuts 1--5). The number of events corresponds to
$200\,\fb^{-1}$.}

%

Finally, we estimated the reach of the GEM detector in $\mzp$ for a
$40\,\tev$ $pp$~collider with integrated luminosity $\sscd$. Assuming that
ten detected $Z' \ra \ee$ events are sufficient for discovery (there is no
background after cuts), the reach was found to be about $8\,\tev$ for both
models considered. The reach for other models may be more or less by
$1-2\,\tev$. This high mass, then, was used to set the upper end of the
dynamic range of the GEM electromagnetic calorimeter.

\subsubsubsec{$Z' \ra \mup \mum$}

The chiral couplings of $Z'$ bosons to quarks and leptons determine
the lepton angular distribution. Thus, the $\cos\theta$ distribution
($\theta$ is the angle between the incoming quark and outgoing
$\mu^-$) can help distinguish alternative models.  The angles and
charges of high energy muons can be measured reliably even at
ultrahigh luminosity. We imposed the criterion that muon charges must
be known with at least $2.5\sigma$ significance. At least 99\% of the
signal events that passed the isolation and invariant mass cuts
described below have the charges of both muons determined. Events with
one mismeasured sign were rejected. The probability of two mismeasured
signs is $\CO (10^{-4})$ and no such events are expected in a data
sample of $\sscd$.

To measure the $\cos\theta$ distribution, one must determine, in addition
to the muons' charges and directions, the direction of the incoming quark
in the lab. The procedure usually followed in these studies has been to
determine the (valence) quark direction by requiring that the $Z'$ be
produced at large rapidity, $\eta_{Z'} \simge 1$.%
\ref\rosner{P.~Langacker, R.~Robinett and J.~Rosner, Proceedings of the
1984 Summer Study on the Design and Utilization of the Superconducting
Super Collider, R.\ Donaldson and J.\ G.\ Morfin, editors; 812 (1984)}
Only about 20\% of the heavy $Z' \ra \mup\mum$ events would pass this cut.
We found it better to use all the events and plot the distribution in
\eqn\thstar{
\cstar = {\rm sgn}(\eta_B) \ts
\tanh\left({\eta_{\mum} - \eta_{\mup} \over {2}}\right) \ts .}
\nref\zpmm{M.~Mohammadi and W.~Orrick, {\it Studies of $Z' \ra \mup\mum$ at
Ultrahigh Luminosity with the GEM Detector}, GEM TN--93--364 (1993).}%
where $\eta_B = \half(\eta_{\mum} + \eta_{\mup})$ is the boost rapidity of
the $\mup\mum$ center-of-mass frame.\cite{\GEMLOI,\response,\zpmm}
For $\rshat > 2\,\tev$, the quark distribution is so much harder than the
antiquark distribution that ${\rm sgn}(\eta_B)$ determines the correct quark
direction at least 75\% of the time. Then $\theta^* \approx \theta$, since
the $p_T$ of the c.m. system is generally much less than $\rshat$.

PYTHIA 5.6 was used to generate 1000 events of $Z' \ra \mup \mum$ with
$\vert \mmm - \mzp \vert < 250\,\gev$ for each of the two models
described in the previous subsection. The cross sections in this
simulation were found to be $\sigma(Z_1 \ra \mup\mum) = 5.86\,\fb$ and
$\sigma(Z_2 \ra \mup\mum) = 4.49\,\fb$. The geometrical acceptance of
the muon system for all events was found to be 63\% for both models;
for muons with $\abseta < 2.46$, it is about 70\%. For these muons,
the trigger was highly efficient.  Thus, before any additional losses,
a data sample of $\sscd$ yields 375 $Z_1 \ra \mup\mum$ and 285 $Z_2
\ra \mup\mum$ events. A study of the muon system's geometrical
acceptance for these $Z'$ events was carried out using a full GEANT
simulation, with results essentially identical to those obtained here
with |gemfast|.

The only physics backgrounds to the $Z'$ signal are continuum Drell-Yan
production and $t \ol t \ra \mup \mum + X$. After momentum smearing is
included, the signal region used for angular distribution studies was taken
to be $3.2\,\tev < \mmm < 4.8\,\tev$. The Drell-Yan background was studied
by generating 4000 $\gamma/Z/Z'$ events with $\mmm > 1.5\,\tev$. Drell-Yan
production contributed six background events to the signal region.

The $\ol t t \ra \mup\mum + X$ background was studied by generating 200K
events with generated $M_{\ol t t} > 1.4\,\tev$ and $\pt(t,\ol t) >
100\,\gev$. The rate for this background is $1.4\,\pb$, so the sample
corresponds to $140\,\fb^{-1}$. (Other potential backgrounds, such as $t W$
and $WW$ production have very much smaller rates.) Detector response to
events having $\mmm > 1\,\tev$ and $\vert \eta_\mu \vert < 2.6$ at the
generator level was simulated with |gemfast|. There were 2900 such events
(for $\sscd$), of which 2700 passed the geometrical acceptance simulated in
|gemfast|. This is higher than the geometrical acceptance for $Z' \ra
\mup\mum$ events because most of the muons from $t$-decay follow their
parent's direction, which is fairly close to the beam, i.e., away from the
dead region near $\eta = 0$. Since these muons are usually accompanied by
hadronic debris, the background can be removed by a combination of
isolation and $\mmm$ cuts.

An isolation cut using a large cone
\eqn\muisol{\sum_{R = 0.7} \et < 0.1\ts \pt(\mu)}
was made. Here, the sum is over energy in the full calorimeter and
$\pt(\mu)$ is the muon momentum corrected for energy loss in the calorimeter.
This cut rejects 96.5\% of the $t \ol t$ background while retaining 94\% of
the $Z'$ events. All these events had identified muon charges. Finally, a
cut on the measured dimuon mass of $\mmm > 3.2\,\tev$ left no $t \ol t$
background events.

A separate GEANT simulation showed that the rate of mismeasured low-$\pt$
muons (e.g., ones with their tracks straightened by scattering) is small
compared to the real high-$\pt$ muon rate.  Therefore, the probability of
getting two mismeasured high-$\pt$ muons with $\pt$ balanced to within
$300\,\gev$ is wholly negligible.

\nfig\LZtwomu{\dofig{85mm}{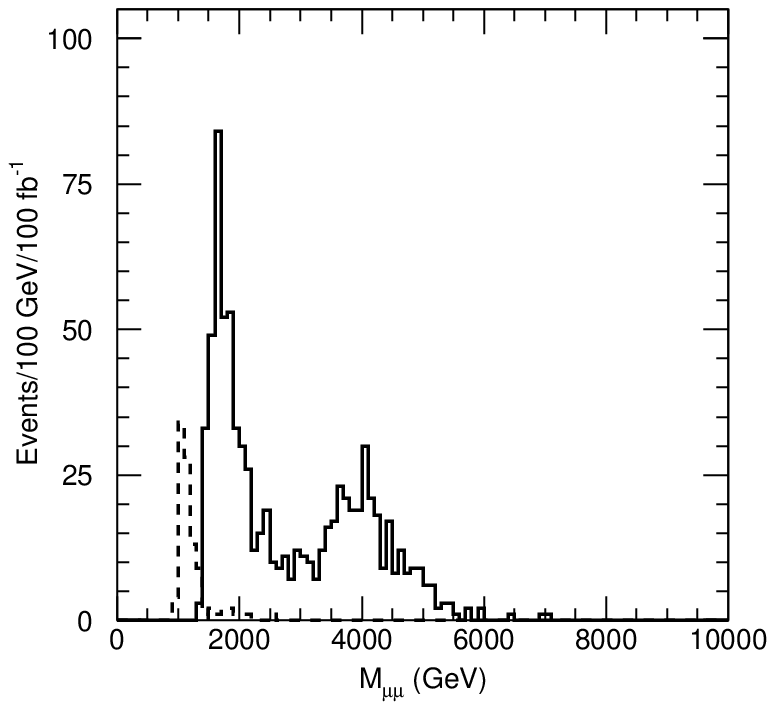}\forceright}{The $\mmm$
distribution for the Drell-Yan
continuum and $Z_2 \ra \mup\mum$. The $\ol t t \ra \mup \mum + X$
background before the invariant mass cut is also shown (dashed).}

A potentially serious loss to the $Z' \ra \mup\mum$ signal comes from
muon-scattering debris that sprays into the muon system, degrading the
reconstruction efficiency. This usually affects the first superlayer of the
muon system.  The average reconstruction efficiency for the high-energy $Z'$
muons was found to be 85\% per muon.  the reconstruction efficiency for $Z'
\ra \mup\mum$ events was raised to 96\% by accepting events with at least
one well-reconstructed muon plus one muon with sufficient hits in two
unaffected superlayers to determine its angle.%
\nref\twmurecon{T.\ Wenaus, {\it A Reconstruction Program for the GEM Muon
Detector}, GEM TN--93--388 (1993).}%
\cite{\twmurecon}
Since the $\pt$ of the poorly-reconstructed muon is expected to match
that of the well-reconstructed one to within $300\,\gev$, its momentum
can be determined with sufficient accuracy. If necessary to do so, it
is reasonable to assume that the charge of the bad muon is opposite
that of the good muon.

In summary, the overall acceptance of the signal (including all
produced events) is 60\%. There are 350 $Z_1$ and 270 $Z_2 \ra
\mup\mum$ events detected by GEM with $\sscd$ of data. The Drell-Yan
dimuon spectrum for the $Z_2$ model is shown in Fig.\ \LZtwomu, which
shows all events that had generated invariant mass greater than
$1.5\,\tev$ and passed all cuts. For comparison, the distribution of
dimuon events from the $\ol t t$ background that passed the isolation
cut is also shown. The Drell-Yan spectra for the two models were fit
with an exponential plus a Gaussian to determine the masses of the
$Z'$ enhancements. These were found to be $M_{Z_1} = 4036 \pm
40\,\gev$ and $M_{Z_2} = 3968 \pm 65\,\gev$. The errors are purely
statistical.

\nfig\LZonec{\dofig{85mm}{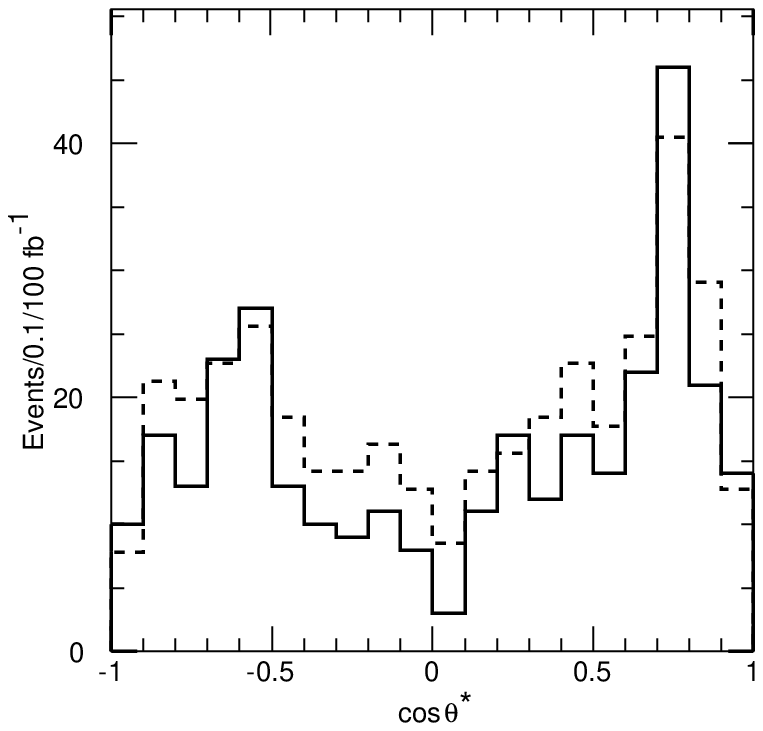}\forceleft}{Generated (dashed) and
reconstructed (solid)
$\cstar$ distributions for $Z_1 \ra \mup\mum$ events passing signal
selections.}

\nfig\LZtwoc{\dofig{85mm}{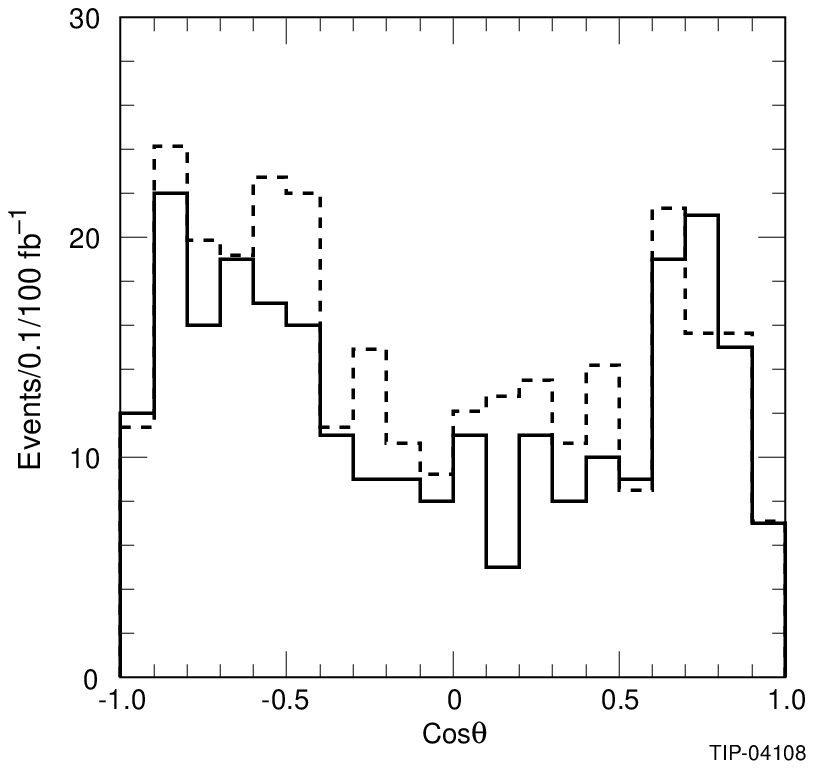}\forceright}{Generated (dashed) and
Reconstructed (solid)
$\cstar$ distributions for $Z_2 \ra \mup\mum$ events passing signal
selections.}

The measured $\cstar$ distributions for Models\ 1 and 2 are shown in Figs.\
\LZonec\ and \LZtwoc. Events in the mass range $3.2 - 4.8\,\tev$ were used
for this analysis. There are 319 $Z_1$ and 257 $Z_2$ events in this sample;
the missing events were lost due to momentum smearing. As noted above, only
six Drell-Yan background events contaminate this sample. For comparison,
the PYTHIA-generated distributions, corresponding to a perfect detector
with GEM's $\eta$-coverage, are shown in these figures. The agreement
between theoretical expectation and simulated measurement is quite good.
The forward-backward asymmetry is defined as
\eqn\FB{\afb = {N(\cstar > 0) - N(\cstar < 0) \over
	       {N(\cstar > 0) + N(\cstar < 0) }} \ts. }
The results for the two models are
\eqn\FBasym{\eqalign{
\afb(Z_1) &= 0.113 \pm 0.056 \quad (\rm GEM) \cr
          &= 0.083 \pm 0.051 \quad (\rm PYTHIA)\cr
\noalign{\smallskip}
\afb(Z_2) &= -0.090 \pm 0.062 \quad (\rm GEM) \cr
          &= -0.115 \pm 0.058 \quad (\rm PYTHIA)\cr}}
The PYTHIA and |gemfast| asymmetries agree well within the statistical
errors and the central values of the two models are separated by about
$3.5\sigma$.

\subsubsec{Studies of Quark-Lepton Substructure in Drell-Yan Processes}

The greatest model-independent sensitivity to quark-lepton substructure is
obtained by measuring the highest possible dilepton and dijet masses. If a
signal is found, the inherent precision of lepton measurement at high
energies will make the Drell-Yan process a much more incisive tool than
high-$\et$ jet production for studying substructure. However, as with the
$Z'$, hundreds of events at very large energies will be required.  The
contact interactions induced by quark-lepton substructure must be
flavor-symmetric if the substructure scale $\Lambda \simle 500\,\tev$.
Thus, the Drell-Yan processes $p p \ra \ellp \ellm$ and $\ellpm \nu_\ell$
will have equal cross sections for electrons and muons.  Signals in $p p
\ra \mup \mum$ are considered here. While it is important to observe the
substructure signal in $pp \ra \ee$ as well, the $\ee$ spectrum cannot help
distinguish between models because a difference in rate can be compensated
for by a change in $\Lambda$. A brief discussion of substructure studies in
the $\mu^\pm \nu_\mu$ channel is presented in Section\ 5.6.3.

Quark-lepton substructure modifies dimuon production in two ways: an excess of
events at high $\mmm = \rshat$~\cite{\ehlq} and a deviation in the angular
distribution of the outgoing $\mum$ relative to the incoming quark.%
\ref\rsclr{R.~S.~Chivukula and L.~Randall, Phys.\ Lett.\ {\bf 202B}, 429
(1988).}
We consider two different chiral forms for the contact interaction
$\CL_{\Lambda}$ arising from substructure at $\Lambda = 25\,\tev$, well
beyond that which can be studied at a 40~TeV hadron collider with
$\hl$. The goal in this study is to
distinguish these two models by their muon angular distribution.

In the first model, only left-handed quarks and leptons are composite and
the contact interaction is assumed to be the product of two weak-isoscalar
currents (the ``ISO'' model):
\eqn\ISO{
\Liso = - {4 \pi \over {\Lambda^2}} \ts \ol Q_{La} \gamma^{\mu}
Q_{La} \ts \ol L_{Lb} \gamma_{\mu} L_{Lb} \ts\ts ,}
where $Q_{La} = (u_a, d_a)_L$ and $L_{La} = (\nu_a, \ell_a)_L$ are
left-handed quark and lepton fields and $a,b = 1,2,3$ label generations.
This interaction, together with the standard Drell-Yan processes, produces
the subprocess cross section
\eqn\sigISO{
{d\sigh(q_i \ol q_i \ra \ellm \ellp) \over {d (\cos \theta)}} =
{\pi \alpha^2 \over {24 \shat}} \ts \Bigl[ A_i(\shat) \ts (1 + \cos
\theta)^2
+ B_i(\shat) \ts (1 - \cos \theta)^2 \Bigr] \ts .}%
The functions $A_i$ and $B_i$ are given by
\eqn\ISOterm{\eqalign{
A_i(\shat) = &\biggl[ {\shat \over {\alpha\Lambda^2}} +
\CQ_i + {4 \over {\sin^2 2 \thw}} \left(T_{3i} - \CQ_i \sin^2 \thw \right)
\left(\textstyle \half - \sin^2 \thw \right)
\left({\shat \over {\shat - M_Z^2}}\right) \biggr]^2 \cr
+&\biggl[ \CQ_i + \CQ_i \tan^2 \thw \left({\shat \over {\shat - M_Z^2}}\right)
\biggr]^2 \ts ; \cr\cr
B_i(\shat) = &\biggl[ \CQ_i - {1 \over{\cos^2 \thw}}
\left(T_{3i} - \CQ_i \sin^2 \thw \right) \left({\shat \over {\shat - M_Z^2}}
\right) \biggr]^2 \cr
+&\biggl[ \CQ_i - {1 \over {\cos^2 \thw}} \CQ_i  \left(\textstyle \half -
\sin^2
\thw \right) \left({\shat \over {\shat - M_Z^2}}\right) \biggr]^2 \ts .\cr}}
Here, $\sin^2 \thw \cong 0.23$, $(T_{3i}, \CQ_i) = (\half, {2 \over {3}})$ for
$q_i = u_i$ and $(-\half, -{1 \over {3}})$ for $q_i = d_i$ At high
energies, $\shat > \alpha \Lambda^2$, $A_i(\shat) \simeq (\shat/\alpha
\ts\Lambda^2)^2$ and the angular distribution of $\ellm$ relative to $q_i$
is approximately $(1 + \cos \theta)^2$.

In the second model the contact interaction is helicity nonconserving
(the ``HNC'' model):
\eqn\HNC{
\Lhnc =  - {4 \pi \over {\Lambda^2}} \ts
\epsilon_{ij} \ol Q_{Lia} u_{Ra} \ts \ol L_{Ljb} \ell_{Rb} + {\rm h.
\ts\ts c.} \ts\ts ,}
where $i,j = 1,2$ label indices in an electroweak doublet and $\epsilon_{12} =
-\epsilon_{21} = 1$.  This interaction, while theoretically unlikely, is
studied here because it generates an angular distribution that becomes
isotropic at large $\shat$. The interaction $\Lhnc$ affects $u_i \ol u_i \ra
\ellm \ellp$ only:
\eqn\sigHNC{
{d\sigh(u_i \ol u_i \ra \ellm \ellp) \over {d (\cos \theta)}} =
{\pi \alpha^2 \over {24 \shat}} \ts \Bigl[ A_u(\shat) (1 + \cos
\theta)^2
+B_u(\shat) (1 - \cos \theta)^2 \Bigr] +
{\pi \shat \over {12 \Lambda^4}} \ts .}
The contributions of $\Lhnc$ and $\gamma/Z^0$ do not interfere because
their chiral structures are different.

PYTHIA 5.5 was used to generate 1500 events each of standard Drell-Yan (DY)
and DY modified with the ISO and HNC contact interactions with $\Lambda =
25\,\tev$. The muons were required to have generated $\mmm > 2\,\tev$. The
cross sections for events in which both muons have $\abseta < 2.46$ are
$2.83\,\fb$, $9.44\,\fb$ and $7.88\,\fb$ for the DY, ISO and HNC cases.
Detector response to events was simulated using |gemfast| with
Gaussian pileup for $\CL = \uhl$. The acceptance was 70\% for events with
both muons in $\abseta < 2.46$, consistent with the acceptance of $Z' \ra
\mup\mum$ events.  Of these, 99.5\% of the standard DY events had the
charges of both muons determined. This dropped to 96\% for the ISO and HNC
models, reflecting the excess of high-energy muons generated by the contact
interactions.

The discussion of backgrounds and muon reconstruction is essentially
the same as in the analysis of $Z' \ra \mup\mum$. The isolation
criterion defined in Eq.\muisol\ and the invariant mass cut $\mmm >
2.5\,\tev$ eliminated the physics backgrounds. The acceptance of these
two cuts for signal events passing previous selections was 41\% for
the DY events, 60\% for the ISO events, and 66\% for HNC. The higher
acceptance in the ISO and HNC cases is due to the excess of high-mass
dimuons. Retaining those events in which at least the momentum of one
muon and the angle of the other are well-measured, the net
reconstruction efficiency was found to be 95\% for the DY and ISO
cases and 92\% for the HNC model.

\nfigwide\LCmmm{\dofig{98.5mm}{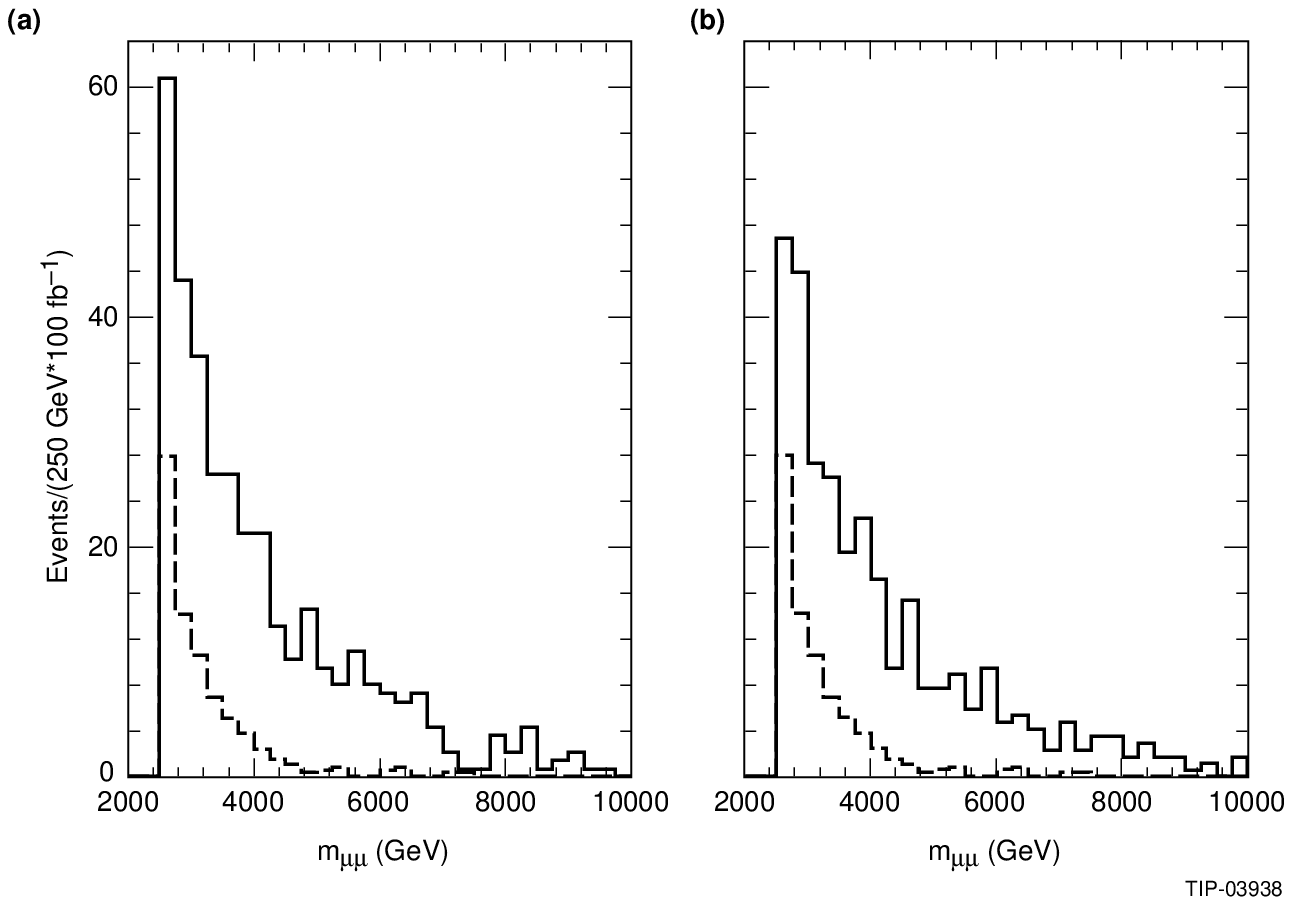}}{The $\mmm$
distributions for the ISO (a) and HNC (b)
substructure models defined in the text. The lower dashed histogram is the
standard Drell-Yan distribution.}

\nfigwide\LCcos{\dofig{96.5mm}{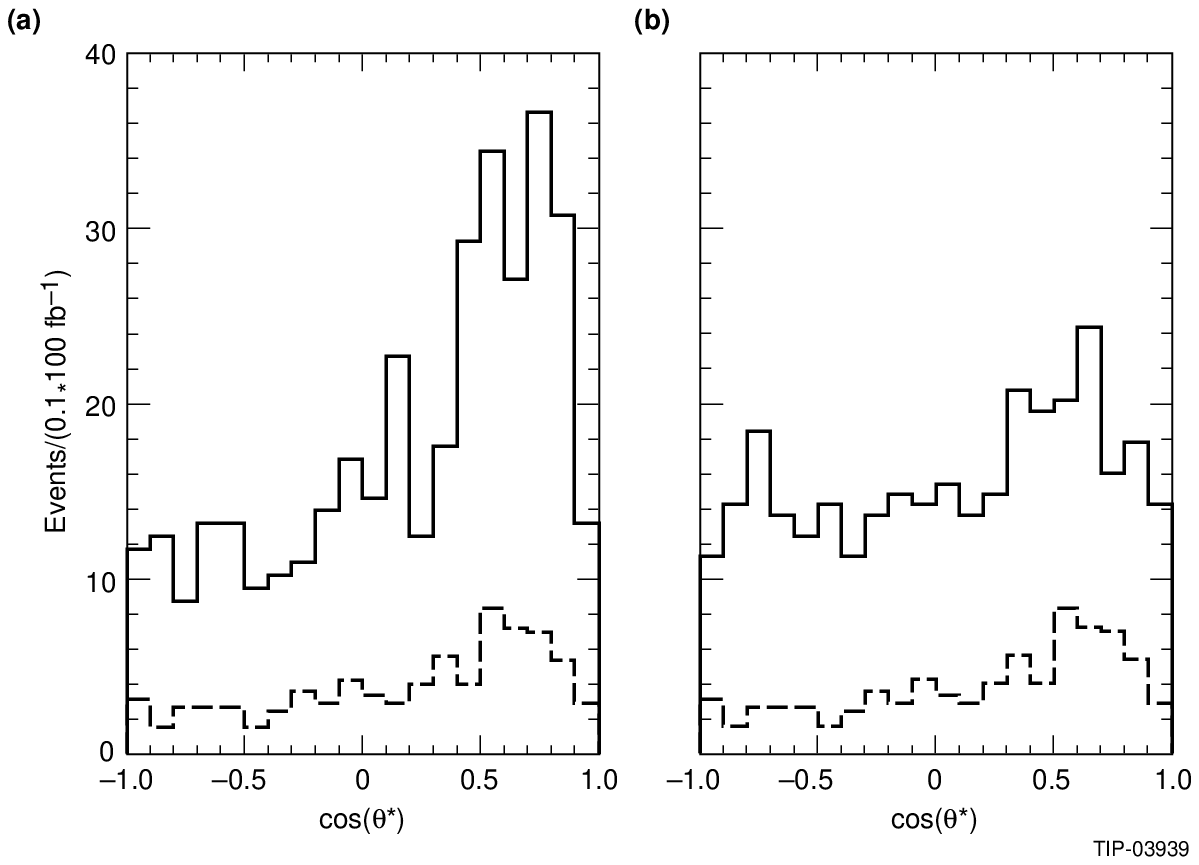}}{The $\cstar$
distributions for the ISO (a) and HNC
(b) substructure models defined in the text. The lower dashed histogram is
the standard Drell-Yan distribution.}

The net acceptances and number of events per $\sscd$ are 23\% and 80
events for DY, 33\% and 360 events for ISO, and 35\% and 315 events
for HNC.  Figure\ \LCmmm\ shows the $\mmm$ distributions for the ISO
and HNC models detected in GEM per $\sscd$. Also shown are the
underlying DY mass distributions. The muon angular distributions were
determined as in the $Z'$ case. The $\cstar$ distribution for the
$\mum$ in the ISO model, compared to DY, is shown in Fig.\ \LCcos\ (a)
and for the HNC model in Fig.\ \LCcos\ (b). The DY background was not
subtracted from the substructure-model distributions in these
figures. The tendency for a $(1 + \cstar)^2$ distribution in the ISO
model and a flat one in the HNC model is clear and the two models are
very well separated. The forward-backward asymmetries are
\eqn\mudist{\eqalign{
\afb({\rm DY})  &= 0.295 \pm 0.108 \cr
\afb({\rm ISO}) &= 0.328 \pm 0.050 \cr
\afb({\rm HNC}) &= 0.122 \pm 0.056 \ts . \cr}}
The errors are statistical only. Since the HNC and Drell-Yan contributions
do not interfere, the Drell-Yan contribution to the asymmetry
can be subtracted. This gives $\afb({\rm HNC}) = 0.065 \pm 0.065$,
consistent with an isotropic distribution.

Finally, we estimate the reach in substructure scale $\Lambda$ that could
be attained with a data sample of $\sscd$ at $\sqrt{s} = 40\,\tev$. We
define the reach to be that value of $\Lambda$ which gives $5\sigma$ over
the Drell-Yan expectation, approximately 125 events. Then, since the
subprocess cross sections above go as $\shat/\Lambda^4$, it is possible to
determine the reach by scaling the numbers of events found above. The reach
of a 40~TeV $pp$ collider estimated in this way is $30-35\,\tev$.

\subsubsec{High Mass and Luminosity Physics Studies of $\ellpm \nu_\ell$
Modes}

The range of physics that can be studied at ultrahigh luminosity can be
greatly extended by searching for isolated, high-$\pt$ leptons accompanied
by large $\etmiss \simeq \pt$. Extensions of the standard gauge group
generally involve $W'$ as well as $Z'$ bosons. Their mass and couplings can
be determined in high-statistics studies of $\wppm \ra e^\pm \nu_e$ and
$\mu^\pm \nu_\mu$, respectively. With $\CL = \uhl$ at a 40~TeV collider, it
should be possible to carry out high-statistics studies up to $M_{W'}
\simeq 5\,\tev$ and to reach as high as $10\,\tev$. If contact interactions
reflecting an underlying quark-lepton substructure exist, they may involve
terms of the form $\ol u d \ol \ell \nu$ and its conjugate. Much can be
learned about the chiral coupling of such interactions by comparing the
rapidity distributions of the outgoing $\mup$ and $\mum$. High-statistics
studies should again be possible up to $\Lambda \simeq 25\,\tev$. This
section contains a brief description of the measurements that can be made
at $\CL = \uhl$ if this new physics exists.

Precise determination of the mass of a $W'$-boson should be possible by
measuring the $\pt$ distribution of the electron in $pp \ra W' \ra e
\nu_e$. The main issue is how well the Jacobian peak determines $M_{W'}$.
For high $M_{W'}$, there are no significant physics backgrounds. For
example, the background $\jet + Z^0$ where the jet fakes an electron and
$Z^0 \ra (\ra \ol \nu \nu)$, is removed by the jet rejection $\CR(e/\jet)
\simeq 3\times 10^{-4}$. The detector-related issues are much the same as
in the $Z' \ra \ee$ study, except that high precision is not as important
here as the resolution on $\etmiss \simge 1\,\tev$.

Information on the $W'$ couplings to quarks and leptons can be obtained by
measuring the angular ($\theta$) distribution in $\wppm \ra \mu^\pm
\nu_\mu$ decay. The $\theta$-distribution of $\mum$ relative to the
incoming $d$-quark will be the same as that of $\mup$ relative to the
incoming $\ol d$. Thus, increased statistics can be obtained by adding data
from both modes. These angular distributions provide the only way in $pp$
colliders to detect the presence of right-handed neutrinos in the decay of
$W'$ bosons. To measure them, one proceeds as follows: (1) Select events
with $\pt(\mu)$ ``near'' the Jacobian peak found in $W' \ra e \nu_e$ and
having balancing $\etmiss$. This enhances the signal relative to any
Drell-Yan continuum or other background. (2) Measure the muon rapidity
$\eta_\mu$. (3) Determine the neutrino rapidity, $\eta_\nu$, by
reconstructing the $W'$. For this, assume $\vec \pt(\nu) = - \vec
\pt(\mu)$. As noted, this is a good approximation for multi-TeV $W'$ bosons
since $\pt(W') \simle 200\,\gev$ about 70\% of the time for $M_{W'} \simge
2\,\tev$. The neutrino 4-momentum and $\eta_\nu$ are determined up to a
quadratic ambiguity by imposing the $W'$ mass constraint. One can either
select the value that minimizes $\eta_{W'} = \eta_B$ or accept each event
twice. The analysis of the muon angular ($\cstar$) distribution can now be
carried exactly as was done for $Z'$ and $\ol q q \ra \mup \mum$. Requiring
good hits in all three muon superlayers, the fake high-$\pt$ background to
the $W' \ra \mu + \etmiss$ signal was found to be negligible.


\nfigwide\LCeta{\dofig{96.5mm}{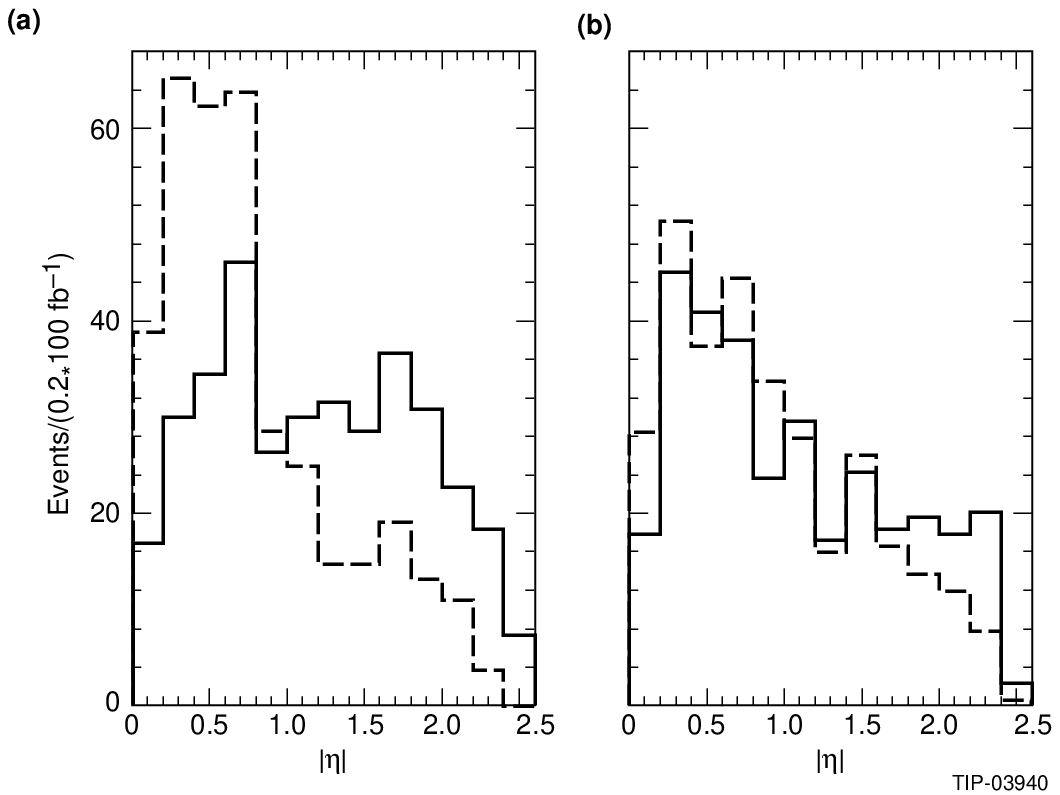}}{$\vert\eta_\mu^-\vert$
(solid) and $\vert\eta_\mu^+\vert$ (dashed) distributions for the ISO (a)
and HNC (b) compositeness models defined in the text.}

Contact interactions of the form $\ol u d \ol \ell \nu$ give an excess of
muons at high-$\pt$. Even though the parton c.m.\ frame cannot be found in
this case, it is still possible to obtain information on the chiral nature
of the contact interaction by comparing the rapidity distributions, $\vert
\eta_{\mu^+} \vert$ and $\vert \eta_{\mu^-} \vert$ of muons with $\pt
\simge 1.5\,\tev$. If, for example, the angular distribution between the
incoming $d$-quark and the outgoing $\mum$ is $(1 + \cos \theta)^2$ in the
parton c.m.\ frame, then $\vert\eta_{\mum}\vert$ is pushed to larger values
because the $d$-quark is harder than the $\ol u$-quark and the $\mum$ tends
to be produced forward. Correspondingly, the $\vert\eta_{\mup}\vert$
distribution is squeezed to smaller values. If the angular distribution is
flat, as in the HNC model, the two rapidity distributions will be
identical. These features are illustrated for the ISO and HNC models in
Figs.\ \LCeta. These plots were made for the surviving dimuon event samples
of Section\ 5.6.2. The two models are well separated by the muons' rapidity
distributions. If quarks and leptons are composite structures, with a
characteristic interaction scale in reach of hadron supercolliders, it is
clear that measurements of the angular distributions in both $\mup \mum$
and $\mu^\pm \nu$ channels will be important for pinning down the chiral
structure of the contact interaction.


\newsec{CONCLUSIONS}

In this article we have presented simulations of the GEM detector's
response for a selection of electroweak, flavor, and new gauge-interaction
physics processes. The physics examples studied were chosen for their
ability to help quantify GEM's capabilities for new physics and to aid in
the detector's design and optimization, as well as for their intrinsic
physics interest. While the examples form a fairly broad range of new
physics within and beyond the standard model, they certainly do not
represent a full survey of the interesting possibilities.

The principal tool used for these studies was the GEANT-based
parameterization of the GEM detector known as |gemfast|. Although there
were a few instances when there was no efficient substitute for full GEANT
simulations of the response of limited portions of the detector --- most
notably in the calorimeter studies of jet backgrounds to $H^0 \ra
\gamma\gamma$ --- the |gemfast| package was essential for our being able to
carry out as broad a range of detailed simulations as we did in a fairly
limited amount of time.  We firmly believe it is necessary that the range
of physics topics studied be as broad as possible. So long as the physics of
the TeV-energy regime remains obscure, detectors built to explore it must
have the capability to find whatever is going on within reach of their
colliders. Thus, we believe that other large collider detector
collaborations will profit from the development of similar tools for their
large-scale simulation studies.

Had the GEM project proceeded, further studies, both in depth and in
breadth, would have been performed and an improved set of tools for
modeling the detector developed. These in turn would have further
influenced the evolution of the experiment. This iteration between design
and physics simulation is an essential part of the creation of any modern
detector. The simulations which were carried out guided the design of GEM
and influenced a number of major technical choices. Examples are:

\item{$\bullet$} The requirements for the energy resolution of the
electromagnetic calorimeter --- both for the stochastic ($\sim 7\%/\sqrt{E}$)
and the constant ($\le 0.4\%$) terms --- were set by the discovery potential
for $\hgg$, $\h \ra ZZ^* \ra \ee\ee$, and $Z' \ra \ee$.

\item{$\bullet$} The pointing capability and angular resolution of the
electromagnetic calorimeter ($(40-50\,{\rm mrad})/\sqrt{E\,(\GeV)} + 0.5
\ts {\rm mrad}$) were also motivated by the need to cleanly identify
and measure $\hgg$ and $Z' \ra \ee$. The pointing ability of the
calorimeter complements the central tracker's determination of the
event vertex, making robust GEM's precise invariant mass resolution for
particles decaying into photons and electrons.

\item{$\bullet$} The segmentations of the calorimeter systems were dictated
by the need to measure EM processes with high precision and the need to
reject hadronic backgrounds. The calorimeter and the tracker combine to
provide clean isolation for photons, electrons and muons.

\item{$\bullet$} The precise momentum resolution of the muon system
is needed to detect $Z' \ra \mup\mum$ and $\h \ra
ZZ^* \ra \mup\mum\mup\mum$ decays with high efficiency.

\item{$\bullet$} The resolution of the compact central tracker permits the
complementary search for supersymmetry in the $\gluino \ts \gluino \ra \ellpm
\ellpm + X$ channel.

\item{$\bullet$} The rapidity coverage of the forward calorimeters,
extending to $\abseta \simeq 5.0$, was dictated by the need to suppress
backgrounds to the $\etmiss$ signatures for $\h \ra \ellp\ellm \ol \nu \nu$
decays and for supersymmetry signals.

\item{$\bullet$} An adequate shielding system was designed to suppress
neutron and photon backgrounds in the muon chambers, even at $\CL
\simeq \uhl$.

\noindent
While the simulation process was by no means complete, the results presented
here demonstrate that GEM could have done outstanding physics at the SSC. We
can only hope that others will have the opportunity to explore this physics
elsewhere.

\vfill\supereject

{\noindent\bf ACKNOWLEDGMENTS}

The results summarized here represent the hard work of a large and
dedicated group of our GEM colleagues: Melynda Brooks, Jim Brau, Rob
Carey, Peter Dingus, Yuri Efremenko, Yuri Fisyak, Geoff Forden, Ray
Frey, Mitch Golden, Yuri Kamyshkov, Mikhail Leltchouk, Hong Ma, Ken
McFarlane, Shawn McKee, Roger McNeil, Geoff Mills, Mohammad Mohammadi,
Steve Mrenna, Harvey Newman, Fred Olness, Sasha Savin, Irwin Sheer,
Sergei Shevchenko, Xiaorong Shi, Mike Shupe, Kostya Shmakov, Liz
Simmons, Jenny Thomas, Henk Uijterwaal, Sasha Vanyashin, Torre Wenaus,
David Worrick, Hiro Yamamoto, Chiaki Yanagisawa, Yanlin Ye, Bing Zhou,
and Ren-yuan Zhu. We also acknowledge the contributions to the
experiment made by the whole GEM collaboration under the leadership of
the spokesmen, Barry Barish and Bill Willis.  Finally, we would like
to thank Holly Durden and the SSCL technical staff for their help in
the preparation of the GEM {\sl Technical Design Report}, from which
many of the plots presented here are drawn.

K.D.L. was supported in part by the U.S. Department of Energy under
Grant No.~DE--FG02--91ER40676 and by the Texas National Research
Laboratory Commission under Grant Nos.~RGFY91--B6, RGFY92--B6, and
RGFY93--278. F.E.P. was supported in part by the U.S. Department of
Energy under Contract DE-AC02-76CH00016. T.S. was supported in part
the U.S. Department of Energy under Contract DE-FG05-92ER40722. T.S.
also thanks the Outstanding Junior Investigator program of DOE and the
SSC Fellowship program of Texas National Research Laboratory
Commission. Fermilab is operated by the Universities Research
Association Inc.\ for the U.S. Department of Energy.

\vfill\supereject
\footatend\vfill\supereject\immediate\closeout\rfile\writestoppt
\baselineskip=14pt\centerline{{\bf References}}\bigskip{\frenchspacing%
\parindent=20pt\escapechar=` \input refs.tmp\vfill\eject}\nonfrenchspacing
\enddoublecolumns

\bye